\documentclass[aps, prx, reprint, amsmath,amssymb,showpacs,floatfix,longbibliography, twocolumn, superscriptaddress,nofootinbib]{revtex4-1}
\usepackage{cmap} 
\usepackage[utf8]{inputenc}
\usepackage{times} 
\usepackage{graphicx}

\usepackage{CJKutf8}
\usepackage{dcolumn}
\usepackage{bm}
\usepackage{color}
\usepackage{xcolor}
\usepackage{hyperref}
\usepackage{enumitem}
\usepackage{cancel}
\usepackage{bbold}
\usepackage{stmaryrd}
\usepackage{mathtools}
\newcommand{\bs}[1]{{\mathbf{#1}}}

\usepackage{bbm}
\usepackage{booktabs}
\usepackage[most]{tcolorbox}
\colorlet{shadecolor}{gray!15!}
\tcbset{
        boxsep=4pt, left=0pt,right=0pt,top=0pt,bottom=0pt,
        colframe=white,colback=shadecolor,  
        highlight math style={enhanced}
        }

\usepackage{comment}

\hypersetup{
    colorlinks,
    linkcolor={red!50!black},
    citecolor={green!60!black},
    urlcolor={blue!60!black}
}

\hypersetup{linktocpage}
\newcolumntype{M}[1]{>{\centering\arraybackslash}m{#1}}
\newcolumntype{N}{@{}m{0pt}@{}}
\usepackage{environ}
\usepackage{booktabs,multirow} 

\bibpunct{[}{]}{,}{n}{}{}

\usepackage{amsfonts,amssymb,amsmath}
\usepackage{amsthm}
\usepackage{makecell}
\DeclareMathOperator*{\argmin}{arg\,min}

\usepackage[T1]{fontenc}
\usepackage{tikz}
\newcommand{\braket}[1]{\left\langle{#1}\right\rangle}
\newcommand{\bra}[1]{\left\langle {#1} \right\vert}
\newcommand{\ket}[1]{ \left\vert {#1} \right\rangle}
\let\originalleft\left
\let\originalright\right
\renewcommand{\left}{\mathopen{}\mathclose\bgroup\originalleft}
\renewcommand{\right}{\aftergroup\egroup\originalright}
\newcommand{\e}{\mathrm{e}}
\newcommand{\T}{\mathrm{T}}

\newcommand{\id}{\mathbb{1}}
\newcommand{\Mod}[1]{\ \mathrm{mod}\ #1}

\newcommand{\HX}{G}
\newcommand{\bfg}{\mathbf{g}}
\newcommand{\HZ}{H}
\newcommand{\bfh}{\mathbf{h}}

\NewEnviron{eqs}{%
\begin{equation}\begin{split}
    \BODY
\end{split}\end{equation}
}

\newtheorem*{defi*}{Definition}

\newenvironment{boxed_defi}
  {\begin{tcolorbox}\begin{defi*}}
  {\end{defi*}\end{tcolorbox}}

\def\prg#1{\paragraph*{{\bf #1}}}

\begin{document}
\begin{CJK*}{UTF8}{gbsn}

\title{Letting the tiger out of its cage: bosonic coding without concatenation}
\author{Yijia Xu (许逸葭)}
\affiliation{Joint Center for Quantum Information and Computer Science, University of Maryland, College Park,
Maryland 20742, USA}
\affiliation{Institute for Physical Science and Technology, University of Maryland, College Park, Maryland 20742, USA}
\author{Yixu Wang (王亦许)}
\affiliation{Institute for Advanced Study, Tsinghua University, Beijing, 100084, China}
\affiliation{Shanghai Institute for Mathematics and Interdisciplinary Sciences, Shanghai, 200433, China}
\affiliation{Joint Center for Quantum Information and Computer Science, University of Maryland, College Park,
Maryland 20742, USA}
\author{Christophe Vuillot}
\affiliation{Université de Lorraine, CNRS, Inria, LORIA, F-54000 Nancy, France}
\author{Victor~V.~Albert}
\affiliation{Joint Center for Quantum Information and Computer Science, University of Maryland, College Park, Maryland 20742, USA}

\date{\today}

\begin{abstract}
Continuous-variable cat codes are encodings into a single photonic or phononic mode that offer a promising avenue for hardware-efficient fault-tolerant quantum computation.
Protecting information in a cat code requires measuring the mode's occupation number modulo two, but this can be relaxed to a linear occupation-number constraint using the alternative two-mode pair-cat encoding.
We construct multimode codes with similar linear constraints
using any two integer matrices
satisfying \textcolor{black}{a CSS-like} homological condition of a quantum rotor code.
Just like the pair-cat code, syndrome extraction can be performed in tandem with stabilizing dissipation using current superconducting-circuit designs.
The framework includes codes with various finite- or infinite-dimensional codespaces, and codes with finite or infinite Fock-state support.
It encompasses two-component cat, pair-cat, dual-rail, two-mode binomial, various bosonic repetition codes, and aspects of chi-squared encodings while also yielding codes from homological products, lattices, generalized coherent states, and algebraic varieties.
Among our examples are analogues of repetition codes, the Shor code, and 
a surface-like code that is not a concatenation of a known cat code with the qubit surface code.
Codewords are coherent states projected into a Fock-state subspace defined by an integer matrix, and their overlaps are governed by Gelfand-Kapranov-Zelevinsky hypergeometric functions.

\end{abstract}
\maketitle
\end{CJK*}

\begin{figure*}[t]
    \centering
    \includegraphics[width=0.9\textwidth]{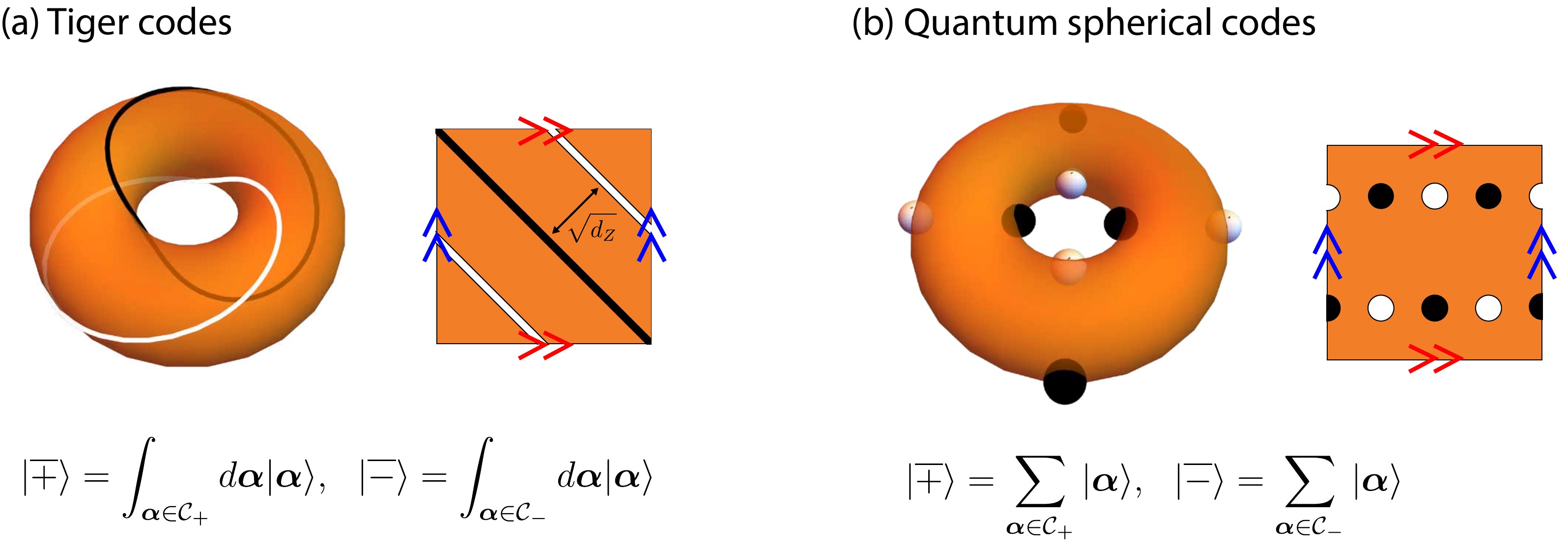}
    \caption{\textcolor{black}{The tiger codewords of this work are superpositions of \textit{continuous} sets of coherent-state "stripes" --- circles or, more generally, tori. In contrast,} quantum spherical codewords are superpositions of a \textit{discrete} set of coherent states.
    The use of continuous superpositions allows for linear Fock-state constraints to define the code, while discrete superpositions correspond to modular and, more generally, representation-theoretic constraints. 
    Two-mode coherent states of values \(\boldsymbol{\alpha}=\alpha(e^{i\phi_1},e^{i\phi_2})\) of fixed coordinate amplitude \(\alpha\) but varying phases $\phi_{1}$ and $ \phi_2$ form a torus, which is depicted either as itself or as a square with opposite boundaries identified.
    Panels (a) and (b) depict the subset of coherent states that form the logical codewords of a tiger code \textcolor{black}{and a quantum spherical code}, respectively. \textcolor{black}{The logical $+$ and $-$ states are defined as equal-weight superpositions of coherent states located at the black and white points. Therefore, the $Z$-distance of tiger code is given by the minimum squared Euclidean distance between points belonging to the constellations $\mathcal{C}_+=\{\alpha(e^{i\phi}, e^{-i\phi})|\forall \phi \in \mathbb{T}\}$ and $\mathcal{C}_{-}=\{\alpha(e^{i\phi}, e^{-i(\phi+\pi)})|\forall \phi \in \mathbb{T}\}$.}
   The tiger code example is the \textcolor{black}{pair-cat codewords defined by matrices \(\HX =\begin{pmatrix} 2 & 2 \end{pmatrix}\) and \(\HZ =\begin{pmatrix} 1 & -1 \end{pmatrix}\), while the quantum spherical code constellations are} $\mathcal{C}_+=\{(\alpha,\alpha), (i\alpha ,- \alpha), (- \alpha, \alpha), (-i \alpha,- \alpha)\}$ and $\mathcal{C}_-=\{(\alpha ,-\alpha), (i\alpha, \alpha), (- \alpha,- \alpha), (-i \alpha,\alpha)\}$.
    }
    \label{fig:spherical_vs_tiger}
\end{figure*}

\section{Introduction}

Quantum mechanical effects offer faster solutions to certain structured problems, secure communication from eavesdropping, and accurate simulation of natural processes \cite{shor1999polynomial,bennett2014quantum,huang2021power,nisqrmp,dalzell2023quantum,yamakawa2024verifiable,jordan2024optimization}.
The needed effects are difficult to preserve in an inherently classical world, and error correction is necessary in order to maintain them and realize useful applications.

The typical building block of a quantum system consists of only two low-lying quantized energy states of a natural system, such as an atom or molecule, or an engineered one, such as an electronic circuit.
These building blocks have to be combined, or "scaled up", to have enough redundancy so as to protect from noise.
Error-correcting codes have been extensively studied for such discrete variable (DV) systems \cite{shor1995scheme,gottesman1997stabilizer,css,calderbank1998quantum}.

However, many quantum systems --- atoms, molecules, and circuits included --- have access to states of higher energy. 
Much research has focused on studying the potential of one or a few such systems, or \textit{modes}, to encode information in their higher-lying state space \cite{gkp,cochrane1999macroscopically,leghtas2013hardware,pairbinomial}. This direction is bootstrapped by dramatic improvements in the control of such bosonic quantum systems (see, e.g., Refs.~\cite{sun2014tracking,leghtas2015confining,fluhmann2019encoding,grimm2020stabilization,putterman2024hardware,gertler2023experimental}, among many other advancements).

It is tantalizing to consider error-correcting codes constructed \textit{intrinsically} in a large number of modes, whose redundancy comes from \textit{both} the large number of building blocks of the system ("breadth") and a high energy per block ("depth").
One can consider concatenating a DV and a single-mode code, i.e., first encoding in a qubit \textcolor{black}{stabilizer} code and then encoding each physical qubit into a mode.
However, since the scaling comes strictly from the DV part of such an encoding, more powerful codes that cannot be expressed as concatenated codes should be possible.

A recent work \cite{jain2024quantum} introduces a quantum spherical code construction for intrinsically multimode codes that cannot be split into a DV and a mode encoding. Spherical codewords are superpositions of a handful of coherent states \(|\boldsymbol{\alpha}\rangle\) --- classical multimode electromagnetic signals whose amplitude and phase are combined into a complex vector \(\boldsymbol{\alpha}\) --- and are generalizations of the single-mode cat \cite{leghtas2013hardware} and two-mode \(2T\)-qutrit code \cite{denys20232} based on polytopes and spherical designs.

In this manuscript, we define "tiger" codes, whose codewords are compact yet \textit{continuous} coherent-state superpositions, with  \(\boldsymbol{\alpha}\) taking values along particular "stripes"  --- circles or, more generally, tori --- in coherent-state configuration space (see Fig.~\ref{fig:spherical_vs_tiger}).
Stabilizing operators are of CSS rotor-code type \cite{vuillotHomologicalQuantumRotor2024}, either linear combinations of occupation-number operators defined by rows of an integer matrix, or products of annihilation/creation operators whose powers form rows of another such matrix.
Any two integer matrices satisfying a CSS-type homological constraint can define a tiger code.

Tiger codes come with several advantages over the cat code and its variants for protecting against loss of energy in each mode, as well as random rotations due to fluctuations of each mode's frequency. 1). The syndrome required to detect losses corresponds to a \textit{linear} combination of occupation-number operators --- a simple multimode generalization of photon counting that relaxes the \textit{nonlinear} constraints required for the cat code and its variants. 2). Protection against dephasing is autonomous, generated by engineered dissipation \cite{poyatos1996quantum,mirrahimi2014dynamically,zanardi2014coherent,albert2016geometry} whose stabilizing jump operators can have a lower degree and fewer terms than those of concatenated-cat and quantum spherical codes.
3). Loss error syndromes can be extracted by appending an ancilla and evolving with an entangling Hamiltonian \textit{without} turning off the autonomous dissipation, in contrast to multi-component cat codes and similar to the pair-cat encoding \cite{albert2019pair}, a special case of our codes.
4). The \textcolor{black}{$Z$-type stabilizers} and \textcolor{black}{$X$-type } dissipators commute with one another when the Fock-state support of the codewords is infinite, making contact with the stabilizer formalism \cite{gottesman1997stabilizer,css}.

Our framework utilizes homology over the integers \cite{vuillotHomologicalQuantumRotor2024,rotorclifford}, giving rise to intrinsically multimode codes that are distinct from \textcolor{black}{qubit stabilizer} codes, which come from homology over \(\mathbb{Z}_2\) \cite{PhysRevLett.77.793,steane1996multiple,PhysRevA.54.1098}. 
The framework includes finite- and infinite-dimensional code spaces consisting of either a finite or infinite number of Fock states.
We unite many seemingly isolated bosonic encodings under one roof, including two-component cat, pair-cat, various bosonic repetition codes \cite{jeong2002efficient, ralph2003quantum}, the dual-rail code \cite{chuang_dualrail,chuang1996dualrail}, two-mode binomial codes \cite{chuang_pairbinomial}, as well as aspects of a \(\chi^{(2)}\) code \cite{niu2018hardware}. At the same time, we obtain new codes from lattices \cite{conway2013sphere}, generalized coherent states \cite{radcliffe1971some,arecchi1972atomic},
and integer hypergraph products \cite{tillich,vuillotHomologicalQuantumRotor2024}.
We show that circular superpositions of coherent states and tiger-code properties are intimately tied to Gelfand-Kapranov-Zelevinsky (GKZ) hypergeometric functions \cite{gel1986general,gel1986generalized,gel1987holonomic,gel1989hypergeometric,adolphson1994hypergeometric,gel1990hypergeometric,gel1992general}, 
yielding codes from algebraic varieties such as Calabi-Yau manifolds \cite{stienstra2007gkz}.

Tiger codes come with intrinsically linear syndromes, sport low-order autonomous protection. They are incredibly diverse, providing a framework for experimentally viable, inherently multimode quantum information processing in the near future.

\section{Summary of results}

Tiger-code logical states, or codewords, consist of a continuous superposition of coherent states along a circle (\(\mathbb{T}\)), or, more generally, a \textit{torus} in coherent-state configuration space.
Any state lying on the torus can be obtained from any other state by applying separate configuration-space rotations (a.k.a. "phase-shifters") on every mode in a particular direction.
The group of such rotations is the first ingredient in our code.

\prg{Projected coherent states}
A single-mode rotation is an exponential of the occupation number operator, \(\hat{n} = \hat{a}^{\dagger}\hat{a}\), where \(\hat{a}\) is the lowering/annihilation operator of a given mode \cite{serafini2017quantum,albert2022bosonic}.
The group is defined by integer vectors \(\mathbf{h}\), which map out directions of rotation \(\mathbf{h}\cdot \hat{\mathbf{n}}\), expressed in terms of the vector of occupation-number operators, \(\mathbf{\hat{n}} = \begin{pmatrix} \hat{n}_1&  \hat{n}_2 &\cdots\end{pmatrix} \).
The vectors \(\mathbf{h}\) generate the group and can be collected as rows of an integer \textit{generator matrix} \(\HZ\).

For example, in the case of the two-mode pair-cat encoding, rotations are generated by the occupation number difference, \(\hat{n}_1 - \hat{n}_2\), corresponding to a generator matrix with one entry, \(\HZ = \mathbf{h} = \begin{pmatrix}1 & -1 \end{pmatrix}\).
The rotation group elements are \(\exp[i\phi (\hat{n}_1 - \hat{n}_2)]\) for any angle \(\phi \in [0, 2\pi)\).

Tiger codewords consist of all possible rotations applied to some fiducial coherent state.
We define \textit{projected coherent states} (cf. \cite{skagerstam1985quasi,drummond2016coherent}),
\begin{eqs}\label{eq:projected-coherent-state-informal}
    |\boldsymbol{\alpha}\rangle_{\boldsymbol{\Delta}}^{\HZ}\propto\int d\boldsymbol{\phi}e^{i\boldsymbol{\phi}(\HZ\hat{\mathbf{n}}-\boldsymbol{\Delta})}|\boldsymbol{\alpha}\rangle~,
\end{eqs}
as continuous superpositions of coherent states \(|\boldsymbol{\alpha}\rangle\) \cite{klauder1985coherent,dodonov2002nonclassical,gazeau2009coherent} whose coefficients are determined by \(\HZ\) and an integer vector \(\boldsymbol{\Delta}\).
Here, the vector of angles \(\boldsymbol{\phi}\) iterates over all possible rotations. 
Pair-coherent \cite{barut1971new,agarwal1986generation,agarwal1988nonclassical,gerry1995nonclassical}, trio-coherent~\cite{an2003even}, two-mode binomial  \cite{radcliffe1971some,arecchi1972atomic},
and \(SU(N)\) generalized coherent states \cite{gitman1993coherent,nemoto2000generalized,calixto2021entanglement} are all examples of such states.

Using the Fock-state expression for an ordinary coherent state, \(|\boldsymbol{\alpha}\rangle\propto \sum_{\mathbf{n}} \boldsymbol{\alpha}^{\mathbf{n}}|\mathbf{n}\rangle/\sqrt{\mathbf{n}!}\) in multi-index notation. 
We see that the application of all possible rotations imposes a linear constraint on the state's Fock states \(|\mathbf{n}\rangle\),
\begin{eqs}\label{eq:constraint-informal}
    \HZ\hat{\mathbf{n}} = \boldsymbol{\Delta}~,
\end{eqs}
and rules out any Fock states that do not satisfy this constraint.
The admissible set of Fock states can be finite or infinite, yielding tiger codes of finite or infinite Fock-state support.
Tiger codewords can be defined for any \textcolor{black}{integer} values of \(\boldsymbol{\Delta}\) for which there exist admissible Fock states.

The normalization function of a projected coherent state,
\begin{eqs}\label{eq:projected-coherent-state-normalization}
    \mathsf{A}_{\boldsymbol{\Delta}}(\mathbf{y}) = \sum_{\HZ\mathbf{n} = \boldsymbol{\Delta}} \frac{\mathbf{y}^{\mathbf{n}}}{\mathbf{n}!}\quad\quad\text{(GKZ)},
\end{eqs}
is a Gelfand-Kapranov-Zelevinsky (GKZ) hypergeometric function \cite{gel1986general,gel1986generalized,gel1987holonomic,gel1989hypergeometric,adolphson1994hypergeometric,gel1990hypergeometric,gel1992general} --- a generalization of many uni- and multi-variate hypergeometric functions.
Projected coherent states are in correspondence with such functions, and we utilize this connection to construct and analyze the resulting tiger codes.

\prg{Error model and protection}
In the context of error correction, the vector \(\boldsymbol{\Delta}\) encodes an error syndrome, and the  constraint~\eqref{eq:constraint-informal} divides the Fock space into sectors delineated by the vector's possible values.
Any loss operator \(\mathbf{\hat{a}}^{\mathbf{p}} = \hat{a}_1^{p_1}\hat{a}_2^{p_2}\cdots\), whose non-negative integer \textit{loss vector} \(\mathbf{p}\) denotes the losses in each mode, will change the value of the syndrome as long as the loss vector is not in the kernel of \(\HZ\).
Detecting such loss errors is done by measuring each row in the vector \(\HZ \hat{\mathbf{n}}\) of linear constraint operators.

The other source of noise protected by tiger codes stems from modal frequency fluctuations \cite{turchette2000decoherence}, which elicit random single-mode rotations. 
Such \textit{dephasing error} can rotate coherent states into each other, inducing logical errors.

Cat codes and more general quantum spherical codes \cite{jain2024quantum} are well protected against dephasing error. 
Coherent-state overlaps are suppressed by the squared Euclidean distance between their configuration-space values,
\begin{eqs}\label{eq:euclidean}
    |\langle\boldsymbol{\alpha}|\boldsymbol{\beta}\rangle|^{2}=\exp(-\|\boldsymbol{\alpha}-\boldsymbol{\beta}\|^{2}),
\end{eqs}
and dephasing errors are suppressed exponentially with the energy and the minimum squared  Euclidean distance between different constellations \(\mathcal{C}\ni \boldsymbol{\alpha}\) participating in each codeword.

We conjecture that projected coherent states with infinite Fock-state support inherit this same degree of suppression.
We prove this conjecture for a large family of tiger codes that contains all of our examples.
We also provide example of codes which protect against dephasing at all non-zero energies, i.e., without requiring a large-energy limit at all!

Tiger codes with finite Fock-state support do not have a continuously tunable energy parameter.
Nevertheless, we present examples of codes where exponential suppression of dephasing errors is achieved in the components of the discrete syndrome vector \(\boldsymbol{\Delta}\), which determine the total energy of the code in this case.

Protection against dephasing error can be implemented autonomously by engineering stabilizing Lindbladians \cite{poyatos1996quantum,mirrahimi2014dynamically,zanardi2014coherent,albert2016geometry} that drive coherent states into the required codeword constellations.
Such Lindbladians are obtained using the fact that coherent states are right eigenstates of lowering operators, \(\mathbf{\hat{a}}^{\mathbf{g}}|\boldsymbol{\alpha}\rangle=\boldsymbol{\alpha}^{\mathbf{g}}|\boldsymbol{\alpha}\rangle\).
Projected coherent states maintain this property as long as \(\mathbf{g}\) is in the kernel of \(\HZ\), yielding Lindbladian \textit{dissipators} of the form \(\mathbf{\hat{a}}^{\mathbf{g}} - \boldsymbol{\alpha}^{\mathbf{g}}\).

The kernel constraint implies that such dissipators commute with the rotations in Eq.~\eqref{eq:projected-coherent-state-informal} for all values of \(\boldsymbol{\phi}\). This means that a syndrome extraction circuit generated by a Hamiltonian in the syndrome generators \(\mathbf{h}\cdot\hat{\mathbf{n}}\) commutes, and can be performed in tandem, with the code's stabilizing dissipation.

Finite-support codes require dissipators whose vectors have negative entries, corresponding to monomials \(\mathbf{\hat{a}}^{\dagger\mathbf{q}}\mathbf{\hat{a}}^{\mathbf{p}}\), for a vector written as a difference of positive vectors, \(\mathbf{g}=\mathbf{p}-\mathbf{q}\).
We can still define dissipators for such codes using the identity \(\hat{a}^{\dagger k}\ket{\alpha}=\left(\hat{n}\right)_{k}\ket{\alpha}/\alpha^{k}\) for a single-mode coherent state \(|\alpha\rangle\) and falling factorial \((u)_v = u(u-1)\cdots(u-v+1)\).
Such dissipators no longer commute with each other, but the codespace is still contained in their joint eigenspace.

Stabilizing dissipation allows one to track the syndrome \(\boldsymbol{\Delta}\) in software without necessarily correcting back to the original codespace.
For codes with infinite support, such tracking can continue indefinitely since there are an infinite number of syndromes.
Tracking is also possible for finite-support codes, but only for a finite number of errors.

The vectors \(\mathbf{g}\) corresponding to dissipators can be \textcolor{black}{arranged} as rows of a generator matrix \(\HX\), \textcolor{black}{whose span lies within the kernel of the “dual” generator matrix} \(\HZ\), \textcolor{black}{which specifies} the linear occupation-number constraints.
The two matrices satisfy the CSS-type \textcolor{black}{condition} 
\begin{eqs}\label{eq:css-homo-constraint}
    \HZ\HX^{\text{T}} = \mathbf{0},
\end{eqs}
and tiger codes can be defined using any such pair.

\begin{boxed_defi}[tiger code]
Let \(\HX\in\mathbb{Z}^{r_x\times N}\) and \(\HZ\in\mathbb{Z}^{r_z\times N}\) be two integer matrices such that \(\HZ \HX^{\textnormal{T}} = \mathbf{0}\).
Let \(\boldsymbol{\Delta}\in\mathbb{Z}^{r_z}\) be \(r_z\) integers and \(\boldsymbol{\alpha}\in\mathbb{C}^N\) be \(N\) complex numbers.

The Tiger Code, \(\mathcal{T}\left(\HX,\HZ,\boldsymbol{\Delta},\boldsymbol{\alpha}\right)\), consists in all rotations of the projected coherent state \(\ket{\boldsymbol{\alpha}}_{\boldsymbol{\Delta}}^{\HZ}\) defined by the kernel of \(\HX\):
\begin{equation}
    \mathcal{T} = \textup{Span}\left\{\e^{i\boldsymbol{\mu}\cdot\hat{\boldsymbol{n}}}\ket{\boldsymbol{\alpha}}_{\boldsymbol{\Delta}}^{\HZ}\;\middle\vert\;\forall\boldsymbol{\mu}\in\mathbb{T}^N,\,\HX\boldsymbol{\mu}=0\!\!\!\!\pmod{2\pi}\right\}.
\end{equation}
The codestates, \(\ket{\overline{\Psi}}\in\mathcal{T}\), furthermore satisfy the following constraints.
For \textcolor{black}{any $1 \leq j\leq r_z$, the $j$-th row of \(\HZ\) and the $j$-th entry of $\mathbf{\Delta}$, denoted by $\mathbf{h}$ and $\Delta_{\mathbf{h}}$ respectively,}
\[\left[\bfh\cdot\hat{\mathbf{n}}-\Delta_{\mathbf{h}}\right]\ket{\overline{\Psi}} = 0.\]
For all linear combination of the rows of \(\HX\), \(\bfg=\mathbf{s}\HX,\,\mathbf{s}\in\mathbb{Z}^{r_x}\), decomposed into \(\bfg=\mathbf{p}-\mathbf{q}\) where \(\mathbf{p}\) and \(\mathbf{q}\) are non-negative integer vectors,
\[\left[\mathbf{\hat{a}}^{\dagger \bs{q}}\mathbf{\hat{a}}^{\bs{p}} - \boldsymbol{\alpha}^{\bfg}\left(\hat{\bs{n}}\right)_{\bs{q}}\right]\ket{\overline{\Psi}} = 0.\]
\end{boxed_defi}
In the definition, \(\mathbb{T}=[0,2\pi)\). We use multi-index notation, e.g., \(\mathbf{u}^{\mathbf{v}} \equiv \prod_j u_j^{v_j}\) \textcolor{black}{for exponentiation} and \((\mathbf{u})_{\mathbf{v}} \equiv \prod_j (u_j)_{v_j}\) \textcolor{black}{for falling factorial}. When the matrix \(\HX\) is non-negative, the second constraint simplifies to \(\left[\mathbf{\hat{a}}^{\bfg} - \boldsymbol{\alpha}^{\bfg}\right]\ket{\overline{\Psi}} = 0\) for any row \(\bfg\) of \(\HX\); this corresponds to the case of infinite Fock-state support (see Secs.~\ref{sec:inf_support}, \ref{sec:examples}, and \ref{sec:surface_tiger}).
We explore the finite-support case in Section~\ref{sec:finitesupport}.

Recognizing the matrix constraint~\eqref{eq:css-homo-constraint} as the defining equation of a chain complex \((\HX,\HZ)\) over the integers \cite{vuillotHomologicalQuantumRotor2024}, we show that the dimension and structure of a tiger codespace
is given by the homology of this complex.
The code space generally consists of finite- and infinite-dimensional factors, meaning that qudit or infinite-dimensional encodings are possible.
Representatives of the homology classes define the logical \(X\) operators, and representatives of the cohomology classes over \(\mathbb{T}=[0,2\pi)\) define logical \(Z\) operators.

Loss and gain errors, \(\mathbf{\hat{a}}^{\dagger\mathbf{q}}\mathbf{\hat{a}}^{\mathbf{p}}\) for \(\mathbf{p}\neq \mathbf{q}\), constitute \(X\)-error for tiger codes.
\textcolor{black}{Undetectable $X$-}errors correspond to vectors \(\mathbf{p} - \mathbf{q}\) that \textcolor{black}{are in the kernel of $H$ (not detectable by $Z$-type stabilizers) but are not in the image of $G$ (are not $X$-type dissipators ) }.
The classes of logical $X$ operators formed by loss-gain monomials correspond to vectors in the quotient, \(\ker \HZ / \text{im}~\HX\), similar to a qubit stabilizer code's logical Pauli operations forming a quotient of a normalizer by the code's stabilizer \cite{gottesman2016surviving}.

\textcolor{black}{The $X$-distance, $d_X$, is determined by the 1-norm of undetectable $X$-error $\mathbf{\hat{a}}^{\dagger\mathbf{q}}\mathbf{\hat{a}}^{\mathbf{p}}$ with the minimal 1-norm of $\mathbf{p}-\mathbf{q}$. }
A code can detect up to \(d_X - 1\) and correct at least \(\lfloor(d_X-1)/2\rfloor\) losses on any mode, granted that dephasing errors are suppressed.
Certain codes can correct more losses due to the structure of the kernel of their \(\HZ\) matrix.

We calculate the \(X\)-distance using \textit{differences} between raising and lowering monomial coordinates in order to remove any dephasing error contribution,  \(\mathbf{\hat{a}}^{\dagger\mathbf{p}}\mathbf{\hat{a}}^{\mathbf{p}}\), which constitutes a \(Z\)-error.
We claim that such dephasing errors are suppressed exponentially with the \(Z\)-distance of the code, which, for the infinite-support case, is the minimal square Euclidean distance between pairs of points in different coherent-state constellations making up each codeword.

Logical \(Z\) gates are tensor products of single-mode rotations which connect different codewords via a Euclidean configuration-space path.
More general quadratic and conditional phase gates can be realized with the same ingredients, yielding a universal gate set (see Table~\ref{tab:pair-cat}).

\begin{table}[t]
\begin{tabular}{ccccc}
\toprule 
Logical & Code & Modes & $d_{X}$ & $d_{Z}$\tabularnewline
\midrule
\multirow{8}{*}{qubit} & two-component cat \cite{mirrahimi2014dynamically} & $1$ & $1$ & $4$\tabularnewline
\cmidrule{2-5}
 & pair-cat \cite{albert2019pair} & $2$ & $2$ & $4$\tabularnewline
\cmidrule{2-5}
 & four-mode tiger $\textcolor{blue}{\star}$ & $4$ & $2$ & $8$\tabularnewline \cmidrule{2-5}
 & coherent-state rep-n \cite{ralph2003quantum} & $N$ & $1$ & $4N$ 
 \tabularnewline 
 & extended pair-cat \cite{albert2019pair} & $N$ & $N$ & $4N\sin^{2}\frac{\pi}{2N}$\tabularnewline 
 & tiger Shor $\textcolor{blue}{\star}$ & $ML$ & $M$ & $4ML\sin^{2}\frac{\pi}{2M}$\tabularnewline 
 & liger surface $\textcolor{blue}{\star}$ & $3r$ & $2$ & ~$4r$~\tabularnewline 
 & tiger surface $\textcolor{blue}{\star}$ & $(2m-1)r$ & $m$ & ~$\geq 4rm\sin^{2}\frac{\pi}{2m}$~\tabularnewline
\midrule 
\multirow{2}{*}{rotor} & pair-coherent state \cite{barut1971new} & $2$ & $1$ & $4\sin^{2}\frac{\varphi}{2}$\tabularnewline
& three-mode tiger $\textcolor{blue}{\star}$ & 3 & 3&  $12\sin^2 \frac{\varphi}{2}$\tabularnewline
\midrule 
mode 
& Fock-state repetition $\textcolor{blue}{\star}$ & $N$ & $N$ & $4N\sin^{2}\frac{\varphi}{2N}$\tabularnewline
\bottomrule
\end{tabular}

\caption{ \label{tab:code_parameters} 
List of distances for several tiger codes with infinite Fock-state support that are described in Sec.~\ref{sec:examples}. The blue stars \textcolor{blue}{$\star$} mark some of the new examples covered in this work.
A code with \(X\)-distance \(d_X\) can detect \textcolor{black}{any $X$-error $\mathbf{\hat{a}}^{\dagger \mathbf{q}} \mathbf{\hat{a}}^{\mathbf{p}}$ with 1-norm $|\mathbf{p}-\mathbf{q}| <d$} and correct at least \(\lfloor (d_X - 1)/2 \rfloor\) losses on any mode, but several codes can correct more errors due to the structure of the kernel of their \(\HZ\) matrix.
For example, the pair-cat code can correct arbitrary losses on any one mode.
The $Z$-distance~\eqref{eq:z-distance} \(d_Z\) for \textcolor{black}{tiger codes encoding a logical qubit} is the minimum squared Euclidean distance between coherent states in different tiger-codeword constellations; this distance governs the code's degree of dephasing suppression.
For codes with infinite-dimensional logical encodings (rotors or modes), \(d_Z\) is still the square of minimum Euclidean distance, but listed as a function of the logical rotation angle \(\varphi\), per Eq.~\eqref{eq:z-distance-continuous}.
The lower bound on the tiger surface code is from Sec.~\eqref{sec:surface_tiger}; it becomes exact when the code is defined on a long strip, becoming the long-tiger, or "liger", code.
}
\end{table}

\begin{table}[h]
\begin{tabular}{ccccc}
\toprule 
Logical & Code & Modes & $d_{X}$ & \makecell{Dephasing\\ suppression}\tabularnewline
\midrule
\multirow{4}{*}{qubit} & 
two-mode binomial
\cite{chuang_pairbinomial} & $2$ & $2$ & exact
\tabularnewline
\cmidrule{2-5}
 & four-mode binomial $\textcolor{blue}{\star}$ & $4$ & $2$ & exact
 \tabularnewline
\cmidrule{2-5}
 & $\chi^{(2)}$-like \cite{niu2018hardware} $\textcolor{blue}{\star}$ & $3$ & $3$ & $\exp(-2\alpha \sqrt{\Delta})$\tabularnewline \cmidrule{2-5}
 & Calabi-Yau $\textcolor{blue}{\star}$ & $4$ & $6$ & $(\sqrt{13}/4)^{\Delta}$ 
 \tabularnewline 
\midrule 
qu$N$it 
& multinomial \textcolor{blue}{$\star$}  & $N$ & $2$ & exact
\tabularnewline 
\bottomrule
\end{tabular}

\caption{ \label{tab:finite_support_code_parameters} 
\textcolor{black}{
List of distances for several tiger codes with finite Fock-state support that are described in Sec.~\ref{sec:finitesupport}. The blue stars \textcolor{blue}{$\star$} mark some of the new examples covered in this work.
The definition of $X$-distance $d_X$ is same as the one for infinite-support case. 
In lieu of the $Z$-distance, we quantify finite-support codes' degree of dephasing supperession. Three examples can exactly detect dephasing errors, satisfying the KL conditions $\langle \overline{\mu} |\mathbf{\hat{a}}^{\dagger \mathbf{p}} \mathbf{\hat{a}}^{ \mathbf{p}} | \overline{\nu}\rangle=0$ for codeword with $\mu \neq \nu$ and $|\mathbf{p}|< \Delta$.
Two of them suppress dephasing errors exponentially with $\Delta$.
}
}
\end{table}

Tiger codes can be defined for any pair of integer matrices satisfying the CSS-type \textcolor{black}{condition} \eqref{eq:css-homo-constraint}.
This exceedingly broad definition allows us to define codes using, e.g., generating matrices of lattices.
The GKZ connection yields codes from matrices defining algebraic varieties.
The contact with integer homology opens the door to using tesselations of topologically nontrivial surfaces to define intrinsically multimode codes.
Altogether, we provide 17 examples of old and new tiger codes.
The \(X\)- and \(Z\)-distances of several infinite-support examples are listed in Table \ref{tab:code_parameters}. \textcolor{black}{The parameters of several finite-support examples are summerized in Table \ref{tab:finite_support_code_parameters}.}

\begin{table*}
\begin{tabular}{ccc}
\toprule 
 & Pair-cat code &  Tiger code (qu\(K\)it) \tabularnewline
\midrule
Generator matrices & $\HX =\begin{pmatrix}2 & 2\end{pmatrix};\quad \HZ =\begin{pmatrix}1 & -1\end{pmatrix}$ & $\HX =\begin{pmatrix}\vdots\\
\bfg\\
\vdots
\end{pmatrix};\quad \HZ =\begin{pmatrix}\vdots\\
\bfh\\
\vdots
\end{pmatrix}$\tabularnewline
\midrule
$\phantom{\displaystyle\int}$$X$-type dissipators$\phantom{\displaystyle\int}$ & $\hat{a}_{1}^{2}\hat{a}_{2}^{2}-\alpha^{4}$ & $\mathbf{\hat{a}}^{\bfg}-\alpha^{|\bfg|}$ for all $\bfg \in \text{row}(\HX) $\tabularnewline
$Z$-type \textcolor{black}{stabilizers} & $\hat{n}_{1}-\hat{n}_{2}$ & $\bfh\cdot\hat{\mathbf{n}}$ for all $\bfh \in \text{row}(\HZ) $\tabularnewline
$\phantom{\displaystyle\int}$\(X\)-type codewords $|\overline \mu\rangle$ $\phantom{\displaystyle\int}$ & pair-coherent states & projected coherent states\tabularnewline
\midrule 
$\phantom{\displaystyle\int}$Detectable losses$\phantom{\displaystyle\int}$ & $\{\hat{a}_{1}^{p_1}\hat{a}_{2}^{p_2}\,|\,p_1\neq p_2\geq0\}$ & $\{\mathbf{\hat{a}}^{\mathbf{p}}\,|\,\mathbf{p}\notin\ker \HZ \}$\tabularnewline
Undetectable $X$-error & $\mathbf{\hat{a}}^{\dagger \mathbf{q}} \mathbf{\hat{a}}^{\mathbf{p}}$ for $\mathbf{p}-\mathbf{q} \in \text{ker} H \backslash \text{im} G$ & $\mathbf{\hat{a}}^{\dagger\mathbf{q}}\mathbf{\hat{a}}^{\mathbf{p}}$ for $\mathbf{p}-\mathbf{q}\in\ker \HZ \backslash \text{im}\HX $
\tabularnewline
$\phantom{\displaystyle\int}$Logical $X$ operator & $\hat{a}_{1}\hat{a}_{2}/ \alpha^2$ & $\mathbf{\hat{a}}^{\mathbf{p}}/\alpha^{|\mathbf{p}|}$ for $\mathbf{p}\in\ker \HZ / \text{im}\HX $
\tabularnewline
$\phantom{\displaystyle\int}$ $X$-distance $d_X$ $\phantom{\displaystyle\int}$& 2& ${\displaystyle \min_{\mathbf{p},\mathbf{q}\neq\mathbf{0}}|\mathbf{p}-\mathbf{q}|}$
\tabularnewline
$\phantom{\displaystyle\int}$Hamiltonian $X$-gate$\phantom{\displaystyle\int}$ & $\hat{a}_{1}\hat{a}_{2}+\text{h.c.}$ & $\mathbf{\hat{a}}^{\dagger\mathbf{q}}\mathbf{\hat{a}}^{\mathbf{p}}+\text{h.c.}$\tabularnewline
Hamiltonian $XX$-gate & $(\hat{a}_{1}\hat{a}_{2})^{\otimes2}+\text{h.c.}$ & $(\mathbf{\hat{a}}^{\dagger\mathbf{q}}\mathbf{\hat{a}}^{\mathbf{p}})^{\otimes2}+\text{h.c.}$\tabularnewline
\midrule 
$\phantom{\displaystyle\int}$Dephasing suppression$\phantom{\displaystyle\int}$ & ~~~$|\langle\overline{\mu}|\hat{a}_{j}^{\dagger p}\hat{a}_{j}^{p}|\overline{\nu}\rangle|^{2}\sim\alpha^{4p}\exp(-4\alpha^{2})$~~~ &  ~~$|\langle\overline{\mu}|\mathbf{\hat{a}}^{\dagger\mathbf{p}}\mathbf{\hat{a}}^{\mathbf{p}}|\overline{\nu}\rangle|^{2}\lesssim\text{poly}(\alpha)\exp(-d_{Z}\alpha^{2})$~~
\tabularnewline
$\phantom{\displaystyle\int}$ Logical $Z$ operators$\phantom{\displaystyle\int}$ & $(-1)^{\hat{n}_{1}}$ & $e^{i\mu\mathbf{z}\cdot\hat{\mathbf{n}}}$ for $\mu\mathbf{z}\in\text{ker}_{2\pi}\HX /\text{im}_{2\pi}\HZ $\tabularnewline
$\phantom{\displaystyle\int}$ $Z$-distance $d_{Z}$$\phantom{\displaystyle\int}$ & $4$ & ${\displaystyle \min_{\mu\in\frac{2\pi}{K}\mathbb{Z}_{K}}\min_{\boldsymbol{\phi}} \left\|\mathbf{1}-e^{i(\boldsymbol{\phi}\HZ +\mu\mathbf{z})} \right\|^{2}}$\tabularnewline
$\phantom{\displaystyle\int}$ Phase gate $\phantom{\displaystyle\int}$ & $(-1)^{\hat{n}_{1}^{2}}$ & $e^{i\frac{2\pi}{K}(\mathbf{z}\cdot\hat{\mathbf{n}})^{2}}$\tabularnewline
SUM gate & $(-1)^{\hat{n}_{1}\otimes\hat{n}_{1}}$ & $e^{i\frac{2\pi}{K}(\mathbf{z}\cdot\hat{\mathbf{n}})\otimes(\mathbf{z}\cdot\hat{\mathbf{n}})}$\tabularnewline
\bottomrule
\end{tabular}
\caption{\label{tab:pair-cat} 
Table comparing the pair-cat qubit to a general qu\(K\)it tiger code that is infinitely supported in Fock space. 
The energy density per mode is \(\alpha^2\) , with \(\alpha>0\) assumed real for simplicity.
Here, \(\HZ\) is an integer matrix, and \(\HX\) is an integer matrix with non-negative entries, respectively. 
Codewords are labeled by indices \(\mu,\nu\in \frac{2\pi}{K}\mathbb{Z}_K = \frac{2\pi}{K}\{0,1,\cdots,K-1\}\).
We conjecture that the ability of a dephasing operator to distinguish the codespace is suppressed exponentially with the product of the code distance \(d_Z\) and energy density, and prove this conjecture for all tiger codes that satisfy \(\mathbf 1= \begin{pmatrix} 1 & 1 &\cdots & 1\end{pmatrix}\in \ker \HZ \) (see Sec.~\ref{sec:dephasingZdist}). 
}
\end{table*}

\prg{Infinite-support examples}
The single-mode two-component cat code is a tiger code, admitting no occupation number constraint (\(\HZ = 0\)) and codewords that are fixed-eigenvalue eigenstates of \(\hat{a}^2\) (\(\HX = 2\)).

Coherent-state repetition codes, whose codewords are tensor products of the same coherent state, correspond to \textcolor{black}{the case where $H=0$ and where $G$ is the check matrix of a repetition code.}
Such codes can encode either a logical rotor or a logical qudit, depending on the \textcolor{black}{parity of} number of physical modes \(N\) --- an interesting feature of homology.
A dual of these codes are the Fock-state repetition codes, whose \(\HZ\) matrix is a generating matrix of the \(A_N\) root lattice.
Two-mode versions of both of these codes are special, yielding infinite-dimensional logical subspaces spanned by various pair-coherent states. 

The two-mode pair-cat qubit is the simplest tiger code with nonzero generator matrices, but it does not improve over the \(Z\)-distance of the two-component cat code.
Its many-mode extension protects against arbitrary losses on all but one mode, but has a \(Z\)-distance that vanishes with the number of modes.

Even small examples of tiger codes, like the three-mode and four-mode tiger codes, show intriguing behavior. For instance, the three-mode tiger code encodes a logical rotor and triples the code distances compared to a bare rotor or a logical rotor encoded in pair-coherent state, respectively. A four-mode code \textcolor{black}{achieves two-fold enhancement in} both \textcolor{black}{$X$- and $Z$-distances compared to the} two-component cat codes.
It belongs to a family of tiger-code analogues of the qubit Shor \cite{shor1995scheme,knill2000efficient,ralph2005loss} (a.k.a. quantum parity check) family.
It's dephasing protection holds for \textit{all} values of the energy and does not require a large-energy limit --- an effect that, to our knowledge, is a first among codes based on coherent states or their concatenations. This implies that codewords for any nonzero \(\Delta\) are exactly orthogonal. Moreover, the code admits a logical \(X\) operation in the form of a quadratic (i.e., beam-splitter) Hamiltonian for such cases.

Our framework yields integer-homology analogues of many established \(\mathbb{Z}_2\)-homology (read: qubit) CSS codes, realized using projected coherent states.
We study the integer-homology analogue of the Kitaev surface code \cite{kitaev1997quantum,kitaev1997quantumimperfect,bravyi1998quantum,dennis2002topological,kitaev2003fault}, whose linear constraints are sums of occupation numbers around a cross of a 2D lattice, and whose dissipators contain products of four lowering operators.
This code inherits the surface code's string-like minimal logical operators, geometric locality of stabilizing operators, and scaling of code distances.
It is less resource intensive than the surface-cat code --- a concatenation of a single-mode cat code with the surface code \cite{chamberland2022building} --- requiring only linear occupation-number measurements, as opposed to nonlinear joint parity measurements, and does not involve any individual mode \textcolor{black}{stabilization}.
Part of this code's dephasing protection does not require a large-energy limit.

\prg{Finite-support examples}

Many infinite-support codes can be converted to finite-support codes by multiplying columns of generator matrices by \(-1\), which corresponds to mathematically rotating the Fock-space sector used to define the code (see Fig.~\ref{fig:paircat_pairbinomial}).

Multiplying the second column of the pair-cat generator matrix by \(-1\), \( \begin{pmatrix} 1 & -1\end{pmatrix}\mapsto \begin{pmatrix} 1& 1 \end{pmatrix}\), yields a constraint on the total occupation number of the two modes.
Pairing this with \(\HX = \begin{pmatrix} 2 & -2 \end{pmatrix}\) yields the two-mode binomial qubit, whose codewords are two-mode binomial states \cite{radcliffe1971some,arecchi1972atomic}.
This code can be extended to a "multinomial" \(N\)-mode qu\(N\)it, whose codewords are \(SU(N)\) generalized coherent states \cite{gitman1993coherent,nemoto2000generalized,calixto2021entanglement}.

A simple three-mode code with \(\HX = \begin{pmatrix} 2& 2& -2 \end{pmatrix}\) and occupation-number constraints on the sums of two mode pairs admits the same Fock-state support as a \(\chi^{(2)}\) code \cite{niu2018hardware}.
We extend this code to a larger family and show that dephasing errors decrease exponentially with the square-root of a component of the syndrome vector \(\boldsymbol{\Delta}\).

Certain GKZ functions encode algebraic curves and, more generally, algebraic varieties \cite{stienstra2007gkz}.
We highlight this connection by defining a code family based on surfaces of Calabi-Yau type.
This family also admits some \(\boldsymbol{\Delta}\)-induced dephasing suppression, and can detect arbitrary losses on any mode.
We utilize the powerful theory of polynomial ideals to define a logical \(X\) operator for this and the \(\chi^{(2)}\)-like codes.

\section{Code construction: infinite support}\label{sec:inf_support}

We go through our construction for the case of infinite Fock-state support, where the two types of stabilizing operators ($X$-type dissipators and $Z$-type \textcolor{black}{stabilizers}) commute. \textcolor{black}{Unless otherwise specified, throughout the remainder of the text, the term \textit{stabilizers} refers collectively to both the $X$-type dissipators and the $Z$-type stabilizers.}
We cover both finite- and infinite-dimensional logical encodings.
For reference, the features of a general qu\(K\)it tiger code are compared side-by-side to the two-mode pait-cat qubit in Table~\ref{tab:pair-cat}.

\subsection{Stabilizers and dissipators}
Tiger codes are defined using $X$- and $Z$-type generator matrices, which we call respectively \(\HX\) and \(\HZ\),
\begin{eqs}
\HX =\begin{pmatrix}\vdots\\
\bfg\\
\vdots
\end{pmatrix}\begin{array}{c}
\uparrow\\
r_{x}\\
\downarrow
\end{array}\quad\quad\text{and}\quad\quad \HZ =\begin{pmatrix}\vdots\\
\bfh\\
\vdots
\end{pmatrix}\begin{array}{c}
\uparrow\\
r_{z}\\
\downarrow
\end{array},
\end{eqs}
where $\HX \in\mathbb{N}^{r_{x}\times N}$ is an $r_{x} \times N$ integer matrix with non-negative entries whose row vectors we generically label by \(\bfg\), and where $\HZ \in\mathbb{Z}^{r_{z}\times N}$ is an $r_{z} \times N$ integer matrix whose row vectors we generically label by \(\bfh\).
We cover the case when \(\HX\) has negative entries in  Section~\ref{sec:finitesupport}.
The two generator matrices are required to satisfy the CSS-type condition
\begin{eqs}\label{eq:css}
    \HX  \HZ ^\T=\mathbf{0}~\quad\Leftrightarrow\quad\bfh\cdot\bfg=0, ~~\forall \bfg\in \text{row}(\HX) , \forall\bfh\in \text{row}(\HZ) ,
\end{eqs} 
where \(\cdot^\T\) is the transpose operation, and where \(\mathbf 0\) is a zero matrix of appropriate dimensions.

The matrix $\HZ $ defines the subspace of Fock states that supports the code.
This subspace consists of all Fock states \(|\mathbf{n}\rangle\) whose labels satisfy the following linear constraints,
\begin{eqs}\label{eq:fock}
    \HZ \mathbf{n}=\boldsymbol{\Delta}\quad\Leftrightarrow\quad\bfh\cdot\mathbf{n}=\Delta_{\bfh}\quad\text{for}\quad\boldsymbol{\Delta}=\begin{pmatrix}\vdots\\
\Delta_{\bfh}\\
\vdots
\end{pmatrix}\begin{array}{c}
\uparrow\\
r_{z}\\
\downarrow
\end{array} ,
\end{eqs}
for all rows \(\bfh\) of \(\HZ \) and for some fixed \(r_{z}\)-dimensional integer vector \(\mathbf\Delta\).
When the explicit value of \(\mathbf \Delta\) does not affect code properties, it will be set to zero, while in some cases it will be beneficial to choose it non-zero.

Since \(\HX \) consists of non-negative entries, the constrained Fock subspace is infinite-dimensional.
Indeed taking any initial vector \(\mathbf{n}\in\mathbb{N}^N\) satisfying the constraints \eqref{eq:fock}, one can add to it any positive integer multiple of any row of \(\HX \), and obtain another valid Fock-state with higher total occupation number, showing there is an infinite number of solutions.
Correspondingly, the codewords will be of infinite Fock-state support.

For example, the two-mode pair-cat code, developed in Sec.~\ref{sec:pair_cat}, corresponds to \(\HZ =\begin{pmatrix} 1 & -1\end{pmatrix}\), with the integer scalar \(\Delta\) fixing the occupation-number difference \(\hat n_1 - \hat n_2\) between the modes.

The operator \(\bfh\cdot\mathbf{\hat{n}}-\Delta_{\bfh}\) for each row \(\bfh\) and each corresponding entry \(\Delta_{\bfh}\) annihilates all codewords and generates a family of unitary \(Z\)-type stabilizers, 
\begin{subequations}
\begin{eqs}\label{eq:ztype}
    \exp\left[i\phi(\bfh\cdot\hat{\mathbf{n}}-\Delta_{\bfh})\right]\quad\quad\text{for}\quad\quad\phi\in\mathbb{T} = [0,2\pi) .
\end{eqs}
Each stabilizer is a tensor product of single-mode Fock-space rotations, and the codespace lies in the \(+1\) eigenspace of all stabilizers for all \(\phi\).

The non-negative-entry matrix $\HX $ corresponds to a non-unitary \(X\)-type "stabilizer" that is a monomial in the annihilation operators $\hat{a}_j$.
We call these dissipators and define them using multi-index notation,
\begin{eqs}\label{eq:xtype}
    \mathbf{\hat{a}}^{\bfg}\equiv\hat{a}_{1}^{g_{1}}\hat{a}_{2}^{g_{2}}\cdots\hat{a}_{N}^{g_{N}} .
\end{eqs}
\end{subequations}

Codewords are eigenstates of such dissipators.
For example, the two-mode pair-cat codewords are eigenstates of \(\hat{a}_1^2 \hat{a}_2^2\), corresponding to \(\HX  = \begin{pmatrix} 2 &2 \end{pmatrix}\).

All \(Z\)-type stabilizers commute with all \(X\)-type dissipators due to the CSS-type condition \eqref{eq:css}, which implies that
\begin{eqs}\label{eq:commutation}
    (\bfh\cdot\mathbf{\hat{n}})(\mathbf{\hat{a}}^{\bfg})=(\mathbf{\hat{a}}^{\bfg})(\bfh\cdot\mathbf{\hat{n}}-\bfh\cdot\bfg)=(\mathbf{\hat{a}}^{\bfg})(\bfh\cdot\mathbf{\hat{n}}) .
\end{eqs}
The above relation holds more generally for any orthogonal pair of vectors \(\bfg\) and \(\bfh\).
It can be obtained by remembering that \(f(\hat n)\hat{a}=\hat{a}f(\hat n - 1)\) for any analytic function $f$ of the occupation-number operator.

\subsection{Projected coherent states}
The \textcolor{black}{projection of a }\(Z\)-type \textcolor{black}{stabilizer} can be expressed in terms of a projection \({\Pi}_{\boldsymbol\Delta}\) onto admissible Fock states \(|\mathbf{n}\rangle\),
which can be defined in terms of the \(Z\)-type stabilizers.
This means that the projection can be thought of as an (integral) averaging over the entire \(Z\)-type stabilizer group.
The integral and sum forms of the projection correspond to the last two equalities below, 
\begin{subequations}\label{eq:Z_projector}
\begin{align}
  {\Pi}_{\boldsymbol\Delta}& \equiv \prod_{\bfh}\int_0^{2\pi}\frac{d\phi}{2\pi}e^{i\phi(\bfh\cdot\hat{\mathbf{n}}-\Delta_{\bfh})},\\
  &=\int_{\boldsymbol \phi \in \mathbb{T}^{r_z}}\frac{d^{r_{z}}\boldsymbol\phi}{(2\pi)^{r_{z}}}e^{i\boldsymbol{\phi}\cdot\left(\HZ \hat{\mathbf{n}}-\boldsymbol{\Delta}\right)},\\
  &=\sum_{\substack{\mathbf{n}\in\mathbb{N}^N:\\\HZ \mathbf{n}=\boldsymbol{\Delta}}}|\mathbf{n}\rangle\langle\mathbf{n}|.
\end{align}
\end{subequations}

In the first expression, \(\bfh\) runs over the rows of \(\HZ \), and its corresponding \(\Delta_{\bfh}\) is the coordinate of the vector \(\boldsymbol\Delta\) at the same position.
For each row \(\bfh\), we integrate over an angle \(\phi\in[0,2\pi)\), i.e., the group \(\mathbb{T}\).
In the second expression, these angles are all combined into one vector \(\boldsymbol{\phi}\) which runs over the stabilizer group, \(\mathbb{T}^{r_{z}}\).
The sum form in the third expression is obtained from the integral by applying the orthogonality relation \(\int_{0}^{2\pi}\frac{d\phi}{2\pi}e^{i\phi(\hat{n}-\Delta)}=\delta_{\hat{n},\Delta}\) for each coordinate of \(\boldsymbol\phi\), where \(\delta\) is the Kronecker delta function.

Projected coherent states, \(|\boldsymbol\alpha\rangle_{\boldsymbol\Delta}^{\HZ}\),
are obtained by applying this projection to the ordinary coherent states \(|\boldsymbol\alpha\rangle\), for \(\boldsymbol{\alpha}\in\mathbb{C}^N\), and normalizing.
Using the integral and sum equivalent forms of the projection yields two complementary representations (cf. \cite{skagerstam1985quasi,drummond2016coherent}),
\begin{subequations}\label{eq:projected-coherent-state}
\begin{align}
|\boldsymbol{\alpha}\rangle_{\boldsymbol\Delta}^{\HZ}& \equiv \frac{e^{\|\boldsymbol{\alpha}\|^{2}/2}}{\sqrt{\mathsf{A}_{\boldsymbol\Delta}(\boldsymbol{\alpha}^{\star}\boldsymbol{\alpha})}}{\Pi}_{\boldsymbol\Delta}|\boldsymbol{\alpha}\rangle\label{eq:proj1} ~,\\
&=\frac{e^{\|\boldsymbol{\alpha}\|^{2}/2}}{\sqrt{\mathsf{A}_{\boldsymbol\Delta}(\boldsymbol{\alpha}^{\star}\boldsymbol{\alpha})}}\int_{\boldsymbol \phi \in \mathbb{T}^{r_z}}\frac{d^{r_{z}}\boldsymbol{\phi}}{(2\pi)^{r_{z}}}e^{-i\boldsymbol{\phi}\cdot\boldsymbol{\Delta}}|\boldsymbol{\alpha}e^{i\boldsymbol{\phi}\HZ }\rangle\label{eq:proj_state_integral_form} ~,\\
&=\frac{1}{\sqrt{\mathsf{A}_{\boldsymbol\Delta}(\boldsymbol{\alpha}^{\star}\boldsymbol{\alpha})}}\sum_{\HZ \mathbf{n}=\boldsymbol{\Delta}}\frac{\boldsymbol{\alpha}^{\mathbf{n}}}{\sqrt{\mathbf{n}!}}|\mathbf{n}\rangle~,\label{eq:proj_state_sum_form}
\end{align}
\end{subequations}
where the coherent-state value in the second expression,
\begin{equation}
    \boldsymbol{\alpha}e^{i\boldsymbol{\phi}\HZ } \equiv \left(\alpha_{1}e^{i(\boldsymbol{\phi}\HZ )_{1}},\alpha_{2}e^{i(\boldsymbol{\phi}\HZ )_{2}},\cdots,\alpha_{N}e^{i(\boldsymbol{\phi}\HZ )_{N}}\right),
\end{equation}
is obtained from the identity \(e^{i\phi\hat{n}}|\alpha\rangle=|\alpha e^{i\phi}\rangle\), and where the notions $\mathbf{n}! \equiv n_1!n_2!\cdots n_N!$, $\boldsymbol{\alpha}^{\mathbf{n}} \equiv \prod_{j=1}^N \alpha_j^{n_j}$, \textcolor{black}{ and $\sum_{H \mathbf{n}=\mathbf{\Delta}} \equiv \sum_{\mathbf{n}\in \mathbb{N}^N: H \mathbf{n}=  \mathbf{\Delta}}$} are used in the third expression.   In Eq.~\eqref{eq:projected-coherent-state}, $\| \boldsymbol{\alpha}\|^2={\sum_{j=1}^N \alpha_j^*\alpha_j}$ denotes the squared 2-norm of the complex vector $\boldsymbol{\alpha}$. \textcolor{black}{The shorthand notations introduced above will be used throughout the remainder of this work.}

The normalization \(\mathsf{A}_{\boldsymbol\Delta}\) is a Gelfand-Kapranov-Zelevinsky (GKZ) hypergeometric (a.k.a. \(\HZ \)-hypergeometric) function.
For any complex \(\mathbf{y} = (y_1,y_2, \cdots,y_N)\), 
\begin{subequations}\label{eq:GKZintegralsum}
\begin{align}
  \mathsf{A}_{\boldsymbol{\Delta}}(\mathbf{y})&\equiv \int_{\boldsymbol \phi \in \mathbb{T}^{r_z}}\frac{d^{r_{z}}\boldsymbol{\phi}}{(2\pi)^{r_{z}}}e^{-i\boldsymbol{\phi}\cdot\boldsymbol{\Delta}}\exp\left({\textstyle \sum_{j=1}^{N}}y_{j}e^{i(\boldsymbol{\phi}\HZ )_{j}}\right)\label{eq:GKZintegral},\\
  &=\sum_{\HZ \mathbf{n}=\boldsymbol{\Delta}}\frac{\mathbf{y}^{\mathbf{n}}}{\mathbf{n}!} ~,\label{eq:GKZsum}
\end{align}
\end{subequations}
with each form obtained from the corresponding expression in Eq.~\eqref{eq:projected-coherent-state} using the formula for the overlap between ordinary coherent states and Fock states, respectively.
This function completely determines the normalization and overlap between projected coherent states,
\begin{eqs}\label{eq:gkz-overlap}
    _{\boldsymbol{\Delta}}^{\HZ}\langle\boldsymbol{\alpha}|\boldsymbol{\beta}\rangle_{\boldsymbol{\Delta}}^{\HZ}=\frac{\mathsf{A}_{\boldsymbol{\Delta}}(\boldsymbol{\alpha}^{\star}\boldsymbol{\beta})}{\sqrt{\mathsf{A}_{\boldsymbol{\Delta}}(\boldsymbol{\alpha}^{\star}\boldsymbol{\alpha})\mathsf{A}_{\boldsymbol{\Delta}}(\boldsymbol{\beta}^{\star}\boldsymbol{\beta})}}~,
\end{eqs}
where the argument \(\boldsymbol{\alpha^{\star}\beta}=\begin{pmatrix}\alpha_{1}^{\star}\beta_{1},\alpha_{2}^{\star}\beta_{2},\cdots,\alpha_{N}^{\star}\beta_{N}\end{pmatrix}\).
We discuss its features in App.~\ref{sec:GKZ}.  Since vectors are used to parameterize both operators and coherent states, we include commas when the vector appears in bra-ket notation or as an argument of GKZ hypergeometric functions, but omit them when the vector represents an operator.

All projected coherent states satisfy the stabilizer conditions
\begin{subequations}\label{eq:stabilizers}
\begin{eqs}
    e^{i\phi(\bfh\cdot\hat{\mathbf{n}}-\Delta_{\bfh})} |\boldsymbol{\alpha}\rangle_{\boldsymbol\Delta}^{\HZ} = |\boldsymbol{\alpha}\rangle_{\boldsymbol\Delta}^{\HZ} ~,\quad\quad\forall \phi\in[0,2\pi)
\end{eqs}
for any row \(\bfh\) of \(\HZ \) since, by definition, they are superpositions of Fock states that satisfy the corresponding constraint.

Projected coherent states also inherit a subset of properties of the original coherent states.
Ordinary coherent states are eigenstates of any dissipator \(\mathbf{\hat{a}}^{\bfg}\), and projected coherent states retain this property for any \(\bfg\) in the kernel of \(\HZ \),
\begin{eqs}
    \mathbf{\hat{a}}^{\bfg}|\boldsymbol{\alpha}\rangle_{\boldsymbol\Delta}^{\HZ}=\boldsymbol{\alpha}^{\bfg}|\boldsymbol{\alpha}\rangle_{\boldsymbol\Delta}^{\HZ} ~,\quad\quad\forall\bfg\in\ker \HZ  .\label{eq:akerHZ}
\end{eqs}
\end{subequations}
This can be derived by using the first expression \eqref{eq:proj1} for the projected coherent state, noting that the dissipator commutes with all stabilizers due to Eq.~\eqref{eq:commutation}, and using \(\hat a|\alpha\rangle=\alpha|\alpha\rangle\). 

More generally, any combination of loss and gain operators that commute with all \(\mathbf{h}\cdot \mathbf{\hat{n}}\) operators act the same way on projected coherent states as they do on ordinary coherent states.
In some cases, this allows for quadratic (i.e., beam-splitter) logical gates for tiger codes.
This also allows for a simple treatment of the damping operator \(\exp(-\hat{n})\), irrespective of the Fock-state support of the code.

Tiger codewords form a subset of the projected coherent states
for certain values of \(\boldsymbol\alpha\).
For simplicity, we chose to always include in this subset the all-ones coherent state,
with \(\boldsymbol{\alpha} \propto \boldsymbol{1}=(1,1,\cdots,1)\), 
\begin{eqs}\label{eq:fiducial}
    \left|\boldsymbol{\alpha}=\alpha\boldsymbol{1}\right\rangle_{\boldsymbol\Delta}^{\HZ}\quad\text{for}\quad\alpha>0\quad\quad\text{(fiducial state)} ,
\end{eqs}
with \textit{total energy} \(N \alpha^2\) and \textit{energy density} \(\alpha^2\).
(We keep \(\alpha\) real to avoid burdensome notation.)
The remaining codewords can be obtained by applying logical operators --- particular tensor products of single-mode rotations --- which preserve the total energy and energy density.

All logical states \(|\overline{\Psi}\rangle\) --- superpositions of tiger codewords --- satisfy the stabilizer-like constraints from Eq.~\eqref{eq:stabilizers}, equivalent to
\begin{subequations}
\label{eq:discrete-stabilizers}
 \begin{align}
     (\bfh\cdot\hat{\mathbf{n}})\ket{\overline{\Psi}} &= \Delta_{\bfh} \ket{\overline{\Psi}} \quad~~~ \forall\bfh\in \textcolor{black}{\text{row}(\HZ)},\\
     \mathbf{\hat{a}}^{\bfg}\ket{\overline{\Psi}} &= \alpha^{|\bfg|}\ket{\overline{\Psi}}\quad~~\forall\bfg\in \textcolor{black}{\text{row} (\HX)},
 \end{align}
\end{subequations}
where \(|\bfg|=\sum_{j=1}^N |g_j|\) denotes the 1-norm of vector $\mathbf{g}$. \textcolor{black}{If $\mathbf{h}$ is the $i$-th row of $\HZ$, then $\Delta_{\mathbf{h}}$ denotes the $i$-th entry of the vector $\mathbf{\Delta}$.}
The joint eigenspace defined by these constraints defines the logical subspace, but its structure can be determined from the properties of \(\HX\) and \(\HZ\) alone.

\subsection{Logical subspace from homology}\label{sec:homology}

With the \(Z\)-type \textcolor{black}{stabilizer} imposed, the \(X\)-type matrix \(\HX \) can be used to determine the structure of the codespace.
As shown in Eq.~\eqref{eq:akerHZ}, the kernel of \(\HZ \) permits to define the annihilation operators which still act as such on the projected coherent states.
The rows of \(\HX \), which are \textcolor{black}{$X$-type} stabilizers, span a strict subspace of this kernel and further constrain the codespace.
The codespace structure is then determined by the remaining part of the kernel of \(\HZ \).
We can write
\begin{equation}
    \ker \HZ  = {\rm im} \HX  + \mathcal{L}_X ,
\end{equation}
where we denote this remaining part \(\mathcal{L}_X\) and also call it the logical-\(X\) group.
Here \(\HZ\) and \(\HX\) acts on vectors with integer entries.
Hence the kernel and the image 
should be thought of as (sub)groups under addition.
The logical-\(X\) group is then the quotient group formed by the kernel of \(\HZ \) modulo the image of \(\HX \),
\begin{subequations}\label{eq:homology}
\begin{eqs}
    \mathcal{L}_X \cong\frac{\ker \HZ }{{\rm im}~\HX }~.
\end{eqs}
This is the homology group of the integer chain complex formed by \(\HX \) and \(\HZ \).
Such a homology group can contain infinite factors, i.e. \(\mathbb{Z}\), as well as finite factors fo the form \(\mathbb{Z}_K = \mathbb{Z}/K\mathbb{Z} = \left\{0, 1, \ldots, K-1\right\}\).
The former are referred to as the \emph{free} factors and the latter as the \emph{torsion} factors.
We can compute a generating matrix for \(\mathcal{L}_X\): It is an integer matrix that we denote by \(L_X\).

In the example of the pair-cat code, where \(\HZ =\begin{pmatrix}1 &-1 \end{pmatrix}\) and \(\HX = \begin{pmatrix} 2 &2 \end{pmatrix}\), the kernel of \(\HZ \) is generated by the vector \( \begin{pmatrix} 1 &1 \end{pmatrix}\), while the image of \(\HX \) consists of only \textit{even-integer} multiples of said vector. 
The quotient group thus corresponds to a qubit,
\begin{equation*}
    {\mathcal{L}}_X=\frac{{\rm im}\begin{pmatrix} 1 & 1 \end{pmatrix}}{{\rm im} \begin{pmatrix} 2 & 2 \end{pmatrix}}\cong\frac{\mathbb{Z}}{2\mathbb{Z}}\cong\mathbb{Z}_{2}\quad\quad\text{(pair-cat)}.
\end{equation*}
This is an example of torsion.

Similar to the dueling sum and integral representations of projected coherent states, the powerful framework of homology yields a "dual" method of calculating the codespace structure.
The homology condition, \(\HX  \HZ ^\T = \mathbf 0\), implies that the rows of \(\HZ \) also span a strict subspace of the kernel of \(\HX \).
The remaining subspace of this kernel, denoted \(\mathcal{L}_Z\) and called the logical-\(Z\) group, provides another characterization of the codespace structure.
The kernel of \(\HX \) splits as 
\begin{equation*}
    \ker_{2\pi} \HX =\text{im}_{2\pi}\HZ  + \mathcal{L}_{Z},
\end{equation*}
where the matrices \(\HX\) and \(\HZ\) are now considered to act on vectors with entries in \(\mathbb{T} = [0,2\pi)\), as signaled by the subscripts.
The logical-\(Z\) group \(\mathcal{L}_Z\) can then be determined by the quotient
\begin{eqs}
    \mathcal{L}_Z \cong\frac{\ker_{2\pi}\HX }{{\rm im}_{2\pi}\HZ }~.
\end{eqs}
\end{subequations}
This is the cohomology group of the \((\HX ,\HZ )\) chain complex, defined over \(\mathbb{T}\) coefficients. 
Because the cohomology is taken over \(\mathbb{T}\), it is guaranteed to have the same structure as that of the homology group \(\mathcal{L}_X\) \cite[Thm.~1]{vuillotHomologicalQuantumRotor2024}.
Namely, for each factor \(\mathbb{Z}\) in \(\mathcal{L}_X\), there is a factor \(\mathbb{T}\) in \(\mathcal{L}_Z\), and for each factor \(\mathbb{Z}_K\) in \(\mathcal{L}_X\), there is a factor \(\frac{2\pi}{K}\mathbb{Z}_K\) in \(\mathcal{L}_Z\),
where \(\frac{2\pi}{K}\mathbb{Z}_K\) is the subgroup of \(\mathbb{T}\) isomorphic to \(\mathbb{Z}_K\):
\begin{equation}
    \frac{2\pi}{K}\mathbb{Z}_K=\left\{0, \frac{2\pi}{K},\ldots,\frac{2\pi(K-1)}{K}\right\}.\label{eq:Zstard}
\end{equation}

In the case of the pair-cat code, the ``\(2\pi\)-kernel'' of \(\HX =\begin{pmatrix} 2 & 2 \end{pmatrix}\) contains all \textit{real} multiples of \( \begin{pmatrix} 1 &-1 \end{pmatrix}\), modulo \(2\pi\).
The ``\(2\pi\)-image'' of \(\HZ \) \textit{precisely} consists of these same multiples, suggesting (incorrectly) that the homology group, and thus the codespace, is trivial.
In reality, the \(2\pi\)-periodicity ensures that there is another vector, namely \(\pi \begin{pmatrix} 1 &0 \end{pmatrix}\), that is in the \(2\pi\)-kernel of \(\HX \) but \textit{not} in the \(2\pi\)-image of \(\HZ \).
Other vectors such as \(\frac{\pi}{2}\begin{pmatrix} 1 &1 \end{pmatrix}\) and \(\pi(0~1)\) are also in \(\ker \HX \), but these can be obtained from \(\pi \begin{pmatrix}1 &0 \end{pmatrix}\) by adding an element in the image of \(\HZ \).
Since doubling \(\pi \begin{pmatrix} 1~0 \end{pmatrix}\) yields the zero vector modulo \(2\pi\),
and since there is no other vector like it, 
we obtain the same result as that of the previous homology calculation,
\begin{eqs}\nonumber
    \mathcal{L}_{Z}=\frac{\mathbb{T}\times\pi\mathbb{Z}_{2}}{\mathbb{T}}\cong\pi\mathbb{Z}_{2}\quad\quad\text{(pair-cat)}~.
\end{eqs}

We can also compute a generating matrix for \(\mathcal{L}_Z\).
It is an integer matrix that we denote by \(L_Z\), and it generates the \(Z\)-logical group by multiplying it with vectors with entries from \(\mathcal{L}_Z\), i.e. \(\mathbb{T}\) or \(\frac{2\pi}{K}\mathbb{Z}_K\).

\prg{Computing (co)homology with torsion}
Torsion occurs every time integer multiples of an integer vector are in the image of \(\HX \) without the vector \textit{itself} being in the image.
In other words, the \(m\)th multiple of a vector \(\mathbf w\) \textit{not} in the image satisfies 
\begin{eqs}\label{eq:torsion}
    m\mathbf w = \mathbf s \HX \quad\quad \text{for some}\quad\quad \mathbf s \in\mathbb{Z}^{r_{x}} .
\end{eqs}
Torsion can occur in homology over rings that are not fields, such as our case of the integers, and allows us to design finite-dimensional logical subspaces.
The presence of torsion is also witnessed by the presence of "discrete vectors", such as that in the pair-cat example, in the \(2\pi\)-kernel of \(\HX \).

Since both the integer homology and \(\mathbb{T}\)-cohomology calculations yield the same answer \cite[Thm.~1]{vuillotHomologicalQuantumRotor2024}, one can choose the more straightforward homology calculation to compute the codespace structure.
Computing \(L_X\) amounts to computing the Smith normal forms of \(\HX\) and \(\HZ\),
\begin{equation}
V_\HX\HX = D_\HX W_\HX ~,\qquad \HZ W_\HZ = V_\HZ D_\HZ~,
\end{equation}
where \(V_{\HX/\HZ}\in\mathbb{Z}^{r_{x/z}\times r_{x/z}}\) and \(W_{\HX/\HZ}\in\mathbb{Z}^{N\times N}\) are some unimodular matrices, and where \(D_{\HX/\HZ}\in\mathbb{Z}^{r_{x/z}\times N}\) only have non-zero integer entries on the diagonal.
The rows of \(W_\HX\) corresponding to non-zero and non-unity diagonal elements of \(D_\HX\) are the generators for the torsion part of \(L_X\), whose order is given by the corresponding diagonal entries.
The generators for the free part are given by the columns of \(W_\HZ\) corresponding to zero entries on the diagonal of \(D_\HZ\) and that are not in the span of the rows of \(W_\HX\) corresponding to non-zero diagonal entries of \(D_\HX\).

Knowing the codespace structure, the computation of \(L_Z\) becomes easier as we can do the computation restricted to coefficients of the homology group.
For instance when we have logical quKits in the torsion part, we just have to do linear algebra over \(\mathbb{Z}_K\) to compute the corresponding generators in \(L_Z\) as rows over \(\mathbb{Z}_K\), and then interpret them as a rows over \(\mathbb{Z}\).
If we have a free part, we can do linear algebra over \(\mathbb{Z}\) to find the corresponding generators.

\subsection{Logical operators \& codewords}\label{sec:logical_codeword}

As described above, the homology (resp. cohomology) group \eqref{eq:homology} determining the codespace is, in general, a product of infinite factors, \(\mathbb{Z}\) (resp. \(\mathbb{T}\)), and, due to torsion, possible finite factors, \(\mathbb{Z}_K\) (resp. \(\frac{2\pi}{K}\mathbb{Z}_K\)), i.e. logical qubit (\(K=2\)) or qu$K$it subsystems.
Each subsystem is generated by its own integer vector \(\mathbf{z}\), and such generating vectors make up the rows of a generating matrix \(L_Z\) of logical $Z$ operators.
Similarily the matrix \(L_X\), of integer vectors \(\mathbf{x}\), determines the  logical $X$ operators,
\begin{eqs}
   L_{X}=\begin{pmatrix}\vdots\\
\mathbf{x}\\
\vdots
\end{pmatrix}\quad\quad\quad\text{and}\quad\quad\quad L_{Z}=\begin{pmatrix}\vdots\\
\mathbf{z}\\
\vdots
\end{pmatrix} ,
\end{eqs}
and the number of rows of both matrices is equal the number of factors in the homology group.
The inner product between a vector \(\mathbf{x}\) and its partner \(\mathbf{z}\) in the same row is one while all other inner products are zero, yielding
\begin{eqs}\label{eq:logicals}
    L_{X}L_{Z}^{\text{T}}=\mathbbm{1}
\end{eqs}
and underpinning the commutation relations between the distinct subsystems.

Both matrices generate subspaces that are in the kernels of their corresponding generator matrices, 
ensuring the their associated logical operators, introduced below, commute with all stabilizers.
Moreover, the generators for the logical operators can always be redefined by adding stabilizers,
\begin{equation}
    L_X^\prime \cong L_X + \mathbf{s}\HX,\qquad L_Z^\prime \cong L_Z + \mathbf{t}\HZ,
\end{equation}
where \(\mathbf{s}\in\mathbb{Z}^{r_{x}}\) and \(\mathbf{t}\in\mathbb{Z}^{r_{z}}\) are any integer vectors.
Notably, since \(\HX\) is non-negative, we can always make \(L_X\) non-negative as well.
When implementing logical-\(Z\) operators, as shown below, we can furthermore use the full stabilizer group, i.e. linear combinations of rows of \(\HZ\) with coefficients in \(\mathbb{T}\). By adding rows of $G$ to $L_X$, one can generate deformed $X$ operator with negative entries. These correspond to operators of the form $\mathbf{\hat{a}}^{\dagger \mathbf{q}} \mathbf{\hat{a}}^{\mathbf{p}}$ which have the $X$ action within the codespace space but do not preserve it, thereby inducing leakage. To determine the $X$-distance of the code, we must account for undetectable $X$-error arising from both monomials solely by $\hat{a}$ and those involving both $\hat{a}^\dagger$ and $\hat{a}$.

Tiger codewords can be obtained by applying logical $Z$ operators corresponding to vectors \(\mathbf{z}\) to the all-ones projected coherent state \(\boldsymbol{\alpha} = \alpha \mathbf{1}\).
Codewords are obtained in this way for both the finite- and infinite-dimensional logical subspace cases.  
From now on, we focus on codes that consist of a single logical subsystem of either finite or infinite dimension.

\prg{Logical qudit}
Suppose that a given row vector \(\mathbf{z}\in\mathbb{Z}^N\) of \(L_Z\) and its partner \(\mathbf{x}\in\mathbb{Z}^N\) in \(L_X\) correspond to a logical qudit encoding of dimension \(K\).
Logical $Z$ operators are tensor-product rotations generated by 
\(\mathbf{z}\cdot\hat{\mathbf{n}}\), multiplied by elements \(\mu\) of the logical-\(Z\) group,
\begin{eqs}\label{eq:zlogicals}
    \overline{Z}(\mu)=\exp\left(-i\mu\mathbf{z}\cdot\hat{\mathbf{n}}\right)\quad\quad\text{for}\quad\quad\mu\in\frac{2\pi}{K}\mathbb{Z}_K .
\end{eqs}
We can add any combination of the rows of \(\HZ\) with arbitrary coefficients and get an equivalent \(Z\)-logical operator, i.e. for \(\boldsymbol{\phi}\in\mathbb{T}^{r_{z}}\),
\begin{equation}
e^{-i\mu\mathbf{z}\cdot\hat{\mathbf{n}}}\cong e^{-i\mu\mathbf{z}\cdot\hat{\mathbf{n}}}e^{-i\boldsymbol{\phi}\cdot(\HZ\hat{\mathbf{n}}-\boldsymbol{\Delta})}=e^{-i\left[\left(\mu\mathbf{z}+\boldsymbol{\phi}\HZ\right)\cdot\hat{\mathbf{n}}-\boldsymbol{\phi}\cdot\boldsymbol{\Delta}\right]},
\end{equation}
where we use "\(\cong\)" to relate operators with equivalent logical action.

Applying the above to the all-ones state~\eqref{eq:fiducial} yields an \(X\)-type codewords of a logical qudit indexed by \(\mu\in\frac{2\pi}{K}\mathbb{Z}_K\),
\begin{subequations}
\label{eq:codeword_formula}
\begin{align}
    |\overline{\mu}\rangle&=\overline{Z}(\mu)|\alpha\mathbf{1}\rangle_{\boldsymbol{\Delta}}^{\HZ} ~,\\
    &=|\alpha e^{-i\mu\mathbf{z}}\rangle_{\boldsymbol{\Delta}}^{\HZ} ~,\\
    &={\textstyle \frac{e^{N\alpha^{2}/2}}{\sqrt{\mathsf{A}_{\boldsymbol{\Delta}}(\alpha^{2}\boldsymbol{1})}}}{\Pi}_{\boldsymbol{\Delta}}|\alpha e^{-i\mu\mathbf{z}}\rangle~,
\end{align}
\end{subequations}
where \(\alpha e^{i\mu\mathbf{z}}=\alpha(e^{i\mu z_{1}},e^{i\mu z_{2}},\cdots,e^{i\mu z_{N}})\),
where the right-most state on the last line is an ordinary coherent state, and where the normalizing GKZ function \(\mathsf{A}\) is \(\mathbf{z}\)-independent since all codewords have the same vector of component norms, \(\boldsymbol{\alpha^{\star}}\boldsymbol{\alpha} = \alpha^2 \boldsymbol{1}\).
The second equality can be obtained by using expression \eqref{eq:proj1} for the projected coherent state and observing that the logicals commute with all stabilizers due to Eq.~\eqref{eq:commutation}.

Logical $X$ operators are monomials of lowering operators whose powers are entries in \(\mathbf{x}\),
\begin{eqs}\label{eq:xlogical-ops}
    \overline{X}=(\mathbf{\hat{a}}/\alpha)^{\mathbf{x}}=\hat{a}_{1}^{x_{1}}\hat{a}_{2}^{x_{2}}\cdots \hat{a}_{N}^{x_{N}}/\alpha^{|\mathbf{x}|} .
\end{eqs}
We normalize each component in the product by \(\alpha\) so that the operator more closely resembles a qudit Pauli operator.
We can append various stabilizing operators without affecting the logical action, and there can exist other variants that contain \(\hat{a}^{\dagger}\) terms.

The two types of logical operators satisfy the qudit commutation relation
\begin{eqs}\label{eq:qudit-commutation}
    \overline{X}\,\overline{Z}=e^{-i\frac{2\pi}{K}}\overline{Z}\,\overline{X} ,
\end{eqs}
where \(\overline{Z}\) without argument is just the generator of the group, \(\overline{Z} = \overline{Z}(\frac{2\pi}{K})\).
Equation~\eqref{eq:qudit-commutation} is obtained from the logical relation~\eqref{eq:logicals} and the commutation relation~\eqref{eq:commutation}.
This commutation relation ensures that the \(X\)-type operator acts on the codewords as \(\overline{X}|\overline{\mu}\rangle=e^{-i\mu}|\overline{\mu}\rangle\).
All logical operators commute with the stabilizing operators since their associated vectors \(\mathbf{x}\) and \(\mathbf{z}\) are in the kernels of \(\HZ \) and \(\HX \), respectively.

A dual basis of \(Z\)-type codewords indexed by \(\ell\in\mathbb{Z}_K\), can be defined by taking the logical Fourier transform,
\begin{subequations}
\begin{align}
   |\overline{\ell}\rangle&\propto\sum_{\mu\in\mathbb{Z}_{K}}e^{i\frac{2\pi}{K}\ell\mu}|\overline{\mu}\rangle ~,\\
   &=\sum_{\mu\in\mathbb{Z}_{K}}e^{i\frac{2\pi}{K}\ell\mu}\overline{Z}^{\mu}|\alpha\boldsymbol{1}\rangle_{\boldsymbol\Delta}^{\HZ}\label{eq:zprojector} ~,\\
   &\propto\sum_{\HZ \mathbf{n}=\boldsymbol{\Delta}\wedge\mathbf{z}\cdot\mathbf{n}=\ell\text{ mod }K}\frac{\alpha^{|\mathbf{n}|}}{\sqrt{\mathbf{n}!}}|\mathbf{n}\rangle~,\label{eq:zprojector-constraint}
\end{align}
\end{subequations}
where $\mathbf{n}!\equiv n_1!n_2!\cdots n_N!$ in multi-index notation, and where \(|\mathbf n| \equiv |n_1|+|n_2|+\cdots+ |n_N|\).
The second constraint in the last sum is modular, dividing the \(\HZ \)-constrained Fock space further into \(K\) sectors corresponding to the \(K\) values of the inner product \(\mathbf{z}\cdot\mathbf n\) modulo \(K\).
Each sector houses a \(Z\)-type codeword, making this basis orthonormal.
These codewords satisfy
\begin{subequations}
\begin{align}
    \overline{Z}(\mu)|\overline{\ell}\rangle &=e^{-i\ell\mu}|\overline{\ell}\rangle ,\\
    \overline{X}|\overline{\ell}\rangle &\propto|\overline{\ell -1\text{ mod }K}\rangle ,
\end{align}
\end{subequations}
which is obtained from the commutation relation~\eqref{eq:qudit-commutation} and Eq.~\eqref{eq:zprojector}.
Note that in the second equation, the normalization of the codestates is not perfectly conserved by \(\overline{X}\). This comes from the non-orthogonality of \(\ket{\overline{\mu}}\) states.

Given this orthonormal basis, we can define the codespace projector as
\begin{eqs}\label{eq:codeword_projector}
    P=\sum_{\ell \in \mathbb{Z}_K} \ket{\overline{\ell}}\bra{\overline{\ell}} .
\end{eqs}

\prg{Infinite-dimensional logical subsystem}
Suppose now that a given row vector \(\mathbf{z}\) of \(L_Z\) and its partner \(\mathbf{x}\) in \(L_X\) correspond to an infinite-dimensional subsystem. \textcolor{black}{In this case, both the logical \textcolor{black}{$Z$} operators and the \textcolor{black}{$X$-type} codewords are parameterized by angles.} Logical $Z$ operators \textcolor{black}{then take the form}
\begin{eqs}
\overline{Z}(\varphi)=\exp\left(-i\varphi\mathbf{z}\cdot\hat{\mathbf{n}}\right),\quad\quad\forall\varphi\in \mathbb{T}~.
\end{eqs}
The \textcolor{black}{$X$-type} codewords are obtained by applying logical $Z$ operators to the fiducial state
\begin{subequations}
\label{eq:continuous-x-basis}
\begin{align}
   |\overline{\varphi}\rangle&=\overline{Z}(\varphi)|\alpha\mathbf{1}\rangle_{\boldsymbol{\Delta}}^{\HZ} ~,\\
   &=|\alpha e^{-i\varphi\mathbf{z}}\rangle_{\boldsymbol{\Delta}}^{\HZ} ~,\\
   &={\textstyle \frac{e^{N\alpha^{2}/2}}{\sqrt{\mathsf{A}_{\boldsymbol{\Delta}}(\alpha^{2}\boldsymbol{1})}}}{\Pi}_{\boldsymbol{\Delta}}|\alpha e^{-i\varphi\mathbf{z}}\rangle~.
\end{align}
\end{subequations}

The logical $X$ operators stay the same as those in Eq.~\eqref{eq:xlogical-ops} for this case, and the commutation relation becomes
\begin{eqs}
    \overline{X}\,\overline{Z}(\varphi)=e^{-i\varphi}\overline{Z}(\varphi)\overline{X} .
\end{eqs}
The logical $X$ operator acts as \(\overline{X}|\overline{\varphi}\rangle=e^{-i\varphi}|\overline{\varphi}\rangle\).

A dual basis of integer-labeled \(Z\)-type codewords 
\(|\overline\ell \rangle\) can be defined by taking the Fourier transform of the codewords \(|\overline\varphi\rangle\),
\begin{subequations}
\begin{align}
    |\overline{\ell}\rangle&\propto\int d\phi e^{i\varphi\ell}|\overline{\varphi}\rangle ~,\\
    &=\int d\phi e^{i\varphi\ell}\overline{Z}(\varphi)|\alpha\mathbf{1}\rangle_{\boldsymbol\Delta}^{\HZ} ~,\\
    &\propto\sum_{\HZ \mathbf{n}=\boldsymbol{\Delta}\wedge\mathbf{z}\cdot\mathbf{n}=\ell}\frac{\alpha^{|\mathbf{n}|}}{\sqrt{\mathbf{n}!}}|\mathbf{n}\rangle~.\label{eq:inf-dim-z-constraint}
\end{align}
\end{subequations} 
The second constraint divides the \(\HZ \)-constrained Fock space into sectors labeled by the values \(\ell\) of the inner product \(\mathbf{z}\cdot\mathbf n\). 
This inner product may yield no solution for some values of \(\ell\), but those sectors that do admit at least one Fock state will house one \(Z\)-type codeword.
Since the codewords are supported on non-overlapping sets of Fock states, this basis is orthonormal.
We provide examples of logical-rotor (\(\ell \in \mathbb{Z}\)) and logical-mode (\(\ell \geq 0\)) encodings in Sec.~\ref{sec:pair_coherent}.

The \(Z\)-type codewords are eigenstates of \(\overline{Z}(\varphi)\) with eigenvalue \(\exp(-i\varphi\ell)\), and action of the logical $X$ operator decreases the value of \(\ell\) by one,
\begin{eqs}
    \overline{X}|\overline{\ell}\rangle\propto|\overline{\ell-1}\rangle .
\end{eqs}
The adjoint of this logical operator can raise the value of $\ell$ by one.
The codespace projector is defined in the same way as that in Eq.~\eqref{eq:codeword_projector}.

\subsection{Logical gates}

The following two types of gates, together with the {logical $X$ and $Z$ gates described above, constitute a universal gate set for tiger codes with infinite Fock-state support and finite-dimensional logical encoding.

Continuous rotations generated by \(\overline{X}\) and its tensor powers can be realized as Hamiltonian perturbations to the engineered dissipation induced by the dissipators \(\mathbf{\hat{a}}^{\mathbf{h}} - \alpha^{|\mathbf{h}|}\).
It is known that the leading-order contribution of a Hamiltonian perturbation to the steady-state subspace of a Lindbladian is the projection of the Hamiltonian into the subspace \cite{albert2016geometry}.
Applying the perturbation long enough relative to the dissipation strength yields the desired rotation \cite{zanardi2014coherent}.
This Zeno-type effect has been outlined in detail for the two-component cat \cite{mirrahimi2014dynamically} and pair-cat \cite{albert2019pair} qubits, and tiger codes follow the same recipe.

Higher-order polynomials \textcolor{black}{of occupation number operators} provide simple realizations of higher-order \(Z\)-type gates for both finite and infinite-dimensional tiger encodings.
Their actions are most easily understood in terms of \(Z\)-type codewords \(|\overline{\ell}\rangle\), which are constrained to consist of Fock states \(|\mathbf{n}\rangle\) satisfying \(\mathbf{z}\cdot\hat{{\bf n}} \equiv \ell\), either modulo \(K\) for the qu\(K\)it case~\eqref{eq:zprojector-constraint} or exacly for the infinite-dimensional case~\eqref{eq:inf-dim-z-constraint}.

A qu\(K\)it phase gate is implemented by the following "Kerr-effect" rotation \cite{girvin2014circuit}
\begin{eqs}
    e^{i\frac{2\pi}{K}(\mathbf{z}\cdot\hat{{\bf n}})^{2}}|\overline{\ell}\rangle=e^{i\frac{2\pi}{K}\ell^{2}}|\overline{\ell}\rangle .
\end{eqs}
This can be proven for the qu\(K\)it case by substituting \(\mathbf{z}\cdot\hat{{\bf n}} = mK + \ell\) for some non-negative integer \(m\).
A similar operation yields a quadratic phase gate \cite{grimsmo2020quantum,rotorclifford} for the infinite-dimensional case.

An entangling SUM gate of similar form acts as
\begin{eqs}
    e^{i\frac{2\pi}{K}(\mathbf{z}\cdot\hat{{\bf n}})\otimes(\mathbf{z}\cdot\hat{{\bf n}})}|\overline{\ell_{1}},\overline{\ell_{2}}\rangle=e^{i\frac{2\pi}{K}\ell_{1}\ell_{2}}|\overline{\ell_{1}},\overline{\ell_{2}}\rangle ,
\end{eqs}
with its cousin for the infinite-dimensional case being a conditional phase gate \cite{rotorclifford}.

\section{Error model and code distance}\label{sec:code_distance}

Tiger codes can protect against occupation-number loss and gain errors, and we consider losses for simplicity in this work.
They also are able to suppress dephasing error, and we observe that the degree of suppression is related to the minimum Euclidean distance between coherent states in different (continuous) tiger constellations.
This distance is the same as that of the quantum spherical codes \cite{jain2024quantum}, up to a factor of the total energy.

\subsection{Loss error detection}

In multi-index notation, occupation-number loss operators on \(N\) modes, 
\begin{eqs}
     \mathbf p \in \mathbb{N}^n\quad\Leftrightarrow\quad \mathbf{\hat{a}}^{\mathbf p}\quad\quad\text{(loss errors),}
\end{eqs}
are parameterized by non-negative integer vectors \(\mathbf p\), which we will use interchangeably with the actual operator.

The \textcolor{black}{$Z$-type stabilizers} \(\bfh \cdot \hat{\mathbf n}\) serve as check operators that can detect loss errors. The code space lies in the Fock-state subspace satisfying the constraint \(\bfh \cdot \hat{\mathbf n} = \Delta_{\bfh}\) for all syndrome generators \(\bfh\) and corresponding entries \(\Delta_{\bfh}\) of the syndrome vector \(\boldsymbol{\Delta}\).

The action of a loss operator on a code state can map the code into a Fock-state subspace satisfying a different constraint.
Permuting the loss operator through a \textcolor{black}{$Z$-type stabilizer} and applying the codespace constraint yields
\begin{eqs}
    (\bfh\cdot\mathbf{\hat{n}})(\mathbf{\hat{a}}^{\mathbf{p}})=(\mathbf{\hat{a}}^{\mathbf{p}})(\bfh\cdot\mathbf{\hat{n}}-\bfh\cdot\mathbf{p})=(\mathbf{\hat{a}}^{\mathbf{p}})(\Delta_{\bfh}-\bfh\cdot\mathbf{p}) .
\end{eqs}
If \(\bfh\cdot\mathbf{p} \neq 0\), then the error has mapped the codespace into an error space inside the Fock-state subspace corresponding to the syndrome \(\Delta_{\bfh} - \bfh\cdot\mathbf{p}\).
In that case, the loss error \(\hat{\mathbf a}^{\mathbf p}\) can be detected by measuring \(\bfh \cdot \hat{\mathbf n}\).

Combining all syndromes and using the definition of the codespace projector~\eqref{eq:Z_projector},
\begin{eqs}\label{eq:projector-permutation}
    \mathbf{\hat{a}}^{\mathbf{p}}{\Pi}_{\boldsymbol{\Delta}}={\Pi}_{\boldsymbol{\Delta}-\HZ \mathbf{p}~}\mathbf{\hat{a}}^{\mathbf{p}} ,
\end{eqs}
we see that the syndrome vector is shifted by \(-\HZ  \mathbf p\).
Any vectors \(\mathbf p\) that are in the kernel of \(\HZ \) will not change the syndrome vector.
This fact yields a numerically tractable condition,
\begin{eqs}\label{eq:detectable}
    \mathbf{p}\notin\ker \HZ \quad\quad\Leftrightarrow\quad\quad\mathbf{\hat{a}}^{\mathbf{p}}\text{ detectable} ,
\end{eqs}
for a loss error to be detectable. More specifically, the Knill-Laflamme condition \cite{kl} for loss error has following property 
\begin{eqs}\label{eq:loss_KL}
    P \mathbf{\hat{a}}^{\mathbf{p}} P\propto \Pi_{\mathbf{\Delta}} \mathbf{\hat{a}}^{\mathbf{p}} \Pi_{\mathbf{\Delta}}=\Pi_{\mathbf{\Delta}} \Pi_{\mathbf{\Delta}-H\mathbf{p}} \mathbf{\hat{a}}^{\mathbf{p}}\propto \delta_{\mathbf{p}\in \text{ker} H} ~,
\end{eqs}
where $P$ is the codespace projector in Eq.~\eqref{eq:codeword_projector} , and $\delta_{\mathbf{p}\in \text{ker} H} \equiv \delta_{H\mathbf{p}, \mathbf{0}}$ \textcolor{black}{vanishes} if $\mathbf{p} \notin \text{ker} H$ . Consequently, the error-detection condition for loss is automatically satisfied for any $\mathbf{\hat{a}}^{\mathbf{p}}$ such that $\mathbf{p} \notin \ker H$.

If a loss vector \(\mathbf p\) is in the kernel of \(\HZ \), then it is either a $X$-type  dissipator or a logical operator.
Dissipators are defined to be those operators that are also in the image of \(\HX \), so \(\mathbf p\) is a dissipator if it is in the span of the generators \(\bfg \in \text{row} (\HX) \).
Errors that are dissipators, like stabilizer Pauli strings for qubit \textcolor{black}{stabilizer} codes, are harmless since they apply the identity gate to the codespace.  \textcolor{black}{For convenience, in the analysis of code distance, we use the term \textit{undetectable $X$-error} specifically for those that induce a logical $X$ action.  }

The last category of loss errors consists of errors that are undetectable and that cause a logical error. By definition, their error vectors are in the kernel of \(\HZ \) and not in the image of \(\HX \),
\begin{eqs}
    \mathbf p \in  \ker \HZ  ~\backslash~ {\rm im}~\HX  \quad\quad \Leftrightarrow \quad\quad \mathbf{\hat{a}}^{\mathbf{p}}~\textcolor{black}{\text{undetectable}} .
\label{eq:uncorrectableloss}\end{eqs}

This set can be organized into distinct logical gates by taking the quotient of the two groups.
Namely, any such loss vector \(\mathbf p\) can be expressed as a product of a coset representative \(\mathbf{x} \in \ker \HZ  / {\rm im}~\HX  \) and a dissipator.
This decomposition is an analogue of the logical Pauli group for stabilizer codes.

For example, the kernel of \(\HZ  = \begin{pmatrix} 1 &-1 \end{pmatrix}\) for the pair-cat code consists of all integer multiples of \( \begin{pmatrix} 1& 1 \end{pmatrix}\).
Vectors \( \begin{pmatrix} p & 0 \end{pmatrix} \) and \( \begin{pmatrix} 0 & p \end{pmatrix} \) for \(p > 0\) are not in the kernel, meaning that the code detects all single-mode losses, \(\hat{a}_1^p\) and \(\hat{a}_2^p\).
The operator \(\hat{a}_1^2 \hat{a}_2^2\) is in the kernel, but it is the code's sole dissipator since its corresponding vector is equal to \(\HX  = \begin{pmatrix} 2 & 2\end{pmatrix} \).
The operator \(\hat{a}_1 \hat{a}_2\) is in the kernel of \(\HZ \) but not in the image of \(\HX \), making it a logical operator.
Any odd power of \(\hat{a}_1 \hat{a}_2\) has the same logical action and can be expressed as a product of \(\hat{a}_1 \hat{a}_2\) and a power of the code's dissipator.

\subsection{Loss error correction \& \texorpdfstring{\(X\)}{}-distance}\label{subsec:loss-correction}

Per the Knill-Laflamme conditions \cite{kl}, correcting two nontrivial loss errors \(\mathbf{\hat{a}}^{\mathbf p}\) and \(\mathbf{\hat{a}}^{\mathbf q}\) requires the ability to detect their product \(\mathbf{\hat{a}}^{\dagger\mathbf{q}}\mathbf{\hat{a}}^{\mathbf{p}}\).
It is useful to treat the adjoint portion as a negative vector and associate the product with the error vector \(\mathbf p - \mathbf q\).

The case \(\mathbf p = \mathbf q\) corresponds to a dephasing error
\begin{eqs}
    \mathbf{\hat{a}}^{\dagger\mathbf{p}}\mathbf{\hat{a}}^{\mathbf{p}} \quad\quad\quad \text{(dephasing errors),}
\end{eqs}
which is undetectable since \(\mathbf p - \mathbf p = \mathbf 0\) is in the kernel of \(\HZ \) for any \(\mathbf p\).
In other words, dephasing errors commute with all \textcolor{black}{$Z$-type stabilizers} \(\bfh \cdot \hat{\mathbf n}\) since they can be expressed as functions of the occupation number operators.
These errors cannot induce logical $X$-errors since they are diagonal in Fock space.
They may induce \(Z\)-errors, which we tackle in the next subsection.

Error products for which \(\mathbf 0 \neq \mathbf p \neq \mathbf q \neq \mathbf 0\) can be organized into two classes, extending the two parts of Eq.~\eqref{eq:detectable},
\begin{align}
\mathbf{p}-\mathbf{q}\notin\ker \HZ \quad&\,\Leftrightarrow\quad\mathbf{\hat{a}}^{\dagger\mathbf{q}}\mathbf{\hat{a}}^{\mathbf{p}}\text{ detectable},\\
\mathbf{p}-\mathbf{q}\in\ker \HZ ~\backslash~{\rm im}~\HX \quad&\Leftrightarrow\quad\mathbf{\hat{a}}^{\dagger\mathbf{q}}\mathbf{\hat{a}}^{\mathbf{p}}\text{~\textcolor{black}{ undetectable}}.\label{eq:undetectable_type2}
\end{align}
Similar to the case of loss errors, a sufficient condition for detectability is for the error vector overlap with at least one dissipator vector \(\bfh \in \text{row} (\HZ) \), thereby mapping the codespace to an error space with a different syndrome \(\boldsymbol \Delta\). Similar to Eq.~\eqref{eq:loss_KL}, we can show that KL condition  holds true for error $\mathbf{\hat{a}}^{\dagger \mathbf{q}} \mathbf{\hat{a}}^{\mathbf{p}}$
\begin{eqs}
    P \mathbf{\hat{a}}^{\dagger \mathbf{q}} \mathbf{\hat{a}}^{\mathbf{p}} P\propto    \Pi_{\mathbf{\Delta}}  \Pi_{\mathbf{\Delta}-H(\mathbf{p}-\mathbf{q})} \mathbf{\hat{a}}^{\dagger \mathbf{q}} \mathbf{\hat{a}}^{\mathbf{p}} \propto \delta_{\mathbf{p}-\mathbf{q}\in\text{ker} H} ~.
\end{eqs}

The second class corresponds to undetectable errors that yield logical errors when projected into the codespace.
Any errors not falling into these classes correspond to operators that can be expressed as a product of dissipators (or their adjoints) and a dephasing error.

For the pair-cat code, the error \(\hat{n}_1 \hat{a}_2\) with \(\mathbf p = \begin{pmatrix} 1 & 1 \end{pmatrix} \) and \(\mathbf q = \begin{pmatrix} 1 & 0\end{pmatrix}\) is detectable since its error vector, \(\mathbf p - \mathbf q = \begin{pmatrix} 0 & 1 \end{pmatrix}\), is not in the kernel of \(\HZ  =  \begin{pmatrix} 1 &-1 \end{pmatrix}\).
This error maps the code space to the same Fock-space sector as \(\hat a_2\) since the dephasing portion of the error does not affect the syndrome.
On the other hand, \(\mathbf p - \mathbf q = \begin{pmatrix} 1 &1 \end{pmatrix}\) with \(\mathbf p = \begin{pmatrix} 2 & 2\end{pmatrix}\) and \(\mathbf q = \begin{pmatrix} 1 &1 \end{pmatrix}\), corresponding to \(\hat{a}_1^{\dagger}\hat{a}_2^{\dagger} \hat{a}_1^2\hat{a}_2^2 = \hat n_1 \hat n_2 \hat a_1 \hat a_2\), is an \textcolor{black}{undetectable} error since it is a product of a dephasing error and a logical operation.

We use the total number of losses occurring during a loss error \(\mathbf{\hat{a}}^{\mathbf p}\) --- the monomial degree or 1-norm \(|\mathbf p|\) --- to define an \(X\)-distance.
Correcting up to \(t\) losses (\(|\mathbf p| \leq t\)) requires detecting operator products \(\mathbf{\hat{a}}^{\dagger\mathbf{q}}\mathbf{\hat{a}}^{\mathbf{p}}\) with twice the total degree, \(|\mathbf p|+|\mathbf q| \leq 2t\).
Since dephasing errors do not cause \textcolor{black}{logical $X$ action},
we exclude them from this distance analysis.
This reduces the set of loss-gain products required for correction to those with \(\mathbf p - \mathbf q \neq \mathbf 0\) and allows us to substitute the 1-norm of the error vector, \(|\mathbf p - \mathbf q|=\sum_{j=1}^N |p_j - q_j|\), for the sum of 1-norms of \(\mathbf p\) and \(\mathbf q\).
The upper bound of \(2t\) remains unchanged since this set includes  "off-set" errors such as \(\hat a_j^{\dagger t} \hat{a}_k^t\) for \(j\neq k\), for which \(|\mathbf p - \mathbf q| = |\mathbf p| + |\mathbf q| = 2t\).

The \(X\)-distance \textcolor{black}{is determined by} the 1-norm of the \textcolor{black}{undetectable} $X$-error \textcolor{black}{with lowest weight},
\begin{eqs}\label{eq:x-distance}
d_{X}=\min_{\mathbf{p},\mathbf{q}}|\mathbf{p}-\mathbf{q}|,\quad\forall~\mathbf{0}\neq\mathbf{p}-\mathbf{q}\in\text{ker} \HZ ~\backslash~\text{im} \HX ~.
\end{eqs}
This ensures that any products with lower 1-norm are combinations of detectable errors,  dissipators, and dephasing errors.
A tiger code with such a distance detects up to \(d_X-1\) losses, i.e., all errors \(\mathbf{\hat a}^{\mathbf p}\) with \(|\mathbf p| < d_X\). 
If \(2t < d_X\) for some \(t\) and if dephasing errors are suppressed in the large-\(\alpha\) limit, then a code can correct \(t\) losses on any mode.
Neither condition is necessary, with codes like the pair-cat code detecting arbitrary single-mode losses at distance two, and with a four-mode code which admits no dephasing errors for \textit{any} \(\alpha\).

Each loss vector \(\mathbf p\) is also characterized by the number of modes which lost at least one particle --- the (Hamming) weight or support, \(\text{wt}~\mathbf p\).
The 1-norm upper-bounds the weight, \(\text{wt}~\mathbf p \leq |\mathbf p|\), with equality occurring when every mode lost at most one particle.
The quantity \(d_X - 1\) provides a tight upper bound on the number of modes that can undergo loss errors while preserving the logical information.
This bound happens to be an equality for all of the examples we study.
It can be shown as follows.
The logical operator \(\mathbf{\hat{a}}^{\dagger\mathbf{q}}\mathbf{\hat{a}}^{\mathbf{p}}\) with \(|\mathbf p - \mathbf q| = d_X\) necessarily has \(\mathbf p\) and \(\mathbf q\) of disjoint support; otherwise, it would contain a dephasing error that could be removed to decrease the distance.
A code that admits such a logical cannot correct both \(\mathbf{\hat{a}}^{\mathbf{p}}\) and \(\mathbf{\hat{a}}^{\mathbf{q}}\). 
Assuming, without loss of generality, that \(\text{wt}~\mathbf{p} \geq \text{wt}~\mathbf{q}\), the code can only correct loss errors acting on at most \(\text{wt}~\mathbf{p}-1\) modes.
Using the claim of disjoint support together with the 1-norm upper bound yields
$\text{wt}~\mathbf{p}\leq\text{wt}(\mathbf{p}-\mathbf{q})\leq|\mathbf{p}-\mathbf{q}|=d_{X}$.

The pair-cat code can detect any single-mode errors, corresponding to vectors of the form \(\mathbf p = \begin{pmatrix} p & 0 \end{pmatrix}\) \textcolor{black}{or $\begin{pmatrix}
    0& p
\end{pmatrix}$ for any $p\in\mathbb{N}$,} whose 1-norm \(p\) \textcolor{black}{can be} arbitrarily high.
But since logical operators such as \(\hat a_1 \hat a_2\) have a 1-norm of two, \(d_X = 2\).
Such a distance implies that this code \textcolor{black}{approximately} corrects a single\textcolor{black}{-mode} loss in either mode in the limit of large \(\alpha\).

\subsection{Dephasing errors \& \texorpdfstring{\(Z\)}{}-distance}
\label{sec:dephasingZdist}
Dephasing errors \(\mathbf{\hat{a}}^{\dagger \mathbf p}\mathbf{\hat{a}}^{\mathbf p}\) cause only logical \(Z\)-errors since they are diagonal in Fock space.
In other words, their expectation values with respect to \(X\)-type codewords are all equal, and such errors manifest as \textit{off-diagonal} matrix elements between the codewords.
We conjecture that these matrix elements are suppressed exponentially in the energy density \(\alpha^2\).
We further conjecture that the coefficient multiplying this density can be lower bounded by the squared Euclidean distance between the coherent-state constellations present in each codeword, which we define to be the \(Z\)-distance of the code.
We verify these behaviors for all qudit tiger codes satisfying a technical constraint, namely, that \(\boldsymbol 1 \in \ker \HZ \).

\prg{Logical qudit}

We consider the dephasing matrix element between qudit codewords~\eqref{eq:codeword_formula}, \(|\overline\mu\rangle\), \(\ket{\overline{\nu}}\) with \(\mu,\nu\in\frac{2\pi}{K}\mathbb{Z}_K\). 
Plugging Eq.~\eqref{eq:projector-permutation} into the projected coherent-state overlap~\eqref{eq:gkz-overlap},
the matrix element reduces to a ratio of GKZ functions with different arguments,
\begin{subequations}\label{eq:dephasing-calc}
\begin{align}
    \langle\overline{\nu}|\mathbf{\hat{a}}^{\dagger\mathbf{p}}\mathbf{\hat{a}}^{\mathbf{p}}|\overline{\mu}\rangle&=\frac{e^{N\alpha^{2}}\langle\alpha e^{i\nu\mathbf{z}}|{\Pi}_{\boldsymbol{\Delta}}\mathbf{\hat{a}}^{\dagger\mathbf{p}}\mathbf{\hat{a}}^{\mathbf{p}}{\Pi}_{\boldsymbol{\Delta}}|\alpha e^{i\mu\mathbf{z}}\rangle}{\mathsf{A}_{\boldsymbol{\Delta}}(\alpha^{2}\boldsymbol{1})} ~,\\
    &=\alpha^{2|\mathbf{p}|}\frac{\mathsf{A}_{\boldsymbol{\Delta}-\HZ\mathbf{p}}(\alpha^{2}e^{i(\mu-\nu)\mathbf{z}})}{\mathsf{A}_{\boldsymbol{\Delta}}(\alpha^{2}\boldsymbol{1})}\,.\label{eq:dephasing_overlap_b}
\end{align}
\end{subequations}
The case \(\mathbf p = \mathbf 0\) corresponds to the overlap between the non-orthogonal states, i.e., an intrinsic memory error.
This element needs to be zero for all \(\mathbf p\) in order for the code to exactly protect against dephasing.
We show that, in the worst case, it goes to zero quickly in the large-\(\alpha\) limit for a large class of codes, and conjecture that this suppression holds for all codes.

Since each tiger codeword is a continuous superposition of a coherent-state constellation, 
this overlap is a “sum” over all possible overlaps between points taken from the two constellations. 
The integral form~\eqref{eq:GKZintegral} of a GKZ function can be easily upper-bounded by the overlap of the two closest points,
\begin{eqs}
  |\mathsf{A}_{\boldsymbol{\Delta}}(\alpha^{2}e^{i\mu\mathbf{z}})|\leq e^{N\alpha^{2}}\int\frac{d^{r_{z}}\boldsymbol{\phi}}{(2\pi)^{r_{z}}}e^{-\frac{1}{2}\alpha^{2}\left\|\boldsymbol{1}-e^{i(\boldsymbol{\phi}\HZ +\mu\mathbf{z})}\right\|^{2}} .
\end{eqs}

The above integrand is the absolute value of the overlap~\eqref{eq:euclidean} between two ordinary coherent states,
the all-ones coherent state and another state whose value depends on \(\boldsymbol{\phi}\). 
It is maximized at the angle vector for which the two coherent states are the closest, i.e., \(\boldsymbol{\phi}_{\star}=\argmin_{\boldsymbol{\phi}}\|\boldsymbol{1}-e^{i(\boldsymbol{\phi}\HZ +\mu\mathbf{z})}\|^{2}\).
Substituting this optimal value yields a bound for the numerator in Eq.~\eqref{eq:dephasing_overlap_b}.

The special case of the above bound with \(\mathbf{z}=\mathbf{0}\) and \(\boldsymbol{\phi}_{\star}=\mathbf{0}\) applies to the denominator in Eq.~\eqref{eq:dephasing_overlap_b}, yielding an upper bound of \(\exp({N\alpha^{2}})\).
We cannot use this to upper-bound the dephasing matrix element since we would need a lower bound for the denominator. 
However, a careful asymptotic analysis of the denominator in the large-\(\alpha\) limit (see App.~\ref{sec:GKZ}) yields an exact expression with the same exponential dependence, provided that \(\boldsymbol 1 \in\ker \HZ \). 
For such cases, we obtain
\begin{eqs}
    \!\! |\langle\overline{\nu}|\mathbf{\hat{a}}^{\dagger\mathbf{p}}\mathbf{\hat{a}}^{\mathbf{p}}|\overline{\mu}\rangle|^2 \overset{\alpha\to\infty}{\lesssim} \text{poly}\ensuremath{(\alpha)}e^{-\alpha^{2}\left\|\boldsymbol{1}-e^{i(\boldsymbol{\phi}_{\star}\HZ +(\mu-\nu)\mathbf{z})}\right\|^{2}},
\end{eqs}
which omits polynomial pre-factors to highlight the exponential suppression.

The above bound shows that all dephasing errors are suppressed at large \(\alpha\), with the \textit{strength} of the suppression depending on the minimal Euclidean distance between codeword constellations.
This motivates a Euclidean definition of the \(Z\)-distance.

Returning to the case of a logical qudit, we have to consider dephasing matrix elements between any two codewords labeled by \(\nu\neq\mu\in \frac{2\pi}{K}\mathbb{Z}_K\).
Global rotations allow us to reduce this to \(K\) matrix elements between \(|\overline \mu\rangle\) and the fiducial state \(|\overline 0\rangle\), with each element bounded by the minimal square Euclidean distance between a pair of constellations,
\begin{subequations}
\begin{align}
   d_{Z}(\mu)&=\min_{\boldsymbol{\phi}\in\mathbb{T}^{r_{z}}}\left\Vert \mathbf{1}-e^{i\left(\boldsymbol{\phi}\HZ+\mu\mathbf{z}\right)}\right\Vert ^{2},\\&=4\min_{\boldsymbol{\phi}\in\mathbb{T}^{r_{z}}}\sum_{j=1}^{N}\sin^{2}{\textstyle \frac{1}{2}}\left(\boldsymbol{\phi}\HZ+\mu\mathbf{z}\right)_{j}\,,\label{eq:z-distance-sines}
\end{align}
\end{subequations}
where the squared Euclidean distance between two complex vectors $\mathbf{u}, \mathbf{v}$ is defined as $\| \mathbf{u}-\mathbf{v}\|^2={\sum_{j=1}^N (u_j^*-v_j^*) (u_j-v_j)}$.
Being a measure of closeness in the configuration space of coherent states, this distance also quantifies the "shortest" logical rotation that one can apply to map one \(X\)-type codeword to another.
The code's $Z$-distance is then the square of minimum Euclidean distance among all paths connecting pairs of distinct $X$-type codewords,
\begin{eqs}\label{eq:z-distance}
    d_Z = \min_{\mu \in \frac{2\pi}{K}\mathbb{Z}_K} d_Z (\mu) .
\end{eqs}
\textcolor{black}{We illustrate the $Z$-distance of the pair-cat code in Fig.~\ref{fig:spherical_vs_tiger}(a), which is the squared Euclidean distance between the black and white lines.}

Plugging this distance into a generic dephasing matrix element for codes satisfying \(\boldsymbol 1 \in \ker \HZ \) and squaring the matrix element to rid ourselves of a factor of one-half yields
\begin{eqs}\label{eq:dephasing-bound}    |\langle\overline{\nu}|\mathbf{\hat{a}}^{\dagger\mathbf{p}}\mathbf{\hat{a}}^{\mathbf{p}}|\overline{\mu}\rangle|^2\overset{\alpha\to\infty}{\lesssim}\text{poly}(\alpha)\exp\left(-d_{Z}\alpha^{2}\right) .
\end{eqs}
For example, in the case of the two-component cat code with codewords \(\left|\pm\alpha\right\rangle\), the distance is $4$ --- the squared Euclidean distance between the two coherent-state values at \(\alpha = 1\).

The code distance yields a tight bound on the dephasing suppression coefficient for all but one of our examples --- the \textcolor{black}{liger} surface code --- where a careful asymptotic analysis of the dephasing matrix elements yields a coefficient that is strictly \textit{larger} than the squared Euclidean distance.
The exact suppression coefficient may be dependent on \(\HZ \), number of modes \(N\), and syndrome vector \(\boldsymbol{\Delta}\).
As such, we conjecture that the code distance serves as a lower bound on the dephasing suppression, with some codes exhibiting greater suppression.
Proving this conjecture requires a derivation of the above asymptotic inequality for codes with \(\mathbf 1 \notin \ker \HZ \).

\prg{Infinite-dimensional logical subsystem}

Infinite dimensional logical encodings are parameterized by a continuous angle parameter \(\varphi\).
There is now a continuum of continuous constellations, one for each value of \(\varphi\).
The analogous rotor-code distance~\cite[Eq.~67]{vuillotHomologicalQuantumRotor2024} was defined as a minimization of a relative distance --- the squared Euclidean distance of the code divided by the squared Euclidean distance of the "bare" single-mode implementation \(e^{i\varphi}\) of the logical rotation.
Here, we keep the Euclidean distance \(\varphi\)-dependent to make contact with the coherent-state overlaps discussed above:
\begin{eqs}\label{eq:z-distance-continuous}
d_{Z}(\varphi)=\min_{\boldsymbol{\phi}\in \mathbb{T}^{r_{z}}}\left\|\mathbf{1}-e^{i\left(\boldsymbol{\phi}\HZ +\varphi\mathbf{z}\right)}\right\|^{2} .
\end{eqs}
We leave the question of relating the two distances a question for future work.

\section{Examples of tiger codes}
\label{sec:examples}

In this section, we provide examples of new tiger codes and recast known codes in our framework. 
All codes here have infinite Fock-state support, but can encode logical qubits, qudits, rotors, or modes.

\subsection{Pair-cat code}
\label{sec:pair_cat}

The pair-cat code \cite{albert2019pair} is a two-mode bosonic code that protects against arbitrary single-mode photon losses.
Having used it as an example throughout the previous section, here we provide only a brief overview.

Its generator matrices are \(\HX =\begin{pmatrix}
2& 2    
\end{pmatrix}\) and \(\HZ =\begin{pmatrix}
 1 & -1   
\end{pmatrix}\),
We can verify that these follow the CSS condition,
\(
    \HX  \HZ ^\T=0
\).
They correspond to one dissipator and one \textcolor{black}{$Z$-type stabilizer}, respectively, 
\begin{eqs}
\hat{a}_{1}^{2}\hat{a}_{2}^{2}-\alpha^{4}\quad\quad\text{and}\quad\quad\hat{n}_{2}-\hat{n}_{1}-\Delta\,,
\end{eqs}
where \(\alpha^2\) is the energy density, and where the integer syndrome \(\Delta\) quantifies the difference between the occupation numbers in the two modes.

A simple homology calculation, done in Sec.~\ref{sec:homology}, shows that the code encodes a qubit with 
logical generators \(\mathbf{x} = \begin{pmatrix} 1 & 1 \end{pmatrix} \) and \(\mathbf{z} = \begin{pmatrix} 1 & 0 \end{pmatrix}\).
These yield logical operator representatives $\overline{X}=\hat{a}_1\hat{a}_2/\alpha^2$ and $\overline{Z}=\exp({i \pi \hat{n}_1})$, respectively.

The \(X\)-type codewords are the coherent states \(|\alpha,\alpha\rangle\) and \(\overline{Z}|\alpha,\alpha\rangle=\left|-\alpha,\alpha\right\rangle\), each projected into a fixed-\(\Delta\) subspace via the projection $\Pi_\Delta=\frac{1}{2\pi}\int d\phi e^{i\phi(\hat{n}_2-\hat{n}_1-\Delta)}$.
Explicitly, for \(\Delta > 0 \),
\begin{eqs}
    \ket{\overline{\pm}}={\frac{1}{\sqrt{I_{\Delta}(2\alpha^{2})}}}\sum_{n\geq0}\frac{(\pm 1)^{n}\alpha^{2n+\Delta}}{\sqrt{n!(n+\Delta)!}}\ket{n,n+\Delta} ,
\end{eqs}
where we have divided out constants upon normalization.
The negative \(\Delta\) case is related to the above by a swap of the modes.
These are instances of pair-coherent \cite{barut1971new,agarwal1986generation,agarwal1988nonclassical,gerry1995nonclassical} or fixed-charge \cite{klauder1985coherent,bhaumik1976charged} states, with the occupation-number difference referred to as the "charge" in this context.

The GKZ function associated with this code is bivariate since the code is on two modes, but the two arguments collapse into one after simplifying,
\begin{eqs}
    \mathsf{A}_{\Delta}(y_{1},y_{2})=\sum_{n\geq0}\frac{y_{1}^{n}y_{2}^{n+\Delta}}{n!(n+\Delta)!}=\Big(\frac{y_2}{y_1}\Big)^{\Delta/2}I_{\Delta}(2\sqrt{y_{1}y_{2}}) ,
\end{eqs}
where \(I\) is the modified Bessel function of the first kind \cite{DLMF}.

Using the definition from Eq.~\eqref{eq:z-distance}, we obtain 
\begin{eqs}
    d_{Z}=\min_{\phi}\Big\|\begin{pmatrix}1 & 1\end{pmatrix}-\begin{pmatrix}-e^{i\phi} &e^{-i\phi} \end{pmatrix}\Big\|^{2}=4 .
\end{eqs}
This matches the exponent of an asymptotic calculation of the dephasing suppression coefficient \cite[Table II]{albert2019pair}, making the bound \eqref{eq:dephasing-bound} tight for this case.

We derived the \(X\)-distance of this code, \(d_X = 2\), in the previous section, indicating that the code can detect arbitrary single-photon losses.
Furthermore, in the large-\(\alpha\) limit, the pair-cat code can approximately correct single-photon losses $\{\hat{a}_1, \hat{a}_2\}$ in each mode.
However, since \(\hat{a}_1^{\dagger} \hat{a}_2\) is not a logical operator, and since all dephasing errors are exponentially suppressed, the code can approximately correct \textit{arbitrary} photon losses in each mode $\{\hat{a}_1^p, \hat{a}_2^p\}_{p>0}$ in the large-$\alpha$ limit. Therefore, the pair-cat code is an approximate error-correcting code and an exact error-detecting code against single-mode losses.

The pair-cat code can be generalized to \(\HZ = \begin{pmatrix}
m_1 & -m_2
\end{pmatrix}\) for any coprime positive integers \(m_1 \neq m_2\).
The kernel of \(\HZ\) is spanned by \(L_X=\begin{pmatrix}
m_2 & m_1    
\end{pmatrix}\), yielding a distance of \(d_X = m_1 + m_2\).
Such a code also detects up to \(m_1+m_2-1\) losses on any mode, in addition to detecting all single-mode losses.

\subsection{Extended pair-cat code}\label{sec:extended-pair-cat}

An \(N\)-mode extension of pair-cat code was shown to detect arbitrary losses in $N-1$ modes~\cite{albert2019pair,albert2018lindbladians} (cf. Ref.~\cite{an2003even} for the \(N=3\) case).
This code is also a tiger code, which we call the extended pair-cat code, with generator matrices
\begin{eqs}
   \HX=\begin{pmatrix}2 & 2 & \cdots & 2\end{pmatrix},\HZ=\begin{pmatrix}1 & -1 & 0 & \cdots & 0 & 0\\
0 & 1 & -1 & \cdots & 0 & 0\\
\vdots & \vdots & \ddots & \ddots & \vdots & \vdots\\
0 & 0 & 0 & \vdots & 1 & -1
\end{pmatrix}=A_{N}~,
\end{eqs}
with the latter matrix generating the \(A_N\) root lattice \cite{conway2013sphere}.
These correspond to the dissipator \(\mathbf{\hat{a}}^{\mathbf 2} - \alpha^{2N}\), where \(\mathbf 2 = \begin{pmatrix}
    2& 2& \cdots &2\end{pmatrix}\), and \(N-1\) pairs of neighboring occupation-number differences, \(\hat n_{j+1}-\hat{n}_{j}-\Delta_{j}\), for its syndromes.

Logical operators are given by
\begin{eqs}
    \overline{X}=(\mathbf{\hat{a}}/\alpha)^{\mathbf 1}\quad\quad\text{and}\quad\quad\overline{Z}=(-1)^{\hat{n}_1} .
\end{eqs}

Logical \(X\) codewords are coherent states projected into the Fock subspace of all occupation-number differences fixed \textcolor{black}{by $\mathbf{\Delta}$}.
We call these \textit{extended pair-coherent states}. For \(\boldsymbol{\Delta}=\mathbf 0\), these are expressed simply as
\begin{eqs}
|\overline{\pm}\rangle=\frac{(\pm 1)^{\hat{n}_1}}{\sqrt{\mathsf{A}_{\mathbf{0}}(\alpha^{2}\mathbf{1})}}\sum_{n\geq0}\left(\frac{\alpha^{n}}{\sqrt{n!}}|n\rangle\right)^{\otimes N} .
\end{eqs}
The GKZ normalization for this case collapses all \(N\) arguments into a product: 
\begin{equation}
\mathsf{A}_{\mathbf 0}(\mathbf{y})=\prescript{}{0}F_{N-1}\left(;\bs{1};y_1y_2\cdots y_N\right)\,,
\end{equation}
where \(\bs{1}\) is the all-ones vector of length \(N-1\), and where \(\prescript{}{p}{F}_q\) is the generalized hypergeometric function \cite{DLMF}.

This code has an \(X\)-distance \(d_X = N\), but has the distinct feature of being able to \textit{approximately correct} (and not just detect) losses on \(N-1\) modes in the large-\(\alpha\) limit.
This is due to the structure of the kernel of \(\HZ \), generated by the logical operator's vector \(\mathbf 1\).
Since vectors \(\mathbf p - \mathbf q\), with \(\mathbf p\) and \(\mathbf q\) supported on up to \(N-1\) modes, are outside of the kernel, their corresponding operators \(\mathbf{\hat{a}}^{\dagger\mathbf{q}}\mathbf{\hat{a}}^{\mathbf{p}}\) are detectable.
Combined with the fact that dephasing errors are suppressed with large \(\alpha\) yields correction of any losses up to one less the code distance.

The large \(X\)-distance of \(N\) comes at the price of a vanishing \(Z\)-distance, $d_Z=4N\sin^2(\frac{\pi}{2N})$.
This is because an alternative logical operator, \(\exp(i\sum_j \pi \hat n_j / N)\), related to the operator \(\exp (i\pi \hat n_1)\) by a \(Z\)-type stabilizer, rotates between codewords by applying a vanishingly small rotation on each mode.

We confirm that the dephasing suppression coefficient of the code is given by the code distance.
Since the estimate from Eq.~\eqref{eq:dephasing-calc} is independent of the power of the dephasing error, we can extract the suppression coefficient from the codeword overlap,
\begin{equation}
 \braket{\overline{-}\vert\overline{+}}=\frac{\mathsf{A}_{\mathbf 0}\left(\alpha^{2}(-1,1,\cdots,1)\right)}{\mathsf{A}_{\mathbf 0}\left(\alpha^{2}\mathbf{1}\right)}=\frac{\prescript{}{0}F_{N-1}\left(;\bs{1};-{\alpha}^{2N}\right)}{\prescript{}{0}F_{N-1}\left(;\bs{1};{\alpha}^{2N}\right)}~.
\end{equation}
Using asymptotic properties of generalized hypergeometric functions \cite{DLMF}, we can determine the coefficient in the large-\(\alpha\) limit,
\begin{eqs}
\left|\frac{\prescript{}{0}F_{N-1}\left(;\bs{1};-{\alpha}^{2N}\right)}{\prescript{}{0}F_{N-1}\left(;\bs{1};{\alpha}^{2N}\right)}\right|^{2}\sim\text{poly}(\alpha)\exp\left(-4N\alpha^{2}\sin^{2}\frac{\pi}{2N}\right)\,.
\end{eqs}

\subsection{Pair-coherent states as tiger codes}\label{sec:pair_coherent}

Two mutually dual variants of the pair-cat code are simple examples of tiger codes with two types of infinite-dimensional logical spaces --- a logical rotor and a logical mode.
These are special cases of the coherent-state repetition code and the Fock-state repetition code, respectively.

\prg{Logical rotor}
Let us pick the generator matrices
\begin{eqs}
    \HX =\begin{pmatrix}
        1 & 1
    \end{pmatrix}\quad\quad\text{and}\quad\quad \HZ =0 ,
\end{eqs}
which correspond to a single dissipator
$\hat{a}_1\hat{a}_2-\alpha^2$.

A $Z$-type codeword for this code consists of pair-coherent states with fixed argument \(\alpha^2\).
Such states are defined for any occupation-number difference, and there is exactly one state in each sector of fixed difference,
\begin{eqs}
\ket{\overline{\ell}}=\frac{1}{\sqrt{I_{\ell}(\alpha)}}\sum_{n\geq0}\frac{\alpha^{2n+\ell}}{\sqrt{n!(n+\ell)!}}\ket{n,n+\ell} ,
\end{eqs}
for positive difference \(\ell\), with the negative-\(\ell\) case related to the above by a swap of the modes.
Since \textcolor{black}{$Z$-type codewords with different $\ell$} lie in separate Fock-state subspaces, they form an orthonormal basis.
Such a codespace is naturally associated with the angular momentum space of (planar) rotor, a.k.a. a particle on a circle.

The \(X\)-type codewords, parameterized by an angle \(\varphi\), can be defined by applying the logical operator \(\overline Z(\varphi) = \exp(i \varphi (\hat{n}_1-\hat{n}_2))\) to the all-ones coherent state,
\begin{eqs}
    |\overline{\varphi}\rangle=\overline{Z}(\varphi)\ket{\boldsymbol{\alpha}=(1,1)}=\ket{\boldsymbol{\alpha}=(e^{i\varphi},e^{-i \varphi})} .
\end{eqs}

The \(X\)-distance \(d_X = 1\) since single\textcolor{black}{-photon} loss vectors are in the kernel of \(\HZ \), which is everything, but not in the image of \(\HX \).

Using the definition from Eq.~\eqref{eq:z-distance-continuous}, we obtain 
\begin{eqs}\label{eq:dz_bare_oscillator}
    d_{Z}= 4\sin^{2}\frac{\varphi}{2}\,,
\end{eqs}
the same as the squared Euclidean distance of the "bare" encoding, \(\|1-e^{i\varphi}\|\).
In other words, the code does no better at protecting against dephasing error then an unencoded mode.

\prg{Logical mode}
The "dual" code is obtained by only imposing the $\HZ $ constraint that generates the kernel of the previous code's \(\HX \) matrix.
The code's generator matrices are
\begin{eqs}
    \HX =0\quad\quad\text{and}\quad\quad \HZ =\begin{pmatrix}
        1 & -1
    \end{pmatrix} ,
\end{eqs}
which correspond to a single \textcolor{black}{$Z$-type stabilizer},
$\hat n_1 - \hat n_2 - \Delta$.
The $X$-type codeword consists of pair-coherent states of arbitrary argument but with fixed difference.

The $Z$-type codeword consists of two-mode fixed-difference Fock states,
\begin{eqs}
    \ket{\overline{\ell}}=\ket{\ell+\Delta,\ell},~~ \forall \ell\geq 0,
\end{eqs}
for positive \(\Delta\), with the negative-\(\Delta\) case related by a swap.
This basis is labeled by non-negative integers, so most naturally corresponds to a logical mode encoding (as opposed to a rotor).
One can also think of this is a rotor whose angular momenta are restricted to be non-negative \cite{rotorclifford}. 

The \(X\)-distance is \(d_X = 2\), since the lowest-degree undetectable $X$-error  is $\hat{a}_1 \hat{a}_2$, which gives the logical $X$ operator.
This code is a special case of the Fock-state repetition code, whose \(Z\)-distance is discussed in the next subsection.

\subsection{Fock-state repetition code}

A close relative of both the logical-mode code from the previous example and the extended pair-cat code is the $N$-mode Fock-state repetition code, with
\begin{eqs}
    \HX =0\quad\quad\text{and}\quad\quad \HZ =A_N~.
\end{eqs}
There are \(N-1\) \textcolor{black}{$Z$-type stabilizers}, \(\hat n_{j+1}-\hat{n}_{j}-\Delta_{j}\) \textcolor{black}{for all $1\leq j \leq N-1$}, the same as those for the extended pair-cat code. The difference is that there is no dissipator, which enlarges the logical space from a qubit to a mode.

The mode's $Z$-type codeword, for \(\boldsymbol{\Delta} = \mathbf 0\), is made up of Fock states \(|\overline{\ell}\rangle^{\otimes N}\).
The mode's logical $X$ operator, \(\overline X = (\mathbf{\hat{a}}/\alpha)^{\mathbf 1}\),
is the same as the logical operator of the extended pair-cat qubit from Sec.~\ref{sec:extended-pair-cat}.

The $X$-type codeword consists of extended pair-coherent states from Sec.~\ref{sec:extended-pair-cat} for arbitrary argument but with all occupation-number differences fixed.
The \(X\)-distance is \(d_X = N\).

As with the extended pair-cat code, the canonical logical $Z$ operator is \(\overline Z(\varphi) = \exp(i \varphi \hat n_1)\), but an alternative logical, \(\exp(i\sum_j \varphi \hat n_j/N)\), applies a rotation that vanishes with \(N\) to each mode.
The \(Z\)-distance from Eq.~\eqref{eq:z-distance-continuous} is 
\begin{eqs}
d_{Z}=4N\sin^{2}\frac{\varphi}{2N}~,
\end{eqs}
being of order \(O(1/N)\) for large \(N\) and any fixed \(\varphi\).

\subsection{Coherent-state repetition code}\label{sec:coherent_repetition}

Dual to the Fock-state repetition code from the previous example is the coherent-state repetition code, whose \(\HX \) matrix is that of a cyclic repetition code,
\begin{eqs}\label{eq:paritycheck_coherentstate_repetition}
    \HX =\begin{pmatrix}
        1 & 1& 0& \cdots &0&0\\
        0 & 1& 1& \cdots &0& 0\\
        \vdots & \vdots & \vdots & \vdots & \vdots& \vdots \\
        1& 0& 0& \cdots & 0& 1
    \end{pmatrix}=R_N \quad\quad\text{and}\quad\quad \HZ =0 .
\end{eqs}
Its $X$-distance is trivial with $d_X=1$ while its $Z$-distance increases as system size increases, \(d_Z = 4N\).

Fixing \(N\) to be odd yields torsion~\eqref{eq:torsion},
namely, a vector \(\bs{e}_1\) which is not in the image of \(\HX \) but whose multiple is.
It is easy to check that the alternating sum of all the rows of \(\HX \) gives \(2\bs{e}_1\),
with \(\bs{e}_1\) itself not in the row-image of \(\HX \):
\begin{subequations}
    \label{eq:a_vector}
\begin{align}
    \bs{f}_N &= \begin{pmatrix}
        1 & -1 & 1 & -1 & \cdots & 1 
    \end{pmatrix}\\
    \bs{e}_1 &=\begin{pmatrix}
        1 & 0 & 0 & \cdots & 0
    \end{pmatrix}\\
    \bs{f}_N\HX  &= 2\bs{e}_1~.
\end{align}
\end{subequations}

The \(X\)-type codewords are
\begin{eqs}
\ket{\overline{\pm}}=\ket{\pm\alpha}^{\otimes N}\,,
\end{eqs}
which can be viewed as a concatenation between the two-component cat code (in the coherent-state basis) and a qubit repetition code \cite{jeong2002efficient, ralph2003quantum}.

A set of logical vectors is \(\mathbf{x}=\bs{e}_1\) and \(\mathbf{z}=\mathbf{1}\),
\begin{eqs}
    \overline{X}=\frac{\hat{a}_{1}}{\alpha}\quad\quad\text{and}\quad\quad\overline{Z}=(-1)^{\sum_{j=1}^{N}\hat{n}_{j}} ,
\end{eqs}
respectively. 
Other logical operators include $\overline{X}=\hat{a}_j/\alpha$ for any mode $j$, and these are related to each other via a dissipator and the condition that any $\hat{a}_i^2 \propto \overline{X}^2$ for all $1 \leq i \leq N$ acts as identity on the codespace.

The strength of dephasing suppression is easily calculated from the codeword overlap,
\begin{eqs}
    |\langle\overline{-}|\overline{+}\rangle|^2=\frac{|\langle\alpha|-\alpha\rangle|^{2N}}{|\langle\alpha|\alpha\rangle|^{2N}}=\exp(-4N \alpha^{2})\,
\end{eqs}
saturating the upper bound in Eq.~\eqref{eq:dephasing-bound} with \(d_Z = 4N\). 
The GKZ function for this code is \(\mathsf{A}(\mathbf{y})=\exp(\sum_{j}y_{j})\).

Other variants of this code arises when  $N$ is even, which only yields a logical rotor despite the closed or open boundary. 
For an odd-length system, one can also obtain a logical rotor by removing the last row of \(\HX \), corresponding to open boundaries on the repetition code.

\subsection{Three-mode tiger code} \label{sec:three-mode}

In this subsection, we introduce a three-mode tiger code inspired by the $U(1)$-covariant three-rotor code proposed in Refs.~\cite{hayden2021error,faist2020continuous} in the context of holographic quantum information. 
The corresponding tiger code encodes a single logical rotor into three physical modes.

The code is defined by the following parity-check matrices:
\begin{eqs}
    G=\begin{pmatrix}
        3& 1& 2
    \end{pmatrix}  \quad\text{and}\quad H=\begin{pmatrix}
        -4 & 6& 3
    \end{pmatrix},
\end{eqs}
which correspond to dissipator $\hat{a}_3^3 \hat{a}_2 \hat{a}_3^2-\alpha^6$ and \textcolor{black}{$Z$-type stabilizer} $-4\hat{n}_1+6\hat{n}_2+3\hat{n}_3-\Delta$.

The $Z$-basis of the logical rotor, for $\Delta=0$, is 
\begin{eqs}
    \ket{\overline{\ell}}\propto \sum_{n\geq \ell}\frac{\alpha^{6n-\ell}}{\sqrt{(3n)! (n+\ell)! (2n-2\ell)! }} \ket{3n, n+\ell, 2(n-\ell)},
\end{eqs}
for all $\ell \in \mathbb{Z}$.

The logical operators are specified by the vectors $\mathbf{x}=\begin{pmatrix}
    3 & 2& 0
\end{pmatrix}$ and $\mathbf{z}= \begin{pmatrix}
    1 & -1& -1
\end{pmatrix}$,
corresponding to logical operators 
\begin{subequations}
\begin{align}
    \overline{X}&= \hat{a}_{1}^{3}\hat{a}_{2}^{2}/\alpha^{5},\\
    \overline{Z}(\varphi)&=e^{i\varphi(\hat{n}_{1}-\hat{n}_{2}-\hat{n}_{3})} .
\end{align} 
\end{subequations}
The $X$-distance of the code as $d_X = 3$ because $\begin{pmatrix}
     0& 1 & -2
\end{pmatrix} \in \text{ker} ~H$ that gives an lowest-degree undetectable $X$-error $\hat{a}^{\dagger}_2 \hat{a}^2_3$. 

To compute the $Z$-distance, we examine the squared Euclidean distance between logical codewords via Eq.~\eqref{eq:z-distance-continuous}. When $\varphi$ is sufficiently small, the distance is 
\begin{eqs}
    d_Z=4 \times 3\sin^2\frac{\varphi}{2}=12 \sin^2\frac{\varphi}{2} ~.
\end{eqs}

Compared to a bare rotor or the logical rotor encoded in pair-coherent states, which have $X$- and $Z$-distances of 1 and $4\sin^2\frac{\varphi}{2}$ respectively, the three-mode tiger code achieves a threefold improvement in both $X$- and $Z$-distances.


\subsection{Four-mode tiger code}\label{sec:four-mode}

Emulating a rotor code whose homology is based on a tesselation of the projective plane \cite{vuillotHomologicalQuantumRotor2024}, we construct a minimal example whose distances, \(d_X =2\) and \(d_Z = 8\), are both \textcolor{black}{twice of} those for the two-component cat code.

This four-mode code is a combination of the logical-rotor pair-coherent-state code from Sec.~\ref{sec:pair_coherent} and the pair-cat code from Sec.~\ref{sec:pair_cat}, and there is no discrete-variable code participating in this construction.
We first define two logical rotors, with each rotor encoded into two modes, and then build a pair-cat qubit on top of the two rotors. 
It can be regarded as a tiger-code (i.e., integer-homology) version of $[[4,1,2]]$ qubit stabilizer code \cite{leung1997approximate}. 
This code exhibits a new feature not previously observed in coherent-state codes --- dephasing errors stemming from detectable losses are detected \textit{exactly} without the need for the large-\(\alpha\) limit.

The code's generator matrices are
\begin{eqs}\label{eq:fourmode_paritycheck}
    \HX =\begin{pmatrix}
        1 & 1& 0& 0\\
        0& 0& 1& 1\\
        0& 2& 0 &2
    \end{pmatrix}\quad\text{and}\quad \HZ =\begin{pmatrix}
        1& -1& -1& 1
    \end{pmatrix} .
\end{eqs}
The first two rows of \(\HX \) correspond to two copies of the logical-rotor pair-coherent-state encoding, while the last row is the \(X\)-type pair-cat constraint.
These yield a set of three dissipators and one \textcolor{black}{$Z$-type stabilizer},
\begin{eqs}\label{eq:current_mirror_stb}
\begin{array}{c}
\hat{a}_{1}\hat{a}_{2}-\alpha^{2}~,\\
\hat{a}_{3}\hat{a}_{4}-\alpha^{2}~,\\
\hat{a}_{2}^{2}\hat{a}_{4}^{2}-\alpha^{4}~,
\end{array}\quad\quad\text{and}\quad\quad\hat{n}_{1}-\hat{n}_{2}-\hat{n}_{3}+\hat{n}_{4}-\Delta\,.
\end{eqs}

A homology calculation yields logical operator vectors \(\mathbf{x}=\begin{pmatrix}0 & 1 & 0 & 1\end{pmatrix}\) and \(\mathbf{z}=\begin{pmatrix}0 & 0 & 1 & 1\end{pmatrix}\), corresponding to logical-qubit operators
\begin{eqs}
\overline{X}=\frac{\hat{a}_{2}\hat{a}_{4}}{\alpha^{2}}\quad\quad\text{and}\quad\quad\overline{Z}=(-1)^{\hat{n}_{3}+\hat{n}_{4}} .
    \end{eqs}

For convenience, we consider codewords for the $\Delta=0$ case.
The \(X\)-type codeword, in integral form, consists of orbits of the two coherent states \(\boldsymbol{\alpha}=\alpha \mathbf 1\) and \(\boldsymbol{\alpha}=\alpha (1,1,-1,-1)\) over the group generated by the \textcolor{black}{$Z$-type stabilizer},
\begin{eqs}
\!\!\!\ket{\overline{\pm}}=\frac{e^{2\alpha^{2}}}{\sqrt{\mathsf{A}_{0}(\alpha^{2}\mathbf{1})}}\int\frac{d\phi}{2\pi}\ket{\alpha(e^{i\phi},e^{-i\phi},\pm e^{-i\phi},\pm e^{i\phi})} .
\end{eqs}

The code's GKZ function simplifies to a modified Bessel function of the first kind \cite{DLMF} with non-trivial argument. We will need it for all values of \(\Delta\).
For positive values, we obtain
\begin{subequations}\begin{align}
\!\!\mathsf{A}_{\Delta}(\mathbf{y})&=\sum_{\mathbf{n}\in \mathbb{N}^4}\delta_{n_{1}-n_{2}-n_{3}+n_{4},\Delta}\frac{\mathbf{y}^{\mathbf{n}}}{\mathbf{n}!} ~,\\&=\sum_{j\geq\Delta}\frac{\left(y_{1}+y_{4}\right)^{j}}{j!}\sum_{n_{2},n_{3}\geq0}\delta_{j,n_{2}+n_{3}+\Delta}\frac{y_{2}^{n_{2}}y_{3}^{n_{3}}}{n_{2}!n_{3}!} ~,\\
&=\left(\frac{y_{1}+y_{4}}{y_{2}+y_{3}}\right)^{\Delta/2}I_{\Delta}\left(2\sqrt{\left(y_{1}+y_{4}\right)\left(y_{2}+y_{3}\right)}\right)~ ,
\end{align}
\end{subequations}
with the case of negative \(\Delta\) related by a swap of variables \(y_1 + y_4 \leftrightarrow y_2+y_3\).
Above, we split the Kronecker delta function with a useful identity,
\begin{eqs}\label{eq:splitting-identity}
    \delta_{n_{1}-n_{2}-n_{3}+n_{4},\Delta}=\sum_{j\geq\Delta}\delta_{n_{1}+n_{4},j}\delta_{n_{2}+n_{3}+\Delta,j}~,
\end{eqs}
and apply the binomial theorem twice.

The code detects any losses \(\mathbf{\hat a}^{\mathbf p}\) with \(\mathbf p \notin \ker \HZ \).
The kernel is spanned by the first two rows of \(\HX \) and by \(\begin{pmatrix} 0 & 1 & 0& 1 \end{pmatrix}\). The two rows are dissipators, while the third is a logical-\(X\) operator.
No loss vectors of weight one are in the kernel, so the code exactly detects up to one loss on any mode --- \(d_X = 2\).
However, there is a weight-two operator in the kernel --- \(\hat{a}_1\hat{a}_4^{\dagger}\) with error vector \(\begin{pmatrix} 1 & 0 &0 &-1 \end{pmatrix}\) --- that is not detectable. 
In other words, 
both \(\hat{a}_1\) and \(\hat{a}_4\) errors yield the same $Z$-type syndrome, and we couldn't distinguish and correct either of them by checking the $Z$-syndrome. 
Therefore, this code cannot correct single-photon losses.

Plugging the GKZ function into the dephasing matrix elements~\eqref{eq:dephasing_overlap_b} and using the fact that \(I_{j}(0) = \delta_{j,0}\) yields 
\begin{eqs}
    \langle\overline{-}|\mathbf{\hat{a}}^{\dagger\mathbf{p}}\mathbf{\hat{a}}^{\mathbf{p}}|\overline{+}\rangle&=\alpha^{2|\mathbf{p}|}\frac{\mathsf{A}_{-\HZ \mathbf{p}}\left(\alpha^{2}(1,1,-1,-1)\right)}{\mathsf{A}_{0}(\alpha^{2}\boldsymbol{1})} ~,\\
    &=\delta_{\mathbf{p}\in\ker \HZ }\frac{\alpha^{2|\mathbf{p}|}}{I_{0}(4\alpha^{2})}~,
\end{eqs}
where the Kronecker delta is 1 when \(\mathbf p\) is in the kernel of \(\HZ \), and zero otherwise.
The matrix element is identically zero for dephasing errors associated with any detectable losses,
meaning that no large-\(\alpha\) limit is needed to detect them!
A similar feature is also present in a thin surface code --- the liger surface code --- described in the next section.

Intrinsic dephasing due to non-orthogonality (\(\mathbf p = \mathbf 0\)) and dephasing associated with a logical operator (e.g., \(\mathbf p = \mathbf{z}\)) remain finite, but are suppressed in said limit with exponent \(4\alpha^2\) since \(I_0(4\alpha^2)\sim \exp(-4\alpha^2)\). The dephasing suppression for arbitrary dephasing is 
\begin{eqs}
    |\langle\overline{-}|\mathbf{\hat{a}}^{\dagger\mathbf{p}}\mathbf{\hat{a}}^{\mathbf{p}}|\overline{+}\rangle|^{2}\sim\delta_{\mathbf{p}\in\ker \HZ } \cdot \alpha^{4|\mathbf{p}|}\exp(-8\alpha^{2}),
\end{eqs}
saturating the bound in Eqs.~\eqref{eq:dephasing-bound} with \(d_Z = 8\).
In other words, not only is the dephasing suppression twice that of the four-mode extended pair-cat code, dephasing errors \(\mathbf{\hat{a}}^{\dagger\mathbf{p}}\mathbf{\hat{a}}^{\mathbf{p}}\) stemming from any nontrivial detectable losses $\mathbf{\hat{a}^{p}}$ (\(\mathbf{p}\neq \mathbf{0}\) and  $\mathbf{p}\notin \text{ker}H$) are corrected exactly.

The above dephasing calculation brings about more surprises. 
First, it guarantees that the logical \(X\)- \textcolor{black}{and $Z$-}bases \textcolor{black}{are both} orthonormal if we pick a nonzero \(\Delta\) parameter to define the codewords. To our knowledge, \textcolor{black}{coherent-state codes with both orthogonal $X$- and $Z$-bases at arbitrary energy, which is a desirable property previously thought to be impossible, have not been reported before.}


Second, such a choice yields a logical \(X\) operator that that can be implemented with a beam splitter,

\begin{equation}
    \overline{X} = \frac{\hat{a}_4^\dagger \hat{a}_1 + \hat{a}_1^\dagger\hat{a}_4 - \hat{a}_3^\dagger \hat{a}_2 - \hat{a}_2^\dagger\hat{a}_3}{\Delta}.
\end{equation}
The technique to obtain this operator is presented in Sec.~\ref{sec:finitesupport}.

\subsection{Tiger Shor code}

Just as the \([[4,1,2]]\) qubit stabilizer code can be extended to the Shor \cite{shor1995scheme,knill2000efficient,ralph2005loss}, a.k.a. \([[m^2,1,m]]\) quantum parity check code, the four-mode code in the previous example is a special case of an analogous tiger family.
We generalize it by combining the higher-mode versions of the logical-rotor pair-coherent-state code and the pair-cat code. 

To increase $d_Z$, we take a length-$L$ coherent state repetition code with generator matrices
\begin{eqs}
    H_{X,1}=T_L \quad\text{and}\quad H_{Z,1}=0, 
\end{eqs}
where $T_L$ is defined as $R_L$ in Eq.~\eqref{eq:paritycheck_coherentstate_repetition} without the last row. 
This code gives logical operators $\mathbf{x}_{1}=\begin{pmatrix}
    1 & 0 & \cdots & 0
\end{pmatrix}, \mathbf{z}_{1}=\begin{pmatrix}
    1 & -1& 1& \cdots & (-1)^{L-1}
\end{pmatrix}$. 

On the other hand, to have an increasing $d_X$, we take a length-$M$ extended pair-cat code with generator matrices
\begin{eqs}
    H_{X,2}=\mathbf{2}_M\quad\quad\text{and}\quad\quad H_{Z,2}=A_M~ ,
\end{eqs}
and logical operators $\mathbf{x}_{2}=\mathbf{1}_M$ and $\mathbf{z}_{2}=\begin{pmatrix}
    1 & 0& 0& \cdots & 0
\end{pmatrix}$. 

Combining those two blocks, we construct the \((L \times M)\)-mode tiger Shor code --- encoding a logical qubit --- with generator matrices
\begin{eqs}
    \HX =\begin{pmatrix}
        H_{X,1}\otimes \mathbb{1}_M\\
        \hline
        \mathbf{x}_{1} \otimes H_{X,2}
    \end{pmatrix}\quad\text{and}\quad \HZ =\begin{pmatrix}
        \mathbf{z}_{1}\otimes H_{Z,2}
    \end{pmatrix}~.
\end{eqs}
Here, \( \mathbb{1}_M\) is the \(M\)-dimensional identity.
Its logical operators are $\mathbf{x}=\mathbf{x}_{1}\otimes \mathbf{1}_M$ and $\mathbf{z}=    \mathbf{z}_{1}\otimes \mathbf{z}_{2}$. 

This code has distances $d_X=|\mathbf{x}|=M$ and $d_Z=4ML \sin^2(\frac{\pi}{2M})$. 
By letting $L\sim M^2$, both distances increase as we increase the system size. However, while all \textcolor{black}{$Z$-type stabilizers} have constant weight, one dissipator is non-local with weight $2M$. This non-local stabilizer also appears in the Shor code.
In the next section, we provide a surface-code construction with constant-degree stabilizers and increasing $X$- and $Z$-distances.

\begin{figure*}[t]
    \centering
    \includegraphics[width=1.0\textwidth]{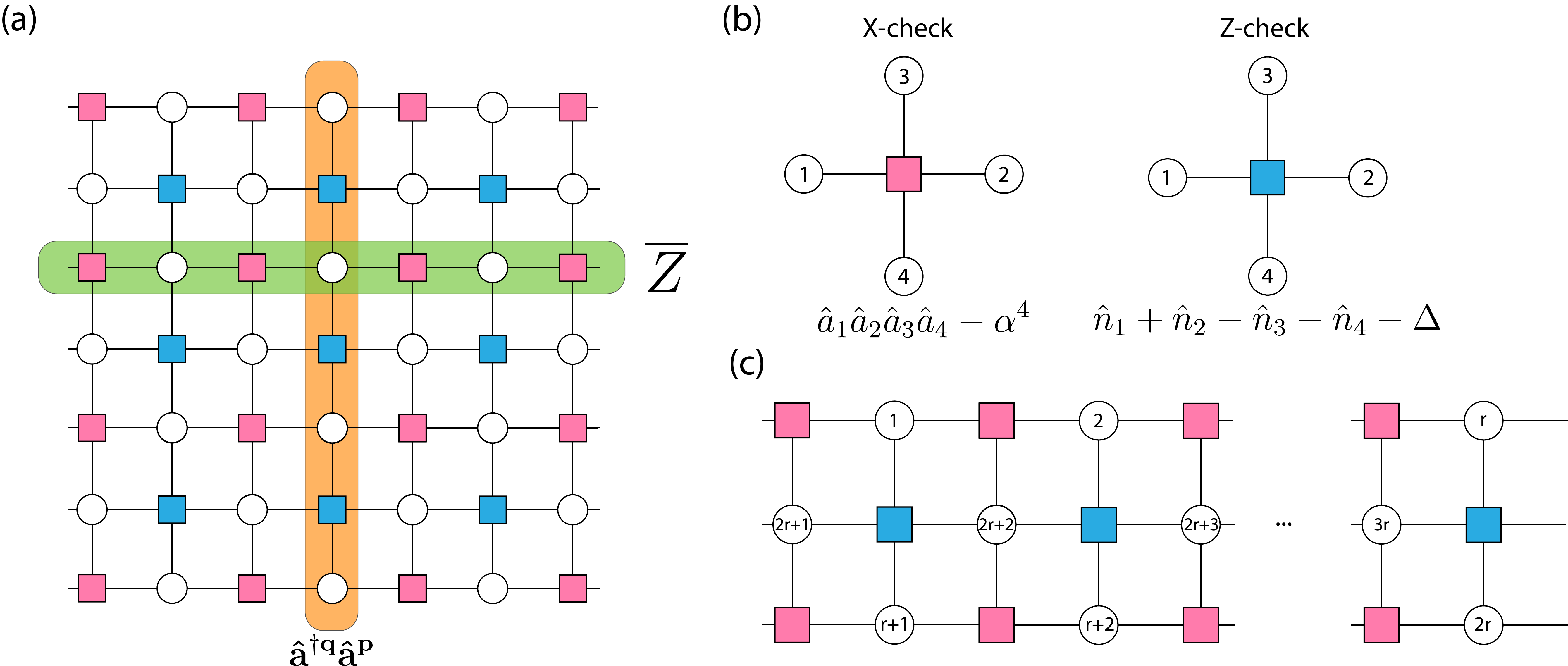}
    \caption{(a) The structure of tiger surface code lattice and its stabilizers, where each white circle represents a physical mode, a pink square represents an $X$-check acting on the physical modes it connects, and a blue square represents a $Z$-check acting on the physical modes it connects. \textcolor{black}{Each row contains $r$ physical modes, and there are $2m-1$ rows in total, forming a 2D lattice with $2mr-r$ physical modes. The undetectable $X$-error $\mathbf{\hat{a}}^{\dagger \mathbf{q}} \mathbf{\hat{a}}^{\mathbf{p}}$, with minimal 1-norm, and the logical $Z$ operator are string operators connecting the top-bottom and left-right boundaries, respectively.}
    (b) The form of the $X$ and $Z$-checks. 
    (c) The liger surface code with $d_X=2$ and $d_Z=4r$, encoding a logical qubit into a chain involving $3r$ physical modes. }
    \label{fig:2d_surface_code}
\end{figure*}

\section{Tiger and liger surface Codes}\label{sec:surface_tiger}

We present a surface-code family of tiger codes with constant-weight stabilizers and $X$- and $Z$-distances that increase with the system size.
This code is distinct from a concatenation of a single-mode cat code with a qubit surface code \cite{chamberland2022building}, and it is obtained by a hypergraph product of two repetition codes over the integers \cite{tillich, vuillotHomologicalQuantumRotor2024}. 
This code inherits the string-like logical operators, geometric locality of stabilizing operators (which we refer to as "checks" in this section), and scaling of code distances of the original surface code.

The code is laid out on a two-dimensional lattice of size $r \times (2m-1)$, encoding a logical qubit into $2rm-r$ physical modes with distances $d_X=m$ and $d_Z \geq 4rm \sin^2\left(\frac{\pi}{2m}\right)$. 
We call the $m=2$ case the \textit{liger} (a.k.a. long-tiger) surface code, whose distances reduce to $d_X=2$ and $d_Z=4r$.
The liger surface code exhibits exact orthogonality in \textcolor{black}{both} logical $X$- \textcolor{black}{and $Z$}-bases for certain \textcolor{black}{choices} of $\boldsymbol{\Delta}$ \textcolor{black}{, in arbitrary energy.}

The generator matrices $R_m, T_m$ for the coherent state repetition code in Sec.~\ref{sec:coherent_repetition} are
\begin{eqs}
    R_m=\begin{pmatrix}
        1& 1& 0& 0&\cdots & 0 \\
        0& 1& 1& 0& \cdots & 0\\
        \vdots & \vdots & \vdots & \vdots& \vdots & \vdots\\
        1& 0& 0& 0& \cdots & 1
    \end{pmatrix},~~ T_m=\begin{pmatrix}
        1& 1& 0& 0&\cdots & 0& 0 \\
        0& 1& 1& 0& \cdots & 0& 0\\
        \vdots & \vdots & \vdots & \vdots& \vdots & \vdots &\vdots\\
        0& 0& 0& 0& \cdots& 1 & 1
    \end{pmatrix},
\end{eqs}
where $R_m$ is a $m\times m$ matrix as Eq.~\eqref{eq:paritycheck_coherentstate_repetition}, and $T_m$ is a $(m-1) \times m$ matrix which is the same as $R_m$ without the last row.

We input $R_r$ and $T_m$ into a hypergraph product, yielding the following generator matrices
\begin{eqs}\label{eq:surface_code_stabilizer}
    \HX  &= \left(\begin{array}{c|c}
        \id_m\otimes R_r  & T_m^\T\otimes \id_r
    \end{array}\right),\\
    \HZ  &= \left(\begin{array}{c|c}
        T_m\otimes\id_r  & -\id_{m-1}\otimes R_r^\T
    \end{array} \right),
\end{eqs}
where $r \geq 3$ is odd, and where $m \geq 2$.
Picking \(r\) odd ensures that the logical space is that of a logical qubit, and one gets an infinite dimensional logical space if \(r\) is even.

As Eq.~\eqref{eq:surface_code_stabilizer} shows, the $R_m$ and $T_m$ only have two non-zero entries per row, and one or two per column.
Hence both $\HX $ and $\HZ $ only have three or four non-zero entries per row, which implies this code is a local stabilizer code consisting of local degree-3 or degree-4 $X$ and $Z$ stabilizing operators, as shown in the parts (a) and (b) of Fig.~\ref{fig:2d_surface_code}. 

The lattice layout of the tiger surface code is depicted in part (a) of Fig.~\ref{fig:2d_surface_code} where top and bottom boundaries of this lattice are open, and the left and right boundaries are periodic. Here the parameter $r$ represents the diameter of the \textcolor{black}{cross section} of the tube, and $m$ represents the length of the tube. 
Part (c) of Fig.~\ref{fig:2d_surface_code} represents the lattice structure of the liger surface code, where three-body $X$-checks are put at the top and bottom rows, and four-body $Z$-checks are in the middle row.

The general construction of logical operators for a hypergraph product is given in \cite[E.2]{vuillotHomologicalQuantumRotor2024}.
When one of the two seed codes has torsion, as in Eq.~\eqref{eq:torsion}, we use it to build the logical operators.
The torsion we have here is described in Sec.~\ref{sec:coherent_repetition}, and we use the vector $\bs{f}_r$ from Eq.~\eqref{eq:a_vector} to obtain
\begin{eqs}
    \mathbf{x}=\left(\begin{array}{c|c}
        \bs{f}_m\otimes\bs{e}_1 & \bs{0}
    \end{array} \right)\quad\text{and}\quad\mathbf{z} = \left(\begin{array}{c|c}
        \bs{e}_1\otimes\bs{f}_r & \bs{0}
    \end{array}\right).
\end{eqs}
One can verify that they indeed \textcolor{black}{commute with stabilizers and have} logical \textcolor{black}{$X$ and $Z$ actions respectively,  \textcolor{black}{using Eq.~\eqref{eq:a_vector} and 
    $T_m \bs{f}_m^\T=\bs{0}$.}}

\begin{figure*}[t]
    \centering
        \includegraphics[width=0.9\textwidth]{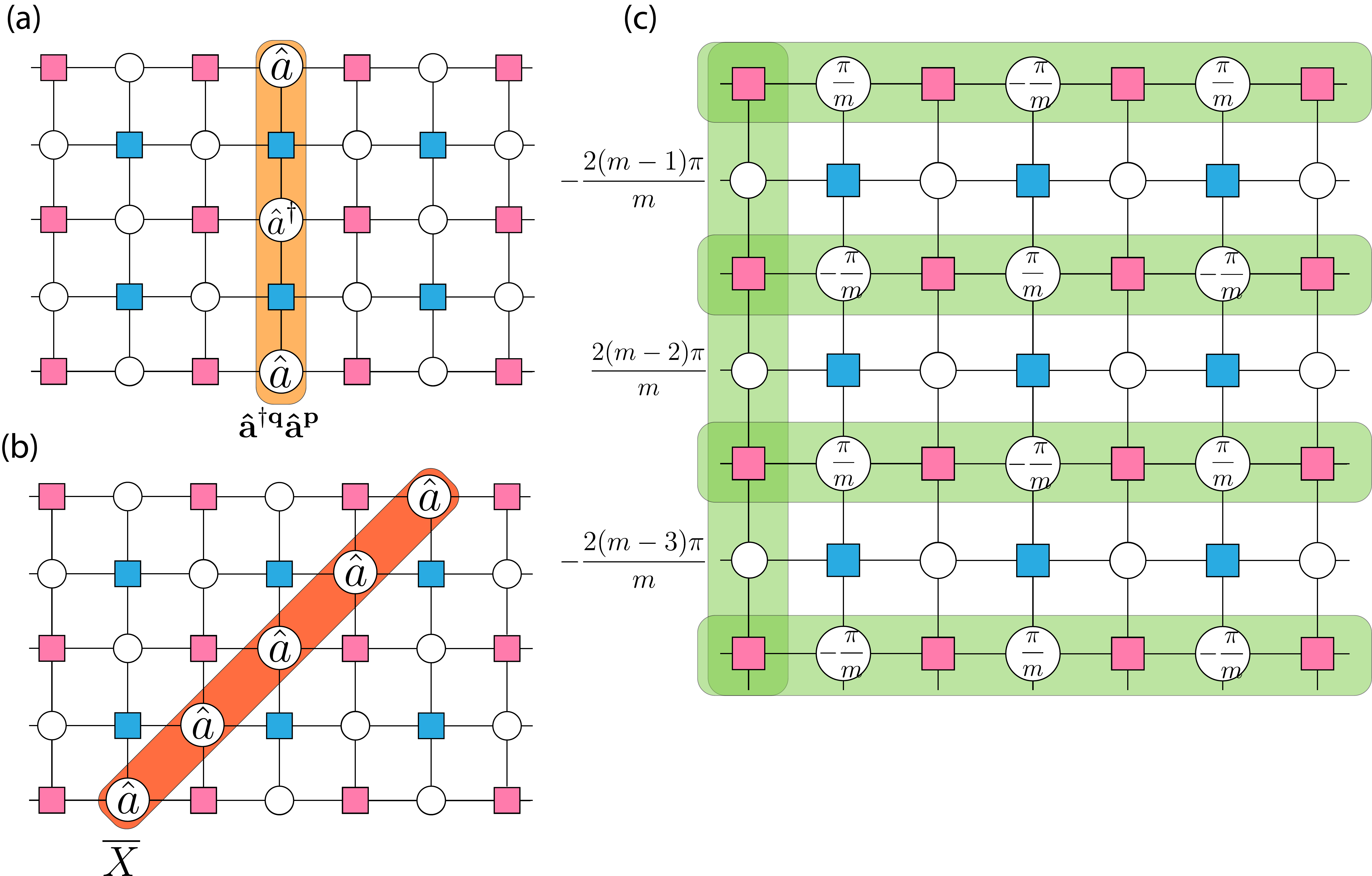}
    \caption{In panel (a), the vertical string operator in the light orange region represents a \textcolor{black}{lowest-weight} undetectable \textcolor{black}{$X$-}error  $\mathbf{\hat{a}}^{\dagger \mathbf{q}} \mathbf{\hat{a}}^{ \mathbf{p}}$. 
    Thus, the tiger surface code can correct loss errors $\mathbf{\hat{a}}^{\mathbf{v}}$ whose 1-norm is at least $\lfloor \frac{m-1}{2} \rfloor$.
    In panel (b), the diagonal string operator in the dark orange region represents the \textcolor{black}{lowest-}weight logical $X$ operator. 
    It implies that tiger surface code can detect loss error $\mathbf{\hat{a}}^{\mathbf{p}}$ one any $(2m-2)$ modes. 
    Panel (c) represents the error pattern that gives upper bound of the squared Euclidean distance between two codewords in Eq.~\eqref{eq:z-distance}, which corresponds to a "smeared" version of the defining logical $Z$ operator of tiger surface code. 
    In this error pattern, each physical mode inside the green horizontal strips \textcolor{black}{undergoes a phase-space rotation} by an \textcolor{black}{angle} of $\textstyle \pm \frac{\pi}{m}$. 
    The physical modes inside the green vertical strip in the first column \textcolor{black}{undergo configuration-space rotations} by angles $\textstyle \frac{-2(m-1)\pi}{m},\frac{2(m-2)\pi}{m},\cdots , \frac{(-1)^{m-1} 2\pi}{m}$, from top to bottom.
    }
    \label{fig:XZ_string}
\end{figure*}

\subsection{Tiger surface code \texorpdfstring{\(X\)}{}-distance}

The logical vector \(\mathbf{x}\) gives the string-like \textcolor{black}{undetectable $X$-error $\mathbf{\hat{a}}^{\dagger \mathbf{q}} \mathbf{\hat{a}}^{ \mathbf{p}}$, where $\mathbf{p}=\begin{pmatrix}
        1& 0& 1& 0& \cdots 
    \end{pmatrix} \otimes \mathbf{e}_1$ and $\mathbf{q}=\begin{pmatrix}
        0& 1& 0& 1 &\cdots
    \end{pmatrix} \otimes \mathbf{e}_1$ \footnote{It is given by the decomposition of logical vector $\mathbf{x}=\mathbf{p}-\mathbf{q}$ as Eq.~\eqref{eq:undetectable_type2}.}, with minimal 1-norm. It consists} alternating $\hat{a}$ and $\hat{a}^\dagger$ acting on neighboring modes along a vertical string connecting the top and bottom boundaries, as shown in Fig.~\ref{fig:XZ_string}(a). \textcolor{black}{The minimal undetectable $X$-error involves}  a non-trivial creation piece $\mathbf{\hat{a}}^{\dagger \mathbf{q}}$ \textcolor{black}{with $\mathbf{q} \neq \mathbf{0}   $ }--- a first example of this kind.
    
The $X$-distance is the $1$-norm of $\mathbf{x}$,
\begin{eqs}\label{eq:surface_code_Xdistance}
    d_X=|\mathbf{x}|=|\bs{f}_m|=m.
\end{eqs}
This code can correct at least $\lfloor \frac{m-1}{2} \rfloor$ losses on any mode.

A longer logical-\(X\) string consisting of only annihilation operators runs diagonally between the top and bottom boundaries, with $\hat{a}$ acting on each of the \(2m-1\) modes inside the string [see Fig.~\ref{fig:XZ_string}(b)].

Since the 1-norm of this minimal pure-loss vector is \(2m-1\), the code can detect up to $2m-2$ losses --- nearly twice the code distance. 
For example, the liger surface code ($m=2$) can detect two losses on any mode, but it cannot correct any losses.

\subsection{Tiger surface code \texorpdfstring{\(Z\)}{}-distance}

In Fig.~\ref{fig:2d_surface_code}(a), the logical $Z$ operator defined by \(\mathbf{z}\), \textcolor{black}{forms} a length-$r$ horizontal string operator consisting of parities $(-1)^{\hat{n}}$ and connecting the left and right boundaries. 
However, the $Z$-distance of a code is determined by the squared Euclidean distance of two neighboring codewords [see Eq.~\eqref{eq:z-distance-sines}], and rotating by the defining logical $Z$ string operator does not produce the shortest Euclidean path. 

We obtain upper and lower bounds on the distance,
\begin{subequations}
    \label{eq:surface_dz_bound}
\begin{align}
   d_{Z}&\geq4rm\sin^{2}\frac{\pi}{2m} ~,\\
   d_{Z}&\leq4rm\sin^{2}\frac{\pi}{2m}+4\sum_{j=1}^{m-1}\sin^{2}\frac{(m-j)\pi}{m} ~,
\end{align}    
\end{subequations}
by choosing other logical \(Z\) operators that act on more modes, but that rotate each mode by a smaller angle.
These bounds differ by an order $O(m)$ term 
in the limit of large system size.

\textcolor{black}{The intuition behind the upper bounds of $d_Z$ is as follows.} Using \(Z\)-type stabilizers, we can "smear" our original logical operator into one that acts on $m$ rows of modes, rotating each mode by $\pi/m$. 
As Fig.~\ref{fig:XZ_string}(c) shows, each physical mode in the green-colored rows is acted by an $\exp(\pm i \pi \hat{n}/m)$ phase shift, and physical modes in each green-colored row contribute $4r \sin^2\frac{\pi}{2m}$ to the squared Euclidean distance.

During the smearing process, we also generate phase shifts in the first column as shown in Fig.~\ref{fig:XZ_string}(c) because of the periodic boundary conditions between the left and right sides. 
The first column contributes a $ 4 \sum_{j=1}^{m-1}\sin^2 \frac{(m-j)\pi}{m}$ term in the upper bound of $d_Z$. Intuitively, for a fixed $m$, we can linearly increase $d_Z$ by increasing $r$ because we add more $m\sin^2 \frac{\pi}{2m}$ terms to the lower and upper bounds.

The lower bound of $d_Z$ is obtained by removing the phase shift acting on the first column induced during the smearing.
The general approach to obtain the $d_Z$ lower bound can be found in Ref.~\cite[Lemma 1]{vuillotHomologicalQuantumRotor2024}.

\subsection{Codewords and dephasing suppression}

Tiger surface codewords can be obtained by applying its logical \(Z\) string operator in Fig.~\ref{fig:2d_surface_code} to the fiducial state,
\begin{eqs}\label{eq:tiger_liger_codewords}
\ket{\overline{\pm}}&=\frac{e^{\frac{(2rm-r) \alpha^2}{2}}}{\sqrt{\mathsf{A}_0(\alpha^2 \mathbf{1})_{\text{tiger}}}} {\Pi_{\boldsymbol{0}}} \ket{\pm\alpha}^{\otimes r} \otimes \ket{\alpha}^{\otimes 2r(m-1)},
\end{eqs}
where $\Pi_{\boldsymbol{0}}$ is the collection of $Z$ projectors given by $\HZ$.

Combining our lower bound~\eqref{eq:surface_dz_bound} on $d_Z$ and Eq.~\eqref{eq:dephasing-bound}, and noting that \(\mathbf{1}\in\ker\HZ\), we can bound the degree of dephasing suppression using the result from App.~\ref{sec:GKZ}, 
\begin{eqs}
    &|\langle \overline{-}|\mathbf{\hat{a}}^{\dagger \mathbf{p}} \mathbf{\hat{a}}^{ \mathbf{p}}|\overline{+} \rangle_{\text{tiger}}|^2\overset{\alpha\to\infty}{\lesssim} \alpha^{4|\mathbf{p}|}\exp\bigg(- 4rm \alpha^2\sin^2\left(\frac{\pi}{2m}\right) \bigg).
\end{eqs}
To achieve the same amount of dephasing suppression, we can either increase $r$ or increase $\alpha^2$. 
To simultaneously increase both the $X$- and $Z$-distances, we can choose $r=m^2$ such that $d_X=m, d_Z\geq 4m^3\sin^2 \frac{\pi}{2m}$ increases as $m$ increases. In general, we can take $r\sim \Omega(m^2)$ to ensure both distances increase with system size.

\subsection{Liger (long-tiger) surface code}

This simpler version of the tiger surface code has a tractable GKZ function with interesting properties.
The \textit{liger} (a.k.a. long tiger) surface code is the tiger surface code on a thin strip with $m=2$. A length-$r$ liger surface code contains $3r$ bosonic modes for odd $r$. 
It encodes a logical qubit with distances $d_X=2$ and $d_Z=4r$, with the latter known exactly because the lower and upper bounds from Eq.~\eqref{eq:dephasing-bound} are equal for this case. 
In other words, the defining logical $Z$ string formed by $(-1)^{\hat{n}}$ in Fig.~\ref{fig:2d_surface_code}(a) exactly gives us the shortest Euclidean path between codewords.

\begin{widetext}
Using codewords shown in Eq.~\eqref{eq:tiger_liger_codewords} and repeatedly applying the splitting identity~\eqref{eq:splitting-identity}, we simplify the liger GKZ function to
\begin{eqs}\label{eq:liger_GKZ}
    \mathsf{A}_{\boldsymbol{\Delta}} (\mathbf{y})_{\text{liger}}=\sum_{\mathbf{n}\in\mathbb{N}^r} \sum_{\mathbf{j}\in \mathbb{N}^r} \prod_{i=1}^r \frac{(y_i+y_{r+i})^{j_i-\min\{0,\Delta_i\}}}{(j_i-\min\{0,\Delta_i\})!} \frac{y_{2r+i}^{n_{i}}}{n_{i}!} \delta_{n_{i}+n_{i+1},j_i+\max\{0,\Delta_i\}} ~,
\end{eqs}
where $y_{3r+1}=y_{2r+1}$ and $n_{r+1}=n_1$ due to the periodic boundary conditions. 
The indices follow the convention of Fig.~\ref{fig:2d_surface_code}(c).

\end{widetext}

Using the GKZ function, we can calculate the liger codeword overlap to be
\begin{eqs}
    \langle \overline{-} | \overline{+} \rangle_{\text{liger}}&=\frac{\mathsf{A}_{0} (-\alpha^2\mathbf{1}_r, \alpha^2 \mathbf{1}_{2r})_{\text{liger}}}{\mathsf{A}_0 (\alpha^2 \mathbf{1})_{\text{liger}}}=\frac{1}{\mathsf{A}_0 (\alpha^2 \mathbf{1})_{\text{liger}}}~,
\end{eqs}
where the numerator is one for arguments $y_i=-y_{r+i}$ for $i=1,2,\cdots,r$ .
In other words, the overlap depends only on the codeword normalization.

The GKZ function of the liger code also yields another property, namely, that $\mathsf{A}_{\boldsymbol{\Delta}} (-\alpha^2 \mathbf{1}_r, \alpha^2 \mathbf{1}_{2r})_{\text{liger}}\propto \varTheta(\boldsymbol{\Delta})$, where the Heaviside step function is zero if the vector \(\boldsymbol{\Delta}\) has at least one negative entry. 

Plugging Eq.~\eqref{eq:liger_GKZ} in Eq.~\eqref{eq:dephasing_overlap_b} yields the explicit expression of the dephasing matrix element 
\begin{eqs}\label{eq:liger_KL}
&|\langle\overline{-}|\mathbf{\hat{a}}^{\dagger\mathbf{p}}\mathbf{\hat{a}}^{\mathbf{p}}|\overline{+}\rangle_{\text{liger}}|^{2}\\
=&{\displaystyle \frac{\varTheta(-H \mathbf{p}) \alpha^{4(|\mathbf{p}|+|H\mathbf{p}|)}}{\mathsf{A}_{0}(\alpha^{2}\mathbf{1})_{\text{liger}}^{2}}}\Big(\sum_{\mathbf{n}\in \mathbb{N}^r} \prod_{i=1}^r \frac{1}{n_i!} \delta_{n_i+n_{i+1},(-H\mathbf{p})_i} \Big)^2,
\end{eqs}
where the dephasing matrix element is exactly zero if $H\mathbf{p}$ has at least one positive entry. The non-zero dephasing matrix elements are exponentially suppressed as $\exp(-6r |\alpha|^2)$, in the large-$\alpha$ limit, where the suppression coefficient $6r$ exceeds the code's $Z$-distance $d_Z=4r$. This is our only example where the dephasing suppression surpasses the \(Z\)-distance.

\subsection{Open boundaries}

Tiger surface codes can also be defined for open left and right boundaries. 
By replacing the last row $R_r$ by $\begin{pmatrix}
    K& 0& 0 &\cdots & 0 \end{pmatrix}$ and keeping $T_m$ unchanged, the hypergraph-product construction will yield a tiger surface code with open boundary condition that encodes a qu$K$it.

This open boundary modification will increase the degree of the boundary $X$-checks and change the coefficient in front of $\hat{n}$ in the boundary $Z$-checks. 
It will not affect any checks in the bulk.
This modification will not change the $X$-distance, and will only slightly decrease the $Z$-distance. 
We leave further exploration of the boundaries of the tiger surface code for future investigation.

\section{Finite-support tiger codes }
\label{sec:finitesupport}

\begin{figure*}[t]
    \centering
    \includegraphics[width=0.8\textwidth]{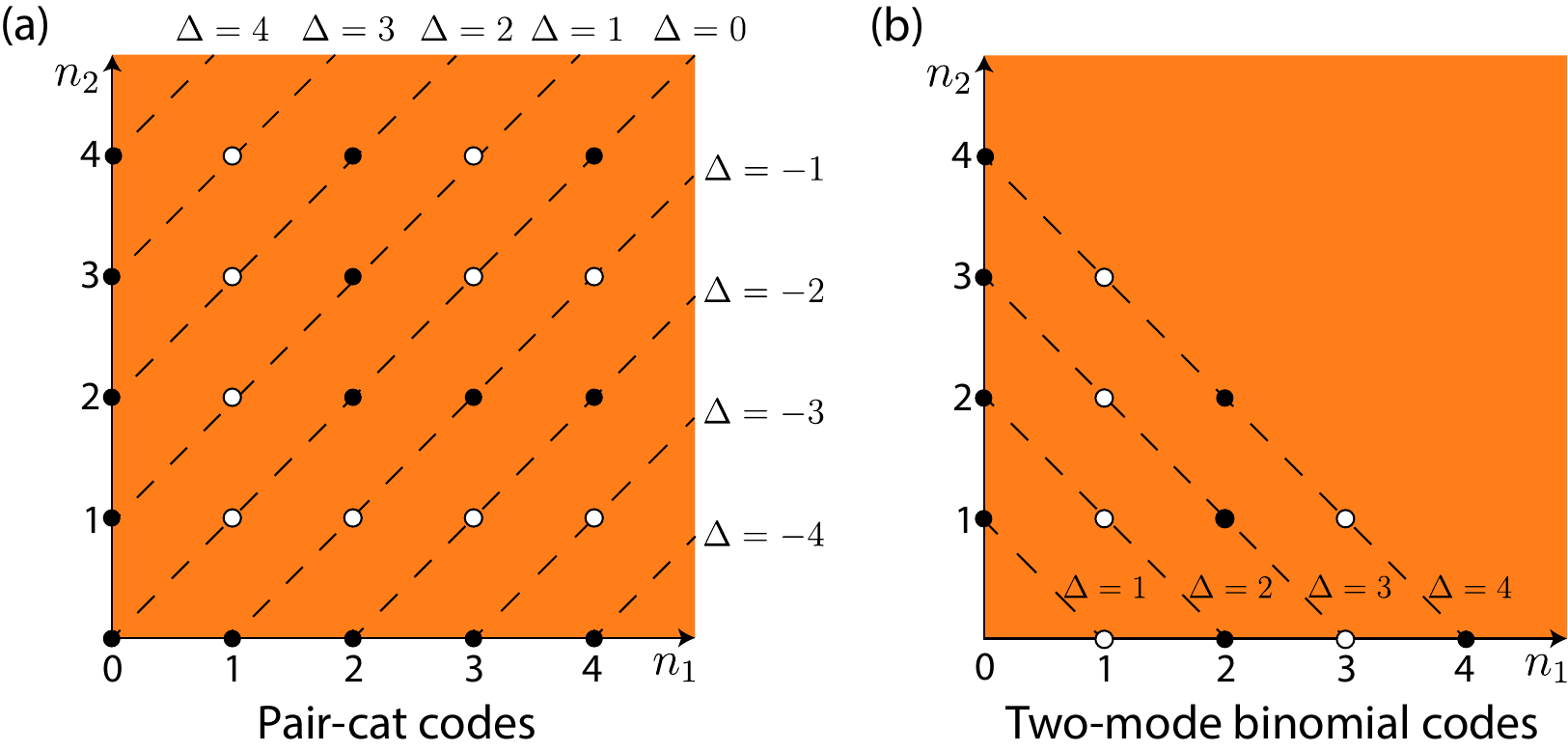}
    \caption{Pictorial comparison between infinite-support pair-cat code and finite-support two-mode binomial code in the two-mode Fock space $(n_1,n_2)$. Panel (a) represents the Fock-state structure of the pair-cat code for different choices of occupation-number difference, $\Delta = n_2-n_1$.
    Each choice specifies a 1D semi-infinite lattice (dashed line with $\Delta$) extending in the upper-right direction in the two-mode Fock space, and the codewords are
    superpositions of the Fock states in the semi-infinite lattice. 
    Inside each lattice, the black (white) circles are in the Fock-state support of codeword $\ket{\overline{0}}_\Delta$ ($\ket{\overline{1}}_\Delta$). 
    Since the 1D lattice with $\Delta$ involves infinite numbers of Fock states, the codewords are superpositions of Fock states involving infinitely many terms. 
    Panel (b) represents the Fock-state structure of the two-mode binomial code for different choices of the total occupation number, $\Delta = n_1+n_2$.
    Each choice specifies a finite 1D lattice in the two-mode Fock space, and the codewords are superpositions of Fock states in the finite lattice.  \textcolor{black}{The finite lattices from left to right correspond to $\Delta=1,2,3,4$.}
    Inside each lattice, the black (white) circles are in the Fock-state support of codeword $\ket{\overline{0}}_\Delta$ ($\ket{\overline{1}}_\Delta$).
    The supports for both $\ket{\overline{0}}_\Delta$ and $\ket{\overline{1}}_\Delta$ are finite. 
    }
    \label{fig:paircat_pairbinomial}
\end{figure*}

In previous sections, we introduce tiger codes with a \textit{non-negative} integer matrix $\HX $ and integer matrix $\HZ $. 
Here, we study codes whose $\HX $ has arbitrary integer entries.
This turns out to imply that the Fock-state subspace fixed by \(\HZ \) will be finite, yielding codewords of finite support and complicating the construction of logical operators.
This also implies that dissipators consist of not only lowering operators, but raising operators, and will thus not commute with each other.
The good news is that homology still yields the codespace, which still lies in a joint eigenspace of all stabilizing operators.

A known code encompassed by our finite-support framework is the two-mode binomial code \cite{chuang_pairbinomial,pairbinomial}, which includes the experimentally relevant \cite{chuang_dualrail,fletcher2008channel,teoh2023dual,kubica2023erasure,koottandavida2024erasure,awsdualrail} dual-rail \(\{|10\rangle,|01\rangle\}\) code as a special case.
We also make contact with a \(\chi^{(2)}\) code \cite{niu2018hardware} by defining a tiger code with the same Fock-state support.

General generator matrices for finite-support codes are both integer-valued,
\begin{eqs}
    \HX \in \mathbb{Z}^{r_{x}\times N},~~ \HZ \in \mathbb{Z}^{r_{z}\times N},~~\HX  \HZ ^\T=\mathbf{0} .
\end{eqs}
To ensure that any negative entries in \(\HX \) are not an artifact of the basis used to define the generator matrix,
we assume that there exists an all-positive vector $\bs{b}$ in the row-image of $\HZ $, i.e., $\bs{b}\in {\rm im}\HZ $ with \(b_j>0\) for all \(j\).
This forces the rows of \(\HX \) to have some negative entries in order to be orthogonal to \(\bs{b}\).
This condition
is equivalent to the condition that there is a finite number of Fock states satisfying the \(\HZ \) constraints.
In other words, only a finite set of Fock states are in any particular eigenspace of the operator \(\mathbf{b}\cdot\hat{\mathbf{n}}\).
Typically, we will have \(\mathbf{b}=\mathbf{1}\), corresponding to a constraint on the total occupation number, \(\mathbf{1}\cdot\hat{\mathbf{n}} = |\hat{\mathbf{n}}|\).

Codewords are defined as projected coherent states, analogous to the infinite-support tiger codes.
The \(Z\)-type \textcolor{black}{stabilizers} are also the same: \(\bfh\cdot \hat{\mathbf{n}} - \Delta_{\bfh}\) for all rows \(\bfh\) of \(\HZ \).
The full Fock-state subspace supported by the code is defined by \(\HZ \hat{\mathbf{n}}= \boldsymbol{\Delta}\) for integer syndrome vector \(\boldsymbol{\Delta}\).

Logical \(Z\) operators, corresponding to vectors \(\mathbf{z}\), are obtained via the same homology calculation.
Codewords can still be obtained by applying logical \(Z\) operators to the all-ones coherent state \(|\alpha\mathbf{1}\rangle\)~\eqref{eq:fiducial}, but the dependence on the energy density \(\alpha^2\) drops out if the total occupation number is fixed by the code.
In other words, finite-support codewords cannot have arbitrary energy since they do not have an infinite runway of Fock states.
Any dephasing protection should thus either be exact or \textcolor{black}{exponentially} suppressed with increasing values of the syndrome vector \(\boldsymbol{\Delta}\). We see both situations in our examples.

The \(X\)-type dissipators now contain raising and lowering operator monomials, i.e., operators \(\mathbf{\hat{a}}^{\dagger\mathbf{q}}\mathbf{\hat{a}}^{\mathbf{p}}\) for any row \(\bfg = \mathbf{p} - \mathbf{q}\) of \(\HX \).
Here, we decompose the row into its positive part \(\bs{p}\in\mathbb{N}^N\) and negative part \(\bs{q}\in\mathbb{N}^N\).
Acting with such operators on coherent states does not produce a constant since
coherent states are not right eigenstates of creation operators.
To develop dissipators which annihilate coherent states, we need an additional formula.

For a non-negative integer \(k\geq 0\), we have that
\begin{align}
     \hat{a}^{\dagger k}\ket{\alpha}=\frac{\left(\hat{n}\right)_{k}}{\alpha^{k}}\ket{\alpha} ,
 \end{align}
where the falling factorial
 \begin{equation}
     \left(\hat{n}\right)_{k} = \hat{a}^{\dagger k}\hat{a}^k = \prod_{j=0}^{k-1}\left(\hat{n}-j\right) = \hat{n}\left(\hat{n}-1\right)\cdots\left(\hat{n}-k+1\right).
     \end{equation}
As a convention, we use \(\left(\hat{n}\right)_0=1\).

Applying the above formula, we can define the dissipators as follows,
\begin{equation}
    \left[\mathbf{\hat{a}}^{\dagger \bs{q}}\mathbf{\hat{a}}^{\bs{p}} - \boldsymbol{\alpha}^{\bfg}\left(\hat{\bs{n}}\right)_{\bs{q}}\right]\ket{\boldsymbol{\alpha}}_{\boldsymbol{\Delta}}^{\HZ} = 0 ,\label{eq:general_a_dissipator}
\end{equation}
where we use multi-index notation, e.g., \((\hat{\mathbf{n}})_{\mathbf{q}}=(\hat{n}_{1})_{q_{1}}(\hat{n}_{2})_{q_{2}}\cdots(\hat{n}_{N})_{q_{N}}\).

These dissipators no longer commute with one another in general.
They still commute with all \(Z\)-type stabilizers since \(\bfg= \mathbf{p} - \mathbf{q}\) is in the kernel of \(\HZ \), and since their second terms are polynomials in the occupation numbers.

Given a logical vector \(\mathbf{x}\) obtained via homology, it is sometimes possible to construct logical \(X\) operators by choosing appropriate compensating occupation-number polynomials.
Let us focus on the case of logical qubit to simplify the presentation, but other logical codespaces work similarly.
For this, we need to pick a logical \(X\) representative \(\bs{x}\in\mathbb{Z}^N\), decomposed into \(\mathbf{v}\in\mathbb{N}^N\) and \(\bs{w}\in\mathbb{N}^N\) as \(\bs{x}=\bs{v}-\bs{w}\), and denote its conjugated logical \(Z\) representative by \(\mathbf{z}\).
We thus have the following two facts,
\begin{subequations}
\begin{align}
    \mathbf{\hat{a}}^{\dagger \mathbf{w}}\mathbf{\hat{a}}^{\mathbf{v}}\ket{\boldsymbol{\alpha}}_{\boldsymbol{\Delta}}^{\HZ}&=\boldsymbol{\alpha}^{\mathbf{x}}\left(\hat{\bs{n}}\right)_{\mathbf{w}}\ket{\boldsymbol{\alpha}}_{\boldsymbol{\Delta}}^{\HZ} ~,\label{eq:almostLX}\\
    (-1)^{\mathbf{z}\cdot\hat{\bs{n}}}\mathbf{\hat{a}}^{\dagger \mathbf{w}}\mathbf{\hat{a}}^{\mathbf{v}}&=-\mathbf{\hat{a}}^{\dagger \mathbf{w}}\mathbf{\hat{a}}^{\mathbf{v}}(-1)^{\mathbf{z}\cdot\hat{\bs{n}}} ~.\label{eq:almostLX-2}
\end{align}
\end{subequations}

If the operator \(\boldsymbol{\alpha}^{\mathbf{x}}\left(\hat{\bs{n}}\right)_{\mathbf{w}}\) was invertible, we would just multiply on the left in Eq.~\eqref{eq:almostLX} by its inverse to obtain the logical \(X\).
This is indeed what happens in the infinite-support case where \(\bs{w}=\bs{0}\), and so \(\boldsymbol{\alpha}^{\mathbf{x}}\left(\hat{\bs{n}}\right)_{\mathbf{w}}=\boldsymbol{\alpha}^\mathbf{x}\).
But here \(\mathbf{x}\) has some negative entries, implying that \(\boldsymbol{\alpha}^{\mathbf{x}}\left(\hat{\bs{n}}\right)_{\mathbf{w}}\) has some zero eigenstates, namely, all Fock states with an occupation number on the \(j\)th mode lower than \(w_j\).
Therefore, \(\boldsymbol{\alpha}^{\mathbf{x}}\left(\hat{\bs{n}}\right)_{\mathbf{w}}\)  is not invertible.

That said, there are several representatives for this logical \(X\), obtained by adding vectors from the row-image of \(\HX \).
In general, any linear combination,
\begin{eqs}
    \sum_{\mathbf{x}=\bs{v}-\bs{w}}\beta_{\mathbf{x}}\mathbf{\hat{a}}^{\dagger\mathbf{w}}\mathbf{\hat{a}}^{\mathbf{v}} ,\label{eq:linearcombinationalmostLX}
\end{eqs}
where the sum is over some set of logical representatives \(\mathbf{x}\), still satisfies Eq.~\eqref{eq:almostLX-2}. 
Crucially, we can promote the coefficient  to \textit{any polynomial} in \(\hat{\bs{n}}\), \(\beta_\mathbf{x}\to \hat{\beta}_\mathbf{x}\), while still maintaining the logical relation.
The operator \(\sum_{\mathbf{x}}\hat{\beta}_\mathbf{x}\boldsymbol{\alpha}^\mathbf{x}\left(\hat{\bs{n}}\right)_{\mathbf{w}}\) on the right-hand side of the analogue of Eq.~\eqref{eq:almostLX} now has more possibilities to be invertible, and can even be \textit{constant} on the code space in some cases, by taking into account \(\HZ \) constraints.
The latter case is more convenient since then we have a logical \(X\) operator which is a polynomial in creation and annihilation operators instead of being a rational function, as in the former case.

Formally, given a chosen logical \(\mathbf{x}\), we can define the polynomial ideal, \(I_\mathbf{x}\subset\mathbb{R}\left[n_1,n_2,\ldots,n_N\right]\) generated by all the \(\left(\bs{n}\right)_{\mathbf{w}}\) for all logical representatives \(\bs{x}^\prime = \bs{v}-\mathbf{w}\) equivalent to \(\mathbf{x}\) and all constraints from \(\HZ \):
\begin{align}
   I_{\mathbf{x}}=\Big\langle&\left\{ \left(\bs{n}\right)_{\bs{w}}\middle\vert\bs{m}\in\mathbb{Z}^{r_{x}},\bs{v},\bs{w}\in\mathbb{N}^{N},\,\bs{x}+\bs{m}\HX=\mathbf{v}-\bs{w}\right\} \nonumber\\&\bigcup\,\,\left\{ \boldsymbol{\lambda}\left(\HZ\cdot\bs{n}-\bs{\Delta}\right)\middle\vert\boldsymbol{\lambda}\in\mathbb{Z}^{r_{z}}\right\} \Big\rangle.\label{eq:ideal}
\end{align}
This ideal, \(I_{\mathbf{x}}\), contains all operators which are polynomials in the \(n_i\) that can be generated on the right-hand side of \eqref{eq:almostLX} by applying on the left hand-side an operator of the type \eqref{eq:linearcombinationalmostLX}.
Besides, since the logical \(X\) has to stabilize the state \(\ket{\boldsymbol{\alpha}}_{\boldsymbol{\Delta}}^{\HZ}\), this right-hand side needs to be \(1\).
Then there exists an operator which is a polynomial of the annihilation and creation operators representing the logical \(X\) given by \(\mathbf{x}\) as long as \(1\in I_\mathbf{x}\), and the coefficient in front of the generators of \(I_{\mathbf{x}}\) generating \(1\) directly give the coefficients \(\beta_{\mathbf{x}}\) after division by \(\boldsymbol{\alpha}^{\mathbf{x}}\).

We present a few examples of interest to illustrate the framework.

\subsection{Two-mode binomial code}

The two-mode binomial code \cite{chuang_pairbinomial,pairbinomial,chuang_dualrail} is an example of a finite-support tiger code.
Our framework reproduces what is known about these codes from the spin-coherent state framework \cite{albert2019pair}, in which the codewords are identified as two-mode binomial states \cite{radcliffe1971some,arecchi1972atomic,calixto2021entanglement} lying on the \(x\)-axis of an effective Bloch sphere.
 In this way, our framework offers another way to look at, and generalize, two-mode binomial states.

The  code  is obtained by moving the sign from one pair-cat generator matrix to the other to yield
 \begin{eqs}
     \HX =\begin{pmatrix}
         2& -2
     \end{pmatrix}\quad\quad\text{and}\quad\quad \HZ =\begin{pmatrix}
         1 & 1
     \end{pmatrix} ,
 \end{eqs}
 where $\HZ $ gives \textcolor{black}{occupation} number constraint \(\hat{n}_1+\hat{n}_2 = \Delta\), 
 and \(\HX \) yields the dissipator
 \begin{align}
\mathbf{\hat{a}}^{\dagger ( 0 ~ 2)}\mathbf{\hat{a}}^{(2~0)}-\alpha^{2-2}(\mathbf{\hat{n}})_{(0~2)}=\hat{a}_{2}^{\dagger2}\hat{a}_{1}^{2}-\hat{n}_{2}\left(\hat{n}_{2}-1\right) ,
\end{align}
Another dissipator for \(-G=\begin{pmatrix}-2 & 2\end{pmatrix}\) gives rise to the above, but with modes swapped.

 A homology calculation yields logical vectors
 \begin{eqs}
     \mathbf{x}=\begin{pmatrix}
         1& -1
     \end{pmatrix}\quad\quad\text{and}\quad\quad \mathbf{z}=\begin{pmatrix}
         1& 0
     \end{pmatrix} .
 \end{eqs}
We can solve for a logical \(X\) operator, which is a linear combination of two representatives of the logical-\(X\) equivalence class,  namely, \(\begin{pmatrix}1 & -1\end{pmatrix}\) and \(\begin{pmatrix}-1 & 1\end{pmatrix}\). 
The logical operators are
\begin{eqs}\label{eq:pairbinomial_logical}
\overline{X}=\frac{\hat{a}_{1}^{\dagger}\hat{a}_{2}+\hat{a}_{2}^{\dagger}\hat{a}_{1}}{\Delta}\quad\quad\text{and}\quad\quad\overline{Z}=(-1)^{\hat{n}_{1}}\cong (-1)^{\hat{n}_{2}+\Delta} ,
 \end{eqs}
 where "\(\cong\)" denotes that two logical operators are equivalent via a \(Z\)-type stabilizer.
Squaring the logical \(X\) gives an additional $X$-stabilizer.

Logical \(X\) codewords are coherent states \(\boldsymbol{\alpha}=(\alpha,~\alpha)\) and \((-\alpha,~\alpha)\), projected into the Fock subspace of fixed total occupation number \(\Delta\),
\begin{subequations}
\begin{align}\label{eq:pairbinomial_codeword}
    \ket{\overline{\pm}}&= \frac{e^{2\alpha^{2}}}{\sqrt{{\mathsf{A}}_{\Delta}(\alpha^{2}\boldsymbol{1})}} \Pi_{\Delta}|\alpha(\pm1,1)\rangle,\\
    &= \frac{1}{\sqrt{{\mathsf{A}}_{\Delta}(\alpha^{2}\boldsymbol{1})}}\sum_{n_{1}+n_{2}=\Delta}\frac{(\pm\alpha)^{n_{1}}\alpha^{n_{2}}}{\sqrt{n_{1}!n_{2}!}}|n_{1},n_{2}\rangle,\\
    &= \frac{1}{\sqrt{2^{\Delta}}} \sum_{n=0}^{\Delta}(\pm1)^{n}\sqrt{\binom{\Delta}{n}}\ket{n,\Delta-n} ,
 \end{align}
 \end{subequations}
with dependence on \(\alpha\) removed after normalizing by the GKZ function, 
\begin{equation}
    \mathsf{A}_{\Delta}(y_1, y_2) = \sum_{n=0}^{\Delta}\frac{y_1^n y_2^{\Delta-n}}{n!(\Delta-n)!}= \frac{(y_1+y_2)^{\Delta}}{\Delta!}~.\label{eq:GKZpairbinomial}
\end{equation}
These \(X\) codewords are orthogonal, as is usually the case for finite-support tiger codes.

Equations~\eqref{eq:dephasing-calc} and \eqref{eq:GKZpairbinomial} imply that this code can detect dephasing errors exactly up to power \(\Delta\),
\begin{equation}\label{eq:binomial-dephasing}
    \left.\bra{\overline{-}}\hat{\bs{a}}^{\dagger \bs{p}}\hat{\bs{a}}^{\bs{p}}\ket{\overline{+}} \propto (y_1-y_2)^{\Delta - p_1 - p_2}\right\vert_{y_1=y_2=\alpha^2}
    =0
\end{equation}
whenever \(\Delta > p_1+p_2\) (cf. \cite{pairbinomial}).

The code can detect \textcolor{black}{arbitrary} number of \textcolor{black}{single-mode} photon losses, just like the pair-cat code.
It cannot correct single-photon losses since their product, \(\hat{a}_1^{\dagger} \hat{a}_2\),
is a logical operator (see  Sec.~\ref{subsec:loss-correction}).

However, losing a photon just moves the system from \(\Delta\) to \(\Delta-1\), hence one can just keep track of the total \textcolor{black}{occupation} number and continue to operate this logical qubit until \(\Delta\) photons have been lost, yielding an erasure.
And indeed, if we pick $\Delta=1$, this reduces to the dual-rail code. 

\subsection{Multinomial code}
The code above can be generalized to three or more modes by defining \(\HX\in\mathbb{Z}^{(N-1)\times N}\) and \(\HZ\in\mathbb{Z}^{1\times N}\) as follows
\begin{equation}
    \HX = \begin{pmatrix}
        1 & -2 & 1 & 0 & \cdots & 0 & 0\\
        0 & 1 & -2 & 1 & \cdots & 0 & 0\\
        0 & 0 & 1 & -2 & \cdots & 0 & 0\\
        \vdots & \vdots & \vdots & \vdots & \ddots & \vdots & \vdots\\
        1 & 0 & 0 & 0 & \cdots & 1 & -2\\
    \end{pmatrix},\; \HZ = \begin{pmatrix}
        1 & 1 &\cdots & 1
    \end{pmatrix}.
\end{equation}
Computing the homology, we find a logical qu\(N\)it on \(N\) modes, with logical vectors \(\mathbf{x}=\begin{pmatrix}
    1& -1 & 0 &\cdots & 0
\end{pmatrix}\) and \(\mathbf{z}=\begin{pmatrix}
    0 & 1 &2 & \cdots &N-1
\end{pmatrix}\).
The corresponding logical operators are given by
\begin{subequations}
\begin{align}
    \overline{X} &= \frac{1}{\Delta}\sum_{j=0}^{N-1}\hat{a}_j^\dagger\hat{a}_{j+1} ~,\\
    \overline{Z} &= \exp\left(i\frac{2\pi}{N}\sum_{j=0}^{N-1}j\hat{n}_j\right) ~.
\end{align}
\end{subequations}

The projected coherent states in this case reduce to the \(SU(N)\) coherent states \cite{gitman1993coherent,nemoto2000generalized,calixto2021entanglement},
\begin{eqs}
    |\boldsymbol{\alpha}\rangle_{{\Delta}}^{\HZ}\propto\sum_{|\mathbf{n}|=\Delta}\sqrt{{\Delta \choose \mathbf{n}}}\boldsymbol{\alpha}^{\mathbf{n}}|\mathbf{n}\rangle ,
\end{eqs}
which are parameterized by the states of a qudit, \(\boldsymbol{\alpha} \in \mathbb{C}P^{N-1} = SU(N)/\mathbb{T}\) (with \(\mathbb{T}\) the group generated by \(\HZ\)) for complex \(\boldsymbol{\alpha}\), and whose Fock-state expansion contains a multinomial coefficient. 
The GKZ function is a multinomial expansion of the coordinate sum of \(\boldsymbol{\alpha}\).
For \(N=2\), these reduce to the two-mode binomial states \cite{radcliffe1971some, arecchi1972atomic, calixto2021entanglement} lying on the Bloch sphere of a qubit.

Any dephasing protection in this case is exact, holding due to the multinomial generalization of Eq.~\eqref{eq:binomial-dephasing}.
The degree of loss protection is also the same.

\subsection{Center-of-mass mode}

Multiplying the second column of the \(\HX \) matrix of the logical-rotor pair-coherent state code by \(-1\) yields
\begin{equation}
    \HX  = \begin{pmatrix}
        1 & -1
    \end{pmatrix}\quad\quad\text{and}\quad\quad
    \HZ  = 0 ,
\end{equation}
corresponding to dissipators
\begin{align}
\hat{a}_{2}^{\dagger}\hat{a}_{1}-\hat{n}_{2}\quad\quad\text{and}\quad\quad\hat{a}_{1}^{\dagger}\hat{a}_{2}-\hat{n}_{1}\,.
\end{align}

The logical space is \textcolor{black}{a single-mode oscillator, which is infinite-dimensional}, with logical vectors \(\mathbf{x}=\begin{pmatrix} 1 & 0\end{pmatrix} \cong \begin{pmatrix} 0 & 1 \end{pmatrix}\) and \(\mathbf{z}= \begin{pmatrix} 1 & 1 \end{pmatrix}\) and related logical $X$ and \(Z\) operators $\overline{X}=\frac{\hat{a}_1+\hat{a}_2}{\sqrt{2}}$, \(\overline{Z}(\varphi)=e^{i\varphi (\hat{n}_{1}+\hat{n}_{2})}\).
We see that the logicals need not commute with the dissipators in the case of finite support.

Applying the logical \(Z\) operator to the all-ones coherent states yields logical \(X\)-type codewords,
\begin{eqs}
   |\overline{\varphi}\rangle=\overline{Z}(\varphi)|\boldsymbol{\alpha}=\alpha\mathbf{1}\rangle=|\alpha e^{i\varphi}\rangle^{\otimes2}\,,
\end{eqs}
which are coherent states of the center-of-mass mode.

Alternatively, adding the dissipators together, we can concoct a \(Z\)-type basis of eigenstates of \(\hat{a}_1^\dagger \hat{a}_2+ \hat{a}_1\hat{a}_2^\dagger \), i.e., two-mode binomial states for each value of the total occupation number \(\ell\),
\begin{eqs}\label{eq:binomial_Fock_state}
\ket{\overline{\ell}}= \frac{1}{\sqrt{2^{\ell}}}\sum_{n=0}^{\ell}\sqrt{\binom{\ell}{n}}\ket{n,\ell-n},~~\forall\ell\geq0~.
\end{eqs}
Each of these can be obtained by applying a 50-50 beamsplitter \(U_{BS}\) to a Fock state,  $\ket{\overline{\ell}}=U_{BS} \ket{\ell,0}$.

The center-of-mass mode admits its own lowering, raising, and occupation-number operators, respectively,
\begin{subequations}\label{eq:com}
\begin{align}
    \overline{a}&=\sqrt{2}\hat{a}_{1}\cong\sqrt{2}\hat{a}_{2}\cong\frac{\hat{a}_{1}+\hat{a}_{2}}{\sqrt{2}}~,\\\overline{a^{\dagger}}&=\frac{\hat{a}_{1}^{\dagger}+\hat{a}_{2}^{\dagger}}{\sqrt{2}}~,\\
    \overline{n}&=\overline{a^{\dagger}}\overline{a}\cong\hat{n}_{1}+\hat{n}_{2}\,,
\end{align}
\end{subequations}
where we use "$\cong$" for operators that are equivalent up to stabilizers or dissipators.
Note that \(\overline{a^{\dagger}} = \overline{a}^{\dagger}\) is satisfied by only one form of the logical lowering operator.

\subsection{Four-mode binomial code}
We can define a finite-support version of the 4-mode tiger code described in Sec.~\ref{sec:four-mode} using generator matrices
\begin{eqs}
    \HX = \begin{pmatrix}
        1 & -1 & 0& 0\\
        0& 0& 1& -1\\
        1& 1& -1& -1
    \end{pmatrix},~~ \HZ =\begin{pmatrix}
        1& 1& 1& 1
    \end{pmatrix},
\end{eqs}
which are obtained by multiplying the second and third columns of the generator matrices in Eq.~\eqref{eq:fourmode_paritycheck} by $-1$ and taking linear combinations.
The logical vectors are $\mathbf{x}=\begin{pmatrix}
    1 & 0& 0& -1
\end{pmatrix},\, \mathbf{z}=\begin{pmatrix}
    1 & 1& 0& 0
\end{pmatrix}$.

We can view this code as a two-mode binomial code constructed out of two center-of-mass modes \(\overline{a}\) and \(\overline{b}\). 
This allows us to succinctly write the codewords as
\begin{eqs}
   \ket{\overline{\pm}}= \frac{1}{\sqrt{2^{\Delta}}} \sum_{n=0}^{\Delta}(\pm1)^{n}\sqrt{\binom{\Delta}{n}}\ket{\overline{n},\overline{\Delta-n}}\,,
\end{eqs}
where the overlined states on the right-hand side are the center-of-mass Fock states from Eq.~\eqref{eq:binomial_Fock_state}.
As with the two-mode binomial code, the codewords are orthogonal.

Similarly, the center-of-mass mode operators in Eq.~\eqref{eq:com} yield four-mode stabilizing operators,
\begin{eqs}
    \Big(\frac{\overline{a^\dagger} \overline{b}+\overline{b^\dagger} \overline{a}}{\Delta}\Big)^2 \quad\quad\text{and}\quad\quad \overline{a^\dagger} \overline{a}+ \overline{b^\dagger} \overline{b}-\Delta ~,
\end{eqs}
and logical operators,
\begin{eqs}
        \overline{X}=\frac{\overline{a^\dagger} \overline{b}+\overline{b^\dagger} \overline{a}}{\Delta}\quad\quad\text{and}\quad\quad \overline{Z}=(-1)^{ \overline{a^\dagger} \overline{a}}\equiv (-1)^{ \overline{b^\dagger} \overline{b}}~.
\end{eqs}

We can construct another logical-\(X\) operator by noticing that \(\begin{pmatrix}1 & 0 & 0 & -1\end{pmatrix}\), \(\begin{pmatrix}-1 & 0 & 0 & 1\end{pmatrix}\), \(\begin{pmatrix}0 & -1 & 1 & 0\end{pmatrix}\) and \(\begin{pmatrix}0 & 1 & -1 & 0\end{pmatrix}\) are all logical \(X\) representatives, meaning that
\begin{subequations}
\begin{align}
&\left(\hat{a}_{4}^{\dagger}\hat{a}_{1}+\hat{a}_{1}^{\dagger}\hat{a}_{4}+\hat{a}_{2}^{\dagger}\hat{a}_{3}+\hat{a}_{3}^{\dagger}\hat{a}_{2}\right)\ket{\overline{\pm}}\\=&\pm\left(\hat{n}_{4}+\hat{n}_{1}+\hat{n}_{2}+\hat{n}_{3}\right)\ket{\overline{\pm}}=\pm\Delta\ket{\overline{\pm}}.
\end{align}
\end{subequations}
Hence, we can use a beam-splitter generator as a logical,
\begin{equation}
 \overline{X}=\frac{\hat{a}_{4}^{\dagger}\hat{a}_{1}+\hat{a}_{1}^{\dagger}\hat{a}_{4}+\hat{a}_{2}^{\dagger}\hat{a}_{3}+\hat{a}_{3}^{\dagger}\hat{a}_{2}}{\Delta}~,
\end{equation}
which contains less terms than the expressions obtained from the concatenation point of view. \textcolor{black}{This code has $d_X=2$ and can exactly suppress dephasing error $\mathbf{\hat{a}}^{\dagger \mathbf{p}} \mathbf{\hat{a}}^{ \mathbf{p}}$ for $|\mathbf{p}| < \Delta$, similar to the dephasing suppression of two-mode binomial code.}

\subsection{\texorpdfstring{\(\chi^{(2)}\)}{}-like Code}

We define a three-mode code with
\begin{equation}
    \HX  = \begin{pmatrix}
        2 & 2 & -2
    \end{pmatrix},\quad \HZ  = \begin{pmatrix}
        0 & 1 & 1\\
        1 & 0 & 1
    \end{pmatrix}, \quad \bs{\Delta} = \begin{pmatrix}
        \Delta_1\\\Delta_2
    \end{pmatrix}.
\end{equation}
Picking \(\Delta_1=\Delta_2=\Delta\) yields a code in the same Fock-state support as one of the \(\chi^{(2)}\) codes from Ref.~\cite{niu2018hardware}.
The general case gives rise to two \textcolor{black}{$Z$-type stabilizers},
\begin{subequations}
\begin{align}
\hat{n}_{2}+\hat{n}_{3}-\Delta_{1}\quad\quad\text{and}\quad\quad\hat{n}_{1}+\hat{n}_{3}-\Delta_{2}~,
\end{align}
and two dissipators,
\begin{align}
\hat{a}_{3}^{\dagger2}\hat{a}_{1}^{2}\hat{a}_{2}^{2}-\alpha^{2}\hat{n}_{3}(\hat{n}_{3}-1) ,&\\\hat{a}_{1}^{\dagger2}\hat{a}_{2}^{\dagger2}\hat{a}_{3}^{2}-\hat{n}_{1}(\hat{n}_{1}-1)\hat{n}_{2}(\hat{n}_{2}-1)/\alpha^{2} .&
\end{align}
\end{subequations}

The codespace is a logical qubit, with logical vectors
\begin{equation}
    \mathbf{x} = \begin{pmatrix}
        1 & 1 & -1
    \end{pmatrix},\quad \mathbf{z} = \begin{pmatrix}
        0 & 1 & 0
    \end{pmatrix}.
\end{equation}
The logical \(Z\) operator is \(\overline{Z} = (-1)^{\hat{n}_2}\).

For the logical \(X\), we need to check if \(1\) is in the following ideal, \(I\subset \mathbb{R}\left[n_1,n_2,n_3\right]\):
\begin{equation}
    I = \langle n_2 + n_3 - \Delta_1, n_1 + n_3 - \Delta_2, n_1n_2, n_3 \rangle.
\end{equation}
This is indeed the case, and we can express the logical \(X\) as
\begin{equation}
    \overline{X} = \frac{\alpha^2\hat{a}_1^\dagger\hat{a}_2^\dagger\hat{a}_3 + \left(\hat{n}_2-\Delta_1-\Delta_2\right)\hat{a}_3^\dagger \hat{a}_1\hat{a}_2}{\alpha\Delta_1\Delta_2}~.
\end{equation}

We can write down codewords for \(\Delta_2 \geq \Delta_1\) as 
\begin{align}
    \ket{\overline{\pm}} &= {\textstyle \frac{\alpha^{\Delta_2}}{\sqrt{\mathsf{A}_{\boldsymbol{\Delta}}(\alpha^2)}}}\sum_{m=0}^{\Delta_1}\frac{(\pm \alpha)^m \ket{\Delta_2-\Delta_1+m, m, \Delta_1 - m}}{\sqrt{(\Delta_2-\Delta_1+m)!m!(\Delta_1-m)!}},
\end{align}
where \(\mathsf{A}_{\boldsymbol{\Delta}}(\mathbf{y})\) the GKZ function is a generalized Laguerre polynomial, \(L_n^{(a)}\) \cite{DLMF},
\begin{subequations}
\begin{align}
    \mathsf{A}_{\boldsymbol{\Delta}}(\mathbf{y}) &= \frac{y_1^{\Delta_2}\left(\frac{y_3}{y_1}\right)^{\Delta_1}}{\Delta_2!}\sum_{m=0}^{\Delta_1}\frac{\left(\frac{y_1y_2}{y_3}\right)^m}{m!}\binom{\Delta_2}{\Delta_1-m},\\
    &=\frac{y_1^{\Delta_2}\left(y_3/y_1\right)^{\Delta_1}}{\Delta_2!}L_{\Delta_1}^{(\Delta_2-\Delta_1)}\left(-\frac{y_1y_2}{y_3}\right).\label{eq:Achi2}
\end{align}
\end{subequations}

The degree of dephasing protection depends on \textit{both} \(\boldsymbol\Delta\) and \(\alpha\) in this case.
Taking the case of large \(\Delta_2\geq\Delta_1\rightarrow\infty\), using Eqs.~\eqref{eq:dephasing-calc} and \eqref{eq:Achi2}, and using known behaviors of Laguerre polynomials \cite{szegoOrthogonalPolynomials1939,DLMF}, we obtain the following asymptotic behavior
    \begin{equation}
    \braket{\overline{-}\vert\mathbf{\hat{a}}^{\dagger\mathbf{p}}\mathbf{\hat{a}}^{\mathbf{p}}\vert\overline{+}}\underset{\Delta_2\geq\Delta_1\rightarrow\infty }{\sim}\exp\left(-2\alpha\sqrt{\Delta_1} + \alpha^2\right){\rm poly}(\alpha,\boldsymbol\Delta),
    \end{equation}
where $\mathbf{p}=\begin{pmatrix} p_1& p_2& p_3 \end{pmatrix}$ is such that \(p_2+p_3\leq\Delta_1\), \(p_1+p_3\leq\Delta_2\) and \(p_2\geq p_1\).
We see that dephasing is exponentially suppressed in \(\sqrt{\Delta_1}\), and that there should be an optimal value for \(\alpha^2\) of the order of \(\Delta_1\).
Besides the polynomial factors not displayed here, there are sweet-spots (cf. \cite{li2017cat,albert2019pair}) when varying \(\alpha^2\) that can be used to further suppress the overlap.

\subsection{Calabi-Yau code}

GKZ functions can be used to define algebraic varieties, including those of the Calabi-Yau type \cite[Table 1]{stienstra2007gkz}.
A simple example is the cubic curve lying in the projective plane, which corresponds to the generator matrix
\begin{eqs}
    \HZ=\begin{pmatrix}1 & 1 & 1 & 1\\
0 & 1 & 0 & -1\\
0 & 0 & 1 & -1
\end{pmatrix} .
\end{eqs}
This matrix constrains the total occupation number, along with differences between the fourth mode and the second and third modes.

The kernel of \(\HZ\) is generated by the vector \(\mathbf{t}=\begin{pmatrix}
    -3& 1 &1 &1
\end{pmatrix}\), and we can define a \textcolor{black}{tiger} code \textcolor{black}{encodes a logical qubit} by setting
\begin{eqs}
    \HX=2\mathbf{t}=\begin{pmatrix}-6 & 2 & 2 & 2\end{pmatrix} .
\end{eqs}
The minimal-weight logical operator corresponds to the vector \(\mathbf{t}\), yielding a code distance \(d_X = |\mathbf{t}| = 6\).
In other words, the code detects at least five losses on any mode.

Similar to the extended pair-cat code, the kernel is quite constrained, and we can say more.
In this case, \textit{all} pure-loss vectors are outside of the kernel.
This means that the code can detect \textit{arbitrary} losses on any mode.

Because the all-one vector is in \({\rm im}\HZ\), the dependency in \(\alpha\) disappears in the codewords.
We can evaluate the behavior of the overlap
\begin{equation}
    \braket{\overline{-}\vert\overline{+}} \overset{\Delta\to\infty}{\sim} \left(\frac{\sqrt{13}}{4}\right)^\Delta{\rm poly}(\Delta).
\end{equation}
Since \(\sqrt{13}/4<1\) we infer that dephasing is exponentially suppressed in \(\Delta\).

Using the method presented at the beginning of Section~\ref{sec:finitesupport}, Eq.~\eqref{eq:ideal}, we can solve for a logical-\(X\) using computer software such as \textsc{SageMath} \cite{sagemath}.
We did not find a general expression, but can find a solution for specific values of \(\Delta\).
For instance, when \(\Delta=3\), we obtain
\begin{align}
    \overline{X} = &\frac{765\hat{n}_4^2 - 1332\hat{n}_4 + 575}{8}\hat{a}_1^3\hat{a}_2^\dagger\hat{a}_3^\dagger\hat{a}_4^\dagger \nonumber\\
    &+ \frac{85\hat{n}_4^2 + 22\hat{n}_4 + 4}{24}\hat{a}_1^{\dagger 3}\hat{a}_2\hat{a}_3\hat{a}_4~ .
    \end{align}

It is simple to generalize the \(\HZ\) matrix to \(N\) modes.
Such a generalization constrains the total occupation number along with differences between a central mode and all but one of the remaining modes. 
This corresponds to a family of hypersurfaces in projective space of dimension \(N-2\) \cite{stienstra2007gkz}.
We leave further study of these exotic codes to future work.


\section{Realizing tiger codes}

\textcolor{black}{Tiger codes are stabilized by $X$-type dissipators and $Z$-type stabilizers. The $X$-type dissipators are monomials of annihilation operators, which can be implemented via engineered dissipation, while the $Z$-type stabilizers are linear constraints on occupation numbers. The pair-cat code is a concrete example within this general framework. 
Experimental techniques developed for implementing pair-cat codes are expected to extend naturally to general tiger codes.}

\subsection{Autonomous QEC against dephasing}

$X$-type dissipators for the infinite-support tiger codes are of the form $\mathbf{\hat{a}}^{\bfg}-\alpha^{|\bfg|}$, where $\bfg$ encodes the powers of the loss operators of each mode in the loss-operator monomial.
These monomial-type jump operators are simpler than the polynomial-type jump operators of spherical codes \cite{jain2024quantum}, and there are several proposals to realize them.

The first idea \cite[Sec.~VI]{albert2019pair} proposes the use of an ancillary Josephson junction and lossy cavity to induce a cascade of four-wave mixing process to realize the pair-cat jump operator, for which $\bfg = \begin{pmatrix} 2 & 2 \end{pmatrix}$. 
Extending this proposal would require a separate process for each nonzero coordinate of $\bfg$.

More recent techniques utilize an asymmetrically threaded SQUID, or ATS \cite{lescanne_exponential_2020}, to realize one or more jump operators with the same ancilla.
These techniques, originally proposed to generate the more difficult polynomial dissipators, are outlined in \cite[Supplement]{lescanne_exponential_2020}\cite[Appx. B.2]{chamberland2022building}\cite[Supplement]{jain2024quantum}.

\subsection{Syndrome-based QEC against loss}

$Z$-type \textcolor{black}{stabilizers} are of the form $\bfh\cdot\hat{\mathbf{n}}$, where $\bfh$ \textcolor{black}{specifies} the integer coefficients in the linear combination of occupation-number operators. The $Z$-type syndrome can be extracted by entangling the physical modes with an ancillary mode via a suitable unitary operation, followed by a measurement of the ancilla.

The entangling unitary between physical modes and ancillary mode is the opto-mechanical interaction $\exp[i(\bfh\cdot \hat{\mathbf{n}})\hat{p}_{\text{anc}}]$, which shifts the position of the ancillary mode by the syndrome value. The syndrome extraction circuit of $Z$-type syndrome $\mathbf{h} \cdot \hat{\mathbf{n}}$ is given by
\begin{eqs}\label{eq:Z_syndrome_extraction}
    &\exp[i(\mathbf{h} \cdot \hat{\mathbf{n}}) \hat{p}_{\text{anc}}] \ket{\psi}_{\mathbf{h}\cdot \hat{\mathbf{n}}=\Delta_\mathbf{h}} \otimes \ket{q=0}_{\text{anc}}\\
    =& \ket{\psi}_{\mathbf{h}\cdot \hat{\mathbf{n}}=\Delta_{\mathbf{h}}} \otimes \ket{q=\Delta_{\mathbf{h}}}_{\text{anc}}~,
\end{eqs}
where $\ket{\psi}_{\mathbf{h}\cdot \hat{\mathbf{n}}=\Delta_{\mathbf{h}}}$ denotes a \textcolor{black}{noisy} code state with syndrome $\Delta_{\mathbf{h}}$, and where $\ket{q=0}_{\text{anc}}$ is the position eigenstate of the ancillary mode. After Eq.~\eqref{eq:Z_syndrome_extraction}, the ancillary mode can be measured in the position basis, with the measurement outcome directly yielding the corresponding $Z$-type syndrome $\Delta_{\mathbf{h}}$.
The continuous-position measurement outcome can be rounded to yield the approximate syndrome value.
This proposal has been detailed for the pair-cat code in \cite[Sec.~VII]{albert2019pair} and is generalizable to other tiger codes.

One could also substitute the ancillary momentum operator $\hat{p}_{\text{anc}}$ with the mode's occupation number $\hat{n}_{\text{anc}}$.
Evolving with this interaction Hamiltonian up to a time $2\pi/q$ and appropriately measuring the ancillary mode would extract the syndrome modulo $q$, which is sufficient if done rapidly so that no more than $q$ photons are lost between measurement rounds.
Engineering this interaction requires tuning each occupation-number interaction, known as a cross-Kerr coupling \cite{krantz2019quantum,blais2021circuit}, to the corresponding coordinate of $\bfh$.
Tuning interactions to fixed values is difficult during fabrication, but there are several proposals for doing so dynamically (e.g., \cite{elliott2018designing,chapple2025balanced}).
A proposal in this family was used to create a pair-coherent state in Ref. \cite{gertler2023experimental}.

It is useful to compare syndrome extraction circuits for cat and tiger codes. 
The $Z$-type syndrome extraction unitary for a four-component cat code is given by $\exp(i \pi \hat{n} \sigma^{\text{anc}}_x)$ , which flips the $Z$ eigenvalue of the ancillary qubit by the syndrome value. This unitary can be implemented by evolving the dispersive interaction $\hat{n} \sigma^{\text{anc}}_x$ for a duration of $\pi$. The corresponding syndrome extraction circuit is:
\begin{eqs}\label{eq:cat_syndrome}
    &\exp(i \pi \hat{n} \sigma^{\text{anc}}_x) \ket{\Phi}_{\hat{n} \Mod 2=\Delta}\otimes \ket{0}_{\text{anc}}\\
    =&\ket{\Phi}_{\hat{n} \Mod 2=\Delta}\otimes \ket{\Delta}_{\text{anc}} ~,
\end{eqs}
where $\ket{\Phi}_{\hat{n} \Mod 2=\Delta}$ denotes a \textcolor{black}{noisy} code state with parity syndrome $\Delta$, and where $\ket{0}_{\text{anc}}$ is the $Z$ eigenstate of the ancillary qubit. After Eq.~\eqref{eq:cat_syndrome}, the ancillary qubit is measured in its $Z$-basis, whose measurement result gives the cat code's syndrome. Note that the dispersive interaction $\hat{n} \sigma^{\text{anc}}_x$ used in the cat code only commutes with the dissipator when the evolution time is an integer multiple of $\pi$. 
This means that the dissipation \(X\)-type protection has to be turned off during the ramp-up time for any cat code with four or more components as well as any of its concatenations. 
In contrast, the time evolution [see Eq.~\eqref{eq:Z_syndrome_extraction}] generated by the linear constraints of a tiger code commutes with all \(X\)-type dissipators for all times, so those dissipators need not be turned off during syndrome extraction. 
This means that tiger codes offer more protection when syndrome extraction is performed using cross-Kerr coupling.

In passing, we note that continuous (i.e., non-syndrome-based) protection against loss has been proposed for the pair-cat code \cite[Sec.~VIII]{albert2019pair} and may be extendable to general tiger codes.

\begin{figure}[t]
    \centering
    \includegraphics[width=0.4\textwidth]{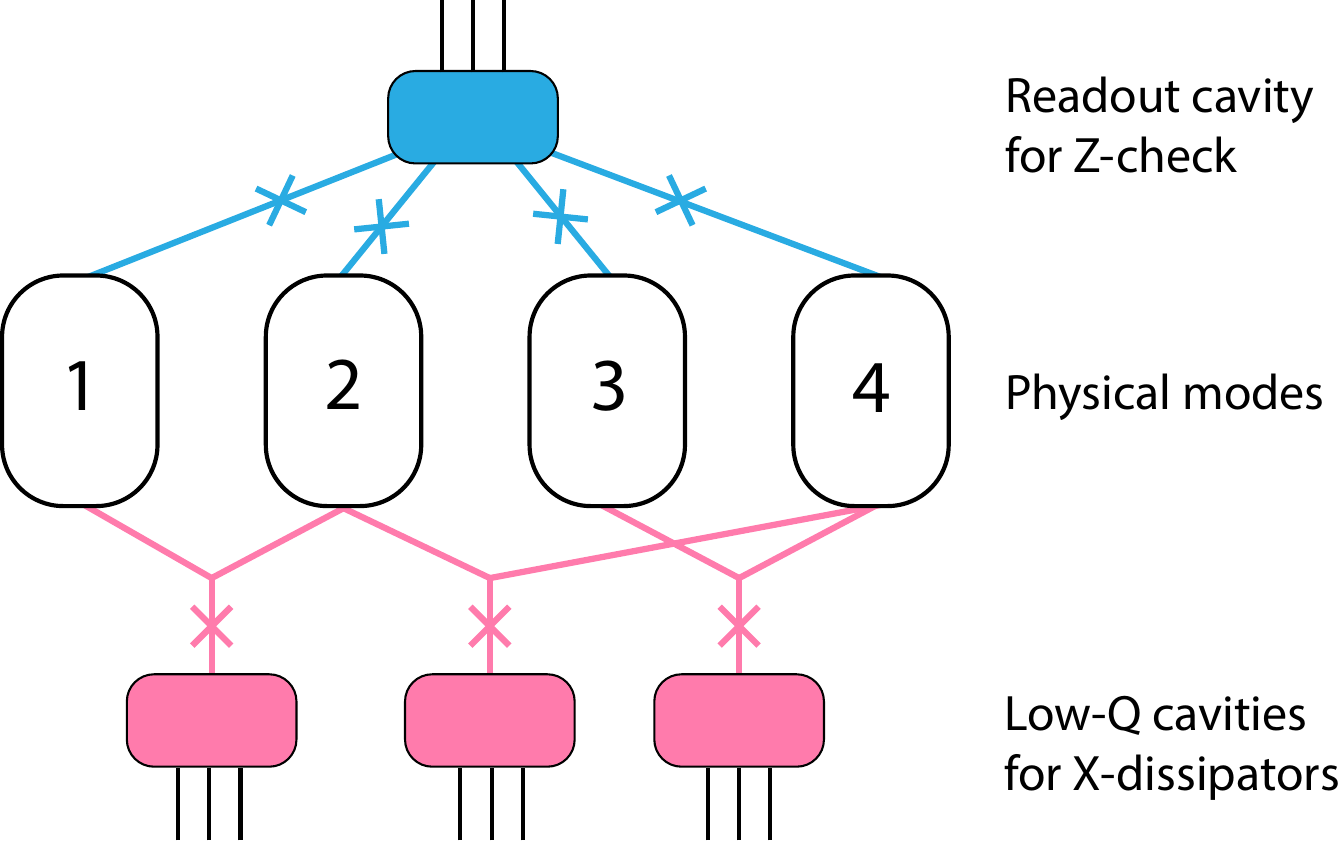}
    \caption{Potential implementation of four-mode tiger codes is analogous to the realization of the pair-cat code in Ref.~\cite{albert2019pair}. A line with a cross-sign represents a Josephson junction ancilla, pink rounded rectangles represent low-Q cavities for entropy extraction, and the blue rounded rectangle at the top represents the readout cavity for $Z$-syndrome extraction. 
    }
    \label{fig:four-mode-tiger}
\end{figure}

\subsection{Example: realizing the four-mode code}

We outline the resource overhead required to realize dissipators and syndrome extraction for the four-mode tiger code.

All $X$-type dissipators in the code are two-body interactions, comprising two degree-2 terms and one degree-4 term. 
Although each dissipator acts on a different subset of modes, the degree-2 terms correspond to the same dissipator that stabilizes pair-coherent states, which is implemented in Ref.~\cite{gertler2023experimental}, while the degree-4 term matches the dissipator used to stabilize pair-cat codes. Each of these dissipators can be implemented by coupling the relevant physical modes to a three-level Josephson junction, driven by a multi-frequency pulse to engineer the desired dissipation \cite[Sec.
VI]{albert2019pair}. Each Josephson junction is coupled to a low-Q cavity that performs entropy extraction. Specifically, realizing the three $X$-type dissipators requires three auxiliary Josephson junctions coupled respectively to the mode pairs $\{\hat{a}_1,\hat{a}_2\}, \{\hat{a}_3,\hat{a}_4\}, \{\hat{a}_2,\hat{a}_4\}$, with appropriate coupling and drive schemes. 

The $Z$-type syndrome measurement, analogous to that in the pair-cat code, can be implemented by coupling each physical mode to a readout cavity via a Josephson junction. The effective interaction induces a conditional displacement of the readout cavity whose amplitude is proportional to the $Z$-syndrome \cite[Sec.~VII]{albert2019pair}.

The hardware cost of realizing each individual $X$-type dissipator and the $Z$-type \textcolor{black}{stabilizer} is comparable to that of the pair-cat code.  The main experimental challenge lies in integrating these components coherently, which demands multiple simultaneous ancilla couplings and driving strengths. The circuit layout is shown in Fig.~\ref{fig:four-mode-tiger}, where the pink parts label ancillary systems for $X$-type dissipators, and where the blue part corresponds to the $Z$-type syndrome extraction system.

\section{Discussion and outlook}

We construct CSS-like intrinsically multimode "tiger codes" from two integer-valued generator matrices, $\HX$ and $\HZ$, satisfying a homological constraint.
The matrix $\HX $ defines code dissipators for autonomous stabilization, while $\HZ $ defines linear constraints on the modes' occupations. 
Our framework yields bosonic codes that are not constructed via concatenation with a qubit \textcolor{black}{stabilizer} code.

Tiger codewords maintain the features of the pair-cat code \cite{albert2019pair} that made it intrinsically amenable to realization.
Namely, code syndromes are linear Fock-space constraints, and extracting them using Hamiltonian evolution is possible without having to turn off any dissipator-induced autonomous stabilization.

Tiger codewords are superpositions of infinitely or finitely many Fock states, and can encode logical qubits, qudits, rotors, or modes. 
Previously known codes, such as the pair-cat code, coherent-state and Fock-state repetition codes, pair-coherent state codes, dual-rail codes, two-mode binomial codes, and aspects of \(\chi^{(2)}\) codes can all be described as instances of a tiger code for specific choices of generator matrices. 
We break new ground with codes based on homology over the integers, the theory of Gelfand-Kapranov-Zelevinsky (GKZ) hypergeometric functions, and algebraic varieties such as those of Calabi-Yau type.

For example, we construct a tiger surface code and its "liger" (long-tiger) version defined in a 2D strip, using a hypergraph product of two repetition codes over the integers.
Its stabilizers are geometrically local and constant-weight, and the code has increasing distance against occupation-number loss and dephasing error. 
This is the first example of an \textit{intrinsically} multimode bosonic code with topological protection, opening a new avenue for constructing topological bosonic codes and dissipative topological phases of matter.
It will be interesting to see what types of phases can be constructed in our framework, and to make contact with phase classification of open systems \cite{lieu2020symmetry,gravina2023critical,de2022symmetry,sa2023symmetry,ma2023average,kawabata2023symmetry}.
It could also be interesting to see how to classify the wide landscape of tiger codes in the language of approximate QEC \cite{yi2024complexity}.

Tiger codewords lie in a subspace of multimode Fock states \(|\mathbf{n}\rangle\) defined by the constraint \(\HZ \mathbf{n} = \boldsymbol{\Delta}\) for some vector \(\boldsymbol{\Delta}\). While the value of $\boldsymbol{\Delta}$ doesn't change the asymptotic behavior of the simpler pair-cat code, it does play an important role for other tiger codes. 
We show that a Shor-like four-mode tiger code as well as the liger surface code, have \textcolor{black}{both} orthogonal $X$- \textcolor{black}{and $Z$}-bases for certain values of $\boldsymbol{\Delta}$, \textcolor{black}{regardless of its energy.} 
Such codes have a counter-intuitive property in that they are based on non-orthogonal coherent states, yet can protect against dephasing without the need of a large-energy limit.

We demonstrate that each tiger code is naturally associated with a corresponding GKZ hypergeometric function \textcolor{black}{which describes the behavior of high-dimensional curves}.
Furthermore, we show that properties (such as code distances) of a given tiger code can be obtained by analyzing its corresponding GKZ hypergeometric functions. Informally, the GKZ hypergeometric function can be viewed as a "generating function" for the tiger code. The Knill-Laflamme condition 
can be readily obtained by taking partial derivatives of the corresponding GKZ function. Therefore, useful tiger codes should be obtainable by searching for GKZ functions with desired properties. GKZ functions may help to draw connections between codes with other fields in physics or mathematics. For example, they are widely discussed in the context of geometric analysis \cite{hosono1996gkz,stienstra2007gkz,hrabowski1985multiple} and Feynman path integrals \cite{nasrollahpoursamami2016periods}.
There may be connections between the integer homology behind tiger codes and related properties of GKZ functions \cite{borisov2013better}, \textcolor{black}{and established formulas for asymptotics of GKZ functions with \(\boldsymbol{\Delta}\) may prove useful~\cite[Thm. 6]{takayama2018hypergeometric}}.
There may also be connections with Grassmanian GKZ functions \cite{gelfand1990gamma,gel1992general,gelfand1993reduction,gelfand1999gg,gelfand1997combinatorics} and Grassmanian coherent states \cite{freidel2011u,calixto2014coherent}.

While we do provide a universal gate set, not all gates are fault tolerant.
However, the homological framework allows for two avenues for developing such gates.
One avenue is to extend qubit gates, e.g., for hypergraph product codes over $\mathbb{Z}_2$~\cite{krishna2021ft, quintavallePartitioningQubitsHypergraph2023} or more general codes \cite{fibrebundle,freedman2021building,gunderman2020local,gunderman2022degenerate,gunderman2024stabilizer,gunderman2025beyond}, to the setting of integer homology.
The second is to extend existing few-mode bosonic protocols, e.g., fault-tolerant schemes for rotation symmetric bosonic codes \cite{grimsmo2020quantum,totey2023performance,marinoff2024explicit} or bias-preserving gates for the pair-cat code \cite{yuan2022construction}.
Our higher-order phase gates in Table~\ref{tab:pair-cat} suggest the presence of a rotor analogue of the Clifford hierarchy \cite{gottesman1999demonstrating,cui2017diagonal,Rengaswamy2019unifying,anderson2024groups,he2024permutation}, whose development and potential connection to the bosonic binary expansion may be interesting directions.

The fault tolerance and threshold properties of the tiger code remain important topics for future research. Prior works \cite{dennis2002topological,chubb2021statistical} have shown that the optimal threshold of discrete-variable stabilizer codes can be determined by analyzing phase transitions in their associated statistical-mechanical models. Extending this statistical-mechanical mapping to the continuous-variable setting, as in \cite{vuillot_toricGKP_2019} for instance, and applying it to investigate the threshold behavior of tiger codes is an interesting future direction.

Hardware-efficient implementations of a concatenated cat codes have been proposed \cite{guillaud2019repetition,darmawan2021practical,chamberland2022building} 
and the repetition-cat code is recently implemented  \cite{putterman2024hardware} in the superconducting-circuit platform. 
Pair-coherent states have also been realized with similar technology \cite{gertler2023experimental}.
While requiring many modes, the hardware requirements \textit{per mode} are arguably less for our codes, as non-concatenated tiger codes avoid the need for individual mode stabilization, making them a viable candidate for future blueprints.

The $t$-design is widely used in the construction of quantum spherical codes \cite{jain2024quantum}, whose codewords are discrete superpositions of coherent states. Its recent extension, the curve $t$-design \cite{ehler2023t}, is a set of curves such that the average of a degree-$t$ polynomial over the curves is equal to that over the entire sphere. 
In the case of tiger codes, whose codewords are continuous superpositions of coherent states along curves, detecting up to degree-$t$ photon loss implies that the expectation values of degree-$t$ polynomials in $\hat{a}$ are indistinguishable between codewords. 
This suggests that curve \(t\)-designs may give rise to new tiger codes, and it remains to determine whether such curves may be stabilized by tiger-code check matrices.

Four- and higher-component cat codes \cite{mirrahimi2014dynamically} and their concatenations are \textit{not} tiger codes since their codespaces are defined by non-linear (parity) occupation-number constraints.
Analogously, single- and two-mode binomial codes which contain a gap, or "spacing", between Fock states in the support of different codewords are not tiger codes.
While codes with linear syndromes are less taxing in terms of resource overhead, they come at a price of requiring a higher number of modes for the same degree of protection.
It will be interesting to develop hybrid codes that are tailored to particular architectures and that strike an appropriate balance between syndrome complexity and degree of protection.

\section*{Acknowledgment}

After completion of this work, we learned of the pre-print \cite{magdalena2024topological} that describes new families of topological bosonic stabilizer
codes (which include analog/Gaussian stabilizer codes and
Gottesman-Kitaev-Preskill codes) using the tools from anyon theory.
The stabilizers of those are discrete or continuous groups of quadrature displacements, and their codewords consist of non-compact superpositions of coherent states. Both our work and the pre-print utilize integer homology \cite{vuillotHomologicalQuantumRotor2024} and develop intrinsically multimode codes with topological protection, but the classes of codes and the primary research focuses are, to our knowledge, completely distinct.

Y.X. thanks En-Jui Kuo for helpful discussions about asymptotic behavior of hypergeometric series.
C.V. thanks Sébastien Draux for helpful discussions about polynomial ideals.
V.V.A.~thanks Howard S.~Cohl for showing us the path to the GKZ hypergeometric function, and  Bonita V.~Saunders for clarifying discussions. 
C.V. acknowledges funding from the Plan France 2030 through the project ANR-22-PETQ-0006.
V.V.A.~acknowledges NSF grants OMA2120757 (QLCI) and CCF2104489. 
Y.W. acknowledges fundings from the National Natural Science Foundation of China Grant No. 12347173, China Postdoctoral Science Foundation Grant No. 2023M742003, and the Shuimu Tsinghua Scholar Program. Y.X.~thanks Yujie Zhang and Yilun Li for hosting the research visits in Tokyo.
V.V.A.~thanks Ryhor Kandratsenia and Olga Albert for providing daycare support throughout this work.

\appendix 
\section{GKZ Hypergeometric functions}\label{sec:GKZ}

Tiger codes have a close connection to the theory of Gelfand-Kapranov-Zelevinsky (GKZ) hypergeometric functions \cite{gel1986general,gel1986generalized,gel1987holonomic,gel1989hypergeometric,adolphson1994hypergeometric,gel1990hypergeometric,gel1992general}, and we collect the various connections and calculations associated with them in this section.
We recommend Refs.~\cite{saito1999hypergeometric,cattani2006three,beukers2011notes,aomoto2011theory,saito2013grobner,Takayama_2020,reichelt2021algebraic} for clear introductions into GKZ theory.

Recall from Eq~\eqref{eq:GKZintegralsum} the integral and sum forms of the GKZ function, reproduced below.
For \(\mathbf{y} = (y_1,y_2, \cdots,y_N)\in\mathbb{C}^N\), 
\begin{subequations}\label{eq:GKZintegralsum2}
\begin{align}
  \mathsf{A}_{\boldsymbol{\Delta}}(\mathbf{y})&=\int\frac{d^{r_{z}}\boldsymbol{\phi}}{(2\pi)^{r_{z}}}e^{-i\boldsymbol{\phi}\cdot\boldsymbol{\Delta}}\exp\left({\textstyle \sum_{j=1}^{N}}y_{j}e^{i(\boldsymbol{\phi}\HZ)_{j}}\right)\label{eq:GKZintegral2},\\
  &=\sum_{\HZ\mathbf{n}=\boldsymbol{\Delta}}\frac{\mathbf{y}^{\mathbf{n}}}{\mathbf{n}!}~,\label{eq:GKZsum2}
\end{align}
where the index \(\mathbf{n}\) goes over non-negative integer vectors, and where we use multi-index notation, e.g., \(\mathbf{y}^{\mathbf{n}}=y_{1}^{n_{1}}y_{2}^{n_{2}}\cdots y_{N}^{n_{N}}\).
The sum can be obtained by expanding the exponential in each \(y_j\) and integrating term by term.
Infinite (finite) sums correspond to projected coherent states supported on an infinite-dimensional (finite-dimensional) Fock-state subspace.

A third form is possible for the infinite-support case if there exists a non-negative integer vector \(\mathbf{s}\) satisfying \(\HZ \mathbf{s} = \boldsymbol{\Delta}\) such that \cite[Lemma~1]{gel1989hypergeometric}\cite{gel1987holonomic}
\begin{eqs}
    \mathsf{A}_{\boldsymbol{\Delta}}(\mathbf{y})=\sum_{\mathbf{n}\in\ker \HZ}\frac{\mathbf{y}^{\mathbf{n}+\mathbf{s}}}{(\mathbf{n}+\mathbf{s})!}\label{eq:GKZkernel}~.
\end{eqs}
\end{subequations}

\subsection{From projected coherent states to GKZ via the Segal-Bargmann mapping}

Both GKZ functions and projected coherent states are defined by the same variables, \(\HZ\) and \(\boldsymbol{\Delta}\).
Moreover, GKZ functions govern overlaps between their corresponding states, as evident by Eq.~\eqref{eq:gkz-overlap}.
This suggests a deep correspondence between appropriate GKZ functions and projected coherent states.

Here, we show that expressing projected coherent states in the Segal-Bargmann representation yields their associated GKZ functions, with arguments scaled by \(\boldsymbol{\alpha}\).
A similar correspondence holds between operator relations of projected coherent states and the defining relations of GKZ functions.

The Segal-Bargmann mapping is a representation of functions in \(L^2(\mathbb{R}^N)\) by holomorphic functions \(F\) in \(N\) complex variables \(z_j\) \cite{bargmann1961hilbert,segal1963mathematical,klauder1985coherent,dodonov2002nonclassical,gazeau2009coherent}.
Each Fock state \(\ket{\bs{n}}\) maps to a monomial term \(\bs{z}^\bs{n}\) such that a general state maps as follows,
\begin{equation}
    \ket{\Psi} = \sum_{\bs{n}\in\mathbb{N}^N}\beta_{\bs{n}}\ket{\bs{n}} \quad\mapsto\quad F_\Psi(\bs{z}) = \sum_{\bs{n}\in\mathbb{N}^N}\frac{\beta_{\bs{n}}}{\sqrt{\bs{n}!}}{\bs{z}}^{\bs{n}}.
\end{equation}

Un-normalized coherent states map to exponentials,
\begin{equation}
   \sum_{\mathbf{n}\in\mathbb{N}^{N}}\frac{\boldsymbol{\alpha}^{\mathbf{n}}}{\sqrt{\mathbf{n}!}}|\mathbf{n}\rangle\quad\mapsto\quad\sum_{\mathbf{n}\in\mathbb{N}^{N}}\frac{\boldsymbol{\alpha}^{\mathbf{n}}}{\mathbf{n}!}\mathbf{z}^{\mathbf{n}}=\exp(\boldsymbol{\alpha}\cdot\mathbf{z})\,.
\end{equation}
Such states correspond to the trivial version of projected coherent states, i.e., with no projection.
More generally, unnormalized projected coherent states map to GKZ functions whose arguments are rescaled by \(\boldsymbol{\alpha}\),
\begin{eqs}
    \sum_{\HZ\mathbf{n}=\boldsymbol{\Delta}}\frac{\boldsymbol{\alpha}^{\mathbf{n}}}{\sqrt{\mathbf{n}!}}|\mathbf{n}\rangle\quad\mapsto\quad\sum_{\HZ\mathbf{n}=\boldsymbol{\Delta}}\frac{\boldsymbol{\alpha}^{\mathbf{n}}}{\mathbf{n}!}\mathbf{z}^{\mathbf{n}}=\mathsf{A}_{\boldsymbol{\Delta}}(\boldsymbol{\alpha}\mathbf{z})~,
\end{eqs}
for argument \(\boldsymbol{\alpha}\mathbf{z}=(\alpha_{1}z_{1},\alpha_{2}z_{2},\cdots,\alpha_{N}z_{N})\).

GKZ functions satisfy certain defining relations, and the Segal-Bargmann representation can be used to obtain those from the coherent-state relations used in this work.
Operators \(\hat{a}_j\), \(\hat{a}^\dagger_j\) and \(\hat{n}_j\) have a simple interpretation,
\begin{equation}
    \hat{a}_j \mapsto {\partial_j},\quad \hat{a}^\dagger_j \mapsto z_j,\quad  \hat{n}\mapsto z_j{\partial_j}~, 
\end{equation}
where the partial derivative \(\partial_j = \partial/\partial z_j\).

Expressing the \(\HZ \) constraints of a tiger code in the Segal-Bargmann representation gives rise to a system of differential equations.
For each row \(\bfh\) of \(\HZ \), we have
\begin{equation}
    \left(\bfh\cdot\hat{\bs{n}} - \Delta_{\bfh}\right)\ket{\Psi}\mapsto\left(\sum_{j=1}^N h_j z_j{\partial_j}-\Delta_{\bfh}\right)F_{\Psi}(\bs{z}) = 0.
\end{equation}
Such a set of equations is one of the two defining constraints of GKZ functions, in which \(\HZ \) is often called \(A\).

We can also put creation and annihilation operator relations, Eq.~\eqref{eq:general_a_dissipator}, into Segal-Bargmann form.
For each row, \(\bfg=(\bs{p}-\bs{q})\in\ker \HZ \), and after a little rewriting 
\begin{subequations}
\begin{align}
\label{eq:diss1}
\left(\mathbf{\hat{a}}^{\dagger\bs{q}}\mathbf{\hat{a}}^{\bs{p}}-\boldsymbol{\alpha}^{\bfg}\left(\hat{\bs{n}}\right)_{\bs{q}}\right)\ket{\Psi}&=0,
\\
\label{eq:diss2}\mathbf{\hat{a}}^{\dagger\bs{q}}\left(\left(\mathbf{\hat{a}}/\boldsymbol{\alpha}\right)^{\bs{p}}-\left(\mathbf{\hat{a}}/\boldsymbol{\alpha}\right)^{\bs{q}}\right)\ket{\Psi}&=0,\\\mapsto\left(\prod_{j}\partial_{j}^{p_{j}}-\prod_{j}\partial_{j}^{q_{j}}\right)F_{\Psi}(\boldsymbol{\alpha}\bs{z})&=0.\label{eq:gkz-defining}
\end{align}
\end{subequations}
Above, prior to applying the mapping, we got rid of the prefactor \(\mathbf{\hat{a}}^{\dagger\bs{q}}\) since it has no right zero eigenstate and so does not change the constraint in itself.
The above mapped constraints form the second defining ingredient of a GKZ system.
The above holds true for any vector in the kernel of \(\HZ \), which, in the case of tiger codes, is partitioned into dissipators generated by \(\HX \) and logicals generated by \(L_X\) (see Sec.~\ref{sec:homology}).

We use the form~\eqref{eq:diss1} for the dissipators in the main text and assume \(\bs{q} = \bs{0}\) for codes with infinite support.
This is not necessary in general, and dissipators of the form \(\mathbf{\hat{a}}^{\bs p} - \mathbf{\hat{a}}^{\bs q}\)~\eqref{eq:diss2} also exist for both finite or infinite support codes.

\subsection{Dephasing matrix elements of projected coherent states}

Matrix elements of dephasing terms, \(\mathbf{\hat{a}}^{\dagger\mathbf{p}}\mathbf{\hat{a}}^{\mathbf{p}}\), with respect to projected coherent states can be calculated using either their integral or sum form, reproducing a non-trivial identity on their corresponding GKZ functions.
This identity leads to relations between expectations values for different \(\mathbf{p}\).

Matrix elements for general projected coherent states \(|\boldsymbol{\alpha}\rangle_{\boldsymbol{\Delta}}^{\HZ}\) and \(|\boldsymbol{\beta}\rangle_{\boldsymbol{\Delta}}^{\HZ}\)  can be obtained by a slight generalization of the calculation in Eq.~\eqref{eq:dephasing-calc}, yielding
\begin{eqs}
    _{\boldsymbol{\Delta}}^{\HZ}\langle\boldsymbol{\alpha}|\mathbf{\hat{a}}^{\dagger\mathbf{p}}\mathbf{\hat{a}}^{\mathbf{p}}|\boldsymbol{\beta}\rangle_{\boldsymbol{\Delta}}^{\HZ}\propto(\boldsymbol{\alpha}^{\star}\boldsymbol{\beta})^{\mathbf{p}}{\mathsf{A}}_{\boldsymbol{\Delta}-\HZ\mathbf{p}}(\boldsymbol{\alpha}^{\star}\boldsymbol{\beta})\,,
\end{eqs}
where the proportionality sign hides normalization factors in the denominator.

The same matrix elements can be obtained by converting Fock-state relations into differential form,
\begin{subequations}
    \begin{align}
        _{\boldsymbol{\Delta}}^{\HZ}
        \langle\boldsymbol{\alpha}|\mathbf{\hat{a}}^{\dagger\mathbf{p}}\mathbf{\hat{a}}^{\mathbf{p}}|\boldsymbol{\beta}\rangle_{\boldsymbol{\Delta}}^{\HZ}&\propto\sum_{\HZ\mathbf{n}=\boldsymbol{\Delta}}\frac{(\boldsymbol{\alpha}^{\star}\boldsymbol{\beta})^{\mathbf{n}}}{\mathbf{n}!}\langle\mathbf{n}|\mathbf{\hat{a}}^{\dagger\mathbf{p}}\mathbf{\hat{a}}^{\mathbf{p}}|\mathbf{n}\rangle,\\
        &=\sum_{\HZ\mathbf{n}=\boldsymbol{\Delta}}\frac{(\boldsymbol{\alpha}^{\star}\boldsymbol{\beta})^{\mathbf{n}}}{\mathbf{n}!}\frac{\mathbf{n}!}{(\mathbf{n}-\mathbf{p})!},\\
        &=\left.\sum_{\HZ\mathbf{n}=\boldsymbol{\Delta}}\mathbf{y}^{\mathbf{p}}\boldsymbol{\partial}^{\mathbf{p}}\frac{\mathbf{y}^{\mathbf{n}}}{\mathbf{n}!}\right|_{\mathbf{y}=\boldsymbol{\alpha}^{\star}\boldsymbol{\beta}},\\
        &=\left.\mathbf{y}^{\mathbf{p}}\boldsymbol{\partial}^{\mathbf{p}}{\mathsf{A}}_{\boldsymbol{\Delta}}(\mathbf{y})\right|_{\mathbf{y}=\boldsymbol{\alpha}^{\star}\boldsymbol{\beta}}
    \end{align}
\end{subequations}
where \(\boldsymbol{\partial}^{\mathbf{p}}=(\partial/\partial y_{1})^{p_{1}}\cdots(\partial/\partial y_{N})^{p_{N}}\).
Equating the above two equations yields an identity between GKZ functions for different \(\boldsymbol{\Delta}\).

The above identity, together with the defining GKZ relation~\eqref{eq:gkz-defining}, \(\boldsymbol{\partial}^{\mathbf{p}}{\mathsf{A}}_{\boldsymbol{\Delta}}(\mathbf{y})=\boldsymbol{\partial}^{\mathbf{q}}{\mathsf{A}}_{\boldsymbol{\Delta}}(\mathbf{y})\) for differences of non-negative vectors, \(\mathbf{p}-\mathbf{q}\in \ker\HZ\), can be used to relate different dephasing matrix elements as
\begin{eqs}
    \mathbf{(\boldsymbol{\alpha}^{\star}\boldsymbol{\beta})}^{\mathbf{q}}{\,}_{\boldsymbol{\Delta}}^{\HZ}\langle\boldsymbol{\alpha}|\mathbf{\hat{a}}^{\dagger\mathbf{p}}\mathbf{\hat{a}}^{\mathbf{p}}|\boldsymbol{\beta}\rangle_{\boldsymbol{\Delta}}^{\HZ}=\mathbf{(\boldsymbol{\alpha}^{\star}\boldsymbol{\beta})}^{\mathbf{p}}{\,}_{\boldsymbol{\Delta}}^{\HZ}\langle\boldsymbol{\alpha}|\mathbf{\hat{a}}^{\dagger\mathbf{q}}\mathbf{\hat{a}}^{\mathbf{q}}|\boldsymbol{\beta}\rangle_{\boldsymbol{\Delta}}^{\HZ}~.
\end{eqs}
This relation reduces the number of matrix elements one needs to calculate in order to determine the degree of dephasing protection of a tiger code.

\subsection{Asymptotics of GKZ integrals}

We can estimate some GKZ integrals for large arguments using the method of steepest descents and the saddle-point approximation, beautifully explained in \cite[Sec.~6.6]{bender2013advanced}.
This yields the degree to which dephasing matrix elements are exponentially suppressed, allowing us to extract the exact value of the exponent and compare it to a given tiger code's \(Z\)-distance. 

Roughly, the method of steepest descents estimates an integral asymptotically in a limit by re-writing it as a complex contour integral, choosing a contour that goes through \textit{saddle points} (at which the phase of the integrand is constant), and approximating the contour integral by Gaussian integrals centered at said points.

To simplify notation, we re-define variables as
\begin{eqs}
    \mathsf{A}(\mathbf{y})=\int\frac{d^{r_{z}}\phi}{(2\pi)^{r_{z}}}\exp\left(-i\sum_{j=1}^{r_{z}}\phi_{j}\Delta_{j}+\sum_{k=1}^{N}|y_{k}|\rho_{k}\right),
\end{eqs}
where $y_k=|y_k|e^{i\theta_k}$, and where
\begin{eqs}
   \rho_{k}\equiv\exp\left[i\left(\sum_{j=1}^{r_{z}}(\HZ)_{jk}\phi_{j}+\theta_{k}\right)\right] .
\end{eqs}

To find a saddle-point value of \(\boldsymbol{\phi}\), we have to solve the following equations,
\begin{equation}\label{eq:saddleofintegral}
\begin{split}
    &\Vec{\nabla}\left(\sum_{k=1}^N |y_k| \rho_k- \sum_{j=1}^{r_{z}} \phi_i \Delta_j \right)=0\\ \implies &\sum_{k=1}^N \rho_k |y_k| (\HZ )_{jk}=\Delta_j~ ,~ \forall j=1,2,\cdots, r_{z} ~.
\end{split}
\end{equation}
We can write $z_j=e^{i \phi_j}$, and the above  becomes
\begin{equation}
 \sum_{k=1}^{N}y_{k}\prod_{j=1}^{r_{z}}z_{j}^{(\HZ)_{jk}}(\HZ)_{lk}=\Delta_{l}~ ,~\forall l=1,2,\cdots,r_{z}~.
\end{equation}

There are $r_{z}$ equations for $r_{z}$ variables above, so solutions exist in principle. 
While it is hard to solve these polynomial equations in general, they can be simplified in some special cases.

The normalization function of tiger codewords, \(\mathsf{A}_{\boldsymbol{\Delta}}(\alpha^2 \mathbf{1})\), is one such case, corresponding to equal GKZ arguments $y_k=\alpha^2$ for real \(\alpha\).
Below, we work in the limit of large energy density, $\alpha^2 \gg \Delta_l$ for all rows $l$, but can also perform the opposite limit for the finite-support cases.
For the former, the saddle point equation can be approximated by 

\begin{equation}
   \sum_{k=1}^{N}\prod_{j=1}^{r_{z}}z_{j}^{(\HZ)_{jk}}(\HZ)_{lk}\approx0,~\forall l=1,2,\cdots,r_{z}~.
\end{equation}

Imposing the additional constraint that the all-ones vector $\mathbf{1} \in \ker \HZ $, we see that $z_j=1$ $(\phi_j=0)$ for all $j$ is a valid solution to the saddle-point equations. In fact, this solution gives the dominant saddle because $N \alpha^2$ is the largest possible exponent. Because the normalization is real, and because $z_j=1~(\phi_j=0)~\forall j$ is the only dominant saddle, we can expand the saddle and perform a Gaussian integration along the real direction of all $\phi_j$'s to obtain the leading-order asymptotic behavior of the tiger-code normalization functions, 
\begin{equation}\label{eq:GKZ_normalization}
    \mathsf{A}(\alpha^{2}\mathbf{1})\sim\frac{\exp(N\alpha^{2})}{(2\pi\alpha^{2})^{r_{z}/2}}\det\left(\HZ\HZ^{\text{T}}\right)^{-1/2} .
\end{equation}

Performing a similar analysis for the more general case of \(\theta_k \neq 0\) requires solving explicitly the saddle-point equations and the tangent direction of the constant phase hyperplane to perform the integration about a possibly complex contour in the enlarged parameter space of \(\boldsymbol{\phi}\).

\bibliography{biblo.bib}

\begin{thebibliography}{148}%
\makeatletter
\providecommand \@ifxundefined [1]{%
 \@ifx{#1\undefined}
}%
\providecommand \@ifnum [1]{%
 \ifnum #1\expandafter \@firstoftwo
 \else \expandafter \@secondoftwo
 \fi
}%
\providecommand \@ifx [1]{%
 \ifx #1\expandafter \@firstoftwo
 \else \expandafter \@secondoftwo
 \fi
}%
\providecommand \natexlab [1]{#1}%
\providecommand \enquote  [1]{``#1''}%
\providecommand \bibnamefont  [1]{#1}%
\providecommand \bibfnamefont [1]{#1}%
\providecommand \citenamefont [1]{#1}%
\providecommand \href@noop [0]{\@secondoftwo}%
\providecommand \href [0]{\begingroup \@sanitize@url \@href}%
\providecommand \@href[1]{\@@startlink{#1}\@@href}%
\providecommand \@@href[1]{\endgroup#1\@@endlink}%
\providecommand \@sanitize@url [0]{\catcode `\\12\catcode `\$12\catcode `\&12\catcode `\#12\catcode `\^12\catcode `\_12\catcode `\%12\relax}%
\providecommand \@@startlink[1]{}%
\providecommand \@@endlink[0]{}%
\providecommand \url  [0]{\begingroup\@sanitize@url \@url }%
\providecommand \@url [1]{\endgroup\@href {#1}{\urlprefix }}%
\providecommand \urlprefix  [0]{URL }%
\providecommand \Eprint [0]{\href }%
\providecommand \doibase [0]{http://dx.doi.org/}%
\providecommand \selectlanguage [0]{\@gobble}%
\providecommand \bibinfo  [0]{\@secondoftwo}%
\providecommand \bibfield  [0]{\@secondoftwo}%
\providecommand \translation [1]{[#1]}%
\providecommand \BibitemOpen [0]{}%
\providecommand \bibitemStop [0]{}%
\providecommand \bibitemNoStop [0]{.\EOS\space}%
\providecommand \EOS [0]{\spacefactor3000\relax}%
\providecommand \BibitemShut  [1]{\csname bibitem#1\endcsname}%
\let\auto@bib@innerbib\@empty
\bibitem [{\citenamefont {Shor}(1999)}]{shor1999polynomial}%
  \BibitemOpen
  \bibfield  {author} {\bibinfo {author} {\bibfnamefont {Peter~W}\ \bibnamefont {Shor}},\ }\bibfield  {title} {\enquote {\bibinfo {title} {Polynomial-time algorithms for prime factorization and discrete logarithms on a quantum computer},}\ }\href {https://doi.org/10.1137/S0036144598347011} {\bibfield  {journal} {\bibinfo  {journal} {SIAM review}\ }\textbf {\bibinfo {volume} {41}},\ \bibinfo {pages} {303--332} (\bibinfo {year} {1999})}\BibitemShut {NoStop}%
\bibitem [{\citenamefont {Bennett}\ and\ \citenamefont {Brassard}(2014)}]{bennett2014quantum}%
  \BibitemOpen
  \bibfield  {author} {\bibinfo {author} {\bibfnamefont {Charles~H}\ \bibnamefont {Bennett}}\ and\ \bibinfo {author} {\bibfnamefont {Gilles}\ \bibnamefont {Brassard}},\ }\bibfield  {title} {\enquote {\bibinfo {title} {Quantum cryptography: Public key distribution and coin tossing},}\ }\href {\doibase 10.1016/j.tcs.2014.05.025} {\bibfield  {journal} {\bibinfo  {journal} {Theoretical computer science}\ }\textbf {\bibinfo {volume} {560}},\ \bibinfo {pages} {7--11} (\bibinfo {year} {2014})}\BibitemShut {NoStop}%
\bibitem [{\citenamefont {Huang}\ \emph {et~al.}(2021)\citenamefont {Huang}, \citenamefont {Broughton}, \citenamefont {Mohseni}, \citenamefont {Babbush}, \citenamefont {Boixo}, \citenamefont {Neven},\ and\ \citenamefont {McClean}}]{huang2021power}%
  \BibitemOpen
  \bibfield  {author} {\bibinfo {author} {\bibfnamefont {Hsin-Yuan}\ \bibnamefont {Huang}}, \bibinfo {author} {\bibfnamefont {Michael}\ \bibnamefont {Broughton}}, \bibinfo {author} {\bibfnamefont {Masoud}\ \bibnamefont {Mohseni}}, \bibinfo {author} {\bibfnamefont {Ryan}\ \bibnamefont {Babbush}}, \bibinfo {author} {\bibfnamefont {Sergio}\ \bibnamefont {Boixo}}, \bibinfo {author} {\bibfnamefont {Hartmut}\ \bibnamefont {Neven}}, \ and\ \bibinfo {author} {\bibfnamefont {Jarrod~R}\ \bibnamefont {McClean}},\ }\bibfield  {title} {\enquote {\bibinfo {title} {Power of data in quantum machine learning},}\ }\href {\doibase 10.1038/s41467-021-22539-9} {\bibfield  {journal} {\bibinfo  {journal} {Nature communications}\ }\textbf {\bibinfo {volume} {12}},\ \bibinfo {pages} {2631} (\bibinfo {year} {2021})}\BibitemShut {NoStop}%
\bibitem [{\citenamefont {Bharti}\ \emph {et~al.}(2022)\citenamefont {Bharti}, \citenamefont {Cervera-Lierta}, \citenamefont {Kyaw}, \citenamefont {Haug}, \citenamefont {Alperin-Lea}, \citenamefont {Anand}, \citenamefont {Degroote}, \citenamefont {Heimonen}, \citenamefont {Kottmann}, \citenamefont {Menke}, \citenamefont {Mok}, \citenamefont {Sim}, \citenamefont {Kwek},\ and\ \citenamefont {Aspuru-Guzik}}]{nisqrmp}%
  \BibitemOpen
  \bibfield  {author} {\bibinfo {author} {\bibfnamefont {Kishor}\ \bibnamefont {Bharti}}, \bibinfo {author} {\bibfnamefont {Alba}\ \bibnamefont {Cervera-Lierta}}, \bibinfo {author} {\bibfnamefont {Thi~Ha}\ \bibnamefont {Kyaw}}, \bibinfo {author} {\bibfnamefont {Tobias}\ \bibnamefont {Haug}}, \bibinfo {author} {\bibfnamefont {Sumner}\ \bibnamefont {Alperin-Lea}}, \bibinfo {author} {\bibfnamefont {Abhinav}\ \bibnamefont {Anand}}, \bibinfo {author} {\bibfnamefont {Matthias}\ \bibnamefont {Degroote}}, \bibinfo {author} {\bibfnamefont {Hermanni}\ \bibnamefont {Heimonen}}, \bibinfo {author} {\bibfnamefont {Jakob~S.}\ \bibnamefont {Kottmann}}, \bibinfo {author} {\bibfnamefont {Tim}\ \bibnamefont {Menke}}, \bibinfo {author} {\bibfnamefont {Wai-Keong}\ \bibnamefont {Mok}}, \bibinfo {author} {\bibfnamefont {Sukin}\ \bibnamefont {Sim}}, \bibinfo {author} {\bibfnamefont {Leong-Chuan}\ \bibnamefont {Kwek}}, \ and\ \bibinfo {author} {\bibfnamefont {Al\'an}\ \bibnamefont {Aspuru-Guzik}},\ }\bibfield  {title} {\enquote
  {\bibinfo {title} {Noisy intermediate-scale quantum algorithms},}\ }\href {\doibase 10.1103/RevModPhys.94.015004} {\bibfield  {journal} {\bibinfo  {journal} {Rev. Mod. Phys.}\ }\textbf {\bibinfo {volume} {94}},\ \bibinfo {pages} {015004} (\bibinfo {year} {2022})}\BibitemShut {NoStop}%
\bibitem [{\citenamefont {Dalzell}\ \emph {et~al.}(2025)\citenamefont {Dalzell}, \citenamefont {McArdle}, \citenamefont {Berta}, \citenamefont {Bienias}, \citenamefont {Chen}, \citenamefont {Gilyén}, \citenamefont {Hann}, \citenamefont {Kastoryano}, \citenamefont {Khabiboulline}, \citenamefont {Kubica},\ and\ \citenamefont {et~al.}}]{dalzell2023quantum}%
  \BibitemOpen
  \bibfield  {author} {\bibinfo {author} {\bibfnamefont {Alexander~M.}\ \bibnamefont {Dalzell}}, \bibinfo {author} {\bibfnamefont {Sam}\ \bibnamefont {McArdle}}, \bibinfo {author} {\bibfnamefont {Mario}\ \bibnamefont {Berta}}, \bibinfo {author} {\bibfnamefont {Przemyslaw}\ \bibnamefont {Bienias}}, \bibinfo {author} {\bibfnamefont {Chi-Fang}\ \bibnamefont {Chen}}, \bibinfo {author} {\bibfnamefont {András}\ \bibnamefont {Gilyén}}, \bibinfo {author} {\bibfnamefont {Connor~T.}\ \bibnamefont {Hann}}, \bibinfo {author} {\bibfnamefont {Michael~J.}\ \bibnamefont {Kastoryano}}, \bibinfo {author} {\bibfnamefont {Emil~T.}\ \bibnamefont {Khabiboulline}}, \bibinfo {author} {\bibfnamefont {Aleksander}\ \bibnamefont {Kubica}}, \ and\ \bibinfo {author} {\bibnamefont {et~al.}},\ }\href {\doibase 10.1017/9781009639651} {\emph {\bibinfo {title} {Quantum Algorithms: A Survey of Applications and End-to-end Complexities}}}\ (\bibinfo  {publisher} {Cambridge University Press},\ \bibinfo {year} {2025})\BibitemShut {NoStop}%
\bibitem [{\citenamefont {Yamakawa}\ and\ \citenamefont {Zhandry}(2024)}]{yamakawa2024verifiable}%
  \BibitemOpen
  \bibfield  {author} {\bibinfo {author} {\bibfnamefont {Takashi}\ \bibnamefont {Yamakawa}}\ and\ \bibinfo {author} {\bibfnamefont {Mark}\ \bibnamefont {Zhandry}},\ }\bibfield  {title} {\enquote {\bibinfo {title} {Verifiable quantum advantage without structure},}\ }\href {\doibase 10.1145/3658665} {\bibfield  {journal} {\bibinfo  {journal} {J. ACM}\ }\textbf {\bibinfo {volume} {71}} (\bibinfo {year} {2024}),\ 10.1145/3658665}\BibitemShut {NoStop}%
\bibitem [{\citenamefont {Jordan}\ \emph {et~al.}(2024)\citenamefont {Jordan}, \citenamefont {Shutty}, \citenamefont {Wootters}, \citenamefont {Zalcman}, \citenamefont {Schmidhuber}, \citenamefont {King}, \citenamefont {Isakov},\ and\ \citenamefont {Babbush}}]{jordan2024optimization}%
  \BibitemOpen
  \bibfield  {author} {\bibinfo {author} {\bibfnamefont {Stephen~P}\ \bibnamefont {Jordan}}, \bibinfo {author} {\bibfnamefont {Noah}\ \bibnamefont {Shutty}}, \bibinfo {author} {\bibfnamefont {Mary}\ \bibnamefont {Wootters}}, \bibinfo {author} {\bibfnamefont {Adam}\ \bibnamefont {Zalcman}}, \bibinfo {author} {\bibfnamefont {Alexander}\ \bibnamefont {Schmidhuber}}, \bibinfo {author} {\bibfnamefont {Robbie}\ \bibnamefont {King}}, \bibinfo {author} {\bibfnamefont {Sergei~V}\ \bibnamefont {Isakov}}, \ and\ \bibinfo {author} {\bibfnamefont {Ryan}\ \bibnamefont {Babbush}},\ }\bibfield  {title} {\enquote {\bibinfo {title} {Optimization by decoded quantum interferometry},}\ }\href {https://arxiv.org/abs/2408.08292} {\bibfield  {journal} {\bibinfo  {journal} {arXiv preprint arXiv:2408.08292}\ } (\bibinfo {year} {2024})}\BibitemShut {NoStop}%
\bibitem [{\citenamefont {Shor}(1995)}]{shor1995scheme}%
  \BibitemOpen
  \bibfield  {author} {\bibinfo {author} {\bibfnamefont {Peter~W.}\ \bibnamefont {Shor}},\ }\bibfield  {title} {\enquote {\bibinfo {title} {Scheme for reducing decoherence in quantum computer memory},}\ }\href {\doibase 10.1103/PhysRevA.52.R2493} {\bibfield  {journal} {\bibinfo  {journal} {Phys. Rev. A}\ }\textbf {\bibinfo {volume} {52}},\ \bibinfo {pages} {R2493--R2496} (\bibinfo {year} {1995})}\BibitemShut {NoStop}%
\bibitem [{\citenamefont {Gottesman}(1997)}]{gottesman1997stabilizer}%
  \BibitemOpen
  \bibfield  {author} {\bibinfo {author} {\bibfnamefont {Daniel}\ \bibnamefont {Gottesman}},\ }\href {https://arxiv.org/abs/quant-ph/9705052} {\emph {\bibinfo {title} {Stabilizer codes and quantum error correction}}}\ (\bibinfo  {publisher} {California Institute of Technology},\ \bibinfo {year} {1997})\BibitemShut {NoStop}%
\bibitem [{\citenamefont {Calderbank}\ \emph {et~al.}(1997)\citenamefont {Calderbank}, \citenamefont {Rains}, \citenamefont {Shor},\ and\ \citenamefont {Sloane}}]{css}%
  \BibitemOpen
  \bibfield  {author} {\bibinfo {author} {\bibfnamefont {A.~R.}\ \bibnamefont {Calderbank}}, \bibinfo {author} {\bibfnamefont {E.~M.}\ \bibnamefont {Rains}}, \bibinfo {author} {\bibfnamefont {P.~W.}\ \bibnamefont {Shor}}, \ and\ \bibinfo {author} {\bibfnamefont {N.~J.~A.}\ \bibnamefont {Sloane}},\ }\bibfield  {title} {\enquote {\bibinfo {title} {Quantum error correction and orthogonal geometry},}\ }\href {\doibase 10.1103/PhysRevLett.78.405} {\bibfield  {journal} {\bibinfo  {journal} {Phys. Rev. Lett.}\ }\textbf {\bibinfo {volume} {78}},\ \bibinfo {pages} {405--408} (\bibinfo {year} {1997})}\BibitemShut {NoStop}%
\bibitem [{\citenamefont {Calderbank}\ \emph {et~al.}(1998)\citenamefont {Calderbank}, \citenamefont {Rains}, \citenamefont {Shor},\ and\ \citenamefont {Sloane}}]{calderbank1998quantum}%
  \BibitemOpen
  \bibfield  {author} {\bibinfo {author} {\bibfnamefont {A~Robert}\ \bibnamefont {Calderbank}}, \bibinfo {author} {\bibfnamefont {Eric~M}\ \bibnamefont {Rains}}, \bibinfo {author} {\bibfnamefont {Peter~M}\ \bibnamefont {Shor}}, \ and\ \bibinfo {author} {\bibfnamefont {Neil~JA}\ \bibnamefont {Sloane}},\ }\bibfield  {title} {\enquote {\bibinfo {title} {Quantum error correction via codes over gf (4)},}\ }\href {https://ieeexplore.ieee.org/document/681315} {\bibfield  {journal} {\bibinfo  {journal} {IEEE Transactions on Information Theory}\ }\textbf {\bibinfo {volume} {44}},\ \bibinfo {pages} {1369--1387} (\bibinfo {year} {1998})}\BibitemShut {NoStop}%
\bibitem [{\citenamefont {Gottesman}\ \emph {et~al.}(2001)\citenamefont {Gottesman}, \citenamefont {Kitaev},\ and\ \citenamefont {Preskill}}]{gkp}%
  \BibitemOpen
  \bibfield  {author} {\bibinfo {author} {\bibfnamefont {Daniel}\ \bibnamefont {Gottesman}}, \bibinfo {author} {\bibfnamefont {Alexei}\ \bibnamefont {Kitaev}}, \ and\ \bibinfo {author} {\bibfnamefont {John}\ \bibnamefont {Preskill}},\ }\bibfield  {title} {\enquote {\bibinfo {title} {Encoding a qubit in an oscillator},}\ }\href {\doibase 10.1103/PhysRevA.64.012310} {\bibfield  {journal} {\bibinfo  {journal} {Phys. Rev. A}\ }\textbf {\bibinfo {volume} {64}},\ \bibinfo {pages} {012310} (\bibinfo {year} {2001})}\BibitemShut {NoStop}%
\bibitem [{\citenamefont {Cochrane}\ \emph {et~al.}(1999)\citenamefont {Cochrane}, \citenamefont {Milburn},\ and\ \citenamefont {Munro}}]{cochrane1999macroscopically}%
  \BibitemOpen
  \bibfield  {author} {\bibinfo {author} {\bibfnamefont {P.~T.}\ \bibnamefont {Cochrane}}, \bibinfo {author} {\bibfnamefont {G.~J.}\ \bibnamefont {Milburn}}, \ and\ \bibinfo {author} {\bibfnamefont {W.~J.}\ \bibnamefont {Munro}},\ }\bibfield  {title} {\enquote {\bibinfo {title} {Macroscopically distinct quantum-superposition states as a bosonic code for amplitude damping},}\ }\href {\doibase 10.1103/PhysRevA.59.2631} {\bibfield  {journal} {\bibinfo  {journal} {Phys. Rev. A}\ }\textbf {\bibinfo {volume} {59}},\ \bibinfo {pages} {2631--2634} (\bibinfo {year} {1999})}\BibitemShut {NoStop}%
\bibitem [{\citenamefont {Leghtas}\ \emph {et~al.}(2013)\citenamefont {Leghtas}, \citenamefont {Kirchmair}, \citenamefont {Vlastakis}, \citenamefont {Schoelkopf}, \citenamefont {Devoret},\ and\ \citenamefont {Mirrahimi}}]{leghtas2013hardware}%
  \BibitemOpen
  \bibfield  {author} {\bibinfo {author} {\bibfnamefont {Zaki}\ \bibnamefont {Leghtas}}, \bibinfo {author} {\bibfnamefont {Gerhard}\ \bibnamefont {Kirchmair}}, \bibinfo {author} {\bibfnamefont {Brian}\ \bibnamefont {Vlastakis}}, \bibinfo {author} {\bibfnamefont {Robert~J.}\ \bibnamefont {Schoelkopf}}, \bibinfo {author} {\bibfnamefont {Michel~H.}\ \bibnamefont {Devoret}}, \ and\ \bibinfo {author} {\bibfnamefont {Mazyar}\ \bibnamefont {Mirrahimi}},\ }\bibfield  {title} {\enquote {\bibinfo {title} {Hardware-efficient autonomous quantum memory protection},}\ }\href {\doibase 10.1103/PhysRevLett.111.120501} {\bibfield  {journal} {\bibinfo  {journal} {Phys. Rev. Lett.}\ }\textbf {\bibinfo {volume} {111}},\ \bibinfo {pages} {120501} (\bibinfo {year} {2013})}\BibitemShut {NoStop}%
\bibitem [{\citenamefont {Michael}\ \emph {et~al.}(2016)\citenamefont {Michael}, \citenamefont {Silveri}, \citenamefont {Brierley}, \citenamefont {Albert}, \citenamefont {Salmilehto}, \citenamefont {Jiang},\ and\ \citenamefont {Girvin}}]{pairbinomial}%
  \BibitemOpen
  \bibfield  {author} {\bibinfo {author} {\bibfnamefont {Marios~H.}\ \bibnamefont {Michael}}, \bibinfo {author} {\bibfnamefont {Matti}\ \bibnamefont {Silveri}}, \bibinfo {author} {\bibfnamefont {R.~T.}\ \bibnamefont {Brierley}}, \bibinfo {author} {\bibfnamefont {Victor~V.}\ \bibnamefont {Albert}}, \bibinfo {author} {\bibfnamefont {Juha}\ \bibnamefont {Salmilehto}}, \bibinfo {author} {\bibfnamefont {Liang}\ \bibnamefont {Jiang}}, \ and\ \bibinfo {author} {\bibfnamefont {S.~M.}\ \bibnamefont {Girvin}},\ }\bibfield  {title} {\enquote {\bibinfo {title} {New class of quantum error-correcting codes for a bosonic mode},}\ }\href {\doibase 10.1103/PhysRevX.6.031006} {\bibfield  {journal} {\bibinfo  {journal} {Phys. Rev. X}\ }\textbf {\bibinfo {volume} {6}},\ \bibinfo {pages} {031006} (\bibinfo {year} {2016})}\BibitemShut {NoStop}%
\bibitem [{\citenamefont {Sun}\ \emph {et~al.}(2014)\citenamefont {Sun}, \citenamefont {Petrenko}, \citenamefont {Leghtas}, \citenamefont {Vlastakis}, \citenamefont {Kirchmair}, \citenamefont {Sliwa}, \citenamefont {Narla}, \citenamefont {Hatridge}, \citenamefont {Shankar}, \citenamefont {Blumoff} \emph {et~al.}}]{sun2014tracking}%
  \BibitemOpen
  \bibfield  {author} {\bibinfo {author} {\bibfnamefont {Luyan}\ \bibnamefont {Sun}}, \bibinfo {author} {\bibfnamefont {Andrei}\ \bibnamefont {Petrenko}}, \bibinfo {author} {\bibfnamefont {Zaki}\ \bibnamefont {Leghtas}}, \bibinfo {author} {\bibfnamefont {Brian}\ \bibnamefont {Vlastakis}}, \bibinfo {author} {\bibfnamefont {Gerhard}\ \bibnamefont {Kirchmair}}, \bibinfo {author} {\bibfnamefont {KM}~\bibnamefont {Sliwa}}, \bibinfo {author} {\bibfnamefont {Aniruth}\ \bibnamefont {Narla}}, \bibinfo {author} {\bibfnamefont {Michael}\ \bibnamefont {Hatridge}}, \bibinfo {author} {\bibfnamefont {Shyam}\ \bibnamefont {Shankar}}, \bibinfo {author} {\bibfnamefont {Jacob}\ \bibnamefont {Blumoff}},  \emph {et~al.},\ }\bibfield  {title} {\enquote {\bibinfo {title} {Tracking photon jumps with repeated quantum non-demolition parity measurements},}\ }\href {\doibase 10.1038/nature13436} {\bibfield  {journal} {\bibinfo  {journal} {Nature}\ }\textbf {\bibinfo {volume} {511}},\ \bibinfo {pages} {444--448} (\bibinfo {year}
  {2014})}\BibitemShut {NoStop}%
\bibitem [{\citenamefont {Leghtas}\ \emph {et~al.}(2015)\citenamefont {Leghtas}, \citenamefont {Touzard}, \citenamefont {Pop}, \citenamefont {Kou}, \citenamefont {Vlastakis}, \citenamefont {Petrenko}, \citenamefont {Sliwa}, \citenamefont {Narla}, \citenamefont {Shankar}, \citenamefont {Hatridge} \emph {et~al.}}]{leghtas2015confining}%
  \BibitemOpen
  \bibfield  {author} {\bibinfo {author} {\bibfnamefont {Zaki}\ \bibnamefont {Leghtas}}, \bibinfo {author} {\bibfnamefont {Steven}\ \bibnamefont {Touzard}}, \bibinfo {author} {\bibfnamefont {Ioan~M}\ \bibnamefont {Pop}}, \bibinfo {author} {\bibfnamefont {Angela}\ \bibnamefont {Kou}}, \bibinfo {author} {\bibfnamefont {Brian}\ \bibnamefont {Vlastakis}}, \bibinfo {author} {\bibfnamefont {Andrei}\ \bibnamefont {Petrenko}}, \bibinfo {author} {\bibfnamefont {Katrina~M}\ \bibnamefont {Sliwa}}, \bibinfo {author} {\bibfnamefont {Anirudh}\ \bibnamefont {Narla}}, \bibinfo {author} {\bibfnamefont {Shyam}\ \bibnamefont {Shankar}}, \bibinfo {author} {\bibfnamefont {Michael~J}\ \bibnamefont {Hatridge}},  \emph {et~al.},\ }\bibfield  {title} {\enquote {\bibinfo {title} {Confining the state of light to a quantum manifold by engineered two-photon loss},}\ }\href {\doibase 10.1126/science.aaa2085} {\bibfield  {journal} {\bibinfo  {journal} {Science}\ }\textbf {\bibinfo {volume} {347}},\ \bibinfo {pages} {853--857} (\bibinfo
  {year} {2015})}\BibitemShut {NoStop}%
\bibitem [{\citenamefont {Fl{\"u}hmann}\ \emph {et~al.}(2019)\citenamefont {Fl{\"u}hmann}, \citenamefont {Nguyen}, \citenamefont {Marinelli}, \citenamefont {Negnevitsky}, \citenamefont {Mehta},\ and\ \citenamefont {Home}}]{fluhmann2019encoding}%
  \BibitemOpen
  \bibfield  {author} {\bibinfo {author} {\bibfnamefont {Christa}\ \bibnamefont {Fl{\"u}hmann}}, \bibinfo {author} {\bibfnamefont {Thanh~Long}\ \bibnamefont {Nguyen}}, \bibinfo {author} {\bibfnamefont {Matteo}\ \bibnamefont {Marinelli}}, \bibinfo {author} {\bibfnamefont {Vlad}\ \bibnamefont {Negnevitsky}}, \bibinfo {author} {\bibfnamefont {Karan}\ \bibnamefont {Mehta}}, \ and\ \bibinfo {author} {\bibfnamefont {JP}~\bibnamefont {Home}},\ }\bibfield  {title} {\enquote {\bibinfo {title} {Encoding a qubit in a trapped-ion mechanical oscillator},}\ }\href {\doibase 10.1038/s41586-019-0960-6} {\bibfield  {journal} {\bibinfo  {journal} {Nature}\ }\textbf {\bibinfo {volume} {566}},\ \bibinfo {pages} {513--517} (\bibinfo {year} {2019})}\BibitemShut {NoStop}%
\bibitem [{\citenamefont {Grimm}\ \emph {et~al.}(2020)\citenamefont {Grimm}, \citenamefont {Frattini}, \citenamefont {Puri}, \citenamefont {Mundhada}, \citenamefont {Touzard}, \citenamefont {Mirrahimi}, \citenamefont {Girvin}, \citenamefont {Shankar},\ and\ \citenamefont {Devoret}}]{grimm2020stabilization}%
  \BibitemOpen
  \bibfield  {author} {\bibinfo {author} {\bibfnamefont {Alexander}\ \bibnamefont {Grimm}}, \bibinfo {author} {\bibfnamefont {Nicholas~E}\ \bibnamefont {Frattini}}, \bibinfo {author} {\bibfnamefont {Shruti}\ \bibnamefont {Puri}}, \bibinfo {author} {\bibfnamefont {Shantanu~O}\ \bibnamefont {Mundhada}}, \bibinfo {author} {\bibfnamefont {Steven}\ \bibnamefont {Touzard}}, \bibinfo {author} {\bibfnamefont {Mazyar}\ \bibnamefont {Mirrahimi}}, \bibinfo {author} {\bibfnamefont {Steven~M}\ \bibnamefont {Girvin}}, \bibinfo {author} {\bibfnamefont {Shyam}\ \bibnamefont {Shankar}}, \ and\ \bibinfo {author} {\bibfnamefont {Michel~H}\ \bibnamefont {Devoret}},\ }\bibfield  {title} {\enquote {\bibinfo {title} {Stabilization and operation of a kerr-cat qubit},}\ }\href {\doibase 10.1038/s41586-020-2587-z} {\bibfield  {journal} {\bibinfo  {journal} {Nature}\ }\textbf {\bibinfo {volume} {584}},\ \bibinfo {pages} {205--209} (\bibinfo {year} {2020})}\BibitemShut {NoStop}%
\bibitem [{\citenamefont {Putterman}\ \emph {et~al.}(2025)\citenamefont {Putterman}, \citenamefont {Noh}, \citenamefont {Hann}, \citenamefont {MacCabe}, \citenamefont {Aghaeimeibodi}, \citenamefont {Patel}, \citenamefont {Lee}, \citenamefont {Jones}, \citenamefont {Moradinejad}, \citenamefont {Rodriguez} \emph {et~al.}}]{putterman2024hardware}%
  \BibitemOpen
  \bibfield  {author} {\bibinfo {author} {\bibfnamefont {Harald}\ \bibnamefont {Putterman}}, \bibinfo {author} {\bibfnamefont {Kyungjoo}\ \bibnamefont {Noh}}, \bibinfo {author} {\bibfnamefont {Connor~T}\ \bibnamefont {Hann}}, \bibinfo {author} {\bibfnamefont {Gregory~S}\ \bibnamefont {MacCabe}}, \bibinfo {author} {\bibfnamefont {Shahriar}\ \bibnamefont {Aghaeimeibodi}}, \bibinfo {author} {\bibfnamefont {Rishi~N}\ \bibnamefont {Patel}}, \bibinfo {author} {\bibfnamefont {Menyoung}\ \bibnamefont {Lee}}, \bibinfo {author} {\bibfnamefont {William~M}\ \bibnamefont {Jones}}, \bibinfo {author} {\bibfnamefont {Hesam}\ \bibnamefont {Moradinejad}}, \bibinfo {author} {\bibfnamefont {Roberto}\ \bibnamefont {Rodriguez}},  \emph {et~al.},\ }\bibfield  {title} {\enquote {\bibinfo {title} {Hardware-efficient quantum error correction via concatenated bosonic qubits},}\ }\href {\doibase 10.1038/s41586-025-08642-7} {\bibfield  {journal} {\bibinfo  {journal} {Nature}\ }\textbf {\bibinfo {volume} {638}},\ \bibinfo {pages}
  {927--934} (\bibinfo {year} {2025})}\BibitemShut {NoStop}%
\bibitem [{\citenamefont {Gertler}\ \emph {et~al.}(2023)\citenamefont {Gertler}, \citenamefont {van Geldern}, \citenamefont {Shirol}, \citenamefont {Jiang},\ and\ \citenamefont {Wang}}]{gertler2023experimental}%
  \BibitemOpen
  \bibfield  {author} {\bibinfo {author} {\bibfnamefont {Jeffrey~M.}\ \bibnamefont {Gertler}}, \bibinfo {author} {\bibfnamefont {Sean}\ \bibnamefont {van Geldern}}, \bibinfo {author} {\bibfnamefont {Shruti}\ \bibnamefont {Shirol}}, \bibinfo {author} {\bibfnamefont {Liang}\ \bibnamefont {Jiang}}, \ and\ \bibinfo {author} {\bibfnamefont {Chen}\ \bibnamefont {Wang}},\ }\bibfield  {title} {\enquote {\bibinfo {title} {Experimental realization and characterization of stabilized pair-coherent states},}\ }\href {\doibase 10.1103/PRXQuantum.4.020319} {\bibfield  {journal} {\bibinfo  {journal} {PRX Quantum}\ }\textbf {\bibinfo {volume} {4}},\ \bibinfo {pages} {020319} (\bibinfo {year} {2023})}\BibitemShut {NoStop}%
\bibitem [{\citenamefont {Jain}\ \emph {et~al.}(2024)\citenamefont {Jain}, \citenamefont {Iosue}, \citenamefont {Barg},\ and\ \citenamefont {Albert}}]{jain2024quantum}%
  \BibitemOpen
  \bibfield  {author} {\bibinfo {author} {\bibfnamefont {Shubham~P}\ \bibnamefont {Jain}}, \bibinfo {author} {\bibfnamefont {Joseph~T}\ \bibnamefont {Iosue}}, \bibinfo {author} {\bibfnamefont {Alexander}\ \bibnamefont {Barg}}, \ and\ \bibinfo {author} {\bibfnamefont {Victor~V}\ \bibnamefont {Albert}},\ }\bibfield  {title} {\enquote {\bibinfo {title} {Quantum spherical codes},}\ }\href {\doibase 10.1038/s41567-024-02496-y} {\bibfield  {journal} {\bibinfo  {journal} {Nature Physics}\ ,\ \bibinfo {pages} {1--6}} (\bibinfo {year} {2024})}\BibitemShut {NoStop}%
\bibitem [{\citenamefont {Denys}\ and\ \citenamefont {Leverrier}(2023)}]{denys20232}%
  \BibitemOpen
  \bibfield  {author} {\bibinfo {author} {\bibfnamefont {Aur{\'e}lie}\ \bibnamefont {Denys}}\ and\ \bibinfo {author} {\bibfnamefont {Anthony}\ \bibnamefont {Leverrier}},\ }\bibfield  {title} {\enquote {\bibinfo {title} {The $2 t $-qutrit, a two-mode bosonic qutrit},}\ }\href {\doibase 10.22331/q-2023-06-05-1032} {\bibfield  {journal} {\bibinfo  {journal} {Quantum}\ }\textbf {\bibinfo {volume} {7}},\ \bibinfo {pages} {1032} (\bibinfo {year} {2023})}\BibitemShut {NoStop}%
\bibitem [{\citenamefont {Vuillot}\ \emph {et~al.}(2024)\citenamefont {Vuillot}, \citenamefont {Ciani},\ and\ \citenamefont {Terhal}}]{vuillotHomologicalQuantumRotor2024}%
  \BibitemOpen
  \bibfield  {author} {\bibinfo {author} {\bibfnamefont {Christophe}\ \bibnamefont {Vuillot}}, \bibinfo {author} {\bibfnamefont {Alessandro}\ \bibnamefont {Ciani}}, \ and\ \bibinfo {author} {\bibfnamefont {Barbara~M.}\ \bibnamefont {Terhal}},\ }\bibfield  {title} {\enquote {\bibinfo {title} {Homological {{Quantum Rotor Codes}}: {{Logical Qubits}} from {{Torsion}}},}\ }\href {\doibase 10.1007/s00220-023-04905-4} {\bibfield  {journal} {\bibinfo  {journal} {Communications in Mathematical Physics}\ }\textbf {\bibinfo {volume} {405}},\ \bibinfo {pages} {53} (\bibinfo {year} {2024})}\BibitemShut {NoStop}%
\bibitem [{\citenamefont {Poyatos}\ \emph {et~al.}(1996)\citenamefont {Poyatos}, \citenamefont {Cirac},\ and\ \citenamefont {Zoller}}]{poyatos1996quantum}%
  \BibitemOpen
  \bibfield  {author} {\bibinfo {author} {\bibfnamefont {J.~F.}\ \bibnamefont {Poyatos}}, \bibinfo {author} {\bibfnamefont {J.~I.}\ \bibnamefont {Cirac}}, \ and\ \bibinfo {author} {\bibfnamefont {P.}~\bibnamefont {Zoller}},\ }\bibfield  {title} {\enquote {\bibinfo {title} {Quantum reservoir engineering with laser cooled trapped ions},}\ }\href {\doibase 10.1103/PhysRevLett.77.4728} {\bibfield  {journal} {\bibinfo  {journal} {Phys. Rev. Lett.}\ }\textbf {\bibinfo {volume} {77}},\ \bibinfo {pages} {4728--4731} (\bibinfo {year} {1996})}\BibitemShut {NoStop}%
\bibitem [{\citenamefont {Mirrahimi}\ \emph {et~al.}(2014)\citenamefont {Mirrahimi}, \citenamefont {Leghtas}, \citenamefont {Albert}, \citenamefont {Touzard}, \citenamefont {Schoelkopf}, \citenamefont {Jiang},\ and\ \citenamefont {Devoret}}]{mirrahimi2014dynamically}%
  \BibitemOpen
  \bibfield  {author} {\bibinfo {author} {\bibfnamefont {Mazyar}\ \bibnamefont {Mirrahimi}}, \bibinfo {author} {\bibfnamefont {Zaki}\ \bibnamefont {Leghtas}}, \bibinfo {author} {\bibfnamefont {Victor~V}\ \bibnamefont {Albert}}, \bibinfo {author} {\bibfnamefont {Steven}\ \bibnamefont {Touzard}}, \bibinfo {author} {\bibfnamefont {Robert~J}\ \bibnamefont {Schoelkopf}}, \bibinfo {author} {\bibfnamefont {Liang}\ \bibnamefont {Jiang}}, \ and\ \bibinfo {author} {\bibfnamefont {Michel~H}\ \bibnamefont {Devoret}},\ }\bibfield  {title} {\enquote {\bibinfo {title} {Dynamically protected cat-qubits: a new paradigm for universal quantum computation},}\ }\href {\doibase 10.1088/1367-2630/16/4/045014} {\bibfield  {journal} {\bibinfo  {journal} {New Journal of Physics}\ }\textbf {\bibinfo {volume} {16}},\ \bibinfo {pages} {045014} (\bibinfo {year} {2014})}\BibitemShut {NoStop}%
\bibitem [{\citenamefont {Zanardi}\ and\ \citenamefont {Campos~Venuti}(2014)}]{zanardi2014coherent}%
  \BibitemOpen
  \bibfield  {author} {\bibinfo {author} {\bibfnamefont {Paolo}\ \bibnamefont {Zanardi}}\ and\ \bibinfo {author} {\bibfnamefont {Lorenzo}\ \bibnamefont {Campos~Venuti}},\ }\bibfield  {title} {\enquote {\bibinfo {title} {Coherent quantum dynamics in steady-state manifolds of strongly dissipative systems},}\ }\href {\doibase 10.1103/PhysRevLett.113.240406} {\bibfield  {journal} {\bibinfo  {journal} {Phys. Rev. Lett.}\ }\textbf {\bibinfo {volume} {113}},\ \bibinfo {pages} {240406} (\bibinfo {year} {2014})}\BibitemShut {NoStop}%
\bibitem [{\citenamefont {Albert}\ \emph {et~al.}(2016)\citenamefont {Albert}, \citenamefont {Bradlyn}, \citenamefont {Fraas},\ and\ \citenamefont {Jiang}}]{albert2016geometry}%
  \BibitemOpen
  \bibfield  {author} {\bibinfo {author} {\bibfnamefont {Victor~V.}\ \bibnamefont {Albert}}, \bibinfo {author} {\bibfnamefont {Barry}\ \bibnamefont {Bradlyn}}, \bibinfo {author} {\bibfnamefont {Martin}\ \bibnamefont {Fraas}}, \ and\ \bibinfo {author} {\bibfnamefont {Liang}\ \bibnamefont {Jiang}},\ }\bibfield  {title} {\enquote {\bibinfo {title} {Geometry and response of lindbladians},}\ }\href {\doibase 10.1103/PhysRevX.6.041031} {\bibfield  {journal} {\bibinfo  {journal} {Phys. Rev. X}\ }\textbf {\bibinfo {volume} {6}},\ \bibinfo {pages} {041031} (\bibinfo {year} {2016})}\BibitemShut {NoStop}%
\bibitem [{\citenamefont {Albert}\ \emph {et~al.}(2019)\citenamefont {Albert}, \citenamefont {Mundhada}, \citenamefont {Grimm}, \citenamefont {Touzard}, \citenamefont {Devoret},\ and\ \citenamefont {Jiang}}]{albert2019pair}%
  \BibitemOpen
  \bibfield  {author} {\bibinfo {author} {\bibfnamefont {Victor~V}\ \bibnamefont {Albert}}, \bibinfo {author} {\bibfnamefont {Shantanu~O}\ \bibnamefont {Mundhada}}, \bibinfo {author} {\bibfnamefont {Alexander}\ \bibnamefont {Grimm}}, \bibinfo {author} {\bibfnamefont {Steven}\ \bibnamefont {Touzard}}, \bibinfo {author} {\bibfnamefont {Michel~H}\ \bibnamefont {Devoret}}, \ and\ \bibinfo {author} {\bibfnamefont {Liang}\ \bibnamefont {Jiang}},\ }\bibfield  {title} {\enquote {\bibinfo {title} {Pair-cat codes: autonomous error-correction with low-order nonlinearity},}\ }\href {\doibase 10.1088/2058-9565/ab1e69} {\bibfield  {journal} {\bibinfo  {journal} {Quantum Science and Technology}\ }\textbf {\bibinfo {volume} {4}},\ \bibinfo {pages} {035007} (\bibinfo {year} {2019})}\BibitemShut {NoStop}%
\bibitem [{\citenamefont {Xu}\ \emph {et~al.}(2024)\citenamefont {Xu}, \citenamefont {Wang},\ and\ \citenamefont {Albert}}]{rotorclifford}%
  \BibitemOpen
  \bibfield  {author} {\bibinfo {author} {\bibfnamefont {Yijia}\ \bibnamefont {Xu}}, \bibinfo {author} {\bibfnamefont {Yixu}\ \bibnamefont {Wang}}, \ and\ \bibinfo {author} {\bibfnamefont {Victor~V.}\ \bibnamefont {Albert}},\ }\bibfield  {title} {\enquote {\bibinfo {title} {Multimode rotation-symmetric bosonic codes from homological rotor codes},}\ }\href {\doibase 10.1103/PhysRevA.110.022402} {\bibfield  {journal} {\bibinfo  {journal} {Phys. Rev. A}\ }\textbf {\bibinfo {volume} {110}},\ \bibinfo {pages} {022402} (\bibinfo {year} {2024})}\BibitemShut {NoStop}%
\bibitem [{\citenamefont {Steane}(1996{\natexlab{a}})}]{PhysRevLett.77.793}%
  \BibitemOpen
  \bibfield  {author} {\bibinfo {author} {\bibfnamefont {A.~M.}\ \bibnamefont {Steane}},\ }\bibfield  {title} {\enquote {\bibinfo {title} {Error correcting codes in quantum theory},}\ }\href {\doibase 10.1103/PhysRevLett.77.793} {\bibfield  {journal} {\bibinfo  {journal} {Phys. Rev. Lett.}\ }\textbf {\bibinfo {volume} {77}},\ \bibinfo {pages} {793--797} (\bibinfo {year} {1996}{\natexlab{a}})}\BibitemShut {NoStop}%
\bibitem [{\citenamefont {Steane}(1996{\natexlab{b}})}]{steane1996multiple}%
  \BibitemOpen
  \bibfield  {author} {\bibinfo {author} {\bibfnamefont {Andrew}\ \bibnamefont {Steane}},\ }\bibfield  {title} {\enquote {\bibinfo {title} {Multiple-particle interference and quantum error correction},}\ }\href {\doibase 10.1098/rspa.1996.0136} {\bibfield  {journal} {\bibinfo  {journal} {Proceedings of the Royal Society of London. Series A: Mathematical, Physical and Engineering Sciences}\ }\textbf {\bibinfo {volume} {452}},\ \bibinfo {pages} {2551--2577} (\bibinfo {year} {1996}{\natexlab{b}})}\BibitemShut {NoStop}%
\bibitem [{\citenamefont {Calderbank}\ and\ \citenamefont {Shor}(1996)}]{PhysRevA.54.1098}%
  \BibitemOpen
  \bibfield  {author} {\bibinfo {author} {\bibfnamefont {A.~R.}\ \bibnamefont {Calderbank}}\ and\ \bibinfo {author} {\bibfnamefont {Peter~W.}\ \bibnamefont {Shor}},\ }\bibfield  {title} {\enquote {\bibinfo {title} {Good quantum error-correcting codes exist},}\ }\href {\doibase 10.1103/PhysRevA.54.1098} {\bibfield  {journal} {\bibinfo  {journal} {Phys. Rev. A}\ }\textbf {\bibinfo {volume} {54}},\ \bibinfo {pages} {1098--1105} (\bibinfo {year} {1996})}\BibitemShut {NoStop}%
\bibitem [{\citenamefont {Jeong}\ and\ \citenamefont {Kim}(2002)}]{jeong2002efficient}%
  \BibitemOpen
  \bibfield  {author} {\bibinfo {author} {\bibfnamefont {H.}~\bibnamefont {Jeong}}\ and\ \bibinfo {author} {\bibfnamefont {M.~S.}\ \bibnamefont {Kim}},\ }\bibfield  {title} {\enquote {\bibinfo {title} {Efficient quantum computation using coherent states},}\ }\href {\doibase 10.1103/PhysRevA.65.042305} {\bibfield  {journal} {\bibinfo  {journal} {Phys. Rev. A}\ }\textbf {\bibinfo {volume} {65}},\ \bibinfo {pages} {042305} (\bibinfo {year} {2002})}\BibitemShut {NoStop}%
\bibitem [{\citenamefont {Ralph}\ \emph {et~al.}(2003)\citenamefont {Ralph}, \citenamefont {Gilchrist}, \citenamefont {Milburn}, \citenamefont {Munro},\ and\ \citenamefont {Glancy}}]{ralph2003quantum}%
  \BibitemOpen
  \bibfield  {author} {\bibinfo {author} {\bibfnamefont {T.~C.}\ \bibnamefont {Ralph}}, \bibinfo {author} {\bibfnamefont {A.}~\bibnamefont {Gilchrist}}, \bibinfo {author} {\bibfnamefont {G.~J.}\ \bibnamefont {Milburn}}, \bibinfo {author} {\bibfnamefont {W.~J.}\ \bibnamefont {Munro}}, \ and\ \bibinfo {author} {\bibfnamefont {S.}~\bibnamefont {Glancy}},\ }\bibfield  {title} {\enquote {\bibinfo {title} {Quantum computation with optical coherent states},}\ }\href {\doibase 10.1103/PhysRevA.68.042319} {\bibfield  {journal} {\bibinfo  {journal} {Phys. Rev. A}\ }\textbf {\bibinfo {volume} {68}},\ \bibinfo {pages} {042319} (\bibinfo {year} {2003})}\BibitemShut {NoStop}%
\bibitem [{\citenamefont {Chuang}\ and\ \citenamefont {Yamamoto}(1995)}]{chuang_dualrail}%
  \BibitemOpen
  \bibfield  {author} {\bibinfo {author} {\bibfnamefont {Isaac~L.}\ \bibnamefont {Chuang}}\ and\ \bibinfo {author} {\bibfnamefont {Yoshihisa}\ \bibnamefont {Yamamoto}},\ }\bibfield  {title} {\enquote {\bibinfo {title} {Simple quantum computer},}\ }\href {\doibase 10.1103/PhysRevA.52.3489} {\bibfield  {journal} {\bibinfo  {journal} {Phys. Rev. A}\ }\textbf {\bibinfo {volume} {52}},\ \bibinfo {pages} {3489--3496} (\bibinfo {year} {1995})}\BibitemShut {NoStop}%
\bibitem [{\citenamefont {Chuang}\ and\ \citenamefont {Yamamoto}(1996)}]{chuang1996dualrail}%
  \BibitemOpen
  \bibfield  {author} {\bibinfo {author} {\bibfnamefont {Isaac~L.}\ \bibnamefont {Chuang}}\ and\ \bibinfo {author} {\bibfnamefont {Yoshihisa}\ \bibnamefont {Yamamoto}},\ }\bibfield  {title} {\enquote {\bibinfo {title} {Quantum bit regeneration},}\ }\href {\doibase 10.1103/PhysRevLett.76.4281} {\bibfield  {journal} {\bibinfo  {journal} {Phys. Rev. Lett.}\ }\textbf {\bibinfo {volume} {76}},\ \bibinfo {pages} {4281--4284} (\bibinfo {year} {1996})}\BibitemShut {NoStop}%
\bibitem [{\citenamefont {Chuang}\ \emph {et~al.}(1997)\citenamefont {Chuang}, \citenamefont {Leung},\ and\ \citenamefont {Yamamoto}}]{chuang_pairbinomial}%
  \BibitemOpen
  \bibfield  {author} {\bibinfo {author} {\bibfnamefont {Isaac~L.}\ \bibnamefont {Chuang}}, \bibinfo {author} {\bibfnamefont {Debbie~W.}\ \bibnamefont {Leung}}, \ and\ \bibinfo {author} {\bibfnamefont {Yoshihisa}\ \bibnamefont {Yamamoto}},\ }\bibfield  {title} {\enquote {\bibinfo {title} {Bosonic quantum codes for amplitude damping},}\ }\href {\doibase 10.1103/PhysRevA.56.1114} {\bibfield  {journal} {\bibinfo  {journal} {Phys. Rev. A}\ }\textbf {\bibinfo {volume} {56}},\ \bibinfo {pages} {1114--1125} (\bibinfo {year} {1997})}\BibitemShut {NoStop}%
\bibitem [{\citenamefont {Niu}\ \emph {et~al.}(2018)\citenamefont {Niu}, \citenamefont {Chuang},\ and\ \citenamefont {Shapiro}}]{niu2018hardware}%
  \BibitemOpen
  \bibfield  {author} {\bibinfo {author} {\bibfnamefont {Murphy~Yuezhen}\ \bibnamefont {Niu}}, \bibinfo {author} {\bibfnamefont {Isaac~L.}\ \bibnamefont {Chuang}}, \ and\ \bibinfo {author} {\bibfnamefont {Jeffrey~H.}\ \bibnamefont {Shapiro}},\ }\bibfield  {title} {\enquote {\bibinfo {title} {Hardware-efficient bosonic quantum error-correcting codes based on symmetry operators},}\ }\href {\doibase 10.1103/PhysRevA.97.032323} {\bibfield  {journal} {\bibinfo  {journal} {Phys. Rev. A}\ }\textbf {\bibinfo {volume} {97}},\ \bibinfo {pages} {032323} (\bibinfo {year} {2018})}\BibitemShut {NoStop}%
\bibitem [{\citenamefont {Conway}\ and\ \citenamefont {Sloane}(2013)}]{conway2013sphere}%
  \BibitemOpen
  \bibfield  {author} {\bibinfo {author} {\bibfnamefont {John~Horton}\ \bibnamefont {Conway}}\ and\ \bibinfo {author} {\bibfnamefont {Neil James~Alexander}\ \bibnamefont {Sloane}},\ }\href {https://link.springer.com/book/10.1007/978-1-4757-6568-7} {\emph {\bibinfo {title} {Sphere packings, lattices and groups}}},\ Vol.\ \bibinfo {volume} {290}\ (\bibinfo  {publisher} {Springer Science \& Business Media},\ \bibinfo {year} {2013})\BibitemShut {NoStop}%
\bibitem [{\citenamefont {Radcliffe}(1971)}]{radcliffe1971some}%
  \BibitemOpen
  \bibfield  {author} {\bibinfo {author} {\bibfnamefont {J~Michael}\ \bibnamefont {Radcliffe}},\ }\bibfield  {title} {\enquote {\bibinfo {title} {Some properties of coherent spin states},}\ }\href {\doibase 10.1088/0305-4470/4/3/009} {\bibfield  {journal} {\bibinfo  {journal} {Journal of Physics A: General Physics}\ }\textbf {\bibinfo {volume} {4}},\ \bibinfo {pages} {313} (\bibinfo {year} {1971})}\BibitemShut {NoStop}%
\bibitem [{\citenamefont {Arecchi}\ \emph {et~al.}(1972)\citenamefont {Arecchi}, \citenamefont {Courtens}, \citenamefont {Gilmore},\ and\ \citenamefont {Thomas}}]{arecchi1972atomic}%
  \BibitemOpen
  \bibfield  {author} {\bibinfo {author} {\bibfnamefont {F.~T.}\ \bibnamefont {Arecchi}}, \bibinfo {author} {\bibfnamefont {Eric}\ \bibnamefont {Courtens}}, \bibinfo {author} {\bibfnamefont {Robert}\ \bibnamefont {Gilmore}}, \ and\ \bibinfo {author} {\bibfnamefont {Harry}\ \bibnamefont {Thomas}},\ }\bibfield  {title} {\enquote {\bibinfo {title} {Atomic coherent states in quantum optics},}\ }\href {\doibase 10.1103/PhysRevA.6.2211} {\bibfield  {journal} {\bibinfo  {journal} {Phys. Rev. A}\ }\textbf {\bibinfo {volume} {6}},\ \bibinfo {pages} {2211--2237} (\bibinfo {year} {1972})}\BibitemShut {NoStop}%
\bibitem [{\citenamefont {Tillich}\ and\ \citenamefont {Zémor}(2014)}]{tillich}%
  \BibitemOpen
  \bibfield  {author} {\bibinfo {author} {\bibfnamefont {Jean-Pierre}\ \bibnamefont {Tillich}}\ and\ \bibinfo {author} {\bibfnamefont {Gilles}\ \bibnamefont {Zémor}},\ }\bibfield  {title} {\enquote {\bibinfo {title} {Quantum ldpc codes with positive rate and minimum distance proportional to the square root of the blocklength},}\ }\href {\doibase 10.1109/TIT.2013.2292061} {\bibfield  {journal} {\bibinfo  {journal} {IEEE Transactions on Information Theory}\ }\textbf {\bibinfo {volume} {60}},\ \bibinfo {pages} {1193--1202} (\bibinfo {year} {2014})}\BibitemShut {NoStop}%
\bibitem [{\citenamefont {Gel'fand}(1986)}]{gel1986general}%
  \BibitemOpen
  \bibfield  {author} {\bibinfo {author} {\bibfnamefont {Izrail~Moiseevich}\ \bibnamefont {Gel'fand}},\ }\bibfield  {title} {\enquote {\bibinfo {title} {General theory of hypergeometric functions},}\ }in\ \href@noop {} {\emph {\bibinfo {booktitle} {Doklady Akademii Nauk}}},\ Vol.\ \bibinfo {volume} {288}\ (\bibinfo {organization} {Russian Academy of Sciences},\ \bibinfo {year} {1986})\ pp.\ \bibinfo {pages} {14--18}\BibitemShut {NoStop}%
\bibitem [{\citenamefont {Gel'fand}\ and\ \citenamefont {Gel'fand}(1986)}]{gel1986generalized}%
  \BibitemOpen
  \bibfield  {author} {\bibinfo {author} {\bibfnamefont {Izrail~Moiseevich}\ \bibnamefont {Gel'fand}}\ and\ \bibinfo {author} {\bibfnamefont {Sergei~Izrail'evich}\ \bibnamefont {Gel'fand}},\ }\bibfield  {title} {\enquote {\bibinfo {title} {Generalized hypergeometric equations},}\ }in\ \href@noop {} {\emph {\bibinfo {booktitle} {Doklady Akademii Nauk}}},\ Vol.\ \bibinfo {volume} {288}\ (\bibinfo {organization} {Russian Academy of Sciences},\ \bibinfo {year} {1986})\ pp.\ \bibinfo {pages} {279--283}\BibitemShut {NoStop}%
\bibitem [{\citenamefont {Gel'fand}\ \emph {et~al.}(1987)\citenamefont {Gel'fand}, \citenamefont {Graev},\ and\ \citenamefont {Zelevinskii}}]{gel1987holonomic}%
  \BibitemOpen
  \bibfield  {author} {\bibinfo {author} {\bibfnamefont {Izrail~Moiseevich}\ \bibnamefont {Gel'fand}}, \bibinfo {author} {\bibfnamefont {Mark~Iosifovich}\ \bibnamefont {Graev}}, \ and\ \bibinfo {author} {\bibfnamefont {Andrei~Vladlenovich}\ \bibnamefont {Zelevinskii}},\ }\bibfield  {title} {\enquote {\bibinfo {title} {Holonomic systems of equations and series of hypergeometric type},}\ }in\ \href@noop {} {\emph {\bibinfo {booktitle} {Doklady Akademii Nauk}}},\ Vol.\ \bibinfo {volume} {295}\ (\bibinfo {organization} {Russian Academy of Sciences},\ \bibinfo {year} {1987})\ pp.\ \bibinfo {pages} {14--19}\BibitemShut {NoStop}%
\bibitem [{\citenamefont {Gel'fand}\ \emph {et~al.}(1989)\citenamefont {Gel'fand}, \citenamefont {Zelevinskii},\ and\ \citenamefont {Kapranov}}]{gel1989hypergeometric}%
  \BibitemOpen
  \bibfield  {author} {\bibinfo {author} {\bibfnamefont {Izrail~Moiseevich}\ \bibnamefont {Gel'fand}}, \bibinfo {author} {\bibfnamefont {Andrei~Vladlenovich}\ \bibnamefont {Zelevinskii}}, \ and\ \bibinfo {author} {\bibfnamefont {Mikhail~Mikhailovich}\ \bibnamefont {Kapranov}},\ }\bibfield  {title} {\enquote {\bibinfo {title} {Hypergeometric functions and toral manifolds},}\ }\href {https://link.springer.com/article/10.1007/BF01078777} {\bibfield  {journal} {\bibinfo  {journal} {Funktsional'nyi Analiz i ego Prilozheniya}\ }\textbf {\bibinfo {volume} {23}},\ \bibinfo {pages} {12--26} (\bibinfo {year} {1989})}\BibitemShut {NoStop}%
\bibitem [{\citenamefont {Adolphson}(1994)}]{adolphson1994hypergeometric}%
  \BibitemOpen
  \bibfield  {author} {\bibinfo {author} {\bibfnamefont {Alan}\ \bibnamefont {Adolphson}},\ }\bibfield  {title} {\enquote {\bibinfo {title} {Hypergeometric functions and rings generated by monomials},}\ }\href {\doibase 10.1215/S0012-7094-94-07313-4} {\bibfield  {journal} {\bibinfo  {journal} {Duke Mathematical Journal}\ }\textbf {\bibinfo {volume} {73}},\ \bibinfo {pages} {269} (\bibinfo {year} {1994})}\BibitemShut {NoStop}%
\bibitem [{\citenamefont {Gel'fand}\ and\ \citenamefont {Graev}(1990)}]{gel1990hypergeometric}%
  \BibitemOpen
  \bibfield  {author} {\bibinfo {author} {\bibfnamefont {Izrail~Moiseevich}\ \bibnamefont {Gel'fand}}\ and\ \bibinfo {author} {\bibfnamefont {Mark~Iosifovich}\ \bibnamefont {Graev}},\ }\bibfield  {title} {\enquote {\bibinfo {title} {Hypergeometric functions associated with the grassmannian g3, 6},}\ }\href {https://iopscience.iop.org/article/10.1070/SM1990v066n01ABEH001931} {\bibfield  {journal} {\bibinfo  {journal} {Mathematics of the USSR-Sbornik}\ }\textbf {\bibinfo {volume} {66}},\ \bibinfo {pages} {1} (\bibinfo {year} {1990})}\BibitemShut {NoStop}%
\bibitem [{\citenamefont {Gel'fand}\ \emph {et~al.}(1992)\citenamefont {Gel'fand}, \citenamefont {Graev},\ and\ \citenamefont {Retakh}}]{gel1992general}%
  \BibitemOpen
  \bibfield  {author} {\bibinfo {author} {\bibfnamefont {Izrail~Moiseevich}\ \bibnamefont {Gel'fand}}, \bibinfo {author} {\bibfnamefont {Mark~Iosifovich}\ \bibnamefont {Graev}}, \ and\ \bibinfo {author} {\bibfnamefont {Vladimir~Solomonovich}\ \bibnamefont {Retakh}},\ }\bibfield  {title} {\enquote {\bibinfo {title} {General hypergeometric systems of equations and series of hypergeometric type},}\ }\href {https://iopscience.iop.org/article/10.1070/RM1992v047n04ABEH000915/pdf} {\bibfield  {journal} {\bibinfo  {journal} {Russian Mathematical Surveys}\ }\textbf {\bibinfo {volume} {47}},\ \bibinfo {pages} {1} (\bibinfo {year} {1992})}\BibitemShut {NoStop}%
\bibitem [{\citenamefont {Stienstra}(2007)}]{stienstra2007gkz}%
  \BibitemOpen
  \bibfield  {author} {\bibinfo {author} {\bibfnamefont {Jan}\ \bibnamefont {Stienstra}},\ }\bibfield  {title} {\enquote {\bibinfo {title} {Gkz hypergeometric structures},}\ }\href {https://link.springer.com/chapter/10.1007/978-3-7643-8284-1_12} {\bibfield  {journal} {\bibinfo  {journal} {Arithmetic and Geometry Around Hypergeometric Functions: Lecture Notes of a CIMPA Summer School held at Galatasaray University, Istanbul, 2005}\ ,\ \bibinfo {pages} {313--371}} (\bibinfo {year} {2007})}\BibitemShut {NoStop}%
\bibitem [{\citenamefont {Serafini}(2017)}]{serafini2017quantum}%
  \BibitemOpen
  \bibfield  {author} {\bibinfo {author} {\bibfnamefont {Alessio}\ \bibnamefont {Serafini}},\ }\href {\doibase 10.1201/9781315118727} {\emph {\bibinfo {title} {Quantum continuous variables: a primer of theoretical methods}}}\ (\bibinfo  {publisher} {CRC press},\ \bibinfo {year} {2017})\BibitemShut {NoStop}%
\bibitem [{\citenamefont {Albert}(2025)}]{albert2022bosonic}%
  \BibitemOpen
  \bibfield  {author} {\bibinfo {author} {\bibfnamefont {Victor~V.}\ \bibnamefont {Albert}},\ }\enquote {\bibinfo {title} {Bosonic coding: introduction and use cases},}\ in\ \href {\doibase 10.3254/ENFI250007} {\emph {\bibinfo {booktitle} {Quantum Fluids of Light and Matter}}},\ \bibinfo {series} {International School of Physics “Enrico Fermi”}, Vol.\ \bibinfo {volume} {209},\ \bibinfo {editor} {edited by\ \bibinfo {editor} {\bibfnamefont {Alberto}\ \bibnamefont {Bramati}}, \bibinfo {editor} {\bibfnamefont {Iacopo}\ \bibnamefont {Carusotto}}, \ and\ \bibinfo {editor} {\bibfnamefont {Cristiano}\ \bibnamefont {Ciuti}}}\ (\bibinfo  {publisher} {IOS Press},\ \bibinfo {address} {NL},\ \bibinfo {year} {2025})\ p.\ \bibinfo {pages} {79–107}\BibitemShut {NoStop}%
\bibitem [{\citenamefont {Skagerstam}(1985)}]{skagerstam1985quasi}%
  \BibitemOpen
  \bibfield  {author} {\bibinfo {author} {\bibfnamefont {B-S}\ \bibnamefont {Skagerstam}},\ }\bibfield  {title} {\enquote {\bibinfo {title} {Quasi-coherent states for unitary groups},}\ }\href {\doibase 10.1088/0305-4470/18/1/011} {\bibfield  {journal} {\bibinfo  {journal} {Journal of Physics A: Mathematical and General}\ }\textbf {\bibinfo {volume} {18}},\ \bibinfo {pages} {1} (\bibinfo {year} {1985})}\BibitemShut {NoStop}%
\bibitem [{\citenamefont {Drummond}\ and\ \citenamefont {Reid}(2016)}]{drummond2016coherent}%
  \BibitemOpen
  \bibfield  {author} {\bibinfo {author} {\bibfnamefont {P.~D.}\ \bibnamefont {Drummond}}\ and\ \bibinfo {author} {\bibfnamefont {M.~D.}\ \bibnamefont {Reid}},\ }\bibfield  {title} {\enquote {\bibinfo {title} {Coherent states in projected hilbert spaces},}\ }\href {\doibase 10.1103/PhysRevA.94.063851} {\bibfield  {journal} {\bibinfo  {journal} {Phys. Rev. A}\ }\textbf {\bibinfo {volume} {94}},\ \bibinfo {pages} {063851} (\bibinfo {year} {2016})}\BibitemShut {NoStop}%
\bibitem [{\citenamefont {Klauder}\ and\ \citenamefont {Skagerstam}(1985)}]{klauder1985coherent}%
  \BibitemOpen
  \bibfield  {author} {\bibinfo {author} {\bibfnamefont {J}~\bibnamefont {Klauder}}\ and\ \bibinfo {author} {\bibfnamefont {B}~\bibnamefont {Skagerstam}},\ }\href {\doibase 10.1142/0096} {\emph {\bibinfo {title} {Coherent states: applications in physics and mathematical physics}}}\ (\bibinfo  {publisher} {WORLD SCIENTIFIC},\ \bibinfo {year} {1985})\BibitemShut {NoStop}%
\bibitem [{\citenamefont {Dodonov}(2002)}]{dodonov2002nonclassical}%
  \BibitemOpen
  \bibfield  {author} {\bibinfo {author} {\bibfnamefont {VV}~\bibnamefont {Dodonov}},\ }\bibfield  {title} {\enquote {\bibinfo {title} {`nonclassical'states in quantum optics: asqueezed'review of the first 75 years},}\ }\href {\doibase 10.1088/1464-4266/4/1/201} {\bibfield  {journal} {\bibinfo  {journal} {Journal of Optics B: Quantum and Semiclassical Optics}\ }\textbf {\bibinfo {volume} {4}},\ \bibinfo {pages} {R1} (\bibinfo {year} {2002})}\BibitemShut {NoStop}%
\bibitem [{\citenamefont {Gazeau}(2009)}]{gazeau2009coherent}%
  \BibitemOpen
  \bibfield  {author} {\bibinfo {author} {\bibfnamefont {Jean-Pierre}\ \bibnamefont {Gazeau}},\ }\enquote {\bibinfo {title} {Coherent states in quantum physics},}\ \ (\bibinfo  {publisher} {John Wiley \& Sons, Ltd},\ \bibinfo {year} {2009})\BibitemShut {NoStop}%
\bibitem [{\citenamefont {Barut}\ and\ \citenamefont {Girardello}(1971)}]{barut1971new}%
  \BibitemOpen
  \bibfield  {author} {\bibinfo {author} {\bibfnamefont {AO}~\bibnamefont {Barut}}\ and\ \bibinfo {author} {\bibfnamefont {L}~\bibnamefont {Girardello}},\ }\bibfield  {title} {\enquote {\bibinfo {title} {New “coherent” states associated with non-compact groups},}\ }\href {\doibase 10.1007/BF01646483} {\bibfield  {journal} {\bibinfo  {journal} {Communications in Mathematical Physics}\ }\textbf {\bibinfo {volume} {21}},\ \bibinfo {pages} {41--55} (\bibinfo {year} {1971})}\BibitemShut {NoStop}%
\bibitem [{\citenamefont {Agarwal}(1986)}]{agarwal1986generation}%
  \BibitemOpen
  \bibfield  {author} {\bibinfo {author} {\bibfnamefont {G.~S.}\ \bibnamefont {Agarwal}},\ }\bibfield  {title} {\enquote {\bibinfo {title} {Generation of pair coherent states and squeezing via the competition of four-wave mixing and amplified spontaneous emission},}\ }\href {\doibase 10.1103/PhysRevLett.57.827} {\bibfield  {journal} {\bibinfo  {journal} {Phys. Rev. Lett.}\ }\textbf {\bibinfo {volume} {57}},\ \bibinfo {pages} {827--830} (\bibinfo {year} {1986})}\BibitemShut {NoStop}%
\bibitem [{\citenamefont {Agarwal}(1988)}]{agarwal1988nonclassical}%
  \BibitemOpen
  \bibfield  {author} {\bibinfo {author} {\bibfnamefont {Govind~S}\ \bibnamefont {Agarwal}},\ }\bibfield  {title} {\enquote {\bibinfo {title} {Nonclassical statistics of fields in pair coherent states},}\ }\href {\doibase 10.1364/JOSAB.5.001940} {\bibfield  {journal} {\bibinfo  {journal} {JOSA B}\ }\textbf {\bibinfo {volume} {5}},\ \bibinfo {pages} {1940--1947} (\bibinfo {year} {1988})}\BibitemShut {NoStop}%
\bibitem [{\citenamefont {Gerry}\ and\ \citenamefont {Grobe}(1995)}]{gerry1995nonclassical}%
  \BibitemOpen
  \bibfield  {author} {\bibinfo {author} {\bibfnamefont {Christopher~C.}\ \bibnamefont {Gerry}}\ and\ \bibinfo {author} {\bibfnamefont {Rainer}\ \bibnamefont {Grobe}},\ }\bibfield  {title} {\enquote {\bibinfo {title} {Nonclassical properties of correlated two-mode schr\"odinger cat states},}\ }\href {\doibase 10.1103/PhysRevA.51.1698} {\bibfield  {journal} {\bibinfo  {journal} {Phys. Rev. A}\ }\textbf {\bibinfo {volume} {51}},\ \bibinfo {pages} {1698--1701} (\bibinfo {year} {1995})}\BibitemShut {NoStop}%
\bibitem [{\citenamefont {An}(2003)}]{an2003even}%
  \BibitemOpen
  \bibfield  {author} {\bibinfo {author} {\bibfnamefont {Nguyen~Ba}\ \bibnamefont {An}},\ }\bibfield  {title} {\enquote {\bibinfo {title} {Even and odd trio coherent states: number distribution, squeezing and realization scheme},}\ }\href {\doibase 10.1016/S0375-9601(03)00652-2} {\bibfield  {journal} {\bibinfo  {journal} {Physics Letters A}\ }\textbf {\bibinfo {volume} {312}},\ \bibinfo {pages} {268--276} (\bibinfo {year} {2003})}\BibitemShut {NoStop}%
\bibitem [{\citenamefont {Gitman}\ and\ \citenamefont {Shelepin}(1993)}]{gitman1993coherent}%
  \BibitemOpen
  \bibfield  {author} {\bibinfo {author} {\bibfnamefont {Dmitri~Maximovitch}\ \bibnamefont {Gitman}}\ and\ \bibinfo {author} {\bibfnamefont {AL}~\bibnamefont {Shelepin}},\ }\bibfield  {title} {\enquote {\bibinfo {title} {Coherent states of su (n) groups},}\ }\href {\doibase 10.1088/0305-4470/26/2/018} {\bibfield  {journal} {\bibinfo  {journal} {Journal of Physics A: Mathematical and General}\ }\textbf {\bibinfo {volume} {26}},\ \bibinfo {pages} {313} (\bibinfo {year} {1993})}\BibitemShut {NoStop}%
\bibitem [{\citenamefont {Nemoto}(2000)}]{nemoto2000generalized}%
  \BibitemOpen
  \bibfield  {author} {\bibinfo {author} {\bibfnamefont {Kae}\ \bibnamefont {Nemoto}},\ }\bibfield  {title} {\enquote {\bibinfo {title} {Generalized coherent states for su (n) systems},}\ }\href {\doibase 10.1088/0305-4470/33/17/307} {\bibfield  {journal} {\bibinfo  {journal} {Journal of Physics A: Mathematical and General}\ }\textbf {\bibinfo {volume} {33}},\ \bibinfo {pages} {3493} (\bibinfo {year} {2000})}\BibitemShut {NoStop}%
\bibitem [{\citenamefont {Calixto}\ \emph {et~al.}(2021)\citenamefont {Calixto}, \citenamefont {Mayorgas},\ and\ \citenamefont {Guerrero}}]{calixto2021entanglement}%
  \BibitemOpen
  \bibfield  {author} {\bibinfo {author} {\bibfnamefont {Manuel}\ \bibnamefont {Calixto}}, \bibinfo {author} {\bibfnamefont {Alberto}\ \bibnamefont {Mayorgas}}, \ and\ \bibinfo {author} {\bibfnamefont {Julio}\ \bibnamefont {Guerrero}},\ }\bibfield  {title} {\enquote {\bibinfo {title} {Entanglement and u (d)-spin squeezing in symmetric multi-qudit systems and applications to quantum phase transitions in lipkin--meshkov--glick d-level atom models},}\ }\href {\doibase 10.1007/s11128-021-03218-6} {\bibfield  {journal} {\bibinfo  {journal} {Quantum Information Processing}\ }\textbf {\bibinfo {volume} {20}},\ \bibinfo {pages} {304} (\bibinfo {year} {2021})}\BibitemShut {NoStop}%
\bibitem [{\citenamefont {Turchette}\ \emph {et~al.}(2000)\citenamefont {Turchette}, \citenamefont {Myatt}, \citenamefont {King}, \citenamefont {Sackett}, \citenamefont {Kielpinski}, \citenamefont {Itano}, \citenamefont {Monroe},\ and\ \citenamefont {Wineland}}]{turchette2000decoherence}%
  \BibitemOpen
  \bibfield  {author} {\bibinfo {author} {\bibfnamefont {Q.~A.}\ \bibnamefont {Turchette}}, \bibinfo {author} {\bibfnamefont {C.~J.}\ \bibnamefont {Myatt}}, \bibinfo {author} {\bibfnamefont {B.~E.}\ \bibnamefont {King}}, \bibinfo {author} {\bibfnamefont {C.~A.}\ \bibnamefont {Sackett}}, \bibinfo {author} {\bibfnamefont {D.}~\bibnamefont {Kielpinski}}, \bibinfo {author} {\bibfnamefont {W.~M.}\ \bibnamefont {Itano}}, \bibinfo {author} {\bibfnamefont {C.}~\bibnamefont {Monroe}}, \ and\ \bibinfo {author} {\bibfnamefont {D.~J.}\ \bibnamefont {Wineland}},\ }\bibfield  {title} {\enquote {\bibinfo {title} {Decoherence and decay of motional quantum states of a trapped atom coupled to engineered reservoirs},}\ }\href {\doibase 10.1103/PhysRevA.62.053807} {\bibfield  {journal} {\bibinfo  {journal} {Phys. Rev. A}\ }\textbf {\bibinfo {volume} {62}},\ \bibinfo {pages} {053807} (\bibinfo {year} {2000})}\BibitemShut {NoStop}%
\bibitem [{\citenamefont {Gottesman}(2016)}]{gottesman2016surviving}%
  \BibitemOpen
  \bibfield  {author} {\bibinfo {author} {\bibfnamefont {Daniel}\ \bibnamefont {Gottesman}},\ }\bibfield  {title} {\enquote {\bibinfo {title} {Surviving as a quantum computer in a classical world},}\ }\href {https://www.cs.umd.edu/class/spring2024/cmsc858G/QECCbook-2024-ch1-11.pdf} {\bibfield  {journal} {\bibinfo  {journal} {Textbook manuscript preprint}\ } (\bibinfo {year} {2016})}\BibitemShut {NoStop}%
\bibitem [{\citenamefont {Knill}\ \emph {et~al.}(2000{\natexlab{a}})\citenamefont {Knill}, \citenamefont {Laflamme},\ and\ \citenamefont {Milburn}}]{knill2000efficient}%
  \BibitemOpen
  \bibfield  {author} {\bibinfo {author} {\bibfnamefont {Emanuel}\ \bibnamefont {Knill}}, \bibinfo {author} {\bibfnamefont {Raymond}\ \bibnamefont {Laflamme}}, \ and\ \bibinfo {author} {\bibfnamefont {G}~\bibnamefont {Milburn}},\ }\bibfield  {title} {\enquote {\bibinfo {title} {Efficient linear optics quantum computation},}\ }\href {https://arxiv.org/abs/quant-ph/0006088} {\bibfield  {journal} {\bibinfo  {journal} {arXiv preprint quant-ph/0006088}\ } (\bibinfo {year} {2000}{\natexlab{a}})}\BibitemShut {NoStop}%
\bibitem [{\citenamefont {Ralph}\ \emph {et~al.}(2005)\citenamefont {Ralph}, \citenamefont {Hayes},\ and\ \citenamefont {Gilchrist}}]{ralph2005loss}%
  \BibitemOpen
  \bibfield  {author} {\bibinfo {author} {\bibfnamefont {T.~C.}\ \bibnamefont {Ralph}}, \bibinfo {author} {\bibfnamefont {A.~J.~F.}\ \bibnamefont {Hayes}}, \ and\ \bibinfo {author} {\bibfnamefont {Alexei}\ \bibnamefont {Gilchrist}},\ }\bibfield  {title} {\enquote {\bibinfo {title} {Loss-tolerant optical qubits},}\ }\href {\doibase 10.1103/PhysRevLett.95.100501} {\bibfield  {journal} {\bibinfo  {journal} {Phys. Rev. Lett.}\ }\textbf {\bibinfo {volume} {95}},\ \bibinfo {pages} {100501} (\bibinfo {year} {2005})}\BibitemShut {NoStop}%
\bibitem [{\citenamefont {Kitaev}(1997{\natexlab{a}})}]{kitaev1997quantum}%
  \BibitemOpen
  \bibfield  {author} {\bibinfo {author} {\bibfnamefont {A~Yu}\ \bibnamefont {Kitaev}},\ }\bibfield  {title} {\enquote {\bibinfo {title} {Quantum computations: algorithms and error correction},}\ }\href {\doibase 10.1070/RM1997v052n06ABEH002155} {\bibfield  {journal} {\bibinfo  {journal} {Russian Mathematical Surveys}\ }\textbf {\bibinfo {volume} {52}},\ \bibinfo {pages} {1191} (\bibinfo {year} {1997}{\natexlab{a}})}\BibitemShut {NoStop}%
\bibitem [{\citenamefont {Kitaev}(1997{\natexlab{b}})}]{kitaev1997quantumimperfect}%
  \BibitemOpen
  \bibfield  {author} {\bibinfo {author} {\bibfnamefont {A~Yu}\ \bibnamefont {Kitaev}},\ }\bibfield  {title} {\enquote {\bibinfo {title} {Quantum error correction with imperfect gates},}\ }in\ \href {\doibase 10.1007/978-1-4615-5923-8_19} {\emph {\bibinfo {booktitle} {Quantum communication, computing, and measurement}}}\ (\bibinfo  {publisher} {Springer},\ \bibinfo {year} {1997})\ pp.\ \bibinfo {pages} {181--188}\BibitemShut {NoStop}%
\bibitem [{\citenamefont {Bravyi}\ and\ \citenamefont {Kitaev}(1998)}]{bravyi1998quantum}%
  \BibitemOpen
  \bibfield  {author} {\bibinfo {author} {\bibfnamefont {Sergey~B}\ \bibnamefont {Bravyi}}\ and\ \bibinfo {author} {\bibfnamefont {A~Yu}\ \bibnamefont {Kitaev}},\ }\bibfield  {title} {\enquote {\bibinfo {title} {Quantum codes on a lattice with boundary},}\ }\href {https://arxiv.org/abs/quant-ph/9811052} {\bibfield  {journal} {\bibinfo  {journal} {arXiv preprint quant-ph/9811052}\ } (\bibinfo {year} {1998})}\BibitemShut {NoStop}%
\bibitem [{\citenamefont {Dennis}\ \emph {et~al.}(2002)\citenamefont {Dennis}, \citenamefont {Kitaev}, \citenamefont {Landahl},\ and\ \citenamefont {Preskill}}]{dennis2002topological}%
  \BibitemOpen
  \bibfield  {author} {\bibinfo {author} {\bibfnamefont {Eric}\ \bibnamefont {Dennis}}, \bibinfo {author} {\bibfnamefont {Alexei}\ \bibnamefont {Kitaev}}, \bibinfo {author} {\bibfnamefont {Andrew}\ \bibnamefont {Landahl}}, \ and\ \bibinfo {author} {\bibfnamefont {John}\ \bibnamefont {Preskill}},\ }\bibfield  {title} {\enquote {\bibinfo {title} {Topological quantum memory},}\ }\href {\doibase 10.1063/1.1499754} {\bibfield  {journal} {\bibinfo  {journal} {Journal of Mathematical Physics}\ }\textbf {\bibinfo {volume} {43}},\ \bibinfo {pages} {4452--4505} (\bibinfo {year} {2002})}\BibitemShut {NoStop}%
\bibitem [{\citenamefont {Kitaev}(2003)}]{kitaev2003fault}%
  \BibitemOpen
  \bibfield  {author} {\bibinfo {author} {\bibfnamefont {A~Yu}\ \bibnamefont {Kitaev}},\ }\bibfield  {title} {\enquote {\bibinfo {title} {Fault-tolerant quantum computation by anyons},}\ }\href {\doibase 10.1016/S0003-4916(02)00018-0} {\bibfield  {journal} {\bibinfo  {journal} {Annals of physics}\ }\textbf {\bibinfo {volume} {303}},\ \bibinfo {pages} {2--30} (\bibinfo {year} {2003})}\BibitemShut {NoStop}%
\bibitem [{\citenamefont {Chamberland}\ \emph {et~al.}(2022)\citenamefont {Chamberland}, \citenamefont {Noh}, \citenamefont {Arrangoiz-Arriola}, \citenamefont {Campbell}, \citenamefont {Hann}, \citenamefont {Iverson}, \citenamefont {Putterman}, \citenamefont {Bohdanowicz}, \citenamefont {Flammia}, \citenamefont {Keller}, \citenamefont {Refael}, \citenamefont {Preskill}, \citenamefont {Jiang}, \citenamefont {Safavi-Naeini}, \citenamefont {Painter},\ and\ \citenamefont {Brand\~ao}}]{chamberland2022building}%
  \BibitemOpen
  \bibfield  {author} {\bibinfo {author} {\bibfnamefont {Christopher}\ \bibnamefont {Chamberland}}, \bibinfo {author} {\bibfnamefont {Kyungjoo}\ \bibnamefont {Noh}}, \bibinfo {author} {\bibfnamefont {Patricio}\ \bibnamefont {Arrangoiz-Arriola}}, \bibinfo {author} {\bibfnamefont {Earl~T.}\ \bibnamefont {Campbell}}, \bibinfo {author} {\bibfnamefont {Connor~T.}\ \bibnamefont {Hann}}, \bibinfo {author} {\bibfnamefont {Joseph}\ \bibnamefont {Iverson}}, \bibinfo {author} {\bibfnamefont {Harald}\ \bibnamefont {Putterman}}, \bibinfo {author} {\bibfnamefont {Thomas~C.}\ \bibnamefont {Bohdanowicz}}, \bibinfo {author} {\bibfnamefont {Steven~T.}\ \bibnamefont {Flammia}}, \bibinfo {author} {\bibfnamefont {Andrew}\ \bibnamefont {Keller}}, \bibinfo {author} {\bibfnamefont {Gil}\ \bibnamefont {Refael}}, \bibinfo {author} {\bibfnamefont {John}\ \bibnamefont {Preskill}}, \bibinfo {author} {\bibfnamefont {Liang}\ \bibnamefont {Jiang}}, \bibinfo {author} {\bibfnamefont {Amir~H.}\ \bibnamefont {Safavi-Naeini}}, \bibinfo {author}
  {\bibfnamefont {Oskar}\ \bibnamefont {Painter}}, \ and\ \bibinfo {author} {\bibfnamefont {Fernando~G.S.L.}\ \bibnamefont {Brand\~ao}},\ }\bibfield  {title} {\enquote {\bibinfo {title} {Building a fault-tolerant quantum computer using concatenated cat codes},}\ }\href {\doibase 10.1103/PRXQuantum.3.010329} {\bibfield  {journal} {\bibinfo  {journal} {PRX Quantum}\ }\textbf {\bibinfo {volume} {3}},\ \bibinfo {pages} {010329} (\bibinfo {year} {2022})}\BibitemShut {NoStop}%
\bibitem [{\citenamefont {Girvin}(2014)}]{girvin2014circuit}%
  \BibitemOpen
  \bibfield  {author} {\bibinfo {author} {\bibfnamefont {S.~M.}\ \bibnamefont {Girvin}},\ }\bibfield  {title} {\enquote {\bibinfo {title} {Circuit qed: superconducting qubits coupled to microwave photons},}\ }in\ \href {\doibase 10.1093/acprof:oso/9780199681181.003.0003} {\emph {\bibinfo {booktitle} {Quantum Machines: Measurement and Control of Engineered Quantum Systems: Lecture Notes of the Les Houches Summer School: Volume 96, July 2011}}}\ (\bibinfo  {publisher} {Oxford University Press},\ \bibinfo {year} {2014})\BibitemShut {NoStop}%
\bibitem [{\citenamefont {Grimsmo}\ \emph {et~al.}(2020)\citenamefont {Grimsmo}, \citenamefont {Combes},\ and\ \citenamefont {Baragiola}}]{grimsmo2020quantum}%
  \BibitemOpen
  \bibfield  {author} {\bibinfo {author} {\bibfnamefont {Arne~L.}\ \bibnamefont {Grimsmo}}, \bibinfo {author} {\bibfnamefont {Joshua}\ \bibnamefont {Combes}}, \ and\ \bibinfo {author} {\bibfnamefont {Ben~Q.}\ \bibnamefont {Baragiola}},\ }\bibfield  {title} {\enquote {\bibinfo {title} {Quantum computing with rotation-symmetric bosonic codes},}\ }\href {\doibase 10.1103/PhysRevX.10.011058} {\bibfield  {journal} {\bibinfo  {journal} {Phys. Rev. X}\ }\textbf {\bibinfo {volume} {10}},\ \bibinfo {pages} {011058} (\bibinfo {year} {2020})}\BibitemShut {NoStop}%
\bibitem [{\citenamefont {Knill}\ \emph {et~al.}(2000{\natexlab{b}})\citenamefont {Knill}, \citenamefont {Laflamme},\ and\ \citenamefont {Viola}}]{kl}%
  \BibitemOpen
  \bibfield  {author} {\bibinfo {author} {\bibfnamefont {Emanuel}\ \bibnamefont {Knill}}, \bibinfo {author} {\bibfnamefont {Raymond}\ \bibnamefont {Laflamme}}, \ and\ \bibinfo {author} {\bibfnamefont {Lorenza}\ \bibnamefont {Viola}},\ }\bibfield  {title} {\enquote {\bibinfo {title} {Theory of quantum error correction for general noise},}\ }\href {\doibase 10.1103/PhysRevLett.84.2525} {\bibfield  {journal} {\bibinfo  {journal} {Phys. Rev. Lett.}\ }\textbf {\bibinfo {volume} {84}},\ \bibinfo {pages} {2525--2528} (\bibinfo {year} {2000}{\natexlab{b}})}\BibitemShut {NoStop}%
\bibitem [{\citenamefont {Bhaumik}\ \emph {et~al.}(1976)\citenamefont {Bhaumik}, \citenamefont {Bhaumik},\ and\ \citenamefont {Dutta-Roy}}]{bhaumik1976charged}%
  \BibitemOpen
  \bibfield  {author} {\bibinfo {author} {\bibfnamefont {Debajyoti}\ \bibnamefont {Bhaumik}}, \bibinfo {author} {\bibfnamefont {Kamales}\ \bibnamefont {Bhaumik}}, \ and\ \bibinfo {author} {\bibfnamefont {Binayak}\ \bibnamefont {Dutta-Roy}},\ }\bibfield  {title} {\enquote {\bibinfo {title} {Charged bosons and the coherent state},}\ }\href {\doibase 10.1088/0305-4470/9/9/011} {\bibfield  {journal} {\bibinfo  {journal} {Journal of Physics A: Mathematical and General}\ }\textbf {\bibinfo {volume} {9}},\ \bibinfo {pages} {1507} (\bibinfo {year} {1976})}\BibitemShut {NoStop}%
\bibitem [{{\relax DLMF}()}]{DLMF}%
  \BibitemOpen
  {\relax DLMF},\ \href {https://dlmf.nist.gov/} {\enquote {\bibinfo {title} {{\it NIST Digital Library of Mathematical Functions}},}\ }\bibinfo {howpublished} {\url{https://dlmf.nist.gov/}, Release 1.2.2 of 2024-09-15},\ \bibinfo {note} {f.~W.~J. Olver, A.~B. {Olde Daalhuis}, D.~W. Lozier, B.~I. Schneider, R.~F. Boisvert, C.~W. Clark, B.~R. Miller, B.~V. Saunders, H.~S. Cohl, and M.~A. McClain, eds.}\BibitemShut {Stop}%
\bibitem [{\citenamefont {Albert}(2018)}]{albert2018lindbladians}%
  \BibitemOpen
  \bibfield  {author} {\bibinfo {author} {\bibfnamefont {Victor~V}\ \bibnamefont {Albert}},\ }\bibfield  {title} {\enquote {\bibinfo {title} {Lindbladians with multiple steady states: theory and applications},}\ }\href {https://arxiv.org/abs/1802.00010} {\bibfield  {journal} {\bibinfo  {journal} {arXiv preprint arXiv:1802.00010}\ } (\bibinfo {year} {2018})}\BibitemShut {NoStop}%
\bibitem [{\citenamefont {Hayden}\ \emph {et~al.}(2021)\citenamefont {Hayden}, \citenamefont {Nezami}, \citenamefont {Popescu},\ and\ \citenamefont {Salton}}]{hayden2021error}%
  \BibitemOpen
  \bibfield  {author} {\bibinfo {author} {\bibfnamefont {Patrick}\ \bibnamefont {Hayden}}, \bibinfo {author} {\bibfnamefont {Sepehr}\ \bibnamefont {Nezami}}, \bibinfo {author} {\bibfnamefont {Sandu}\ \bibnamefont {Popescu}}, \ and\ \bibinfo {author} {\bibfnamefont {Grant}\ \bibnamefont {Salton}},\ }\bibfield  {title} {\enquote {\bibinfo {title} {Error correction of quantum reference frame information},}\ }\href {\doibase 10.1103/PRXQuantum.2.010326} {\bibfield  {journal} {\bibinfo  {journal} {PRX Quantum}\ }\textbf {\bibinfo {volume} {2}},\ \bibinfo {pages} {010326} (\bibinfo {year} {2021})}\BibitemShut {NoStop}%
\bibitem [{\citenamefont {Faist}\ \emph {et~al.}(2020)\citenamefont {Faist}, \citenamefont {Nezami}, \citenamefont {Albert}, \citenamefont {Salton}, \citenamefont {Pastawski}, \citenamefont {Hayden},\ and\ \citenamefont {Preskill}}]{faist2020continuous}%
  \BibitemOpen
  \bibfield  {author} {\bibinfo {author} {\bibfnamefont {Philippe}\ \bibnamefont {Faist}}, \bibinfo {author} {\bibfnamefont {Sepehr}\ \bibnamefont {Nezami}}, \bibinfo {author} {\bibfnamefont {Victor~V.}\ \bibnamefont {Albert}}, \bibinfo {author} {\bibfnamefont {Grant}\ \bibnamefont {Salton}}, \bibinfo {author} {\bibfnamefont {Fernando}\ \bibnamefont {Pastawski}}, \bibinfo {author} {\bibfnamefont {Patrick}\ \bibnamefont {Hayden}}, \ and\ \bibinfo {author} {\bibfnamefont {John}\ \bibnamefont {Preskill}},\ }\bibfield  {title} {\enquote {\bibinfo {title} {Continuous symmetries and approximate quantum error correction},}\ }\href {\doibase 10.1103/PhysRevX.10.041018} {\bibfield  {journal} {\bibinfo  {journal} {Phys. Rev. X}\ }\textbf {\bibinfo {volume} {10}},\ \bibinfo {pages} {041018} (\bibinfo {year} {2020})}\BibitemShut {NoStop}%
\bibitem [{\citenamefont {Leung}\ \emph {et~al.}(1997)\citenamefont {Leung}, \citenamefont {Nielsen}, \citenamefont {Chuang},\ and\ \citenamefont {Yamamoto}}]{leung1997approximate}%
  \BibitemOpen
  \bibfield  {author} {\bibinfo {author} {\bibfnamefont {Debbie~W.}\ \bibnamefont {Leung}}, \bibinfo {author} {\bibfnamefont {M.~A.}\ \bibnamefont {Nielsen}}, \bibinfo {author} {\bibfnamefont {Isaac~L.}\ \bibnamefont {Chuang}}, \ and\ \bibinfo {author} {\bibfnamefont {Yoshihisa}\ \bibnamefont {Yamamoto}},\ }\bibfield  {title} {\enquote {\bibinfo {title} {Approximate quantum error correction can lead to better codes},}\ }\href {\doibase 10.1103/PhysRevA.56.2567} {\bibfield  {journal} {\bibinfo  {journal} {Phys. Rev. A}\ }\textbf {\bibinfo {volume} {56}},\ \bibinfo {pages} {2567--2573} (\bibinfo {year} {1997})}\BibitemShut {NoStop}%
\bibitem [{\citenamefont {Fletcher}\ \emph {et~al.}(2008)\citenamefont {Fletcher}, \citenamefont {Shor},\ and\ \citenamefont {Win}}]{fletcher2008channel}%
  \BibitemOpen
  \bibfield  {author} {\bibinfo {author} {\bibfnamefont {Andrew~S.}\ \bibnamefont {Fletcher}}, \bibinfo {author} {\bibfnamefont {Peter~W.}\ \bibnamefont {Shor}}, \ and\ \bibinfo {author} {\bibfnamefont {Moe~Z.}\ \bibnamefont {Win}},\ }\bibfield  {title} {\enquote {\bibinfo {title} {Channel-adapted quantum error correction for the amplitude damping channel},}\ }\href {\doibase 10.1109/TIT.2008.2006458} {\bibfield  {journal} {\bibinfo  {journal} {IEEE Transactions on Information Theory}\ }\textbf {\bibinfo {volume} {54}},\ \bibinfo {pages} {5705--5718} (\bibinfo {year} {2008})}\BibitemShut {NoStop}%
\bibitem [{\citenamefont {Teoh}\ \emph {et~al.}(2023)\citenamefont {Teoh}, \citenamefont {Winkel}, \citenamefont {Babla}, \citenamefont {Chapman}, \citenamefont {Claes}, \citenamefont {de~Graaf}, \citenamefont {Garmon}, \citenamefont {Kalfus}, \citenamefont {Lu}, \citenamefont {Maiti} \emph {et~al.}}]{teoh2023dual}%
  \BibitemOpen
  \bibfield  {author} {\bibinfo {author} {\bibfnamefont {James~D}\ \bibnamefont {Teoh}}, \bibinfo {author} {\bibfnamefont {Patrick}\ \bibnamefont {Winkel}}, \bibinfo {author} {\bibfnamefont {Harshvardhan~K}\ \bibnamefont {Babla}}, \bibinfo {author} {\bibfnamefont {Benjamin~J}\ \bibnamefont {Chapman}}, \bibinfo {author} {\bibfnamefont {Jahan}\ \bibnamefont {Claes}}, \bibinfo {author} {\bibfnamefont {Stijn~J}\ \bibnamefont {de~Graaf}}, \bibinfo {author} {\bibfnamefont {John~WO}\ \bibnamefont {Garmon}}, \bibinfo {author} {\bibfnamefont {William~D}\ \bibnamefont {Kalfus}}, \bibinfo {author} {\bibfnamefont {Yao}\ \bibnamefont {Lu}}, \bibinfo {author} {\bibfnamefont {Aniket}\ \bibnamefont {Maiti}},  \emph {et~al.},\ }\bibfield  {title} {\enquote {\bibinfo {title} {Dual-rail encoding with superconducting cavities},}\ }\href {\doibase 10.1073/pnas.2221736120} {\bibfield  {journal} {\bibinfo  {journal} {Proceedings of the National Academy of Sciences}\ }\textbf {\bibinfo {volume} {120}},\ \bibinfo {pages}
  {e2221736120} (\bibinfo {year} {2023})}\BibitemShut {NoStop}%
\bibitem [{\citenamefont {Kubica}\ \emph {et~al.}(2023)\citenamefont {Kubica}, \citenamefont {Haim}, \citenamefont {Vaknin}, \citenamefont {Levine}, \citenamefont {Brand\~ao},\ and\ \citenamefont {Retzker}}]{kubica2023erasure}%
  \BibitemOpen
  \bibfield  {author} {\bibinfo {author} {\bibfnamefont {Aleksander}\ \bibnamefont {Kubica}}, \bibinfo {author} {\bibfnamefont {Arbel}\ \bibnamefont {Haim}}, \bibinfo {author} {\bibfnamefont {Yotam}\ \bibnamefont {Vaknin}}, \bibinfo {author} {\bibfnamefont {Harry}\ \bibnamefont {Levine}}, \bibinfo {author} {\bibfnamefont {Fernando}\ \bibnamefont {Brand\~ao}}, \ and\ \bibinfo {author} {\bibfnamefont {Alex}\ \bibnamefont {Retzker}},\ }\bibfield  {title} {\enquote {\bibinfo {title} {Erasure qubits: Overcoming the ${T}_{1}$ limit in superconducting circuits},}\ }\href {\doibase 10.1103/PhysRevX.13.041022} {\bibfield  {journal} {\bibinfo  {journal} {Phys. Rev. X}\ }\textbf {\bibinfo {volume} {13}},\ \bibinfo {pages} {041022} (\bibinfo {year} {2023})}\BibitemShut {NoStop}%
\bibitem [{\citenamefont {Koottandavida}\ \emph {et~al.}(2024)\citenamefont {Koottandavida}, \citenamefont {Tsioutsios}, \citenamefont {Kargioti}, \citenamefont {Smith}, \citenamefont {Joshi}, \citenamefont {Dai}, \citenamefont {Teoh}, \citenamefont {Curtis}, \citenamefont {Frunzio}, \citenamefont {Schoelkopf},\ and\ \citenamefont {Devoret}}]{koottandavida2024erasure}%
  \BibitemOpen
  \bibfield  {author} {\bibinfo {author} {\bibfnamefont {Akshay}\ \bibnamefont {Koottandavida}}, \bibinfo {author} {\bibfnamefont {Ioannis}\ \bibnamefont {Tsioutsios}}, \bibinfo {author} {\bibfnamefont {Aikaterini}\ \bibnamefont {Kargioti}}, \bibinfo {author} {\bibfnamefont {Cassady~R.}\ \bibnamefont {Smith}}, \bibinfo {author} {\bibfnamefont {Vidul~R.}\ \bibnamefont {Joshi}}, \bibinfo {author} {\bibfnamefont {Wei}\ \bibnamefont {Dai}}, \bibinfo {author} {\bibfnamefont {James~D.}\ \bibnamefont {Teoh}}, \bibinfo {author} {\bibfnamefont {Jacob~C.}\ \bibnamefont {Curtis}}, \bibinfo {author} {\bibfnamefont {Luigi}\ \bibnamefont {Frunzio}}, \bibinfo {author} {\bibfnamefont {Robert~J.}\ \bibnamefont {Schoelkopf}}, \ and\ \bibinfo {author} {\bibfnamefont {Michel~H.}\ \bibnamefont {Devoret}},\ }\bibfield  {title} {\enquote {\bibinfo {title} {Erasure detection of a dual-rail qubit encoded in a double-post superconducting cavity},}\ }\href {\doibase 10.1103/PhysRevLett.132.180601} {\bibfield  {journal} {\bibinfo
  {journal} {Phys. Rev. Lett.}\ }\textbf {\bibinfo {volume} {132}},\ \bibinfo {pages} {180601} (\bibinfo {year} {2024})}\BibitemShut {NoStop}%
\bibitem [{\citenamefont {Levine}\ \emph {et~al.}(2024)\citenamefont {Levine}, \citenamefont {Haim}, \citenamefont {Hung}, \citenamefont {Alidoust}, \citenamefont {Kalaee}, \citenamefont {DeLorenzo}, \citenamefont {Wollack}, \citenamefont {Arrangoiz-Arriola}, \citenamefont {Khalajhedayati}, \citenamefont {Sanil}, \citenamefont {Moradinejad}, \citenamefont {Vaknin}, \citenamefont {Kubica}, \citenamefont {Hover}, \citenamefont {Aghaeimeibodi}, \citenamefont {Alcid}, \citenamefont {Baek}, \citenamefont {Barnett}, \citenamefont {Bawdekar}, \citenamefont {Bienias}, \citenamefont {Carson}, \citenamefont {Chen}, \citenamefont {Chen}, \citenamefont {Chinkezian}, \citenamefont {Chisholm}, \citenamefont {Clifford}, \citenamefont {Cosmic}, \citenamefont {Crisosto}, \citenamefont {Dalzell}, \citenamefont {Davis}, \citenamefont {D'Ewart}, \citenamefont {Diez}, \citenamefont {D'Souza}, \citenamefont {Dumitrescu}, \citenamefont {Elkhouly}, \citenamefont {Fang}, \citenamefont {Fang}, \citenamefont {Flammia}, \citenamefont
  {Fling}, \citenamefont {Garcia}, \citenamefont {Gharzai}, \citenamefont {Gorshkov}, \citenamefont {Gray}, \citenamefont {Grimberg}, \citenamefont {Grimsmo}, \citenamefont {Hann}, \citenamefont {He}, \citenamefont {Heidel}, \citenamefont {Howell}, \citenamefont {Hunt}, \citenamefont {Iverson}, \citenamefont {Jarrige}, \citenamefont {Jiang}, \citenamefont {Jones}, \citenamefont {Karabalin}, \citenamefont {Karalekas}, \citenamefont {Keller}, \citenamefont {Lasi}, \citenamefont {Lee}, \citenamefont {Ly}, \citenamefont {MacCabe}, \citenamefont {Mahuli}, \citenamefont {Marcaud}, \citenamefont {Matheny}, \citenamefont {McArdle}, \citenamefont {McCabe}, \citenamefont {Merton}, \citenamefont {Miles}, \citenamefont {Milsted}, \citenamefont {Mishra}, \citenamefont {Moncelsi}, \citenamefont {Naghiloo}, \citenamefont {Noh}, \citenamefont {Oblepias}, \citenamefont {Ortuno}, \citenamefont {Owens}, \citenamefont {Pagdilao}, \citenamefont {Panduro}, \citenamefont {Paquette}, \citenamefont {Patel}, \citenamefont {Peairs},
  \citenamefont {Perello}, \citenamefont {Peterson}, \citenamefont {Ponte}, \citenamefont {Putterman}, \citenamefont {Refael}, \citenamefont {Reinhold}, \citenamefont {Resnick}, \citenamefont {Reyna}, \citenamefont {Rodriguez}, \citenamefont {Rose}, \citenamefont {Rubin}, \citenamefont {Runyan}, \citenamefont {Ryan}, \citenamefont {Sahmoud}, \citenamefont {Scaffidi}, \citenamefont {Shah}, \citenamefont {Siavoshi}, \citenamefont {Sivarajah}, \citenamefont {Skogland}, \citenamefont {Su}, \citenamefont {Swenson}, \citenamefont {Sylvia}, \citenamefont {Teo}, \citenamefont {Tomada}, \citenamefont {Torlai}, \citenamefont {Wistrom}, \citenamefont {Zhang}, \citenamefont {Zuk}, \citenamefont {Clerk}, \citenamefont {Brand\~ao}, \citenamefont {Retzker},\ and\ \citenamefont {Painter}}]{awsdualrail}%
  \BibitemOpen
  \bibfield  {author} {\bibinfo {author} {\bibfnamefont {H.}~\bibnamefont {Levine}}, \bibinfo {author} {\bibfnamefont {A.}~\bibnamefont {Haim}}, \bibinfo {author} {\bibfnamefont {J.~S.~C.}\ \bibnamefont {Hung}}, \bibinfo {author} {\bibfnamefont {N.}~\bibnamefont {Alidoust}}, \bibinfo {author} {\bibfnamefont {M.}~\bibnamefont {Kalaee}}, \bibinfo {author} {\bibfnamefont {L.}~\bibnamefont {DeLorenzo}}, \bibinfo {author} {\bibfnamefont {E.~A.}\ \bibnamefont {Wollack}}, \bibinfo {author} {\bibfnamefont {P.}~\bibnamefont {Arrangoiz-Arriola}}, \bibinfo {author} {\bibfnamefont {A.}~\bibnamefont {Khalajhedayati}}, \bibinfo {author} {\bibfnamefont {R.}~\bibnamefont {Sanil}}, \bibinfo {author} {\bibfnamefont {H.}~\bibnamefont {Moradinejad}}, \bibinfo {author} {\bibfnamefont {Y.}~\bibnamefont {Vaknin}}, \bibinfo {author} {\bibfnamefont {A.}~\bibnamefont {Kubica}}, \bibinfo {author} {\bibfnamefont {D.}~\bibnamefont {Hover}}, \bibinfo {author} {\bibfnamefont {S.}~\bibnamefont {Aghaeimeibodi}}, \bibinfo {author}
  {\bibfnamefont {J.~A.}\ \bibnamefont {Alcid}}, \bibinfo {author} {\bibfnamefont {C.}~\bibnamefont {Baek}}, \bibinfo {author} {\bibfnamefont {J.}~\bibnamefont {Barnett}}, \bibinfo {author} {\bibfnamefont {K.}~\bibnamefont {Bawdekar}}, \bibinfo {author} {\bibfnamefont {P.}~\bibnamefont {Bienias}}, \bibinfo {author} {\bibfnamefont {H.~A.}\ \bibnamefont {Carson}}, \bibinfo {author} {\bibfnamefont {C.}~\bibnamefont {Chen}}, \bibinfo {author} {\bibfnamefont {L.}~\bibnamefont {Chen}}, \bibinfo {author} {\bibfnamefont {H.}~\bibnamefont {Chinkezian}}, \bibinfo {author} {\bibfnamefont {E.~M.}\ \bibnamefont {Chisholm}}, \bibinfo {author} {\bibfnamefont {A.}~\bibnamefont {Clifford}}, \bibinfo {author} {\bibfnamefont {R.}~\bibnamefont {Cosmic}}, \bibinfo {author} {\bibfnamefont {N.}~\bibnamefont {Crisosto}}, \bibinfo {author} {\bibfnamefont {A.~M.}\ \bibnamefont {Dalzell}}, \bibinfo {author} {\bibfnamefont {E.}~\bibnamefont {Davis}}, \bibinfo {author} {\bibfnamefont {J.~M.}\ \bibnamefont {D'Ewart}}, \bibinfo {author}
  {\bibfnamefont {S.}~\bibnamefont {Diez}}, \bibinfo {author} {\bibfnamefont {N.}~\bibnamefont {D'Souza}}, \bibinfo {author} {\bibfnamefont {P.~T.}\ \bibnamefont {Dumitrescu}}, \bibinfo {author} {\bibfnamefont {E.}~\bibnamefont {Elkhouly}}, \bibinfo {author} {\bibfnamefont {M.~T.}\ \bibnamefont {Fang}}, \bibinfo {author} {\bibfnamefont {Y.}~\bibnamefont {Fang}}, \bibinfo {author} {\bibfnamefont {S.}~\bibnamefont {Flammia}}, \bibinfo {author} {\bibfnamefont {M.~J.}\ \bibnamefont {Fling}}, \bibinfo {author} {\bibfnamefont {G.}~\bibnamefont {Garcia}}, \bibinfo {author} {\bibfnamefont {M.~K.}\ \bibnamefont {Gharzai}}, \bibinfo {author} {\bibfnamefont {A.~V.}\ \bibnamefont {Gorshkov}}, \bibinfo {author} {\bibfnamefont {M.~J.}\ \bibnamefont {Gray}}, \bibinfo {author} {\bibfnamefont {S.}~\bibnamefont {Grimberg}}, \bibinfo {author} {\bibfnamefont {A.~L.}\ \bibnamefont {Grimsmo}}, \bibinfo {author} {\bibfnamefont {C.~T.}\ \bibnamefont {Hann}}, \bibinfo {author} {\bibfnamefont {Y.}~\bibnamefont {He}}, \bibinfo {author}
  {\bibfnamefont {S.}~\bibnamefont {Heidel}}, \bibinfo {author} {\bibfnamefont {S.}~\bibnamefont {Howell}}, \bibinfo {author} {\bibfnamefont {M.}~\bibnamefont {Hunt}}, \bibinfo {author} {\bibfnamefont {J.}~\bibnamefont {Iverson}}, \bibinfo {author} {\bibfnamefont {I.}~\bibnamefont {Jarrige}}, \bibinfo {author} {\bibfnamefont {L.}~\bibnamefont {Jiang}}, \bibinfo {author} {\bibfnamefont {W.~M.}\ \bibnamefont {Jones}}, \bibinfo {author} {\bibfnamefont {R.}~\bibnamefont {Karabalin}}, \bibinfo {author} {\bibfnamefont {P.~J.}\ \bibnamefont {Karalekas}}, \bibinfo {author} {\bibfnamefont {A.~J.}\ \bibnamefont {Keller}}, \bibinfo {author} {\bibfnamefont {D.}~\bibnamefont {Lasi}}, \bibinfo {author} {\bibfnamefont {M.}~\bibnamefont {Lee}}, \bibinfo {author} {\bibfnamefont {V.}~\bibnamefont {Ly}}, \bibinfo {author} {\bibfnamefont {G.}~\bibnamefont {MacCabe}}, \bibinfo {author} {\bibfnamefont {N.}~\bibnamefont {Mahuli}}, \bibinfo {author} {\bibfnamefont {G.}~\bibnamefont {Marcaud}}, \bibinfo {author} {\bibfnamefont
  {M.~H.}\ \bibnamefont {Matheny}}, \bibinfo {author} {\bibfnamefont {S.}~\bibnamefont {McArdle}}, \bibinfo {author} {\bibfnamefont {G.}~\bibnamefont {McCabe}}, \bibinfo {author} {\bibfnamefont {G.}~\bibnamefont {Merton}}, \bibinfo {author} {\bibfnamefont {C.}~\bibnamefont {Miles}}, \bibinfo {author} {\bibfnamefont {A.}~\bibnamefont {Milsted}}, \bibinfo {author} {\bibfnamefont {A.}~\bibnamefont {Mishra}}, \bibinfo {author} {\bibfnamefont {L.}~\bibnamefont {Moncelsi}}, \bibinfo {author} {\bibfnamefont {M.}~\bibnamefont {Naghiloo}}, \bibinfo {author} {\bibfnamefont {K.}~\bibnamefont {Noh}}, \bibinfo {author} {\bibfnamefont {E.}~\bibnamefont {Oblepias}}, \bibinfo {author} {\bibfnamefont {G.}~\bibnamefont {Ortuno}}, \bibinfo {author} {\bibfnamefont {J.~C.}\ \bibnamefont {Owens}}, \bibinfo {author} {\bibfnamefont {J.}~\bibnamefont {Pagdilao}}, \bibinfo {author} {\bibfnamefont {A.}~\bibnamefont {Panduro}}, \bibinfo {author} {\bibfnamefont {J.-P.}\ \bibnamefont {Paquette}}, \bibinfo {author} {\bibfnamefont {R.~N.}\
  \bibnamefont {Patel}}, \bibinfo {author} {\bibfnamefont {G.}~\bibnamefont {Peairs}}, \bibinfo {author} {\bibfnamefont {D.~J.}\ \bibnamefont {Perello}}, \bibinfo {author} {\bibfnamefont {E.~C.}\ \bibnamefont {Peterson}}, \bibinfo {author} {\bibfnamefont {S.}~\bibnamefont {Ponte}}, \bibinfo {author} {\bibfnamefont {H.}~\bibnamefont {Putterman}}, \bibinfo {author} {\bibfnamefont {G.}~\bibnamefont {Refael}}, \bibinfo {author} {\bibfnamefont {P.}~\bibnamefont {Reinhold}}, \bibinfo {author} {\bibfnamefont {R.}~\bibnamefont {Resnick}}, \bibinfo {author} {\bibfnamefont {O.~A.}\ \bibnamefont {Reyna}}, \bibinfo {author} {\bibfnamefont {R.}~\bibnamefont {Rodriguez}}, \bibinfo {author} {\bibfnamefont {J.}~\bibnamefont {Rose}}, \bibinfo {author} {\bibfnamefont {A.~H.}\ \bibnamefont {Rubin}}, \bibinfo {author} {\bibfnamefont {M.}~\bibnamefont {Runyan}}, \bibinfo {author} {\bibfnamefont {C.~A.}\ \bibnamefont {Ryan}}, \bibinfo {author} {\bibfnamefont {A.}~\bibnamefont {Sahmoud}}, \bibinfo {author} {\bibfnamefont
  {T.}~\bibnamefont {Scaffidi}}, \bibinfo {author} {\bibfnamefont {B.}~\bibnamefont {Shah}}, \bibinfo {author} {\bibfnamefont {S.}~\bibnamefont {Siavoshi}}, \bibinfo {author} {\bibfnamefont {P.}~\bibnamefont {Sivarajah}}, \bibinfo {author} {\bibfnamefont {T.}~\bibnamefont {Skogland}}, \bibinfo {author} {\bibfnamefont {C.-J.}\ \bibnamefont {Su}}, \bibinfo {author} {\bibfnamefont {L.~J.}\ \bibnamefont {Swenson}}, \bibinfo {author} {\bibfnamefont {J.}~\bibnamefont {Sylvia}}, \bibinfo {author} {\bibfnamefont {S.~M.}\ \bibnamefont {Teo}}, \bibinfo {author} {\bibfnamefont {A.}~\bibnamefont {Tomada}}, \bibinfo {author} {\bibfnamefont {G.}~\bibnamefont {Torlai}}, \bibinfo {author} {\bibfnamefont {M.}~\bibnamefont {Wistrom}}, \bibinfo {author} {\bibfnamefont {K.}~\bibnamefont {Zhang}}, \bibinfo {author} {\bibfnamefont {I.}~\bibnamefont {Zuk}}, \bibinfo {author} {\bibfnamefont {A.~A.}\ \bibnamefont {Clerk}}, \bibinfo {author} {\bibfnamefont {F.~G. S.~L.}\ \bibnamefont {Brand\~ao}}, \bibinfo {author} {\bibfnamefont
  {A.}~\bibnamefont {Retzker}}, \ and\ \bibinfo {author} {\bibfnamefont {O.}~\bibnamefont {Painter}},\ }\bibfield  {title} {\enquote {\bibinfo {title} {Demonstrating a long-coherence dual-rail erasure qubit using tunable transmons},}\ }\href {\doibase 10.1103/PhysRevX.14.011051} {\bibfield  {journal} {\bibinfo  {journal} {Phys. Rev. X}\ }\textbf {\bibinfo {volume} {14}},\ \bibinfo {pages} {011051} (\bibinfo {year} {2024})}\BibitemShut {NoStop}%
\bibitem [{\citenamefont {Szeg{\"o}}(1939)}]{szegoOrthogonalPolynomials1939}%
  \BibitemOpen
  \bibfield  {author} {\bibinfo {author} {\bibfnamefont {G{\'a}bor}\ \bibnamefont {Szeg{\"o}}},\ }\href@noop {} {\emph {\bibinfo {title} {Orthogonal Polynomials}}},\ \bibinfo {edition} {online-ausg}\ ed.,\ \bibinfo {series} {Colloquium {{Publications}}}\ No.~\bibinfo {number} {23}\ (\bibinfo  {publisher} {American mathematical society},\ \bibinfo {address} {New York city},\ \bibinfo {year} {1939})\BibitemShut {NoStop}%
\bibitem [{\citenamefont {Li}\ \emph {et~al.}(2017)\citenamefont {Li}, \citenamefont {Zou}, \citenamefont {Albert}, \citenamefont {Muralidharan}, \citenamefont {Girvin},\ and\ \citenamefont {Jiang}}]{li2017cat}%
  \BibitemOpen
  \bibfield  {author} {\bibinfo {author} {\bibfnamefont {Linshu}\ \bibnamefont {Li}}, \bibinfo {author} {\bibfnamefont {Chang-Ling}\ \bibnamefont {Zou}}, \bibinfo {author} {\bibfnamefont {Victor~V.}\ \bibnamefont {Albert}}, \bibinfo {author} {\bibfnamefont {Sreraman}\ \bibnamefont {Muralidharan}}, \bibinfo {author} {\bibfnamefont {S.~M.}\ \bibnamefont {Girvin}}, \ and\ \bibinfo {author} {\bibfnamefont {Liang}\ \bibnamefont {Jiang}},\ }\bibfield  {title} {\enquote {\bibinfo {title} {Cat codes with optimal decoherence suppression for a lossy bosonic channel},}\ }\href {\doibase 10.1103/PhysRevLett.119.030502} {\bibfield  {journal} {\bibinfo  {journal} {Phys. Rev. Lett.}\ }\textbf {\bibinfo {volume} {119}},\ \bibinfo {pages} {030502} (\bibinfo {year} {2017})}\BibitemShut {NoStop}%
\bibitem [{\citenamefont {{The Sage Developers}}(2022)}]{sagemath}%
  \BibitemOpen
  \bibfield  {author} {\bibinfo {author} {\bibnamefont {{The Sage Developers}}},\ }\href@noop {} {\emph {\bibinfo {title} {{S}ageMath, the {S}age {M}athematics {S}oftware {S}ystem ({V}ersion 9.5)}}} (\bibinfo {year} {2022}),\ \bibinfo {note} {{\tt https://www.sagemath.org}}\BibitemShut {NoStop}%
\bibitem [{\citenamefont {Lescanne}\ \emph {et~al.}(2020)\citenamefont {Lescanne}, \citenamefont {Villiers}, \citenamefont {Peronnin}, \citenamefont {Sarlette}, \citenamefont {Delbecq}, \citenamefont {Huard}, \citenamefont {Kontos}, \citenamefont {Mirrahimi},\ and\ \citenamefont {Leghtas}}]{lescanne_exponential_2020}%
  \BibitemOpen
  \bibfield  {author} {\bibinfo {author} {\bibfnamefont {Raphaël}\ \bibnamefont {Lescanne}}, \bibinfo {author} {\bibfnamefont {Marius}\ \bibnamefont {Villiers}}, \bibinfo {author} {\bibfnamefont {Théau}\ \bibnamefont {Peronnin}}, \bibinfo {author} {\bibfnamefont {Alain}\ \bibnamefont {Sarlette}}, \bibinfo {author} {\bibfnamefont {Matthieu}\ \bibnamefont {Delbecq}}, \bibinfo {author} {\bibfnamefont {Benjamin}\ \bibnamefont {Huard}}, \bibinfo {author} {\bibfnamefont {Takis}\ \bibnamefont {Kontos}}, \bibinfo {author} {\bibfnamefont {Mazyar}\ \bibnamefont {Mirrahimi}}, \ and\ \bibinfo {author} {\bibfnamefont {Zaki}\ \bibnamefont {Leghtas}},\ }\bibfield  {title} {\enquote {\bibinfo {title} {Exponential suppression of bit-flips in a qubit encoded in an oscillator},}\ }\href {\doibase 10.1038/s41567-020-0824-x} {\bibfield  {journal} {\bibinfo  {journal} {Nat. Phys.}\ }\textbf {\bibinfo {volume} {16}},\ \bibinfo {pages} {509--513} (\bibinfo {year} {2020})},\ \bibinfo {note} {\_eprint: 1907.11729}\BibitemShut
  {NoStop}%
\bibitem [{\citenamefont {Krantz}\ \emph {et~al.}(2019)\citenamefont {Krantz}, \citenamefont {Kjaergaard}, \citenamefont {Yan}, \citenamefont {Orlando}, \citenamefont {Gustavsson},\ and\ \citenamefont {Oliver}}]{krantz2019quantum}%
  \BibitemOpen
  \bibfield  {author} {\bibinfo {author} {\bibfnamefont {P.}~\bibnamefont {Krantz}}, \bibinfo {author} {\bibfnamefont {M.}~\bibnamefont {Kjaergaard}}, \bibinfo {author} {\bibfnamefont {F.}~\bibnamefont {Yan}}, \bibinfo {author} {\bibfnamefont {T.~P.}\ \bibnamefont {Orlando}}, \bibinfo {author} {\bibfnamefont {S.}~\bibnamefont {Gustavsson}}, \ and\ \bibinfo {author} {\bibfnamefont {W.~D.}\ \bibnamefont {Oliver}},\ }\bibfield  {title} {\enquote {\bibinfo {title} {A quantum engineer's guide to superconducting qubits},}\ }\href {\doibase 10.1063/1.5089550} {\bibfield  {journal} {\bibinfo  {journal} {Applied Physics Reviews}\ }\textbf {\bibinfo {volume} {6}},\ \bibinfo {pages} {021318} (\bibinfo {year} {2019})},\ \Eprint {http://arxiv.org/abs/https://pubs.aip.org/aip/apr/article-pdf/doi/10.1063/1.5089550/16667201/021318\_1\_online.pdf} {https://pubs.aip.org/aip/apr/article-pdf/doi/10.1063/1.5089550/16667201/021318\_1\_online.pdf} \BibitemShut {NoStop}%
\bibitem [{\citenamefont {Blais}\ \emph {et~al.}(2021)\citenamefont {Blais}, \citenamefont {Grimsmo}, \citenamefont {Girvin},\ and\ \citenamefont {Wallraff}}]{blais2021circuit}%
  \BibitemOpen
  \bibfield  {author} {\bibinfo {author} {\bibfnamefont {Alexandre}\ \bibnamefont {Blais}}, \bibinfo {author} {\bibfnamefont {Arne~L.}\ \bibnamefont {Grimsmo}}, \bibinfo {author} {\bibfnamefont {S.~M.}\ \bibnamefont {Girvin}}, \ and\ \bibinfo {author} {\bibfnamefont {Andreas}\ \bibnamefont {Wallraff}},\ }\bibfield  {title} {\enquote {\bibinfo {title} {Circuit quantum electrodynamics},}\ }\href {\doibase 10.1103/RevModPhys.93.025005} {\bibfield  {journal} {\bibinfo  {journal} {Rev. Mod. Phys.}\ }\textbf {\bibinfo {volume} {93}},\ \bibinfo {pages} {025005} (\bibinfo {year} {2021})}\BibitemShut {NoStop}%
\bibitem [{\citenamefont {Elliott}\ \emph {et~al.}(2018)\citenamefont {Elliott}, \citenamefont {Joo},\ and\ \citenamefont {Ginossar}}]{elliott2018designing}%
  \BibitemOpen
  \bibfield  {author} {\bibinfo {author} {\bibfnamefont {Matthew}\ \bibnamefont {Elliott}}, \bibinfo {author} {\bibfnamefont {Jaewoo}\ \bibnamefont {Joo}}, \ and\ \bibinfo {author} {\bibfnamefont {Eran}\ \bibnamefont {Ginossar}},\ }\bibfield  {title} {\enquote {\bibinfo {title} {Designing kerr interactions using multiple superconducting qubit types in a single circuit},}\ }\href {\doibase 10.1088/1367-2630/aa9243} {\bibfield  {journal} {\bibinfo  {journal} {New Journal of Physics}\ }\textbf {\bibinfo {volume} {20}},\ \bibinfo {pages} {023037} (\bibinfo {year} {2018})}\BibitemShut {NoStop}%
\bibitem [{\citenamefont {Chapple}\ \emph {et~al.}(2025)\citenamefont {Chapple}, \citenamefont {Benhayoune-Khadraoui}, \citenamefont {Richer},\ and\ \citenamefont {Blais}}]{chapple2025balanced}%
  \BibitemOpen
  \bibfield  {author} {\bibinfo {author} {\bibfnamefont {Alex~A}\ \bibnamefont {Chapple}}, \bibinfo {author} {\bibfnamefont {Othmane}\ \bibnamefont {Benhayoune-Khadraoui}}, \bibinfo {author} {\bibfnamefont {Simon}\ \bibnamefont {Richer}}, \ and\ \bibinfo {author} {\bibfnamefont {Alexandre}\ \bibnamefont {Blais}},\ }\bibfield  {title} {\enquote {\bibinfo {title} {Balanced cross-kerr coupling for superconducting qubit readout},}\ }\href {https://arxiv.org/abs/2501.09010} {\bibfield  {journal} {\bibinfo  {journal} {arXiv preprint arXiv:2501.09010}\ } (\bibinfo {year} {2025})}\BibitemShut {NoStop}%
\bibitem [{\citenamefont {Lieu}\ \emph {et~al.}(2020)\citenamefont {Lieu}, \citenamefont {Belyansky}, \citenamefont {Young}, \citenamefont {Lundgren}, \citenamefont {Albert},\ and\ \citenamefont {Gorshkov}}]{lieu2020symmetry}%
  \BibitemOpen
  \bibfield  {author} {\bibinfo {author} {\bibfnamefont {Simon}\ \bibnamefont {Lieu}}, \bibinfo {author} {\bibfnamefont {Ron}\ \bibnamefont {Belyansky}}, \bibinfo {author} {\bibfnamefont {Jeremy~T.}\ \bibnamefont {Young}}, \bibinfo {author} {\bibfnamefont {Rex}\ \bibnamefont {Lundgren}}, \bibinfo {author} {\bibfnamefont {Victor~V.}\ \bibnamefont {Albert}}, \ and\ \bibinfo {author} {\bibfnamefont {Alexey~V.}\ \bibnamefont {Gorshkov}},\ }\bibfield  {title} {\enquote {\bibinfo {title} {Symmetry breaking and error correction in open quantum systems},}\ }\href {\doibase 10.1103/PhysRevLett.125.240405} {\bibfield  {journal} {\bibinfo  {journal} {Phys. Rev. Lett.}\ }\textbf {\bibinfo {volume} {125}},\ \bibinfo {pages} {240405} (\bibinfo {year} {2020})}\BibitemShut {NoStop}%
\bibitem [{\citenamefont {Gravina}\ \emph {et~al.}(2023)\citenamefont {Gravina}, \citenamefont {Minganti},\ and\ \citenamefont {Savona}}]{gravina2023critical}%
  \BibitemOpen
  \bibfield  {author} {\bibinfo {author} {\bibfnamefont {Luca}\ \bibnamefont {Gravina}}, \bibinfo {author} {\bibfnamefont {Fabrizio}\ \bibnamefont {Minganti}}, \ and\ \bibinfo {author} {\bibfnamefont {Vincenzo}\ \bibnamefont {Savona}},\ }\bibfield  {title} {\enquote {\bibinfo {title} {Critical schr\"odinger cat qubit},}\ }\href {\doibase 10.1103/PRXQuantum.4.020337} {\bibfield  {journal} {\bibinfo  {journal} {PRX Quantum}\ }\textbf {\bibinfo {volume} {4}},\ \bibinfo {pages} {020337} (\bibinfo {year} {2023})}\BibitemShut {NoStop}%
\bibitem [{\citenamefont {de~Groot}\ \emph {et~al.}(2022)\citenamefont {de~Groot}, \citenamefont {Turzillo},\ and\ \citenamefont {Schuch}}]{de2022symmetry}%
  \BibitemOpen
  \bibfield  {author} {\bibinfo {author} {\bibfnamefont {Caroline}\ \bibnamefont {de~Groot}}, \bibinfo {author} {\bibfnamefont {Alex}\ \bibnamefont {Turzillo}}, \ and\ \bibinfo {author} {\bibfnamefont {Norbert}\ \bibnamefont {Schuch}},\ }\bibfield  {title} {\enquote {\bibinfo {title} {Symmetry protected topological order in open quantum systems},}\ }\href {\doibase 10.22331/q-2022-11-10-856} {\bibfield  {journal} {\bibinfo  {journal} {Quantum}\ }\textbf {\bibinfo {volume} {6}},\ \bibinfo {pages} {856} (\bibinfo {year} {2022})}\BibitemShut {NoStop}%
\bibitem [{\citenamefont {S\'a}\ \emph {et~al.}(2023)\citenamefont {S\'a}, \citenamefont {Ribeiro},\ and\ \citenamefont {Prosen}}]{sa2023symmetry}%
  \BibitemOpen
  \bibfield  {author} {\bibinfo {author} {\bibfnamefont {Lucas}\ \bibnamefont {S\'a}}, \bibinfo {author} {\bibfnamefont {Pedro}\ \bibnamefont {Ribeiro}}, \ and\ \bibinfo {author} {\bibfnamefont {Toma\ifmmode \check{z}\else~\v{z}\fi{}}\ \bibnamefont {Prosen}},\ }\bibfield  {title} {\enquote {\bibinfo {title} {Symmetry classification of many-body lindbladians: Tenfold way and beyond},}\ }\href {\doibase 10.1103/PhysRevX.13.031019} {\bibfield  {journal} {\bibinfo  {journal} {Phys. Rev. X}\ }\textbf {\bibinfo {volume} {13}},\ \bibinfo {pages} {031019} (\bibinfo {year} {2023})}\BibitemShut {NoStop}%
\bibitem [{\citenamefont {Ma}\ and\ \citenamefont {Wang}(2023)}]{ma2023average}%
  \BibitemOpen
  \bibfield  {author} {\bibinfo {author} {\bibfnamefont {Ruochen}\ \bibnamefont {Ma}}\ and\ \bibinfo {author} {\bibfnamefont {Chong}\ \bibnamefont {Wang}},\ }\bibfield  {title} {\enquote {\bibinfo {title} {Average symmetry-protected topological phases},}\ }\href {\doibase 10.1103/PhysRevX.13.031016} {\bibfield  {journal} {\bibinfo  {journal} {Phys. Rev. X}\ }\textbf {\bibinfo {volume} {13}},\ \bibinfo {pages} {031016} (\bibinfo {year} {2023})}\BibitemShut {NoStop}%
\bibitem [{\citenamefont {Kawabata}\ \emph {et~al.}(2023)\citenamefont {Kawabata}, \citenamefont {Kulkarni}, \citenamefont {Li}, \citenamefont {Numasawa},\ and\ \citenamefont {Ryu}}]{kawabata2023symmetry}%
  \BibitemOpen
  \bibfield  {author} {\bibinfo {author} {\bibfnamefont {Kohei}\ \bibnamefont {Kawabata}}, \bibinfo {author} {\bibfnamefont {Anish}\ \bibnamefont {Kulkarni}}, \bibinfo {author} {\bibfnamefont {Jiachen}\ \bibnamefont {Li}}, \bibinfo {author} {\bibfnamefont {Tokiro}\ \bibnamefont {Numasawa}}, \ and\ \bibinfo {author} {\bibfnamefont {Shinsei}\ \bibnamefont {Ryu}},\ }\bibfield  {title} {\enquote {\bibinfo {title} {Symmetry of open quantum systems: Classification of dissipative quantum chaos},}\ }\href {\doibase 10.1103/PRXQuantum.4.030328} {\bibfield  {journal} {\bibinfo  {journal} {PRX Quantum}\ }\textbf {\bibinfo {volume} {4}},\ \bibinfo {pages} {030328} (\bibinfo {year} {2023})}\BibitemShut {NoStop}%
\bibitem [{\citenamefont {Yi}\ \emph {et~al.}(2024)\citenamefont {Yi}, \citenamefont {Ye}, \citenamefont {Gottesman},\ and\ \citenamefont {Liu}}]{yi2024complexity}%
  \BibitemOpen
  \bibfield  {author} {\bibinfo {author} {\bibfnamefont {Jinmin}\ \bibnamefont {Yi}}, \bibinfo {author} {\bibfnamefont {Weicheng}\ \bibnamefont {Ye}}, \bibinfo {author} {\bibfnamefont {Daniel}\ \bibnamefont {Gottesman}}, \ and\ \bibinfo {author} {\bibfnamefont {Zi-Wen}\ \bibnamefont {Liu}},\ }\bibfield  {title} {\enquote {\bibinfo {title} {Complexity and order in approximate quantum error-correcting codes},}\ }\href {\doibase 10.1038/s41567-024-02621-x} {\bibfield  {journal} {\bibinfo  {journal} {Nature Physics}\ ,\ \bibinfo {pages} {1--6}} (\bibinfo {year} {2024})}\BibitemShut {NoStop}%
\bibitem [{\citenamefont {Hosono}\ \emph {et~al.}(1996)\citenamefont {Hosono}, \citenamefont {Lian},\ and\ \citenamefont {Yau}}]{hosono1996gkz}%
  \BibitemOpen
  \bibfield  {author} {\bibinfo {author} {\bibfnamefont {Shinobu}\ \bibnamefont {Hosono}}, \bibinfo {author} {\bibfnamefont {Bong~H}\ \bibnamefont {Lian}}, \ and\ \bibinfo {author} {\bibfnamefont {S~T}\ \bibnamefont {Yau}},\ }\bibfield  {title} {\enquote {\bibinfo {title} {Gkz-generalized hypergeometric systems in mirror symmetry of calabi-yau hypersurfaces},}\ }\href {\doibase 10.1007/BF02506417} {\bibfield  {journal} {\bibinfo  {journal} {Communications in Mathematical Physics}\ }\textbf {\bibinfo {volume} {182}},\ \bibinfo {pages} {535--577} (\bibinfo {year} {1996})}\BibitemShut {NoStop}%
\bibitem [{\citenamefont {Hrabowski}(1985)}]{hrabowski1985multiple}%
  \BibitemOpen
  \bibfield  {author} {\bibinfo {author} {\bibfnamefont {Jan}\ \bibnamefont {Hrabowski}},\ }\bibfield  {title} {\enquote {\bibinfo {title} {Multiple hypergeometric functions and simple lie groups sl and sp},}\ }\href {\doibase 10.1137/0516066} {\bibfield  {journal} {\bibinfo  {journal} {SIAM journal on mathematical analysis}\ }\textbf {\bibinfo {volume} {16}},\ \bibinfo {pages} {876--886} (\bibinfo {year} {1985})}\BibitemShut {NoStop}%
\bibitem [{\citenamefont {Nasrollahpoursamami}(2016)}]{nasrollahpoursamami2016periods}%
  \BibitemOpen
  \bibfield  {author} {\bibinfo {author} {\bibfnamefont {Emad}\ \bibnamefont {Nasrollahpoursamami}},\ }\bibfield  {title} {\enquote {\bibinfo {title} {Periods of feynman diagrams and gkz d-modules},}\ }\href {https://arxiv.org/abs/1605.04970} {\bibfield  {journal} {\bibinfo  {journal} {arXiv preprint arXiv:1605.04970}\ } (\bibinfo {year} {2016})}\BibitemShut {NoStop}%
\bibitem [{\citenamefont {Borisov}\ and\ \citenamefont {Paul~Horja}(2013)}]{borisov2013better}%
  \BibitemOpen
  \bibfield  {author} {\bibinfo {author} {\bibfnamefont {Lev~A}\ \bibnamefont {Borisov}}\ and\ \bibinfo {author} {\bibfnamefont {R}~\bibnamefont {Paul~Horja}},\ }\bibfield  {title} {\enquote {\bibinfo {title} {On the better behaved version of the gkz hypergeometric system},}\ }\href {\doibase 10.1007/s00208-013-0913-6} {\bibfield  {journal} {\bibinfo  {journal} {Mathematische annalen}\ }\textbf {\bibinfo {volume} {357}},\ \bibinfo {pages} {585--603} (\bibinfo {year} {2013})}\BibitemShut {NoStop}%
\bibitem [{\citenamefont {Takayama}\ \emph {et~al.}(2018)\citenamefont {Takayama}, \citenamefont {Kuriki},\ and\ \citenamefont {Takemura}}]{takayama2018hypergeometric}%
  \BibitemOpen
  \bibfield  {author} {\bibinfo {author} {\bibfnamefont {Nobuki}\ \bibnamefont {Takayama}}, \bibinfo {author} {\bibfnamefont {Satoshi}\ \bibnamefont {Kuriki}}, \ and\ \bibinfo {author} {\bibfnamefont {Akimichi}\ \bibnamefont {Takemura}},\ }\bibfield  {title} {\enquote {\bibinfo {title} {A-hypergeometric distributions and newton polytopes},}\ }\href@noop {} {\bibfield  {journal} {\bibinfo  {journal} {Advances in Applied Mathematics}\ }\textbf {\bibinfo {volume} {99}},\ \bibinfo {pages} {109--133} (\bibinfo {year} {2018})}\BibitemShut {NoStop}%
\bibitem [{\citenamefont {Gelfand}\ \emph {et~al.}(1990)\citenamefont {Gelfand}, \citenamefont {Graev},\ and\ \citenamefont {Retakh}}]{gelfand1990gamma}%
  \BibitemOpen
  \bibfield  {author} {\bibinfo {author} {\bibfnamefont {IM}~\bibnamefont {Gelfand}}, \bibinfo {author} {\bibfnamefont {MI}~\bibnamefont {Graev}}, \ and\ \bibinfo {author} {\bibfnamefont {VS}~\bibnamefont {Retakh}},\ }\bibfield  {title} {\enquote {\bibinfo {title} {Gamma-series and general hypergeometric functions on the manifold of $k\times n$ matrices},}\ }\href@noop {} {\bibfield  {journal} {\bibinfo  {journal} {Inst. Mat. Akad. Nauk SSSR, preprint 1990, no. 64}\ } (\bibinfo {year} {1990})}\BibitemShut {NoStop}%
\bibitem [{\citenamefont {Gelfand}\ \emph {et~al.}(1993)\citenamefont {Gelfand}, \citenamefont {Graev},\ and\ \citenamefont {Retakh}}]{gelfand1993reduction}%
  \BibitemOpen
  \bibfield  {author} {\bibinfo {author} {\bibfnamefont {IM}~\bibnamefont {Gelfand}}, \bibinfo {author} {\bibfnamefont {MI}~\bibnamefont {Graev}}, \ and\ \bibinfo {author} {\bibfnamefont {VS}~\bibnamefont {Retakh}},\ }\bibfield  {title} {\enquote {\bibinfo {title} {Reduction formulas for hypergeometric functions associated with the grassmannian gnk and description of these functions on strata of small codimension in gnk},}\ }\href@noop {} {\bibfield  {journal} {\bibinfo  {journal} {Russian J. Math. Phys}\ }\textbf {\bibinfo {volume} {1}},\ \bibinfo {pages} {19--56} (\bibinfo {year} {1993})}\BibitemShut {NoStop}%
\bibitem [{gel(1997)}]{gelfand1999gg}%
  \BibitemOpen
  \bibfield  {title} {\enquote {\bibinfo {title} {Gg-functions and their relation to general hypergeometric functions},}\ }\href {https://iopscience.iop.org/article/10.1070/RM1997v052n04ABEH002055/meta} {\bibfield  {journal} {\bibinfo  {journal} {Russian Mathematical Surveys}\ }\textbf {\bibinfo {volume} {52}},\ \bibinfo {pages} {639} (\bibinfo {year} {1997})}\BibitemShut {NoStop}%
\bibitem [{\citenamefont {Gelfand}\ \emph {et~al.}(1997)\citenamefont {Gelfand}, \citenamefont {Graev},\ and\ \citenamefont {Postnikov}}]{gelfand1997combinatorics}%
  \BibitemOpen
  \bibfield  {author} {\bibinfo {author} {\bibfnamefont {Israel~M}\ \bibnamefont {Gelfand}}, \bibinfo {author} {\bibfnamefont {Mark~I}\ \bibnamefont {Graev}}, \ and\ \bibinfo {author} {\bibfnamefont {Alexander}\ \bibnamefont {Postnikov}},\ }\bibfield  {title} {\enquote {\bibinfo {title} {Combinatorics of hypergeometric functions associated with positive roots},}\ }in\ \href {https://link.springer.com/chapter/10.1007/978-1-4612-4122-5_10} {\emph {\bibinfo {booktitle} {The Arnold-Gelfand mathematical seminars}}}\ (\bibinfo {organization} {Springer},\ \bibinfo {year} {1997})\ pp.\ \bibinfo {pages} {205--221}\BibitemShut {NoStop}%
\bibitem [{\citenamefont {Freidel}\ and\ \citenamefont {Livine}(2011)}]{freidel2011u}%
  \BibitemOpen
  \bibfield  {author} {\bibinfo {author} {\bibfnamefont {Laurent}\ \bibnamefont {Freidel}}\ and\ \bibinfo {author} {\bibfnamefont {Etera~R}\ \bibnamefont {Livine}},\ }\bibfield  {title} {\enquote {\bibinfo {title} {U (n) coherent states for loop quantum gravity},}\ }\href {\doibase 10.1063/1.3587121} {\bibfield  {journal} {\bibinfo  {journal} {Journal of mathematical physics}\ }\textbf {\bibinfo {volume} {52}} (\bibinfo {year} {2011}),\ 10.1063/1.3587121}\BibitemShut {NoStop}%
\bibitem [{\citenamefont {Calixto}\ and\ \citenamefont {P{\'e}rez-Romero}(2014)}]{calixto2014coherent}%
  \BibitemOpen
  \bibfield  {author} {\bibinfo {author} {\bibfnamefont {M}~\bibnamefont {Calixto}}\ and\ \bibinfo {author} {\bibfnamefont {E}~\bibnamefont {P{\'e}rez-Romero}},\ }\bibfield  {title} {\enquote {\bibinfo {title} {Coherent states on the grassmannian u (4)/u (2) 2: Oscillator realization and bilayer fractional quantum hall systems},}\ }\href {\doibase 10.1088/1751-8113/47/11/115302} {\bibfield  {journal} {\bibinfo  {journal} {Journal of Physics A: Mathematical and Theoretical}\ }\textbf {\bibinfo {volume} {47}},\ \bibinfo {pages} {115302} (\bibinfo {year} {2014})}\BibitemShut {NoStop}%
\bibitem [{\citenamefont {Krishna}\ and\ \citenamefont {Poulin}(2021)}]{krishna2021ft}%
  \BibitemOpen
  \bibfield  {author} {\bibinfo {author} {\bibfnamefont {Anirudh}\ \bibnamefont {Krishna}}\ and\ \bibinfo {author} {\bibfnamefont {David}\ \bibnamefont {Poulin}},\ }\bibfield  {title} {\enquote {\bibinfo {title} {Fault-tolerant gates on hypergraph product codes},}\ }\href {\doibase 10.1103/PhysRevX.11.011023} {\bibfield  {journal} {\bibinfo  {journal} {Phys. Rev. X}\ }\textbf {\bibinfo {volume} {11}},\ \bibinfo {pages} {011023} (\bibinfo {year} {2021})}\BibitemShut {NoStop}%
\bibitem [{\citenamefont {Quintavalle}\ \emph {et~al.}(2023)\citenamefont {Quintavalle}, \citenamefont {Webster},\ and\ \citenamefont {Vasmer}}]{quintavallePartitioningQubitsHypergraph2023}%
  \BibitemOpen
  \bibfield  {author} {\bibinfo {author} {\bibfnamefont {Armanda~O.}\ \bibnamefont {Quintavalle}}, \bibinfo {author} {\bibfnamefont {Paul}\ \bibnamefont {Webster}}, \ and\ \bibinfo {author} {\bibfnamefont {Michael}\ \bibnamefont {Vasmer}},\ }\bibfield  {title} {\enquote {\bibinfo {title} {Partitioning qubits in hypergraph product codes to implement logical gates},}\ }\href {\doibase 10.22331/q-2023-10-24-1153} {\bibfield  {journal} {\bibinfo  {journal} {Quantum}\ }\textbf {\bibinfo {volume} {7}},\ \bibinfo {pages} {1153} (\bibinfo {year} {2023})}\BibitemShut {NoStop}%
\bibitem [{\citenamefont {Hastings}\ \emph {et~al.}(2021)\citenamefont {Hastings}, \citenamefont {Haah},\ and\ \citenamefont {O'Donnell}}]{fibrebundle}%
  \BibitemOpen
  \bibfield  {author} {\bibinfo {author} {\bibfnamefont {Matthew~B.}\ \bibnamefont {Hastings}}, \bibinfo {author} {\bibfnamefont {Jeongwan}\ \bibnamefont {Haah}}, \ and\ \bibinfo {author} {\bibfnamefont {Ryan}\ \bibnamefont {O'Donnell}},\ }\bibfield  {title} {\enquote {\bibinfo {title} {Fiber bundle codes: breaking the n1/2 polylog(n) barrier for quantum ldpc codes},}\ }in\ \href {\doibase 10.1145/3406325.3451005} {\emph {\bibinfo {booktitle} {Proceedings of the 53rd Annual ACM SIGACT Symposium on Theory of Computing}}},\ \bibinfo {series and number} {STOC 2021}\ (\bibinfo  {publisher} {Association for Computing Machinery},\ \bibinfo {address} {New York, NY, USA},\ \bibinfo {year} {2021})\ p.\ \bibinfo {pages} {1276–1288}\BibitemShut {NoStop}%
\bibitem [{\citenamefont {Freedman}\ and\ \citenamefont {Hastings}(2021)}]{freedman2021building}%
  \BibitemOpen
  \bibfield  {author} {\bibinfo {author} {\bibfnamefont {Michael}\ \bibnamefont {Freedman}}\ and\ \bibinfo {author} {\bibfnamefont {Matthew}\ \bibnamefont {Hastings}},\ }\bibfield  {title} {\enquote {\bibinfo {title} {Building manifolds from quantum codes},}\ }\href {\doibase 10.1007/s00039-021-00567-3} {\bibfield  {journal} {\bibinfo  {journal} {Geometric and Functional Analysis}\ }\textbf {\bibinfo {volume} {31}},\ \bibinfo {pages} {855--894} (\bibinfo {year} {2021})}\BibitemShut {NoStop}%
\bibitem [{\citenamefont {Gunderman}(2020)}]{gunderman2020local}%
  \BibitemOpen
  \bibfield  {author} {\bibinfo {author} {\bibfnamefont {Lane~G.}\ \bibnamefont {Gunderman}},\ }\bibfield  {title} {\enquote {\bibinfo {title} {Local-dimension-invariant qudit stabilizer codes},}\ }\href {\doibase 10.1103/PhysRevA.101.052343} {\bibfield  {journal} {\bibinfo  {journal} {Phys. Rev. A}\ }\textbf {\bibinfo {volume} {101}},\ \bibinfo {pages} {052343} (\bibinfo {year} {2020})}\BibitemShut {NoStop}%
\bibitem [{\citenamefont {Gunderman}(2022)}]{gunderman2022degenerate}%
  \BibitemOpen
  \bibfield  {author} {\bibinfo {author} {\bibfnamefont {Lane~G.}\ \bibnamefont {Gunderman}},\ }\bibfield  {title} {\enquote {\bibinfo {title} {Degenerate local-dimension-invariant stabilizer codes and an alternative bound for the distance preservation condition},}\ }\href {\doibase 10.1103/PhysRevA.105.042424} {\bibfield  {journal} {\bibinfo  {journal} {Phys. Rev. A}\ }\textbf {\bibinfo {volume} {105}},\ \bibinfo {pages} {042424} (\bibinfo {year} {2022})}\BibitemShut {NoStop}%
\bibitem [{\citenamefont {Gunderman}(2024)}]{gunderman2024stabilizer}%
  \BibitemOpen
  \bibfield  {author} {\bibinfo {author} {\bibfnamefont {Lane~G.}\ \bibnamefont {Gunderman}},\ }\bibfield  {title} {\enquote {\bibinfo {title} {Stabilizer {C}odes with {E}xotic {L}ocal-dimensions},}\ }\href {\doibase 10.22331/q-2024-02-12-1249} {\bibfield  {journal} {\bibinfo  {journal} {{Quantum}}\ }\textbf {\bibinfo {volume} {8}},\ \bibinfo {pages} {1249} (\bibinfo {year} {2024})}\BibitemShut {NoStop}%
\bibitem [{\citenamefont {Gunderman}(2025)}]{gunderman2025beyond}%
  \BibitemOpen
  \bibfield  {author} {\bibinfo {author} {\bibfnamefont {Lane~G}\ \bibnamefont {Gunderman}},\ }\bibfield  {title} {\enquote {\bibinfo {title} {Beyond integral-domain stabilizer codes},}\ }\href@noop {} {\bibfield  {journal} {\bibinfo  {journal} {arXiv preprint arXiv:2501.04888}\ } (\bibinfo {year} {2025})}\BibitemShut {NoStop}%
\bibitem [{\citenamefont {Totey}\ \emph {et~al.}(2023)\citenamefont {Totey}, \citenamefont {Kyle}, \citenamefont {Liu}, \citenamefont {Barge}, \citenamefont {Lordi},\ and\ \citenamefont {Combes}}]{totey2023performance}%
  \BibitemOpen
  \bibfield  {author} {\bibinfo {author} {\bibfnamefont {Saurabh}\ \bibnamefont {Totey}}, \bibinfo {author} {\bibfnamefont {Akira}\ \bibnamefont {Kyle}}, \bibinfo {author} {\bibfnamefont {Steven}\ \bibnamefont {Liu}}, \bibinfo {author} {\bibfnamefont {Pratik~J}\ \bibnamefont {Barge}}, \bibinfo {author} {\bibfnamefont {Noah}\ \bibnamefont {Lordi}}, \ and\ \bibinfo {author} {\bibfnamefont {Joshua}\ \bibnamefont {Combes}},\ }\bibfield  {title} {\enquote {\bibinfo {title} {The performance of random bosonic rotation codes},}\ }\href {https://arxiv.org/abs/2311.16089} {\bibfield  {journal} {\bibinfo  {journal} {arXiv preprint arXiv:2311.16089}\ } (\bibinfo {year} {2023})}\BibitemShut {NoStop}%
\bibitem [{\citenamefont {Marinoff}\ \emph {et~al.}(2024)\citenamefont {Marinoff}, \citenamefont {Bush},\ and\ \citenamefont {Combes}}]{marinoff2024explicit}%
  \BibitemOpen
  \bibfield  {author} {\bibinfo {author} {\bibfnamefont {Benjamin}\ \bibnamefont {Marinoff}}, \bibinfo {author} {\bibfnamefont {Miles}\ \bibnamefont {Bush}}, \ and\ \bibinfo {author} {\bibfnamefont {Joshua}\ \bibnamefont {Combes}},\ }\bibfield  {title} {\enquote {\bibinfo {title} {Explicit error-correction scheme and code distance for bosonic codes with rotational symmetry},}\ }\href {\doibase 10.1103/PhysRevA.109.032436} {\bibfield  {journal} {\bibinfo  {journal} {Phys. Rev. A}\ }\textbf {\bibinfo {volume} {109}},\ \bibinfo {pages} {032436} (\bibinfo {year} {2024})}\BibitemShut {NoStop}%
\bibitem [{\citenamefont {Yuan}\ \emph {et~al.}(2022)\citenamefont {Yuan}, \citenamefont {Xu},\ and\ \citenamefont {Jiang}}]{yuan2022construction}%
  \BibitemOpen
  \bibfield  {author} {\bibinfo {author} {\bibfnamefont {Ming}\ \bibnamefont {Yuan}}, \bibinfo {author} {\bibfnamefont {Qian}\ \bibnamefont {Xu}}, \ and\ \bibinfo {author} {\bibfnamefont {Liang}\ \bibnamefont {Jiang}},\ }\bibfield  {title} {\enquote {\bibinfo {title} {Construction of bias-preserving operations for pair-cat codes},}\ }\href {\doibase 10.1103/PhysRevA.106.062422} {\bibfield  {journal} {\bibinfo  {journal} {Phys. Rev. A}\ }\textbf {\bibinfo {volume} {106}},\ \bibinfo {pages} {062422} (\bibinfo {year} {2022})}\BibitemShut {NoStop}%
\bibitem [{\citenamefont {Gottesman}\ and\ \citenamefont {Chuang}(1999)}]{gottesman1999demonstrating}%
  \BibitemOpen
  \bibfield  {author} {\bibinfo {author} {\bibfnamefont {Daniel}\ \bibnamefont {Gottesman}}\ and\ \bibinfo {author} {\bibfnamefont {Isaac~L}\ \bibnamefont {Chuang}},\ }\bibfield  {title} {\enquote {\bibinfo {title} {Demonstrating the viability of universal quantum computation using teleportation and single-qubit operations},}\ }\href {\doibase 10.1038/46503} {\bibfield  {journal} {\bibinfo  {journal} {Nature}\ }\textbf {\bibinfo {volume} {402}},\ \bibinfo {pages} {390--393} (\bibinfo {year} {1999})}\BibitemShut {NoStop}%
\bibitem [{\citenamefont {Cui}\ \emph {et~al.}(2017)\citenamefont {Cui}, \citenamefont {Gottesman},\ and\ \citenamefont {Krishna}}]{cui2017diagonal}%
  \BibitemOpen
  \bibfield  {author} {\bibinfo {author} {\bibfnamefont {Shawn~X.}\ \bibnamefont {Cui}}, \bibinfo {author} {\bibfnamefont {Daniel}\ \bibnamefont {Gottesman}}, \ and\ \bibinfo {author} {\bibfnamefont {Anirudh}\ \bibnamefont {Krishna}},\ }\bibfield  {title} {\enquote {\bibinfo {title} {Diagonal gates in the clifford hierarchy},}\ }\href {\doibase 10.1103/PhysRevA.95.012329} {\bibfield  {journal} {\bibinfo  {journal} {Phys. Rev. A}\ }\textbf {\bibinfo {volume} {95}},\ \bibinfo {pages} {012329} (\bibinfo {year} {2017})}\BibitemShut {NoStop}%
\bibitem [{\citenamefont {Rengaswamy}\ \emph {et~al.}(2019)\citenamefont {Rengaswamy}, \citenamefont {Calderbank},\ and\ \citenamefont {Pfister}}]{Rengaswamy2019unifying}%
  \BibitemOpen
  \bibfield  {author} {\bibinfo {author} {\bibfnamefont {Narayanan}\ \bibnamefont {Rengaswamy}}, \bibinfo {author} {\bibfnamefont {Robert}\ \bibnamefont {Calderbank}}, \ and\ \bibinfo {author} {\bibfnamefont {Henry~D.}\ \bibnamefont {Pfister}},\ }\bibfield  {title} {\enquote {\bibinfo {title} {Unifying the clifford hierarchy via symmetric matrices over rings},}\ }\href {\doibase 10.1103/PhysRevA.100.022304} {\bibfield  {journal} {\bibinfo  {journal} {Phys. Rev. A}\ }\textbf {\bibinfo {volume} {100}},\ \bibinfo {pages} {022304} (\bibinfo {year} {2019})}\BibitemShut {NoStop}%
\bibitem [{\citenamefont {Anderson}(2024)}]{anderson2024groups}%
  \BibitemOpen
  \bibfield  {author} {\bibinfo {author} {\bibfnamefont {Jonas~T}\ \bibnamefont {Anderson}},\ }\bibfield  {title} {\enquote {\bibinfo {title} {On groups in the qubit clifford hierarchy},}\ }\href {\doibase 10.22331/q-2024-06-13-1370} {\bibfield  {journal} {\bibinfo  {journal} {Quantum}\ }\textbf {\bibinfo {volume} {8}},\ \bibinfo {pages} {1370} (\bibinfo {year} {2024})}\BibitemShut {NoStop}%
\bibitem [{\citenamefont {He}\ \emph {et~al.}(2024)\citenamefont {He}, \citenamefont {Robitaille},\ and\ \citenamefont {Tan}}]{he2024permutation}%
  \BibitemOpen
  \bibfield  {author} {\bibinfo {author} {\bibfnamefont {Zhiyang}\ \bibnamefont {He}}, \bibinfo {author} {\bibfnamefont {Luke}\ \bibnamefont {Robitaille}}, \ and\ \bibinfo {author} {\bibfnamefont {Xinyu}\ \bibnamefont {Tan}},\ }\bibfield  {title} {\enquote {\bibinfo {title} {Permutation gates in the third level of the clifford hierarchy},}\ }\href {https://arxiv.org/abs/2410.11818} {\bibfield  {journal} {\bibinfo  {journal} {arXiv preprint arXiv:2410.11818}\ } (\bibinfo {year} {2024})}\BibitemShut {NoStop}%
\bibitem [{\citenamefont {Chubb}\ and\ \citenamefont {Flammia}(2021)}]{chubb2021statistical}%
  \BibitemOpen
  \bibfield  {author} {\bibinfo {author} {\bibfnamefont {Christopher~T}\ \bibnamefont {Chubb}}\ and\ \bibinfo {author} {\bibfnamefont {Steven~T}\ \bibnamefont {Flammia}},\ }\bibfield  {title} {\enquote {\bibinfo {title} {Statistical mechanical models for quantum codes with correlated noise},}\ }\href {\doibase 10.4171/AIHPD/105} {\bibfield  {journal} {\bibinfo  {journal} {Annales de l’Institut Henri Poincar{\'e} D}\ }\textbf {\bibinfo {volume} {8}},\ \bibinfo {pages} {269--321} (\bibinfo {year} {2021})}\BibitemShut {NoStop}%
\bibitem [{\citenamefont {Vuillot}\ \emph {et~al.}(2019)\citenamefont {Vuillot}, \citenamefont {Asasi}, \citenamefont {Wang}, \citenamefont {Pryadko},\ and\ \citenamefont {Terhal}}]{vuillot_toricGKP_2019}%
  \BibitemOpen
  \bibfield  {author} {\bibinfo {author} {\bibfnamefont {Christophe}\ \bibnamefont {Vuillot}}, \bibinfo {author} {\bibfnamefont {Hamed}\ \bibnamefont {Asasi}}, \bibinfo {author} {\bibfnamefont {Yang}\ \bibnamefont {Wang}}, \bibinfo {author} {\bibfnamefont {Leonid~P.}\ \bibnamefont {Pryadko}}, \ and\ \bibinfo {author} {\bibfnamefont {Barbara~M.}\ \bibnamefont {Terhal}},\ }\bibfield  {title} {\enquote {\bibinfo {title} {Quantum error correction with the toric gottesman-kitaev-preskill code},}\ }\href {\doibase 10.1103/PhysRevA.99.032344} {\bibfield  {journal} {\bibinfo  {journal} {Phys. Rev. A}\ }\textbf {\bibinfo {volume} {99}},\ \bibinfo {pages} {032344} (\bibinfo {year} {2019})}\BibitemShut {NoStop}%
\bibitem [{\citenamefont {Guillaud}\ and\ \citenamefont {Mirrahimi}(2019)}]{guillaud2019repetition}%
  \BibitemOpen
  \bibfield  {author} {\bibinfo {author} {\bibfnamefont {J\'er\'emie}\ \bibnamefont {Guillaud}}\ and\ \bibinfo {author} {\bibfnamefont {Mazyar}\ \bibnamefont {Mirrahimi}},\ }\bibfield  {title} {\enquote {\bibinfo {title} {Repetition cat qubits for fault-tolerant quantum computation},}\ }\href {\doibase 10.1103/PhysRevX.9.041053} {\bibfield  {journal} {\bibinfo  {journal} {Phys. Rev. X}\ }\textbf {\bibinfo {volume} {9}},\ \bibinfo {pages} {041053} (\bibinfo {year} {2019})}\BibitemShut {NoStop}%
\bibitem [{\citenamefont {Darmawan}\ \emph {et~al.}(2021)\citenamefont {Darmawan}, \citenamefont {Brown}, \citenamefont {Grimsmo}, \citenamefont {Tuckett},\ and\ \citenamefont {Puri}}]{darmawan2021practical}%
  \BibitemOpen
  \bibfield  {author} {\bibinfo {author} {\bibfnamefont {Andrew~S.}\ \bibnamefont {Darmawan}}, \bibinfo {author} {\bibfnamefont {Benjamin~J.}\ \bibnamefont {Brown}}, \bibinfo {author} {\bibfnamefont {Arne~L.}\ \bibnamefont {Grimsmo}}, \bibinfo {author} {\bibfnamefont {David~K.}\ \bibnamefont {Tuckett}}, \ and\ \bibinfo {author} {\bibfnamefont {Shruti}\ \bibnamefont {Puri}},\ }\bibfield  {title} {\enquote {\bibinfo {title} {Practical quantum error correction with the xzzx code and kerr-cat qubits},}\ }\href {\doibase 10.1103/PRXQuantum.2.030345} {\bibfield  {journal} {\bibinfo  {journal} {PRX Quantum}\ }\textbf {\bibinfo {volume} {2}},\ \bibinfo {pages} {030345} (\bibinfo {year} {2021})}\BibitemShut {NoStop}%
\bibitem [{\citenamefont {Ehler}\ and\ \citenamefont {Gr{\"o}chenig}(2023)}]{ehler2023t}%
  \BibitemOpen
  \bibfield  {author} {\bibinfo {author} {\bibfnamefont {Martin}\ \bibnamefont {Ehler}}\ and\ \bibinfo {author} {\bibfnamefont {Karlheinz}\ \bibnamefont {Gr{\"o}chenig}},\ }\bibfield  {title} {\enquote {\bibinfo {title} {t-design curves and mobile sampling on the sphere},}\ }in\ \href {\doibase 10.1017/fms.2023.106} {\emph {\bibinfo {booktitle} {Forum of Mathematics, Sigma}}},\ Vol.~\bibinfo {volume} {11}\ (\bibinfo {organization} {Cambridge University Press},\ \bibinfo {year} {2023})\ p.\ \bibinfo {pages} {e105}\BibitemShut {NoStop}%
\bibitem [{\citenamefont {de~la Fuente}\ \emph {et~al.}(2024)\citenamefont {de~la Fuente}, \citenamefont {Ellison}, \citenamefont {Cheng},\ and\ \citenamefont {Williamson}}]{magdalena2024topological}%
  \BibitemOpen
  \bibfield  {author} {\bibinfo {author} {\bibfnamefont {Julio C.~Magdalena}\ \bibnamefont {de~la Fuente}}, \bibinfo {author} {\bibfnamefont {Tyler~D.}\ \bibnamefont {Ellison}}, \bibinfo {author} {\bibfnamefont {Meng}\ \bibnamefont {Cheng}}, \ and\ \bibinfo {author} {\bibfnamefont {Dominic~J.}\ \bibnamefont {Williamson}},\ }\href {https://arxiv.org/abs/2411.04993} {\enquote {\bibinfo {title} {Topological stabilizer models on continuous variables},}\ } (\bibinfo {year} {2024}),\ \Eprint {http://arxiv.org/abs/2411.04993} {arXiv:2411.04993 [quant-ph]} \BibitemShut {NoStop}%
\bibitem [{\citenamefont {Saito}\ \emph {et~al.}(1999)\citenamefont {Saito}, \citenamefont {Sturmfels},\ and\ \citenamefont {Takayama}}]{saito1999hypergeometric}%
  \BibitemOpen
  \bibfield  {author} {\bibinfo {author} {\bibfnamefont {Mutsumi}\ \bibnamefont {Saito}}, \bibinfo {author} {\bibfnamefont {Bernd}\ \bibnamefont {Sturmfels}}, \ and\ \bibinfo {author} {\bibfnamefont {Nobuki}\ \bibnamefont {Takayama}},\ }\bibfield  {title} {\enquote {\bibinfo {title} {Hypergeometric polynomials and integer programming},}\ }\href {\doibase 10.1023/A:1000609524994} {\bibfield  {journal} {\bibinfo  {journal} {Compositio Mathematica}\ }\textbf {\bibinfo {volume} {115}},\ \bibinfo {pages} {231--240} (\bibinfo {year} {1999})}\BibitemShut {NoStop}%
\bibitem [{\citenamefont {Cattani}(2006)}]{cattani2006three}%
  \BibitemOpen
  \bibfield  {author} {\bibinfo {author} {\bibfnamefont {EDUARDO}\ \bibnamefont {Cattani}},\ }\bibfield  {title} {\enquote {\bibinfo {title} {Three lectures on hypergeometric functions},}\ }\href@noop {} {\bibfield  {journal} {\bibinfo  {journal} {Notes for a course}\ } (\bibinfo {year} {2006})}\BibitemShut {NoStop}%
\bibitem [{\citenamefont {Beukers}(2011)}]{beukers2011notes}%
  \BibitemOpen
  \bibfield  {author} {\bibinfo {author} {\bibfnamefont {Frits}\ \bibnamefont {Beukers}},\ }\bibfield  {title} {\enquote {\bibinfo {title} {Notes on a-hypergeometric functions},}\ }\href {https://webspace.science.uu.nl/~beuke106/AHGcourse.pdf} {\bibfield  {journal} {\bibinfo  {journal} {Arithmetic and Galois theories of differential equations, S{\'e}min. Congr}\ }\textbf {\bibinfo {volume} {23}},\ \bibinfo {pages} {25--61} (\bibinfo {year} {2011})}\BibitemShut {NoStop}%
\bibitem [{\citenamefont {Aomoto}(2011)}]{aomoto2011theory}%
  \BibitemOpen
  \bibfield  {author} {\bibinfo {author} {\bibfnamefont {Kazuhiko}\ \bibnamefont {Aomoto}},\ }\href {\doibase 10.1007/978-4-431-53938-4} {\emph {\bibinfo {title} {Theory of hypergeometric functions}}}\ (\bibinfo  {publisher} {Springer},\ \bibinfo {year} {2011})\BibitemShut {NoStop}%
\bibitem [{\citenamefont {Saito}\ \emph {et~al.}(2013)\citenamefont {Saito}, \citenamefont {Sturmfels},\ and\ \citenamefont {Takayama}}]{saito2013grobner}%
  \BibitemOpen
  \bibfield  {author} {\bibinfo {author} {\bibfnamefont {Mutsumi}\ \bibnamefont {Saito}}, \bibinfo {author} {\bibfnamefont {Bernd}\ \bibnamefont {Sturmfels}}, \ and\ \bibinfo {author} {\bibfnamefont {Nobuki}\ \bibnamefont {Takayama}},\ }\href {\doibase 10.1007/978-3-662-04112-3} {\emph {\bibinfo {title} {Gr{\"o}bner deformations of hypergeometric differential equations}}},\ Vol.~\bibinfo {volume} {6}\ (\bibinfo  {publisher} {Springer Science \& Business Media},\ \bibinfo {year} {2013})\BibitemShut {NoStop}%
\bibitem [{\citenamefont {Takayama}(2020)}]{Takayama_2020}%
  \BibitemOpen
  \bibfield  {author} {\bibinfo {author} {\bibfnamefont {N.}~\bibnamefont {Takayama}},\ }\enquote {\bibinfo {title} {A-hypergeometric functions},}\ in\ \href {\doibase 10.1017/9780511777165.005} {\emph {\bibinfo {booktitle} {Encyclopedia of Special Functions: The Askey-Bateman Project}}},\ \bibinfo {editor} {edited by\ \bibinfo {editor} {\bibfnamefont {Tom~H.}\ \bibnamefont {Koornwinder}}\ and\ \bibinfo {editor} {\bibfnamefont {Jasper~V.Editors}\ \bibnamefont {Stokman}}}\ (\bibinfo  {publisher} {Cambridge University Press},\ \bibinfo {year} {2020})\ p.\ \bibinfo {pages} {101–121}\BibitemShut {NoStop}%
\bibitem [{\citenamefont {Reichelt}\ \emph {et~al.}(2021)\citenamefont {Reichelt}, \citenamefont {Schulze}, \citenamefont {Sevenheck},\ and\ \citenamefont {Walther}}]{reichelt2021algebraic}%
  \BibitemOpen
  \bibfield  {author} {\bibinfo {author} {\bibfnamefont {Thomas}\ \bibnamefont {Reichelt}}, \bibinfo {author} {\bibfnamefont {Mathias}\ \bibnamefont {Schulze}}, \bibinfo {author} {\bibfnamefont {Christian}\ \bibnamefont {Sevenheck}}, \ and\ \bibinfo {author} {\bibfnamefont {Uli}\ \bibnamefont {Walther}},\ }\bibfield  {title} {\enquote {\bibinfo {title} {Algebraic aspects of hypergeometric differential equations},}\ }\href {\doibase 10.1007/s13366-020-00560-1} {\bibfield  {journal} {\bibinfo  {journal} {Beitr{\"a}ge zur Algebra und Geometrie/Contributions to Algebra and Geometry}\ }\textbf {\bibinfo {volume} {62}},\ \bibinfo {pages} {137--203} (\bibinfo {year} {2021})}\BibitemShut {NoStop}%
\bibitem [{\citenamefont {Bargmann}(1961)}]{bargmann1961hilbert}%
  \BibitemOpen
  \bibfield  {author} {\bibinfo {author} {\bibfnamefont {Valentine}\ \bibnamefont {Bargmann}},\ }\bibfield  {title} {\enquote {\bibinfo {title} {On a hilbert space of analytic functions and an associated integral transform part i},}\ }\href {\doibase 10.1002/cpa.3160140303} {\bibfield  {journal} {\bibinfo  {journal} {Communications on pure and applied mathematics}\ }\textbf {\bibinfo {volume} {14}},\ \bibinfo {pages} {187--214} (\bibinfo {year} {1961})}\BibitemShut {NoStop}%
\bibitem [{\citenamefont {Segal}(1963)}]{segal1963mathematical}%
  \BibitemOpen
  \bibfield  {author} {\bibinfo {author} {\bibfnamefont {Irving~Ezra}\ \bibnamefont {Segal}},\ }\href@noop {} {\emph {\bibinfo {title} {Mathematical problems of relativistic physics}}},\ Vol.~\bibinfo {volume} {2}\ (\bibinfo  {publisher} {American Mathematical Soc.},\ \bibinfo {year} {1963})\BibitemShut {NoStop}%
\bibitem [{\citenamefont {Bender}\ and\ \citenamefont {Orszag}(2013)}]{bender2013advanced}%
  \BibitemOpen
  \bibfield  {author} {\bibinfo {author} {\bibfnamefont {Carl~M}\ \bibnamefont {Bender}}\ and\ \bibinfo {author} {\bibfnamefont {Steven~A}\ \bibnamefont {Orszag}},\ }\href {\doibase 10.1007/978-1-4757-3069-2} {\emph {\bibinfo {title} {Advanced mathematical methods for scientists and engineers I: Asymptotic methods and perturbation theory}}}\ (\bibinfo  {publisher} {Springer Science \& Business Media},\ \bibinfo {year} {2013})\BibitemShut {NoStop}%
\end{thebibliography}%
\end{document}